\documentclass[dvipsnames, 11pt]{article}
\pdfoutput=1
\usepackage[utf8]{inputenc}

\usepackage{amsmath}
\usepackage{amssymb}
\usepackage{tikz}
\usetikzlibrary{calc}
\usepackage[a4paper, total={6in,9in}, footskip=60pt, footnotesep = 26pt]{geometry}

\usepackage[toc,page]{appendix}
\usepackage{enumerate}
\usepackage{tensor}
\usepackage{booktabs}
\usepackage{bbm}
\usepackage{youngtab}
\usepackage{slashed}
\usepackage{multirow}
\usepackage{makecell}
\usepackage{float}
\usepackage{verbatim}
\usepackage{xcolor}
\usepackage[framemethod=tikz]{mdframed}
\usepackage{soul}
\usepackage[font={small,it}]{caption}

\newcommand{\ab}{|}
\newcommand{\der}{\partial}
\newcommand{\de}{\mathrm{d}}
\newcommand{\detr}[1]{\mathrm{det} \, #1}
\newcommand{\ket}[1]{\ab #1 \rangle}

\newcommand{\al}{\sigma_s}
\newcommand{\vevc}{\gamma}
\newcommand{\sdil}{\phi}

\newcommand{\Vzero}{\mathcal{V}_{(0)}}
\newcommand{\Vw}{\mathcal{V}_{w}}
\newcommand{\Vzerofour}{\mathcal{V}^{\Sigma_4}_{\!(0)}}
\newcommand{\Vwfour}{\mathcal{V}^{\Sigma_4}_{w}}
\newcommand{\Vzerotwo}{\mathcal{V}^{\mathrm{T}^2}_{\!(0)}}
\newcommand{\Vwtwo}{\mathcal{V}^{\mathrm{T}^2}_{w}}

\newcommand{\Ds}{\sigma^3}
\newcommand{\bDs}{\bar{\sigma}^3}
\newcommand{\Dt}{\varphi^3}
\newcommand{\bDt}{\bar{\varphi}^3}

\newcommand{\s}{s}
\newcommand{\g}{g_{X \bar{X}}}

\allowdisplaybreaks

\usepackage{titlesec} 
\titleformat{\paragraph}
{\normalfont \normalsize \itshape}{\theparagraph}{2ex}{}
\titlespacing*{\paragraph}
{0pt}{3.25ex plus 1ex minus .2ex}{1.5ex plus .2ex}

\usepackage{chngcntr}

\numberwithin{equation}{section}

\usepackage{abstract}

\title{\textbf{Non-supersymmetric String Models \\[4pt] from Anti-D3-/D7-branes \\[4pt] in Strongly Warped Throats}}
\author{\\[4pt] Susha Parameswaran, Flavio Tonioni \\[12pt] {\normalsize \emph{Department of Mathematical Sciences, University of Liverpool}} \\ {\normalsize \emph{Liverpool, L69 7ZL, United Kingdom}}}
\date{\normalsize \href{mailto:susha@liv.ac.uk}{susha@liv.ac.uk}, \href{mailto:flavio.tonioni@liv.ac.uk}{flavio.tonioni@liv.ac.uk}}

\usepackage{hyperref}
\hypersetup{
    colorlinks=true,
    linkcolor=blue,
    citecolor=cyan,
    filecolor=cyan,      
    urlcolor=cyan,
}

\usepackage{cite}
\usepackage{mcite}

\begin{document}

\pagenumbering{roman}

\maketitle

\vfill

\vspace{50pt}

\begin{abstract}

This article discusses model-building scenarios including anti-D3-/D7-branes, in which supersymmetry is broken spontaneously, despite having no scale at which sparticles appear and standard supersymmetry is restored. If the branes are placed on singularities at the tip of warped throats in Calabi-Yau orientifold flux compactifications, they may give rise to realistic particle spectra, closed- and open-string moduli stabilisation with a Minkowski/de Sitter uplift, and a geometrical origin for the scale hierarchies. The paper derives the low-energy effective field theory description for such scenarios, i.e. a non-linear supergravity theory for standard and constrained supermultiplets, including soft supersymmetry-breaking matter couplings. The effect of closed-string moduli stabilisation on the open-string matter sector is worked out, incorporating non-perturbative and perturbative effects, and the mass and coupling hierarchies are computed with a view towards phenomenology.

\end{abstract}

\vfill

\thispagestyle{empty}

\newpage
\thispagestyle{empty}

{\hypersetup{linkcolor=black}
\tableofcontents
}

\newpage
\pagenumbering{arabic}

\section{Introduction}

Supersymmetry has been a key ingredient of string theory model building and a leading candidate for a solution to the long-standing gauge hierarchy problem \cite{martinSUSY, Quevedo:2010ui}. However, the present-day absence of supersymmetric partners at the LHC \cite{Tanabashi:2018oca}, together with the failure of supersymmetry to explain the even bigger cosmological problem, suggests that the nature of supersymmetry breaking has not yet been understood. Recently, the fact that anti-D-branes in type II Calabi-Yau orientifold compactifications \cite{Giddings:2001yu, Grana:2005jc} spontaneously break supersymmetry has received a great deal of attention \cite{Ferrara:2014kva, Kallosh:2014via, Kallosh:2014wsa, Bergshoeff:2015jxa, Kallosh:2015nia, Bertolini:2015hua, Bandos:2015xnf, Aparicio:2015psl, Garcia-Etxebarria:2015lif, Dasgupta:2016prs,Vercnocke:2016fbt, Kallosh:2016aep, Bandos:2016xyu, Aalsma:2017ulu, Kallosh:2017wnt, GarciadelMoral:2017vnz, Cribiori:2017laj, Kitazawa:2018zys, Krishnan:2018udc, Aalsma:2018pll, Bena:2018fqc, Cribiori:2018dlc, Kallosh:2019axr, Cribiori:2019hod, Cribiori:2019bfx, Kallosh:2019zgd, Cribiori:2019drf, Randall:2019ent, Dudas:2019pls, Cribiori:2020bgt} (for earlier analyses, see Refs. \cite{Antoniadis:1999xk, *Angelantonj:1999jh, *Aldazabal:1999jr, *Angelantonj:1999ms, *Uranga:1999ib, *Dudas:2000nv, *Pradisi:2001yv}).  Together with fluxes, non-perturbative, and perturbative effects, whose interplay can address the moduli stabilisation problem, the positive-definite energy density of anti-D-branes may help to obtain a (quasi-)de Sitter vacuum corresponding to the observed Universe \cite{Kachru:2003aw, Conlon:2005ki}. Whilst the consistency of these de Sitter constructions is still under debate (for an incomplete list, see Refs. \cite{Kachru:2002gs, Frey:2003dm, *Brown:2009yb, *Bena:2009xk, *Bena:2010pr, *Bena:2010ze, *Bena:2011hz, *Bena:2011wh, *Bena:2012bk, *Bena:2012ek, *Bena:2012tx, *Bena:2012vz, *Bena:2013hr, *Danielsson:2014yga, *Gautason:2015ola, *Bena:2015kia, *Gautason:2015tla, *Cohen-Maldonado:2015ssa, *Bena:2015lkx, *Bena:2016fqp, *Moritz:2017xto, *Sethi:2017phn, *Brennan:2017rbf, *Danielsson:2018ztv, *Moritz:2018sui, *Obied:2018sgi, *Garg:2018reu, *Cicoli:2018kdo, *Kallosh:2018nrk, *Kallosh:2018wme, *Kallosh:2018psh, *Gautason:2018gln, *Armas:2018rsy, *Bena:2019sxm, *Qiu:2020los, Hamada:2019ack, Carta:2019rhx, Gautason:2019jwq, Blumenhagen:2019qcg, Blaback:2019ucp, Bena:2019mte, Kachru:2019dvo}), the spontaneous breaking of supersymmetry by anti-D-branes means that these objects can be used in string model building whilst maintaining the powerful machinery of supersymmetry. 

In more detail, there is a precise identification between the anti-D3-brane action in flat space placed on an orientifold plane and the Volkov-Akulov theory of non-linearly realised supersymmetry \cite{Volkov:1973ix, Kallosh:2014wsa}.  Moreover, all the degrees of freedom on an anti-D3-brane can be described using the tools familiar from linear supergravity by  placing the low-energy fields in constrained supermultiplets \cite{Vercnocke:2016fbt, Kallosh:2016aep, Cribiori:2020bgt}, where the constraints ensure that either only the bosonic or only the fermionic component is an independent degree of freedom \cite{Komargodski:2009rz, DallAgata:2016syy}.  In particular, the anti-D3-brane gaugino plays the role of the goldstino, and falls in a nilpotent superfield, $X$, where the constraint, $\smash{X^2=0}$, fixes the scalar component in terms of the fermion component and auxiliary field as $\smash{\varphi^X = \psi^X \psi^X/F^X}$, and $F^X$ is non-zero by assumption. The standard non-linear supersymmetry transformation for the goldstino, $\smash{\sqrt{2} \, \delta_\epsilon \lambda \sim \epsilon / l^2}$, can be seen after the field redefinition $\smash{\lambda \sim \psi^X/(2 l^2 F^X)}$, where $l$ is the scale where the massive string states come into play.

This progress has made it possible to describe how the anti-D3-brane couples to bulk fields in type IIB Calabi-Yau orientifold flux compactifications, including the closed-string moduli, and to determine the mutual interplay between the closed- and open-string sectors \cite{Ferrara:2014kva, Kallosh:2015nia, Bergshoeff:2015jxa, Garcia-Etxebarria:2015lif, Aparicio:2015psl, GarciadelMoral:2017vnz, Bena:2018fqc, Cribiori:2019hod, Dudas:2019pls}. The low-energy effective field theory corresponds to a non-linear supergravity theory, including standard and constrained superfields, with the anti-D3-brane uplift corresponding to an $F^X$-term contribution to the scalar potential. In particular, Ref. \cite{Cribiori:2019hod} has derived the complete action for an anti-D3-brane in the KKLT-scenario by means of constrained superfields, and Ref. \cite{Dudas:2019pls} has considered the coupling of the anti-D3-brane goldstino to the complex structure modulus controlling the warp factor in a Klebanov-Strassler throat \cite{Klebanov:2000hb}.  Non-linear supersymmetry strongly constrains the theory; for example, the well-known non-renormalisation theorems fulfilled by low-energy effective linearly realised supergravities descending from string theory extend to the non-linear supergravity theories \cite{GarciadelMoral:2017vnz}. \\

Given the null results thus far in sparticle searches, the recent insights into anti-D-brane supersymmetry breaking, and the potential importance of the latter in cosmological model building, this paper develops the idea that quasi-realistic particle physics models, with non-standard realisations of supersymmetry, may be obtained using anti-D3-branes.  Anti-D3-/D7-brane systems placed at orbifold singularities are known to lead to interesting low-energy particle spectra, comprising non-Abelian gauge groups, adjoint fermions, bifundamental scalar and bifundamental fermions \cite{Aldazabal:2000sk, Aldazabal:2000sa, Berenstein:2001nk, Angulo:2002wf, Cascales:2003wn, Marchesano:2004yn, Cicoli:2012vw, Cicoli:2013mpa} (for reviews, see e.g. Refs. \cite{Malyshev:2007zz, Maharana:2012tu}). Intriguingly, as a consequence of the orbifold projection, the $\bar{3}7$- and $7\bar{3}$-sector intersecting fermions and scalars fall into distinct bifundamental representations of the gauge groups, and so the low-energy spectrum does not fulfil the usual superpartner pairing.  It is natural to consider such systems at the tip of a strongly warped throat, which may be dynamically obtained since anti-D3-branes minimize their energy there.  Depending on the warping, volume and mass-sourcing fluxes, both closed- and open-string sectors may localise either in the highly-redshifted region or in the bulk, and hierarchical mass scales are explained via geometrical warping \cite{Randall:1999ee, Giddings:2001yu, DeWolfe:2002nn, Giddings:2005ff, Burgess:2006mn}. This article focuses on strongly-warped scenarios in which most of the degrees of freedom, from both the closed- and the open-string sectors, tend to localise in the highly-redshifted region of the internal space \cite{Burgess:2006mn}, but the results could easily be extended to any model with intersecting anti-D3-/D7-branes.  Interesting bottom-up particle physics models may thus plausibly be embedded into complete string compactifications, with in principle all closed- and open-string moduli stabilised via fluxes, perturbative and non-perturbative effects. 

Towards this objective, this article computes the low-energy effective field theory describing an anti-D3-/D7-brane system at an orbifold singularity at the tip of a strongly warped throat, within a supersymmetric type IIB Calabi-Yau orientifold flux compactification \cite{Giddings:2001yu, DeWolfe:2002nn, Giddings:2005ff, Burgess:2006mn, Frey:2008xw}. Whilst the closed-string and $77$-sector degrees of freedom fulfil a linear supersymmetry, and fall into standard supermultiplets \cite{Ibanez:1998rf, Giddings:2001yu, Camara:2003ku, Grana:2003ek, Grimm:2004uq, Lust:2004cx, Lust:2004fi, Camara:2004jj, Jockers:2004yj, Lust:2005bd}, the $\bar{3}\bar{3}$- and $\bar{3}7$-/$7\bar{3}$-sector degrees of freedom have non-linear supersymmetry transformations, and fall into constrained supermultiplets \cite{Camara:2004jj, Vercnocke:2016fbt, Kallosh:2016aep, GarciadelMoral:2017vnz, Cribiori:2019hod, Cribiori:2020bgt}. By a dimensional reduction of the ten-dimensional and worldvolume actions, and by exploiting how the internal spacetime symmetries transform the intersecting states (for which no action is known), one can infer the non-linear supergravity action, encapsulated as usual in a K\"{a}hler potential, a superpotential, the gauge kinetic functions and the Fayet-Iliopoulos terms.  This non-linear supergravity theory allows one to infer the interactions related by supersymmetry, both linear and non-linear, and to work out the consequences of closed string moduli stabilisation, including perturbative and non-perturbative effects, on the open-string sectors.  Previous studies on the supersymmetry-breaking effects of anti-D3-branes in the KKLT setup considered the possibility in which the matter sector originates from D3- and D7-branes \cite{Aparicio:2015psl, Choi:2005ge, Choi:2005uz, Falkowski:2005ck, Lebedev:2006qq}, while in this work the anti-D3-brane sectors provide both the uplift energy and matter.

It is interesting to compare the effective field theory description of anti-D3-brane supersymmetry breaking with the standard hidden-sector supersymmetry breaking via some non-zero closed-string field F-term. For this purpose, pure anti-D3-brane breaking may be assumed, though in the main text setups with both open- and closed-string breaking active will also be considered. Similarly to the standard procedure, one considers a vacuum that spontaneously breaks supersymmetry via a non-zero $F^X$-term and expands the action around this F-term, to obtain a set of soft-breaking terms in the Lagrangian. The anti-D3-/D7-brane systems give rise to several further low-energy fields - beyond the goldstino - which also lie in constrained superfields without physical superpartners and which can acquire soft-breaking terms (some of the constraints used not only fix the would-be superpartner, but also the auxiliarly field in terms of the goldstino; in this case, the supergravity expansions are different to the standard case and have been computed in appendix \ref{LEEFT SUGRA in IIB compactifications with non-linearly realised SUSY}). As in standard gravity-mediated hidden-sector supersymmetry breaking scenarios, the scale of the soft-breaking masses is $m_{\mathrm{soft}} \sim f_X/m_{P}$, where $f_X$ sets the uplift energy of the anti-D3-brane provided by the $F^X$-term. Whereas, in a standard supersymmetry breaking scenario, the light fields would fall in constrained superfields below the scale $m_{\mathrm{soft}}$, for the anti-D3-brane, constrained superfields are necessary even above $m_{\mathrm{soft}}$, and there is no scale at which superpartners appear.  Instead, the structure that gives the remarkable finiteness properties of string theory is expected to involve the entire spectrum of string states, which appear at the warped string scale $m_s^w$ for anti-D3-branes at the tip of strongly warped throats. The article discusses the scales that emerge for anti-D3-/D7-brane systems embedded in KKLT-like moduli stabilisation, after the interplay between open- and closed-string F-terms.  \\

The article is organised as follows. Section \ref{warped IIB closed-string sector} reviews strongly warped scenarios in type IIB string theory, highlighting the hierarchies among the string, Kaluza-Klein and flux-induced energy scales as well as the conditions for a low-energy supergravity formulation to be valid, with focus on the role of anti-D3-branes. As a helpful example, section \ref{warped D3- and D7-branes} discusses the supergravity description of models with intersecting D3-/D7-branes in strongly warped regimes, including possible supersymmetry breaking by fluxes. Then, section \ref{warped anti-D3- and D7-branes} extends to intersecting anti-D3-/D7-branes models, making use of the tools of constrained superfields, and embeds them into scenarios where the closed-string sector moduli are stabilised. Section \ref{overview on the extension to non-Abelian theories} discusses the supergravity description of quasi-realistic standard-like models on anti-D3-/D7-brane models at orbifold singularities. Finally, a summary of possible mass scales in these setups is provided in section \ref{analysis of the mass hierarchies} and section \ref{conclusions} outlines the main conclusions. The appendices provide useful elements for the dimensional reduction of type IIB theories, a review of hidden-sector supersymmetry breaking and supergravity soft-breaking terms, and an extension of the latter in the presence of constrained superfields.

\section{Warped IIB Closed-String Sector} \label{warped IIB closed-string sector}
Focussing on strongly warped type IIB compactifications, this section introduces the appropriate 10-dimensional metric, shows the hierarchies between the mass scales and discusses the conditions for well-defined 4-dimensional supergravity formulations.

\subsection{Warped Metric and Closed-String Sector Supergravity}  
In warped type IIB compactifications, the 10-dimensional metric takes the form \cite{Giddings:2005ff, Frey:2008xw}
\begin{equation} \label{metric}
    ds_{10}^2 = \dfrac{ \vevc^{3/2} \, e^{2 \Omega [c]}}{[e^{-4A} + c]^{1/2}} \, \bigl[ g_{\mu \nu} \, \de x^\mu \de x^\nu + 2 \der_\mu c \, \der_m b \, \de x^\mu \de y^m \bigr] + [e^{-4A} + c]^{1/2} \, g_{mn} \, \de y^m \de y^n,
\end{equation}
where the coordinates $x^\mu$ and $y^m$ describe the non-compact 4-dimensional spacetime $X_{1,3}$ and the compact 6-dimensional space $Y_6$, respectively, $e^{2 \Omega} [c(x)]$ is a Weyl rescaling factor to the 4-dimensional Einstein frame, defined as
\begin{equation} \label{Weyl factor}
    e^{2 \Omega [c]} = \dfrac{\displaystyle \int_{Y_6} \de^6 y \, \sqrt{g_{6}}}{\displaystyle \int_{Y_6} \de^6 y \, \sqrt{g_{6}} \; \bigl[ e^{-4A} + c \bigr]},
\end{equation}
$\vevc$ is an extra arbitrary constant, and $b=b(y)$ is a compensator field needed to solve the Einstein equations \cite{Frey:2008xw} but ignored in the following as it is sources only derivative couplings with the open-string excitations. The warp factor $e^{-4A}=e^{-4A(y)}$ and the volume-controlling real K\"{a}hler modulus $c=c(x)$ combine together into the generalised warp factor
\begin{equation} \label{general warp factor}
e^{-4 A[c(x),y]} = e^{-4A(y)} + c(x).
\end{equation} 

From the metric above, the physical internal volume in the Einstein frame is
\begin{equation*}
    \mathrm{vol}_6 = \int_{Y_6} \de^6 y \, \sqrt{g_{6}} \; [e^{-4A} + c]^{3/2},
\end{equation*}
whilst the dimensionless unwarped and warped internal volumes are defined respectively as
\begin{equation*}
l_s^6 \Vzero = \int_{Y_6} \de^6 y \, \sqrt{g_{6}}, \qquad \qquad \qquad \qquad l_s^6 \Vw = \int_{Y_6} \de^6 y \, \sqrt{g_{6}} \; e^{-4A}.
\end{equation*}
Moreover, the dimensionless physical internal volume is defined as $\smash{\mathcal{V} = \mathrm{vol}_6/\Vzero l_s^6}$, in units of the unwarped volume.  Given the 10-dimensional gravitational coupling $2 \hat{\kappa}_{10}^2 = g_s^2l_s^8 / 2 \pi$, with the string coupling $g_s$ and the string length $l_s$, the 4-dimensional reduced Planck length $\kappa_4$ turns out to be
\begin{equation}
    2 \kappa_4^2 = \dfrac{2 \hat{\kappa}_{10}^2}{\vevc^{3/2} l_s^6 \Vzero} =  \dfrac{g_s^2 l_s^2}{2 \pi \vevc^{3/2} \Vzero},
\end{equation}
with the reduced Planck mass $m_P$ being defined as the inverse $m_P = 1 / \kappa_4$.  In the large volume limit, where warping becomes negligible, one can identify the field $c$ as $c = e^{4u} = \mathcal{V}^{2/3}$ and the Weyl factor as $e^{2\Omega} = 1/c = e^{-4u}$, and fixing the constant $\vevc = \langle c \rangle$ ensures that the string and Planck scales are related by the physical internal volume \cite{Giddings:2001yu, Conlon:2005ki}.  

In a Calabi-Yau orientifold compactification with Hodge number $h_+^{1,1} = 1$, one can reproduce the 4-dimensional effective action corresponding to the axio-dilaton $\smash{\tau = C_0 + i \, e^{-\sdil}}$, the complex structure moduli $u^\alpha$, with $\alpha=1,\dots,h_-^{2,1}$, and the K\"{a}hler modulus $\rho = \chi + i c$ by means of the K\"{a}hler and superpotential \cite{Frey:2008xw,DeWolfe:2002nn,Chemtob:2019hwb}
\begin{subequations}
\begin{align}
    \kappa_4^2 \hat{K} & = - \mathrm{ln} \, [- i (\tau - \bar{\tau})] - \mathrm{ln} \, \biggl[ - i \int_{Y_6} e^{-4A} \, \Omega \wedge \bar{\Omega} \biggr] - 3 \, \mathrm{ln} \, \bigl[ 2 \, e^{-2 \Omega} \bigr] + \mathrm{ln} \, \biggl[ \dfrac{2}{\pi} \dfrac{\Vw}{[\Vzero]^3} \biggr], \label{closed-strings - Kaehler potential} \\
    \kappa_4^3 \hat{W} & = \dfrac{g_s}{l_s^2} \int_{Y_6} G_3 \wedge \Omega. \label{closed-strings - superpotential}
\end{align}
\end{subequations}
Note that $e^{-2 \Omega} = \mathrm{Im} \, \rho + c_0$, with $c_0 = \Vw / \Vzero$, gives a K\"{a}hler potential for the volume modulus of the usual no-scale form.\footnote{For future reference, it is immediate to show the identity $\der_\rho e^{2 \Omega} = i \, e^{4 \Omega} / 2$; then one finds the derivatives
    \begin{equation*}
    \kappa_4^2 \hat{K}_\rho = \dfrac{3i}{2} \, e^{2 \Omega}, \qquad \qquad \qquad \kappa_4^2 \hat{K}_{\rho \bar{\rho}} = \dfrac{3}{4} \, e^{4 \Omega}, \qquad \qquad \qquad \kappa_4^3 \nabla_\rho \hat{W} = \dfrac{3i g_s}{2 l_s^2} \, e^{2 \Omega} \int_{Y_6} G_3 \wedge \Omega.
    \end{equation*}
    Notice that the no-scale structure is preserved as a consequence of the identity $\kappa_4^2 \, \hat{K}^{\rho \bar{\rho}} \hat{K}_\rho \hat{K}_{\bar{\rho}} = 3$.}
    
    Some more details of these results are reviewed in appendix \ref{appendix: warped dimensional reduction}. The focus in the current work is on local configurations of intersecting anti-D3-/D7-branes within such warped geometries, and it will be assumed throughout that the global configuration of fluxes, branes and O-planes within the Calabi-Yau orientifold compactifications considered satisfy the RR-tadpole cancellation conditions necessary for an overall consistency.

\subsection{Field Localisation and 4-dimensional Supergravity Conditions} \label{field localisation}
In the presence of a highly warped throat, there can be non-trivial localisation effects for the closed-string sector fields; further, there are interesting hierarchies between mass scales in the bulk and in the redshifted region.  These scenarios are studied in detail by Ref. \cite{Burgess:2006mn} and, because they are relevant in the model-building setups considered in this article, a review of their main features is provided below. For brevity, the normalisation $\smash{\Vzero = 1}$ is assumed in the rest of the subsection.

\subsubsection{Closed-String Sector Field Localisation}
As a guiding example for the closed-string sector fields with a flux-induced mass, one can study the behaviour of the axio-dilaton $\tau$. The linearised field equation for the axio-dilaton wavefunction, labelled as $\tau=\tau(y)$, takes the form \cite{Giddings:2005ff, Frey:2006wv}
\begin{equation*}
    \dfrac{e^{2 \Omega}}{[e^{-4A} + c]} \, \Delta_{6} \, \tau (y) + \dfrac{m^2}{\vevc^{3/2}} \, \tau (y) = \dfrac{1}{12 \, \mathrm{Im} \, \tau} \, \dfrac{e^{2 \Omega}}{[e^{-4A} + c]^2} \, G_{mnp} \bar{G}^{mnp} \, \tau (y),
\end{equation*}
where $m^2$ is the 4-dimensional axio-dilaton mass, with the Laplacian $\Delta_{6}$ and the 3-form terms sourcing the Kaluza-Klein tower and the flux-induced mass, respectively. By estimating these terms, one can qualitatively understand the non-trivial localisation effects.
\begin{itemize}
    \item In the bulk, the unwarped metric $g_{mn}$ is of order one and the 3-form flux is of the order of its quantisation integer $n_f$, that is $G_{mnp} \sim n_f / l_s$. The background warp factor is negligible compared to the volume modulus, that is
    \begin{equation*}
        e^{-4 A} \ll c \sim \mathcal{V}^{2/3}.
    \end{equation*}
    Following these estimates, and assuming without loss of generality that integrals are dominated by the bulk, the order of magnitude of the flux-induced moduli masses in the bulk is
    \begin{equation} \label{bulk flux-mass}
        m^2_{\mathrm{flux}} = \dfrac{\vevc^{3/2}}{12} \, \dfrac{e^{2 \Omega}}{[e^{-4A} + c]^2} \, G_{mnp} \bar{G}^{mnp} \sim \dfrac{n_f^2}{\mathcal{V}^2} \, \dfrac{\vevc^{3/2}}{l_s^2} \sim \dfrac{g_s^2 n_f^2}{\mathcal{V}^2} \, \dfrac{1}{\kappa_4^2}.
    \end{equation}
    Also, given the characteristic length scale of the bulk $\lambda$ as measured in terms of the unwarped metric $g_{mn}$ (with $\lambda^6 \sim \Vzero$ in general), the bulk Kaluza-Klein scale is
    \begin{equation} \label{bulk KK-mass}
        m^2_{\mathrm{KK}} \sim \dfrac{e^{2\Omega}}{[e^{-4A} + c]} \dfrac{\vevc^{3/2}}{\lambda^2 l_s^2} \sim \dfrac{1}{\lambda^2\mathcal{V}^{4/3}} \, \dfrac{\vevc^{3/2}}{l_s^2} \sim \dfrac{g_s^2}{\lambda^2 \mathcal{V}^{4/3}} \, \dfrac{1}{\kappa_4^2}.
    \end{equation}
    From the dimensional reduction of the Einstein-Hilbert action one can observe that the string mass is $m_s^2 = g_s^2 M_P^2 / 4 \pi \vevc^{3/2}$, so the bulk string scale must be defined as
    \begin{equation} \label{bulk string-mass}
        m_s^2 = \dfrac{g_s^2}{4 \pi \mathcal{V}} \dfrac{1}{\kappa_4^2}.
    \end{equation}
    \item At the tip of a highly warped throat, where $e^{-4A} \gg c$, the scenario changes drastically. Let $n^0_f$ be the order of the 3-form flux units therein. For example, for a Klebanov-Strassler throat threaded by $M$ units of $F_3$-flux on the 3-sphere and $K$ units of $H_3$-flux on the dual 3-cycle of the deformed conifold, $n^0_f \sim M, K$. In the vicinity of the would-be conifold singularity, the 10-dimensional Einstein-frame metric takes the form \cite{Klebanov:2000hb, Giddings:2001yu}
    \begin{equation*}
        ds_{10}^2 = e^{2 A_0} \, \breve{g}_{\mu \nu} \, \de x^\mu \de x^\nu + r_0^2 \, \biggl[ \dfrac{1}{2} \, \de \tau^2 + d \Omega_3^2 + \dfrac{1}{4} \, \tau^2 \, d \Omega_2^2 \biggr],
    \end{equation*}
    where $\tau$ is the radial coordinate of the deformed conifold, the tip being located at $\tau = 0$, while the other line elements describe the 3- and 2-sphere of the conifold base, and $r_0$ is the radius of the 3-sphere at the tip of the throat, such that $r_0^2 \sim n^0_f$. This indicates that the internal metric at the tip of the throat has the behaviour
    \begin{equation} \label{tip metric}
        g^{0}_{mn} \sim n^0_f \, e^{2 A_0},
    \end{equation}
    where $A_0$ is the warp factor at the tip of the throat, with the 3-form flux scaling as $G^0_{mnp} \sim n^0_f / l_s$. In this way, the characteristic scale of the closed-string sector flux-induced mass evaluated at the tip of the throat is
    \begin{equation} \label{throat flux-mass}
        (m^{w}_{\mathrm{flux}})^2 = \dfrac{\vevc^{3/2}}{12} \, e^{2\Omega + 8A_0} \, G^0_{mnp} \bar{G}_0^{mnp} \sim \dfrac{e^{2 A_0}}{n^0_f \mathcal{V}^{2/3}} \dfrac{\vevc^{3/2}}{l_s^2} \sim \dfrac{g_s^2}{n^0_f \mathcal{V}^{2/3}} \, \dfrac{1}{\kappa_4^2} \, e^{2 A_0}.
    \end{equation}
    On the other hand, according to the definition of the metric, the generic throat Kaluza-Klein scale is
    \begin{equation} \label{throat KK-mass}
        (m^{w}_{\mathrm{KK}})^2 \sim \dfrac{e^{2\Omega + 4A_0}}{\lambda_0^2} \dfrac{\vevc^{3/2}}{l_s^2} \sim \dfrac{e^{2 A_0}}{n^0_f \chi^2 \mathcal{V}^{2/3}} \dfrac{\vevc^{3/2}}{l_s^2} \sim \dfrac{g_s^2}{n^0_f \chi^2 \mathcal{V}^{2/3}} \, \dfrac{1}{\kappa_4^2} \, e^{2 A_0},
    \end{equation}
    where the length scale of a cycle at the tip of the throat, measured by $g_{mn}^0$, has been written as $\lambda_0^2 \sim n^0_f \,  e^{2 A_0} \, \chi^2$, with $\chi$ a parameter independent of the warp factor.
    By observing the dimensionally-reduced Einstein-Hilbert term, one may also infer that the warped string scale can be defined as
    \begin{equation} \label{throat string-mass}
        (m^{w}_{s})^2 = \dfrac{g_s^2}{4 \pi \mathcal{V}^{2/3}} \, \dfrac{1}{\kappa_4^2} \, e^{2 A_0}.
    \end{equation}
    Notice that the factor controlling the size of the throat is preferably taken to be $\chi > 1$, so that the warped Kaluza-Klein scale is smaller than the warped string scale.
\end{itemize}

In particular, if the warped mass of eqn. (\ref{throat flux-mass}) is smaller than the bulk mass of eqn. (\ref{bulk flux-mass}), then it is energetically favourable for the closed-string sector fields to be mostly localised at the tip of the throat. Roughly, the condition for this to happen is therefore
\begin{equation} \label{localisation condition}
    \dfrac{\mathcal{V}^{2/3}}{n_f (n^0_f)^{1/2}} \lesssim e^{-A_0}.
\end{equation}
Noticeably, the warped flux-induced and warped Kaluza-Klein scales $m^{w}_{\mathrm{flux}}$ and $m^{w}_{\mathrm{KK}}$ are comparable. Because the cutoff for the 4-dimensional effective theory has to be at most the warped Kaluza-Klein scale, most of the degrees of freedom from the closed-string sector fall above the 4-dimensional threshold. Fields surviving the cutoff include the K\"{a}hler volume modulus, which does not have a flux-induced mass, and potentially some complex structure moduli associated to the geometry at the infrared end of the throat.

The flux integers $n_f$ and $n_f^0$ will be dropped in most of the remaining sections as they are irrelevant in fixing the order of magnitude of the energy scales.

\subsubsection{Conditions for a 4-dimensional Supergravity Formulation} \label{generalities on warped 4-dimensional SUGRAs}
Whilst  below the warped Kaluza-Klein scale the effective theory is 4-dimensional, an $\mathcal{N}_4 = 1$ supergravity formulation is not always possible. In particular, in the presence of supersymmetry breaking, the gravitino gauging the broken supersymmetry becomes massive and may happen to be localised by warping in the infrared end of the throat. In this case, it would have stronger couplings than those expected from a supergravity description, since they would be suppressed by the warped Planck scale rather than by the actual Planck scale \cite{Burgess:2006mn}.  This will now be discussed in more detail, beginning with supersymmetry breaking by fluxes, and followed by comments on supersymmetry breaking with an anti-D3-brane.

The 4-dimensional gravitino corresponding to the least broken supersymmetry (i.e. broken at the smallest scale) is identified with the lightest Kaluza-Klein mode, which becomes massless as the supersymmetry breaking parameter is taken to zero. Taking this 4-dimensional gravitino $\psi_\mu$ to be embedded in the 10-dimensional gravitino as $\Psi_\mu(x,y) = \psi_\mu(x) \otimes \eta(y)$, the qualitative behaviour of the gravitino wavefunction $\eta$ in the extra dimensions can be determined from the 10-dimensional gravitino field equation, which implies a flux-induced mass for $\psi_\mu$ that is of order
\begin{equation*}
    \dfrac{m_{3/2}}{\vevc^{3/4}} \sim \dfrac{e^{\Omega}}{[e^{-4A} + c]} \, G_{mnp} \gamma^{mnp},
\end{equation*}
where $\gamma^m$ are the Dirac matrices representing the Clifford algebra $\lbrace \gamma^m, \gamma^n \rbrace = 2 g^{mn}$ and $G_{mnp}$ is the supersymmetry-breaking 3-from flux. Similarly to the case of the axio-dilaton described above, this mass gives rise to two possible scales across the internal manifold:
\begin{enumerate}[(i)]
    \item a 3-form flux of order $G_{mnp} \sim n_f \theta / l_s$ in the bulk generates a gravitino mass
    \begin{equation} \label{bulk gravitino-mass}
        m_{3/2} \sim \dfrac{e^{\Omega} \vevc^{3/4}}{[e^{-4A} + c]} \, G_{mnp} \gamma^{mnp} \sim \dfrac{n_f \theta}{\mathcal{V}} \, \dfrac{\vevc^{3/4}}{l_s} \sim \dfrac{g_s n_f \theta}{\mathcal{V}} \, \dfrac{1}{\kappa_4};
    \end{equation}
    \item a 3-form flux of order $G_{mnp} \sim n^0_f \theta_0 / l_s$ in the throat generates a gravitino mass
    \begin{equation} \label{throat gravitino-mass}
        m^{w}_{3/2} \sim e^{\Omega + 4 A_0} \, G^0_{mnp} \gamma_0^{mnp} \vevc^{3/4} \sim \dfrac{\theta_0 \, e^{A_0}}{(n^0_f)^{1/2} \mathcal{V}^{1/3}} \, \dfrac{\vevc^{3/4}}{l_s} \sim \dfrac{g_s \theta_0}{(n^0_f)^{1/2} \mathcal{V}^{1/3}} \, \dfrac{1}{\kappa_4} \, e^{A_0}.
    \end{equation}
\end{enumerate}
These are also the expected orders of magnitude of the mass splittings among the fields of any supermultiplet, depending on where the fields are localised. For supersymmetry-breaking flux parameters such that $\smash{m^{w}_{3/2} \ll m_{3/2}}$, which is expected from condition (\ref{localisation condition}), it is energetically favourable for the lightest gravitino to localise at the infrared end of the throat. Its interactions are then suppressed by the warped Kaluza-Klein scale, in contrast to the Planck-suppressed graviton interactions, making a standard supergravity description difficult. However, when the flux parameters satisfy
\begin{equation} \label{supergravity condition}
    \dfrac{\theta}{\theta_0} \ll \dfrac{e^{A_0} \mathcal{V}^{2/3}}{{n_f (n^0_f)^{1/2}}}, 
\end{equation}
which is fulfilled in particular as $\theta \rightarrow 0$, the gravitino mass scales in eqns. (\ref{bulk gravitino-mass}) and (\ref{throat gravitino-mass}) are such that $\smash{m_{3/2} \ll m^{w}_{3/2}}$. In this case, the 4-dimensional gravitino does not localise in the throat, allowing it to have standard $m_P$-suppressed interactions. Nevertheless, the gravitino mass is still warped-down, that is $\smash{\hat{m}_{3/2}^w = e^{A_0} m_{3/2}}$, as the supersymmetry-breaking scale is set at the tip of the throat where the super-Higgs mechanism is triggered.

This is the framework considered in the article and it is thus sensible to formulate an $\mathcal{N}_4 = 1$ supergravity theory below a cutoff scale set as the warped Kaluza-Klein scale $m^{w}_{\mathrm{KK}}$ if the supergravity condition in eqn. (\ref{supergravity condition}) holds, in the regime set by the localisation condition in eqn. (\ref{localisation condition}). In particular, one can reproduce the supergravity description of a highly warped theory by means of a K\"{a}hler potential with the structure
\begin{equation} \label{warped Kaehler potential}
    \kappa_4^2 \mathcal{K} = 2 A_0 + \kappa_4^2 K,
\end{equation}
where $K$ is the K\"{a}hler potential that one would define in the absence of the extremely strong warping effects discussed above and $A_0$ is the warp factor at the tip of the throat, with the superpotential $W$ (and the gauge kinetic functions $f_{AB}$) unchanged.\footnote{In general, purely closed-string contributions to $K$, $W$ and $f_{AB}$ are then independent of $A_0$, but note that the open-string terms (or local geometric closed-string moduli terms) may have a dependence on $A_0$ if they are located in a region of strong warping.} Indeed, such a formulation manifestly provides redshifted energy scales and, in particular, all the masses are warped down. This includes the warped-down gravitino mass, $\smash{\hat{m}_{3/2}^w = e^{A_0} m_{3/2}}$, where the redshift is induced by the $2 A_0$-shift and the unwarped mass is $\smash{m_{3/2} = e^{\kappa_4^2 K/2} W}$, as given by eqn. (\ref{bulk gravitino-mass}). From now on, $\smash{\mathcal{F}^M}$ and $\smash{\mathcal{V}_F}$ also denote the F-terms and the F-term potentials associated to a highly warped scenario.

To summarise, some fields are localised in the bulk region, like the graviton and the gravitino, while others are localised at the tip of the warped throat, like the K\"{a}hler modulus and possible open-string states, which provide the degrees of freedom for the standard-like models of interest in this article. In particular:
\begin{itemize}
    \item fields that are localised at the tip of the throat have redshifted mass scales and are part of the low-energy effective theory, including the K\"{a}hler modulus and the local open-string states;
    \item fields localised in the bulk typically have masses above the cutoff scale (like bulk complex structure moduli) and/or highly suppressed couplings with the throat degrees of freedom (like bulk branes, which could provide massless degrees of freedom), and therefore they can be neglected.
\end{itemize}
In Ref. \cite{Burgess:2006mn}, this discussion is applied to the spontaneous supersymmetry breaking by fluxes. In this article, supersymmetry breaking by anti-D3-branes at the tip of a throat is also considered. Although the way anti-D3-branes break supersymmetry is conceptually different to flux supersymmetry breaking, the arguments on the localisation of the gravitino in the bulk, for small bulk supersymmetry-breaking fluxes, follow through in the same way. Hence, the following sections show how to incorporate open-string degrees of freedom in a description with the $2A_0$-shift in the K\"{a}hler potential as in eqn. (\ref{warped Kaehler potential}).

\subsubsection*{K\"{a}hler Modulus Localisation}
In KKLT-like constructions, in which the K\"{a}hler modulus is stabilised by non-perturbative effects such as D7-brane gaugino condensates \cite{Derendinger:1985kk, Dine:1985rz, Burgess:1995aa, Blumenhagen:2009qh} or Euclidean D3-brane instantons \cite{Witten:1996bn}, the K\"{a}hler potential shift in eqn. (\ref{warped Kaehler potential}) implies that the scalar potential sourced by non-perturbative effects is redshifted by the warp factor, even though the non-perturbative effects are not necessarily localised near the throat.  
    
To understand this redshifting, one should consider the localisation of the K\"{a}hler modulus $\rho$. The field $\rho$ is massless before the compactification, so naively one expects it to be not localised. However, an explicit analysis is performed in Ref. \cite{Frey:2008xw} and reveals that:
\begin{enumerate}[(i)]
    \item the wavefunction of the 4-dimensional graviton $g_{\mu \nu}$ is strongly peaked in the bulk region, both in the presence and in the absence of strong warping;
    \item the wavefunction of the K\"{a}hler modulus $\rho$ tends to be more and more peaked in the throat as the warping becomes stronger.
\end{enumerate}

Notice that even with non-perturbative effects, the K\"{a}hler modulus is very light and well below the warped KK-scale cutoff, suggesting that its wavefunction is perturbed only slightly and in particular that it is still peaked in the throat.  Then, $\rho$ should feel any non-perturbative effects localised in the bulk via a redshifted mediation to the tip of the throat.  Consistently with this picture, one can observe that with a warped-down non-perturbative contribution to the scalar potential, the stronger the warping is - i.e. the longer the throat is - the less efficient the stabilisation becomes. Another challenge is that any supersymmetry-breaking $(0,3)$-flux  localises around the gaugino condensate usually in the bulk \cite{Dymarsky:2010mf, Baumann:2010sx}, which could result in the gravitino localising at the throat tip, making a supergravity description difficult.

\section{Warped D3- and D7-branes} \label{warped D3- and D7-branes}
This section considers D3- and D7-branes in strongly warped Calabi-Yau orientifold compactifications, as a warm up before the anti-D3-/D7-brane constructions. As D3-/D7-brane systems preserve the same $\mathcal{N}_4 = 1$ supersymmetry as the closed-string sector, the only sources of supersymmetry breaking considered here are $(0,3)$-fluxes. An $\mathcal{N}_4 = 1$ supergravity description can be derived by matching with the operators that are obtained from the dimensional reduction. Appendix \ref{D-brane actions} reviews the field content and the generic form of the worldvolume actions. In terms of $\mathcal{N}_4=1$ supersymmetry, the low-energy degrees of freedom from the D-branes, composing the matter sector, are the following.
\begin{itemize}
    \item D3-branes contain three complex scalars $\varphi^a$ parametrising the position of the brane in the internal space and three spinors $\psi^a$ in an $\mathrm{SU}(3)$-triplet with respect to the internal tangent space group, which form three chiral multiplets, as well as one Abelian gauge vector $A_\mu$ and a spinor $\lambda$ in an $\mathrm{SU}(3)$-singlet, which form a vector multiplet.
    \item D7-branes wrapping a 4-cycle in the internal space contain one complex scalar $\sigma^3$ parametrising the position of the brane in the internal space and a spinor $\eta$, which together form a chiral multiplet, as well as one Abelian gauge vector $B_\mu$ and a spinor $\zeta$, which form a vector multiplet. Extra degrees of freedom associated to the Wilson lines are absent if the wrapped cycle has no non-contractible 1-cycles.
    \item When D3 and D7-branes overlap, the intersecting $37$- and $73$-states correspond to two complex scalars $\varphi$ and $\tilde{\varphi}$ and two spinors $\psi$ and $\tilde{\psi}$, which form two chiral multiplets in conjugate representations of the gauge groups.  Specifically, the chiral multiplets $\varphi$ and $\tilde{\varphi}$ have charges $q_{\mathrm{D3}}=+1,-1$ and $q_{\mathrm{D7}}=-1,+1$, respectively, under the D3- and D7-brane $\mathrm{U}(1)$ gauge groups.
\end{itemize}

A summary of the supergravity expansions for models with matter and supersymmetry-breaking hidden sectors, the latter including bulk moduli, is given in appendix \ref{LEEFT SUGRA in IIB compactifications}. In the following subsections, the specific form of these interactions from the dimensional reduction of D3-/D7-branes in warped flux compactifications will be derived and intersecting states will also be discussed. The total K\"{a}hler potential and the total superpotential will be found to take the form
\begin{subequations}
\begin{gather}
    \begin{split}
        \mathcal{K} = \dfrac{2 A_0}{\kappa_4^2} + \hat{K} + Z_{\Ds \bDs} \Ds \bDs + \dfrac{1}{2} \, \Bigl[ H_{\Ds \Ds} \Ds \Ds + \mathrm{c.c.} & \Bigr] \\
        + Z_{\varphi^a \bar{\varphi}^b} \varphi^a \bar{\varphi}^b + Z_{\varphi \bar{\varphi}} \varphi \bar{\varphi} + Z_{\tilde{\varphi} \tilde{\bar{\varphi}}} \varphi \tilde{\bar{\varphi}} &,
    \end{split} \label{Kaehler potential - D3/D7} \\
    W = \hat{W} + \dfrac{1}{2} \tilde{\mu}_{\Ds \Ds} \Ds \Ds + \tilde{y} (\beta \Ds - \Dt) \, \varphi \tilde{\varphi},  \label{superpotential - D3/D7}
\end{gather}
\end{subequations}
where $\hat{K}$ and $\hat{W}$ are the pure closed-string potentials of eqns. (\ref{closed-strings - Kaehler potential}, \ref{closed-strings - superpotential}) and all the other terms represent the open-string couplings. The gauge kinetic functions, D-term potentials and -- in the case of supersymmetry-breaking fluxes -- soft terms will also be worked out. The $2A_0$-shift will be inserted if working under the conditions (\ref{localisation condition}) and (\ref{supergravity condition}), that is all masses redshifted by the warp factor.

The details of the open-string sector terms depend on the brane configuration, with two main constructions considered. The D3-brane will be placed at the tip of a highly warped throat, whereas the D7-brane will wrap a 4-cycle either located at the tip of the throat or extending from the tip into the bulk. Detailed global constructions are deferred for future work and, when explicit, the wrapped 4-cycle will be assumed to be a torus orbifold for simplicity; throats with such cycles have been constructed e.g. in \cite{Cascales:2003wn}.  Unless otherwise stated, only a pure $(2,1)$-flux is assumed to exist at the tip of the throat. The dimensional reduction will not capture the complex structure moduli couplings, but the supergravity formulation will correctly account for them. Stabilisation of the volume modulus $\rho$ will be considered in subsection \ref{moduli stabilisation + anomaly mediation} focussing on the main case of interest, which is the presence of KKLT-like non-perturbative corrections and uplifting anti-D3-branes. Further, notice that worldvolume fluxes will not be considered in this work.

\subsection{Pure D3- and D7-brane States} \label{single D3- and D7-brane SUGRA}
This subsection overviews the analysis of D3- and D7-branes in type IIB Calabi-Yau orientifold compactifications, adapting it to the strongly warped metric of eqn. (\ref{metric}). In the following, superscripts and subscripts '$0$' denote quantities evaluated at the tip of the throat.

\subsubsection{Warped D3-branes} \label{single D3-brane SUGRA}
As discussed in appendix \ref{D-brane actions}, it is possible to express the action of the D3-brane degrees of freedom by adapting the results of the dimensional reductions from Refs. \cite{Camara:2003ku, Grana:2003ek, McGuirk:2012sb, Bergshoeff:2015jxa}.

\paragraph{D3-brane Chiral Superfields}\label{D3-brane chiral superfields}
The pure kinetic action for the D3-brane scalars takes the form (see also Refs. \cite{Burgess:2006mn, Chen:2009zi, Cownden:2016hpf}, which work directly in the regime of strong warping)
\begin{equation*}
S_\textrm{kin}^{\textrm{D3-scalars}} = - \dfrac{1}{2 \pi g_s} \int_{X_{1,3}} \! \de^4 x \, \sqrt{- \detr{g_4}} \; e^{2 \Omega} \, g^{0}_{a \bar{b}} \, g^{\mu \nu} \, \nabla_\mu \varphi^a \nabla_\nu \bar{\varphi}^b.
\end{equation*}
Therefore, one can include this term within the K\"{a}hler potential of the K\"{a}hler modulus as
\begin{equation*}
\kappa_4^2 K = - 3 \, \mathrm{ln} \, \biggl[ 2 \, e^{-2 \Omega} - \dfrac{\kappa_4^2}{3 \pi g_s} \, g^{0}_{a \bar{b}} \, \varphi^a \bar{\varphi}^b \biggr].
\end{equation*}
This logarithmic no-scale structure, with $K$ of the form $K=-3\log\left[f_{\textrm{hid}}(\rho, \bar\rho)+ f_{\textrm{vis}}(\varphi, \bar\varphi) \right]$, is a common feature of D-brane supergravity and suggests the possibility of sequestering \cite{Randall:1998uk, DeWolfe:2002nn} (see also Ref. \cite{Grana:2003ek}).  Indeed, it implies that the brane scalars do not feel hidden-sector supersymmetry breaking at tree-level, and it turns out that brane fermions also stay massless at tree-level. From the expression above, it follows that the K\"{a}hler matter metric reads
\begin{equation} \label{D3-brane - Z}
    Z_{\varphi^a \bar{\varphi}^b} = \dfrac{1}{2 \pi g_s} \, e^{2 \Omega} \, g_{a \bar{b}}^{0}.
\end{equation}
Due to supersymmetry, the D3-brane modulini are also captured by these couplings. Since, the chiral multiplet $\varphi^a$ is massless in an imaginary self-dual flux background, this K\"{a}hler potential is enough to account for the D3-brane chiral field couplings.

As discussed in subsection \ref{field localisation}, for a low-energy effective field theory describing fields at the tip of a highly warped throat, the K\"{a}hler potential is shifted by the constant $\smash{2 A_0}$. This clearly does not change the K\"{a}hler matter metric for the D3-brane fields.

\paragraph{D3-brane Gauge Sector}
The Weyl scaling from the 10- to the 4-dimensional Einstein frame does not affect the D3-brane gauge kinetic terms in the action, so one has
\begin{equation*}
    S_{\textrm{kin}}^{\textrm{D3-vector}} = - \dfrac{1}{4 \pi g_s} \int_{X_{1,3}} e^{-\sdil} \, F_2 \wedge * F_2 + \dfrac{1}{4 \pi g_s} \int_{X_{1,3}} C_0 \, F_2 \wedge F_2
\end{equation*}
and the gauge kinetic function is as usual
\begin{equation} \label{D3-brane - f}
    f_{\mathrm{D3}} = - \dfrac{i \tau}{2 \pi g_s}.
\end{equation}
This does not depend on the warp factor due to the cancellation happening in the metric-dependent factors. The dimensional reduction of the gaugino is not performed as the action can be reproduced by supersymmetry arguments.

\subsubsection{D7-branes Extending from the Tip of a Warped Throat into the Bulk} \label{single bulk D7-brane SUGRA}
This subsubsection describes a D7-brane wrapping a 4-cycle $\Sigma_4$ which extends from the tip of a warped throat up into the bulk region. Details of the dimensional reduction of the D7-brane worldvolume action can be found in Refs. \cite{Camara:2004jj, Jockers:2004yj, Marchesano:2008rg} (see also Refs. \cite{Lust:2004cx, Lust:2004fi, Lust:2005bd}) and are briefly overviewed in subsection \ref{D-brane actions}. A toy model is described below, including the geometric configuration and the corresponding dimensional reduction. In particular, the warp factor is assumed to be only a function of the directions parallel to the 4-cycle.

\begin{figure}[H]
\centering
\begin{tikzpicture}[scale=0.5]

\draw[thick, color=cyan] (0,1.5) ellipse (7 and 2);
\begin{scope}
    \clip (0,3.7) ellipse (7 and 3);
    \draw[thick, color=cyan, fill=white] (0,-0.7) ellipse (7 and 3);
\end{scope}
\begin{scope}
    \clip(0,-0.3) ellipse (7 and 3);
    \draw[thick, color=cyan] (0,3.7) ellipse (7 and 3);
\end{scope}

\draw[thick, color=purple, fill=white!90!red] (-9.5,6.5) ellipse (1.5 and 5);
\begin{scope}
    \clip (-8.3,6.5) ellipse (1.8 and 5);
    \draw[thick, color=purple,fill=white] (-10.7,6.5) ellipse (1.8 and 5);
\end{scope}
\begin{scope}
    \clip (-10.5,6.5) ellipse (1.8 and 5);
    \draw[thick, color=purple] (-8.3,6.5) ellipse (1.8 and 5);
\end{scope}

\draw[thick, color=purple, fill=white!90!red] (-9.3,11.4) -- (-9.5,12.5) -- (-9.7,11.4);
\node at (-8.1,11.3) {$r_{\mathrm{UV}}$};
\draw[thick, fill=green] (-9.5,12.65) circle (0.1);
\begin{scope}
    \clip (-9.3,11) rectangle (-9.7,11.4);
    \draw[thick, red] (-9.5,11.7) circle (0.4);
\end{scope}

\draw[thick, color=cyan] (0,12) ellipse (2.5 and 1.0);
\begin{scope}
    \clip (0,14.2) ellipse (3 and 2.5);
    \draw[thick, color=cyan, fill=white] (0,9.8) ellipse (3 and 2.5);
\end{scope}
\begin{scope}
    \clip(0,10.2) ellipse (3 and 2.5);
    \draw[thick, color=cyan] (0,14.2) ellipse (3 and 2.5);
\end{scope}

\draw[->, thin] (-6.15,-0.3) to [out=-14, in=194] (6.15,-0.3) node[below]{$\theta$};
\draw[->, thin] (8.5,11.5) -- (8.5,1.5) node[above right]{$r$};

\node at (-9.5,6.5) {$\Sigma_4$};
\node at (0,1.5) {$\Sigma_2$};
\node at (3,10.5) {$\Sigma_2$};

\node at (-9.75,0.5) {\setulcolor{pink} \setul{}{1pt} \ul{$\textrm{D7-brane}$}};
\node at (-13.75,12.75) {\setulcolor{green} \setul{}{1pt} \ul{$\textrm{(anti-)D3-brane}$}};

\end{tikzpicture}
\vspace{-24pt}\caption{A sketch of the toy configuration under consideration, with the D7-brane wrapping the 4-space at $\smash{\theta=0}$ and some throat being glued to the bulk at $r = r_{\mathrm{UV}}$. The D3- or anti-D3-brane provides extra open-string states, as discussed in sections \ref{warped D3- and D7-branes} and \ref{warped anti-D3- and D7-branes}, respectively.}
\label{brane setup - extended 4-cycle}
\end{figure}
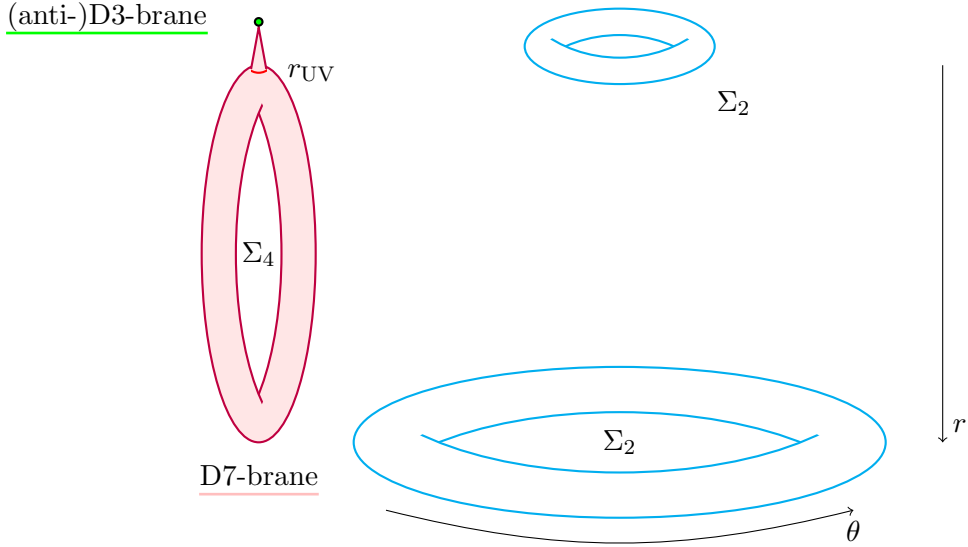

\paragraph{D7-brane Configuration and Field Localisation Conditions}
It is assumed that the internal space, locally in the neighbourhood of the wrapped D7-brane, takes the form  $\Sigma_4 \ltimes \Sigma_2$. Let the coordinates $y^{m'}$ span the 4-space $\Sigma_4$, for $m'=4,\dots,7$, with $z^1, z^2$ the corresponding complexified directions, and let the coordinates $y^{\dot{m}}$, for $\dot{m}=8,9$, parametrise the transverse 2-space $\Sigma_2$, with $z^3$ the associated complex coordinate. Given some convenient coordinates  $r^{m'} = r^{m'}(y^{n'})$ and $\theta^{\dot{m}} = \theta^{\dot{m}} (y^{\dot{n}})$, the metric is of the form
\begin{equation*}
    ds^2_6 = e^{-2 A} \, g_{mn} \de y^m \de y^n = e^{-2 A(r)} \, \Bigr(g_{m'n'} (r) \, \de y^{m'} \de y^{n'} +  g_{3 \bar{3}} (r, \theta) \, \de z^3 \de \bar{z}^3 \Bigl).
\end{equation*}
At some $\smash{r^2 = r_{m'} r^{m'} = r_{\mathrm{UV}}^2}$, the bulk is glued to a warped throat, which ends at its tip with a tiny warp factor $e^{2 A} (r=0) = e^{2 A_0}$. The D7-brane wraps the slice corresponding to the coordinates $\theta_i = 0$. See Fig. \ref{brane setup - extended 4-cycle}.

\vspace{-36pt}

In order to be able to perform explicit calculations, the warp factor is assumed to be a function of only the 4-space coordinates. Further, the 4-cycle is assumed to be the orbifold $\Sigma_4 = \mathrm{T}^4/\mathbb{Z}_2$ and locally the orthogonal directions are the 2-torus $\mathrm{T}^2$, i.e. the metric is such that
\begin{equation*}
    g_{m'n'} (r \in \Sigma_4) = g^{(\mathrm{T}^4 / \mathbb{Z}_2)}_{m'n'}, \qquad \qquad \qquad g_{3 \bar{3}} (r \in \Sigma_4, \theta) = g^{(\mathrm{T}^2)}_{3 \bar{3}}.
\end{equation*}
Finally, in analogy with the Klebanov-Strassler throat, it is assumed that at the throat tip the metric scales with the constant $e^{2 A_0}$, as in eqn. (\ref{tip metric}), that is
\begin{equation*}
    g_{m'n'} (r < r_{\mathrm{UV}}) \overset{r \sim 0}{\sim} e^{2 A_0}, \qquad \qquad \qquad g_{3 \bar{3}} (r < r_{\mathrm{UV}}, \theta) \overset{r \sim 0}{\sim}  e^{2 A_0}.
\end{equation*}

\paragraph*{Localisation Scenarios}
In analogy with what happens for the closed-string sector, one might guess that the open-string moduli of the wrapped D7-brane can become localised at the tip of the throat too. The conditions under which this occurs will now be worked out.

One can analyse the internal wavefunction of the D7-brane scalar fields by dimensionally reducing the real fields $\sigma^{\dot{m}} = \sigma^{\dot{m}}(x,y)$, with $\dot{m} = 8,9$, in a similar way to Refs. \cite{Marchesano:2008rg, Camara:2009xy}. The D7-brane 8-dimensional scalar action can be written in terms of the 4-dimensional Einstein frame metric as
\begin{equation*}
    \begin{split}
        S_{\mathrm{D7}}^{\textrm{scalar}} =  - \tau_{\mathrm{D7}} \al^2 \int_{X_{1,3}} \!\! \de^4 x \, \sqrt{- \detr{g_4}} \int_{\Sigma_4} \!\! \de^4 y \, \sqrt{ \detr{g_{\Sigma_4}}} \; \biggl[ e^{2 \Omega + \sdil} \,[e^{-4A} + c] \, g_{\dot{r} \dot{s}} \, g^{\mu \nu} \, \nabla_\mu \sigma^{\dot{r}} \nabla_\nu \sigma^{\dot{s}} \\
        + \vevc^{3/2} \, e^{4 \Omega + \sdil} \, g_{\dot{r} \dot{s}} \, g^{m'n'} \! \nabla_{m'} \sigma^{\dot{r}} \nabla_{n'} \sigma^{\dot{s}} + \dfrac{1}{2} \, \dfrac{\vevc^{3/2} \, e^{4 \Omega + 2 \sdil}}{e^{-4A} +c} \, G_{m'n'\dot{r}} \tensor{\bar{G}}{^{m'} ^{n'} _{\!\dot{s}}} \, \sigma^{\dot{r}} \sigma^{\dot{s}} & \biggr],
    \end{split}
\end{equation*}
where it is understood that only some of the 3-form fluxes contribute, determined by the interference of the DBI- and CS-actions \cite{Camara:2004jj}.  For constant K\"{a}hler modulus and axio-dilaton backgrounds, one finds the field equation
\begin{equation*}
    \vevc^{-3/2} \, \Delta_{4} \sigma^{\dot{r}} + \dfrac{e^{2 \Omega}}{[e^{-4A} + c]} \Delta_{\Sigma_4} \sigma^{\dot{r}} - \dfrac{1}{2} \, \dfrac{e^{2 \Omega + \sdil}}{[e^{-4A} + c]^2} \, G^{m' n' \dot{r}} \bar{G}_{m' n' \dot{s}} \, \sigma^{\dot{s}} = 0.
\end{equation*}
Then, defining the Kaluza-Klein decomposition of the field as
\begin{equation*}
    \sigma^{\dot{r}} (x, y) = \sum_\omega \sigma_\omega^{\dot{r}} (x) \, \zeta_\omega^{\dot{r}} (y)
\end{equation*}
and imposing the Klein-Gordon equations $\Delta_{4} \sigma_\omega^{\dot{r}} = m_\omega^2 \sigma_\omega^{\dot{r}}$, one eventually obtains the internal wavefunction field equation
\begin{equation*}
    \dfrac{e^{2\Omega}}{[e^{-4A}+c]} \Delta_{\Sigma_4} \zeta_\omega^{\dot{r}} + \dfrac{m_\omega^2}{\vevc^{3/2}} \, \zeta_\omega^{\dot{r}} = \dfrac{1}{2} \, \dfrac{e^{2\Omega + \sdil}}{[e^{-4A} + c]^2} \, G^{m' n' \dot{r}} \bar{G}_{m' n' \dot{s}} \, \zeta_\omega^{\dot{s}}.
\end{equation*}
This is the same equation as the one defining the axio-dilaton wavefunction, with the only difference that the wavefunction is 4- rather than 6-dimensional.

Following subsection \ref{field localisation}, the compactification volume can be sufficiently large so that warped-down masses are still greater than bulk masses (cf. eqn. (\ref{localisation condition})), and fields tend to localise in the bulk.  However, the D7-brane chiral superfield is localised near the tip of the throat whenever the warped-down mass $m_{\mathrm{D7}}^w$ is smaller than the unwarped bulk mass $m_{\mathrm{D7}}$, that is if (in analogy with eqn. (\ref{localisation condition}))
\begin{equation} \label{D7 localisation condition}
    \dfrac{e^{A_0} \mathcal{V}^{2/3}}{n_f (n^0_f)^{1/2}} \lesssim \dfrac{\theta'}{\theta'_0},
\end{equation}
where the fluxes sourcing the D7-brane field masses have been taken to be $G_{mnp} \sim \theta' n_f/l_s$ in the bulk and $G_{mnp} \sim \theta'_0 n^0_f/l_s$ near the tip.  For generic flux parameters, $\theta'$ and $\theta'_0$, the warped mass is of the same order as the warped flux-induced axio-dilaton mass $m_{\mathrm{flux}}^w$ of eqn. (\ref{throat flux-mass}) and the warped Kaluza-Klein scale $m_{\mathrm{KK}}^w$ of eqn. (\ref{throat KK-mass}), i.e.
\begin{equation*}
    (m_{\mathrm{D7}}^w)^2 \sim \dfrac{g_s^2 {\theta'_0}^{2}}{n^0_f \mathcal{V}^{2/3}} \, \dfrac{1}{\kappa_4^2} \, e^{2 A_0},
\end{equation*}
so that these fields are too heavy to stay in the low-energy theory.  However, if $\theta'_0$ is small enough, it may be that fluxes sourcing the D7-brane masses allow both $m_{\mathrm{D7}}^w \lesssim m_{\mathrm{D7}}$, so fields are localised, and also $m_{\mathrm{D7}}^w \ll m_{\mathrm{KK}}^w$, so fields stay in the low-energy theory. It may also happen that $\theta'$ is small enough that the hierarchy is $m_{\mathrm{D7}}^w \gtrsim m_{\mathrm{D7}}$, so it is energetically favourable for the D7-brane fields to be localised in the bulk, and yet fluxes at the tip of a highly warped throat source a warped-down mass, analogously to what happens to the gravitino mass in eqn. (\ref{supergravity condition}). These three scenarios will now be discussed in detail.

\paragraph{D7-brane Chiral Superfield in the Bulk} \label{extended 4-cycle: case i}
For large enough internal volumes that do not satisfy the localisation condition of eqn. (\ref{D7 localisation condition}), $m_{\mathrm{D7}} \lesssim m_{\mathrm{D7}}^w$ and D7-brane fields generally extend along the throat from the tip into the bulk.  Before the compactification over the wrapped 4-cycle, the kinetic term for the D7-brane transverse complexified scalar $\sigma^3$ reads
\begin{equation*}
    S_\textrm{kin}^{\textrm{D7-scalar}} = - \dfrac{1}{2 \pi g_s l_s^4} \int_{X_{1,3}} \! \de^4 x \, \sqrt{- g_{4}} \int_{\Sigma_4} \de^4 y \, \sqrt{ g_{\Sigma_4}} \; [e^{-4A} + c] \, e^{2\Omega + \sdil} \, g_{3 \bar{3}} \, g^{\mu \nu} \, \nabla_\mu \sigma^{3} \nabla_\nu \bar{\sigma}^{3}.
\end{equation*}
Since the warp factor varies only longitudinally with respect to the brane, one can define the dimensionless unwarped and warped 4-dimensional volumes
\begin{equation*}
    l_s^4 \Vzerofour = \int_{\Sigma_4} \de^4 y \, \sqrt{ \detr{g_{\Sigma_4}}}, \qquad \qquad \qquad l_s^4 \Vwfour = \int_{\Sigma_4} \de^4 y \, \sqrt{\detr{g_{\Sigma_4}}} \; e^{-4A}.
\end{equation*}
In particular, the internal metric, being that of a torus, is independent of the 4-cycle coordinates and, following the definition of the Weyl factor in eqn. (\ref{Weyl factor}), it is apparent that the kinetic term becomes
\begin{equation*}
S_\textrm{kin}^{\textrm{D7-scalar}} = - \dfrac{\Vzerofour}{2\pi g_s} \int_{X_{1,3}} \! \de^4 x \, \sqrt{- g_{4}} \; e^{\sdil} \, g_{3 \bar{3}} \, g^{\mu \nu} \, \nabla_\mu \sigma^{3} \nabla_\nu \bar{\sigma}^{3}.
\end{equation*}
One can reproduce this within a supergravity action by modifying the axio-dilaton K\"{a}hler potential as
\begin{equation*}
\kappa_4^2 K = - \mathrm{ln} \, \biggl[ -i (\tau - \bar{\tau}) - \dfrac{\kappa_4^2}{\pi g_s} \, \Vzerofour \, g_{3 \bar{3}} \, \sigma^{3} \bar{\sigma}^{3} \biggr],
\end{equation*}
or equivalently by defining the K\"{a}hler matter metric
\begin{equation} \label{bulk D7-brane - Z}
    Z_{\sigma^3 \bar{\sigma}^3} = \dfrac{\Vzerofour}{\pi g_s} \, \dfrac{g_{3 \bar{3}}}{[-i (\tau - \bar{\tau})]}.
\end{equation}

As far as the mass term is concerned, from the dimensional reduction, in real notation one finds an action of the form
\begin{equation*}
    S_{\textrm{mass}}^{\textrm{D7-scalar}} = - \dfrac{1}{2 \pi g_s l_s^4} \int_{X_{1,3}} \!\!\! \de^4 x \sqrt{- g_{4}} \int_{\Sigma_4} \!\!\! \de^4 y \sqrt{g_{\Sigma_4}} \; \dfrac{1}{8 \pi \Vzero} \, \dfrac{g_s^2}{\kappa_4^2} \, \dfrac{e^{4\Omega + 2 \sdil}}{e^{-4A} + c} \, l_s^2 \, G_{m'n'\dot{r}} \, \tensor{\bar{G}}{^{m'} ^{n'} _{\!\dot{s}}} \, \sigma^{\dot{r}} \sigma^{\dot{s}}.
\end{equation*}
D7-branes have a supersymmetric mass sourced by a $(2,1)$-flux. In the toy model under consideration, in the vicinity of the brane it is possible to decompose forms in the 6-dimensional space into products of forms in the 4- and 2-dimensional spaces, $\Sigma_4=\mathrm{T}^4/{\mathbb{Z}_2}$ and $\mathrm{T}^2$ (see appendix \ref{appendix: throat geometry}). In particular, the specific mass-sourcing  $(2,1)$-flux can be written as \cite{Camara:2004jj} (the hat denotes the specific component)
\begin{equation*}
    \hat{G}_3 (r, \theta = 0) = f (r, \theta = 0) \, \chi_\vartheta,
\end{equation*}  
where the $(2,1)$-form $\chi_\vartheta=\eta \wedge \de \bar{w}^3$ is defined in terms of the  $(2,0)$-form of the 4-cycle as $\eta = \de z^1 \wedge \de z^2$ and $\de \bar{w}^3$, with $w^3 = z^3/l_s$ a dimensionless coordinate, and $f = f(r, \theta)$ is a function representing the near-brane dependences. For definiteness, let the integrals be dominated by the throat region, where $e^{-4A} \gg \langle c\rangle$. As $e^{4A} \hat{G}_3$ is a harmonic form, one can express the 2-form component $g_2 = f(r,\theta=0) \,\eta$ in terms of the harmonic $(2,0)$-form $\eta$ as
\begin{equation*}
    e^{4A} g_2 = \dfrac{1}{\omega_w^{\Sigma_4}} \, \eta \int_{\Sigma_4} g_2 \wedge \bar{\eta},
\end{equation*}
with $\omega_w^{\Sigma_4} = \int_{\Sigma_4} e^{-4A} \eta \wedge \bar{\eta}$. Now, starting from the general action above, the supersymmetric mass term can be expressed as
\begin{equation*}
    S_{\textrm{mass}}^{\textrm{D7-scalar}} = - \dfrac{1}{2 \pi g_s l_s^4} \int_{X_{1,3}} \!\! \de^4 x \sqrt{- g_{4}} \int_{\Sigma_4} \!\! \de^2 z \, d^2 \bar{z} \, \sqrt{ g_{\Sigma_4}} \; \dfrac{1}{8 \pi \Vzero} \, \dfrac{g_s^2}{\kappa_4^2} \, e^{4\Omega + 4 A + 2 \sdil} \, (g_2 \, \cdot \, \bar{g}_2) \, \sigma^{3} \bar{\sigma}^{3}
\end{equation*}
where, because $g_2$ is automatically self-dual, i.e. $*_4 g_2 = g_2$, the 4-cycle integral is
\begin{equation*}
   \int_{\Sigma_4} \!\! \de^2 z \, d^2 \bar{z} \sqrt{ g_{\Sigma_4}} \; e^{4A} \, g_2 \, \cdot \, \bar{g}_2 = \int_{\Sigma_4} e^{4A} \, g_2 \wedge \bar{g}_2 = \dfrac{1}{(\omega_w^{\Sigma_4})^2} \int_{\Sigma_4} \! e^{-4A} \, \eta \wedge \bar{\eta} \int_{\Sigma_4} \! g_2 \wedge \bar{\eta} \int_{\Sigma_4} \! \bar{g}_2 \wedge \eta.
\end{equation*}
The first integral factor can be written as
\begin{equation*}
   \lambda_{\Sigma_4} = \int_{\Sigma_4} e^{-4A} \, \eta \wedge \bar{\eta} = \omega_w^{\Sigma_4} \sim \omega_w^{\Sigma_4} \dfrac{\Vwfour}{\Vzerofour} e^{2\Omega},
\end{equation*}
where an approximate unit factor has been introduced in the final relation for convenience in the comparison of the dimensionally reduced action with the supergravity. In the end the scalar mass term becomes
\begin{equation*}
    S_{\textrm{mass}}^{\textrm{D7-scalar}} = - \dfrac{1}{2 \pi g_s} \int_{X_{1,3}} \!\! \de^4 x \sqrt{- g_{4}} \;  \dfrac{1}{8 \pi \Vzero} \, \dfrac{g_s^2}{\kappa_4^2} \, \dfrac{e^{6\Omega + 2 \sdil}}{\omega_w^{\Sigma_4}} \, \dfrac{\Vwfour}{\Vzerofour} \, \dfrac{1}{l_s^4} \int_{\Sigma_4} \bar{g}_2 \wedge \bar{\eta} \int_{\Sigma_4} g_2 \wedge \eta \; \sigma^{3} \bar{\sigma}^{3}.
\end{equation*}
The opposite approximation to that used above, where integrals are dominated by the bulk region, can be obtained easily by taking formally $e^{4A} = 1$ everywhere, and $e^{2\Omega} = 1/c$.

In view of Ref. \cite{Lust:2005bd}, to generate the $\smash{(2,1)}$-flux-induced mass one introduces the holomorphic superpotential bilinear coupling
\begin{equation} \label{bulk D7-brane - mu}
    \begin{split}
        \tilde{\mu}_{\sigma^{3} \sigma^{3}} & = - \dfrac{\Vzero}{\pi} \, \dfrac{1}{\kappa_4 l_s^2} \der_\tau \der_{u^\vartheta} \int_{Y_6} \Bigl[ G_3 \wedge \Omega \Bigr] \, \delta^{(2)} (\theta) \\
        & = \biggl[ \dfrac{\Vzero}{\pi [-i(\tau - \bar{\tau})] \kappa_4 l_s^2} \int_{Y_6} (G_3 - \bar{G}_3) \wedge \biggl( \dfrac{i}{\omega_w} \, (\der_{u^\vartheta} \omega_w) \, \Omega - \chi_{u^\vartheta} \biggr) \, \delta^{(2)} (\theta) \biggr]
    \end{split}
\end{equation}
where use has been made of the identity $\der_{u^\alpha} \Omega = [\der_{u^\alpha} \mathrm{ln} \, \omega_w] \, \Omega + i \chi_{\alpha}$. Indeed, in the specific case in which the background is pure $(2,1)$-flux, this is
\begin{equation*}
    \bigl[\tilde{\mu}_{\sigma^{3} \sigma^{3}}\bigr]_{(2,1)} = \biggl[ \dfrac{\Vzero}{\pi \, [-i(\tau - \bar{\tau})] \kappa_4 l_s^2} \int_{\Sigma_4} \bar{g}_2 \wedge \eta \biggr] \, \delta_{33}.
\end{equation*}
As required, the effective coupling $\smash{\mu_{\sigma^3 \sigma^3} = e^{\kappa_4^2 \hat{K}/2} [\tilde{\mu}_{\sigma^3 \sigma^3}]_{(2,1)}}$, reproduces a supersymmetric mass $m^2_{\sigma^{3} \bar{\sigma}^{3}} = Z^{\sigma^{3} \bar{\sigma}^{3}} \mu_{\sigma^{3} \sigma^{3}} \bar{\mu}_{\bar{\sigma}^{3} \bar{\sigma}^{3}}$ that corresponds precisely to the one inferred from the dimensional reduction. The identification takes place if $\smash{e^{\kappa_4^2 \hat{K}_{\mathrm{cs}}} \Vw = \Vwfour / \omega_w^{\Sigma_4}}$, otherwise the bilinear coupling $\tilde{\mu}_{\sigma^{3} \sigma^{3}}$ should be rescaled by an order one factor $\smash{\left(\Vw/\omega_w\right)^{-1/2} \left(\Vwfour / \omega_w^{\Sigma_4}\right)^{1/2}}$, in which the apparent non-holomorphicity is expected to cancel. For the canonically normalised field, one recognises the mass
\begin{equation*}
    m^2_{\mathrm{D7}} \sim \dfrac{g_s^2}{\mathcal{V}^2} \dfrac{1}{\kappa_4^2}.
\end{equation*}
As will be seen from all the dimensional reductions, all the couplings of the theory have 4-dimensional scales which are defined in terms of the reduced Planck length with, depending on the interactions, various suppressions from the string coupling, the volume and/or the warp factor, while the string length factor precisely accounts for the integrations over the compact space.  Notice that $m_{\mathrm{D7}}$ is below the cutoff $m_{\mathrm{KK}}^w$.

\paragraph*{Comment on Generic Flux Backgrounds}
For a generic flux background, one can again take advantage of the results of Refs. \cite{Camara:2004jj, Lust:2005bd} and a similar dimensional reduction follows as above: one obtains the same supersymmetric mass just found, plus some soft-breaking scalar mass terms.

In particular, Ref. \cite{Lust:2005bd} considered unwarped toroidal orbifold compactifications, and showed that all these terms can be generated by the holomorphic bilinear coupling $\tilde{\mu}_{\sigma^{3} \sigma^{3}}$ of eqn. (\ref{bulk D7-brane - mu}) and a non-vanishing K\"{a}hler potential $H$-term, which, together with the axio-dilaton and complex structure moduli F-terms, give the same effective $\mu$-coupling as above (see eqn. (\ref{effective mu})), along with the soft-breaking terms (see eqns. (\ref{soft masses}, \ref{B-term})).

For less isotropic scenarios, where for instance only the wrapped cycle is a toroidal orbifold $O_4 = \mathrm{T}^4 / \mathbb{Z}_2$, some difficulties may arise. The complex structure moduli K\"ahler potential includes $\kappa_4^2 \hat{K}(u, \bar{u}) = - \mathrm{ln} \, \omega_w$ with $\smash{\omega_w = i \, \Vwfour \Vzerotwo \, \prod_{a=1}^3 [-i(u^a - \bar{u}^a)]}$, where $u^3 = u^\vartheta$ is the modulus associated to the $(2,1)$-form $\chi^\vartheta$, and the $\smash{H}$-coupling should be
\begin{equation} \label{bulk D7-brane - H}
    H_{\sigma^{3} \sigma^{3}} = - \dfrac{1}{\pi g_s} \,  \dfrac{\Vzerofour}{[-i (\tau -\bar{\tau})][-i (u^{3} - \bar{u}^{3})]} \, \delta_{33}.
\end{equation}
The interplay between the various terms in eqn. (\ref{effective mu}) can take place here only if the closed-string sector factors are also defined by integrations over the 4-cycle. This is true only if the 3-form flux is constant over the whole transverse space. Similar considerations hold for the soft-breaking masses of eqn. (\ref{soft masses}). The $B$-term also follows from eqn. (\ref{B-term}).

\paragraph{Strongly Warped Throats with D7-brane Chiral Superfield at Tip} \label{extended 4-cycle: case ii}
If the internal volume is sufficiently small as to satisfy the condition of eqn. (\ref{localisation condition}) and in particular the D7-brane mass flux parameters satisfy eqn. (\ref{D7 localisation condition}), $m_{\mathrm{D7}}^w \lesssim m_{\mathrm{D7}}$ and the D7-brane chiral superfield field localises at the tip of the throat.

One can impose the localisation of the D7-brane scalar at the level of the dimensional reduction by means of a delta-function that accompanies the superfield $\sigma^{3}$, meaning the substitution $\sigma^{3} \bar{\sigma}^{3} \to l_s^4 \delta^{(4)} (y - y_0) \, \sigma^{3} \bar{\sigma}^{3}$. Adapting the previous results (in particular, the integration over the 4-cycle gives a factor $e^{2A_0}$ originating from the metric terms, which at the tip depend on $g_{m'n'} \sim e^{2 A_0}$), one finds the action
\begin{equation*}
    \begin{split}
        S_{\textrm{D7-scalar}} = & - \dfrac{1}{2 \pi g_s} \int_{X_{1,3}} \! \de^4 x \, \sqrt{- \detr{g_4}} \; e^{2\Omega + 2 A_0 + \sdil} \, g^{\mu \nu} \, \nabla_\mu \sigma^{3} \nabla_\nu \bar{\sigma}^{3} \\
        & \quad - \dfrac{1}{2 \pi g_s} \int_{X_{1,3}} \! \de^4 x \, \sqrt{- \detr{g_4}} \; \dfrac{1}{8 \pi \Vzero} \, \dfrac{g_s^2}{\kappa_4^2} \, e^{4\Omega + 8 A_0 + 2 \sdil} \, (g_2^{0} \, \cdot \, \bar{g}_2^{0}) \, \sigma^{3} \bar{\sigma}^{3}.
    \end{split}
\end{equation*}
The 2-form $g_2^{0}$ is the component of the mass-sourcing flux precisely at the tip of the throat, with $G_3^{0} = g_2^{0} \wedge \de w^3$. It is convenient to absorb the warp factors into the scalar $\dot{\sigma}^3 = e^{ A_0} \sigma^{3}$, for which the kinetic action becomes
\begin{equation*}
    \begin{split}
        S_{\textrm{D7-scalar}} = - \int_{X_{1,3}} \! \de^4 x \, \sqrt{- \detr{g_4}} \; \biggl[ \dfrac{1}{\pi g_s} \, \dfrac{e^{2\Omega}}{[-i (\tau -\bar{\tau})]} \, g^{\mu \nu} \, \nabla_\mu \dot{\sigma}^{3} \nabla_\nu \dot{\bar{\sigma}}^{3} \\
        + \dfrac{g_s}{4 \pi^2 \Vzero} \dfrac{e^{4\Omega + 6 A_0}}{[-i (\tau - \bar{\tau})]^2} \, (g_2^{0} \, \cdot \, \bar{g}_2^{0}) \, \dfrac{1}{\kappa_4^2} \, \dot{\sigma}^{3} \dot{\bar{\sigma}}^{3} & \biggr].
    \end{split}
\end{equation*}
The action can be reproduced by means of the K\"{a}hler matter metric
\begin{equation} \label{localised bulk D7-brane - Z}
Z_{\dot{\sigma}^3 \dot{\bar{\sigma}}^3} = \dfrac{1}{\pi g_s} \, \dfrac{e^{2 \Omega}}{[-i (\tau - \bar{\tau})]}
\end{equation}
and, in the presence of only $\smash{(2,1)}$-flux at the tip, the superpotential bilinear coupling
\begin{equation} \label{localised bulk D7-brane - mu}
\bigl[\tilde{\mu}_{\dot{\sigma}^3 \dot{\sigma}^3}\bigr]_{(2,1)} = \dfrac{\Vzerotwo \bigl[ \Vzerofour \bigr]^{1/2}}{\pi [-i(\tau - \bar{\tau})] \kappa_4} \, g_{12}^{0}.
\end{equation}
Notice that the bilinear coupling is holomorphic since it can be seen to arise from the GVW-superpotential deformation
\begin{equation*}
    \delta W = \dfrac{1}{2 \pi} \, \bigl[ \Vzerofour \bigr]^{1/2} \, \der_{\tau} \der_{u^\vartheta} \int_{Y_6} \Bigl[ G_3 \wedge \Omega \Bigr] \, \delta^{(4)} (r)  \, \delta^{(2)} (\theta) \, \dot{\sigma}^3 \dot{\sigma}^3 \equiv \dfrac{1}{2} \, \tilde{\mu}_{\sigma^3 \sigma^3} \dot{\sigma}^3 \dot{\sigma}^3.
\end{equation*}
This reproduces the mass term when the total K\"{a}hler potential contains the $2A_0$-shift, namely when the theory is formulated as in eqn. (\ref{warped Kaehler potential}). In particular, as expected, the canonically normalised mass reads
\begin{equation*}
    (m_{\mathrm{D7}}^w)^2 \sim \dfrac{g_s^2}{\mathcal{V}^{2/3}} \, \dfrac{1}{\kappa_4^2} \, e^{2 A_0}.
\end{equation*}

The structure in the K\"{a}hler and superpotential couplings for the D7-brane chiral superfields here is identical to the case in which the D7-brane wraps a 4-cycle localised at the tip of the throat, as discussed in subsubsection \ref{single throat D7-brane SUGRA}, after replacing the flux evaluated at the warped end of the 4-cycle with the integral of the flux in the 4-cycle at the tip. Therefore, the case discussed above will not be treated separately in the following.

\paragraph{Strongly Warped Scenarios with D7-brane fields in the Bulk} \label{extended 4-cycle: case iii}

An interesting scenario arises in the presence fluxes at the tip of the throat that would give a warped-down mass for the D7-brane fields, $m_{\mathrm{D7}}^w$, that is still heavier than flux-induced masses in the bulk, $m_{\mathrm{D7}}$. In this case, the D7-brane fields minimise their energy by localising in the bulk, so the D7-brane couplings are those in eqns. (\ref{bulk D7-brane - Z}, \ref{bulk D7-brane - mu}).  However, as discussed above, strongly warped scenarios fulfilling eqn. (\ref{localisation condition}), which allow a supergravity description thanks to eqn. (\ref{supergravity condition}), have a K\"{a}hler potential with the structure in eqn. (\ref{warped Kaehler potential}).  So, similarly to what happens with the gravitino when eqn. (\ref{supergravity condition}) is satisfied, in the 4-dimensional effective field theory the canonically normalised D7-brane scalar mass then scales as
\begin{equation*}
    e^{A_0} \, m_{\mathrm{D7}} \sim \dfrac{\theta' g_s}{\mathcal{V}} \, \dfrac{1}{\kappa_4} e^{A_0}.
\end{equation*}

\paragraph{D7-brane Gauge Sector}
From the DBI-action of a stack of D7-branes one can observe the kinetic action for the 4-dimensional gauge field to be
\begin{equation*}
\begin{split}
    S_{\textrm{kin}}^{\textrm{D7-vector}} = - \dfrac{\tau_{\mathrm{D7}} \al^2}{4} \int_{X_{1,3}} \! \de^4 x \, \sqrt{- \detr{g_4}} \int_{\Sigma_4} \de^4 y \, \sqrt{ \detr{g_{\Sigma_4}}} \; [e^{-4A} + c] \, g^{\mu \rho} g^{\nu \sigma} \, F_{\mu \nu} F_{\rho \sigma}.
\end{split}
\end{equation*}
It is thus possible to recognise the inverse of the Weyl factor and write
\begin{equation*}
S_{\textrm{kin}}^{\textrm{D7-vector}} = - \dfrac{\Vzerofour}{8 \pi g_s} \int_{X_{1,3}} \! \de^4 x \, \sqrt{- \detr{g_4}} \; e^{-2\Omega} \, g^{\mu \rho} g^{\nu \sigma} \, F_{\mu \nu} F_{\rho \sigma},
\end{equation*}
so that from the Yang-Mills coupling condition
\begin{equation*}
\dfrac{4 \pi}{g^2_{\mathrm{YM}}} = \mathrm{Im} \, \tau_{\mathrm{YM}} = \dfrac{1}{g_s} \, e^{-2\Omega} \, \Vzerofour = \dfrac{1}{g_s} \, \biggl[ - \dfrac{i}{2} \, (\rho - \bar{\rho}) + \dfrac{\Vwfour}{\Vzerofour} \biggr] \, \Vzerofour,
\end{equation*}
together with holomorphicity, one concludes that the gauge kinetic function has to be
\begin{equation} \label{bulk D7-brane - f}
f_{\mathrm{D7}} = - \dfrac{i \Vzerofour}{2 \pi g_s} \, \bigl[ \rho + i c_0 \bigr],
\end{equation}
with the constant $\smash{c_0 = \Vwfour / \Vzerofour}$. So, for strong warping, it preserves the usual structure, provided the inclusion of the shift suggested by Ref. \cite{Frey:2008xw}. In the limit where integrals are dominated by the bulk region, the gauge-kinetic function becomes  $\smash{f_{\mathrm{D7}} = - i \Vzerofour \rho / 2 \pi g_s}$.

It would be interesting to study localisation effects such as those that can take place in the chiral sector.  The gaugino soft-breaking mass is provided by $(0,3)$-fluxes, following eqn. (\ref{gaugino mass}). Meanwhile, similar mechanisms seem to be prevented for the gauge field, since the vectors do not have flux-induced masses.

\subsubsection{D7-branes at the Tip of Warped Throats} \label{single throat D7-brane SUGRA}
This subsubsection describes the dimensional reduction and the supergravity formulation of a D7-brane wrapping a 4-cycle $\Sigma_4$ at the tip of a warped throat, assuming that the warp factor varies only transversally with respect to the brane. A toy model is described below, including the geometric configuration and the corresponding dimensional reduction.

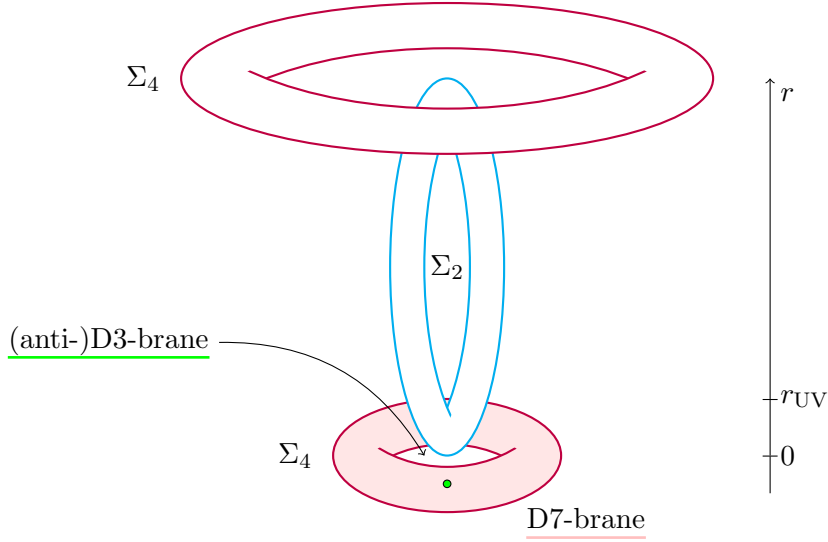
\begin{figure}[H]
\centering
\begin{tikzpicture}[scale=0.5]

\draw[thick, color=purple, fill=white!90!red] (0,0) ellipse (3 and 1.5);
\begin{scope}
    \clip (0,2.2) ellipse (3 and 2.5);
    \draw[thick, color=purple, fill=white] (0,-2.2) ellipse (3 and 2.5);
\end{scope}
\begin{scope}
    \clip(0,-1.8) ellipse (3 and 2.5);
    \draw[thick, color=purple] (0,2.2) ellipse (3 and 2.5);
\end{scope}

\draw[thick, color=cyan,fill=white] (0,5) ellipse (1.5 and 5);
\begin{scope}
    \clip (1.2,5) ellipse (1.8 and 5);
    \begin{scope}
        \clip (-1.2,5) ellipse (1.8 and 5);
        \draw[thick, color=purple, fill=white!90!red] (0,0) ellipse (3 and 1.5);
    \end{scope}
    \draw[thick, color=cyan] (-1.2,5) ellipse (1.8 and 5);
\end{scope}
\begin{scope}
    \clip (-1,5) ellipse (1.8 and 5);
    \draw[thick, color=cyan] (1.2,5) ellipse (1.8 and 5);
\end{scope}

\draw[thick, color=purple, fill=white] (0,10) ellipse (7 and 2);
\begin{scope}
    \clip (0,12.2) ellipse (7 and 3);
    \draw[thick, color=purple, fill=white] (0,7.8) ellipse (7 and 3);
    \draw[thick, color=cyan] (0,5) ellipse (1.5 and 5);
\end{scope}
\begin{scope}
    \clip(0,8.2) ellipse (7 and 3);
    \draw[thick, color=purple] (0,12.2) ellipse (7 and 3);
\end{scope}

\draw[-|] (8.5,-1) -- (8.5,0) node[right] {$0$};
\draw[-|] (8.5,0) -- (8.5,1.5) node [right] {$r_{\mathrm{UV}}$};
\draw[->] (8.5,1.5) -- (8.5,10) node[below right]{$r$};

\draw[fill=green] (0,-0.75) circle (0.1);
\draw [->, bend left, thin] (-6,3) node[left] {\setulcolor{green} \setul{}{1pt} \ul{$\textrm{(anti-)D3-brane}$}} to (-0.6,0);

\node at (0,5) {$\Sigma_2$};
\node at (-4,0) {$\Sigma_4$};
\node at (-8,10) {$\Sigma_4$};
\node at (3.4,-1.8) {\setulcolor{pink} \setul{}{1pt} \ul{$\textrm{D7-brane}$}};

\end{tikzpicture}
\vspace{-28pt}\caption{A sketch of the toy configuration under consideration, with the D7-brane wrapping the 4-space at $\smash{r=0}$. The D3- or anti-D3-brane provides extra open-string states, as discussed in sections \ref{warped D3- and D7-branes} and \ref{warped anti-D3- and D7-branes}, respectively.}
\label{brane setup - localised 4-cycle}
\end{figure}

\paragraph{D7-brane Configuration}
Let the internal 6-dimensional space in the vicinity of the D7-brane wrapped at the tip of the warped throat take the form $\Sigma_4 \rtimes \Sigma_2$. Let the coordinates $y^{m'}$ span a 4-space, for $m'=4,\dots,7$, with $\smash{z^1, z^2}$ their complex version, and let $y^{\dot{m}}$ parametrise the transverse 2-space, for $\dot{m}=8,9$, with $z^3$ the corresponding complex direction. Given some convenient coordinates $\psi^{m'} = \psi^{m'}(y^{n'})$ and $r^{\dot{m}} = r^{\dot{m}}(y^{\dot{n}})$ for the 4- and 2-dimensional spaces, respectively, the internal metric near the throat tip is
\begin{equation*}
    ds^2_6 = e^{-2 A} \, g_{mn} \de y^m \de y^n = e^{-2 A(r)} \, \left(g_{m'n'} (\psi, r) \, \de y^{m'} \de y^{n'} +  g_{3 \bar{3}} (r) \, \de z^3 \de \bar{z}^3\right).
\end{equation*}
The D7-brane is assumed to wrap the 4-dimensional slice corresponding to the position $r = 0$ at the tip and this 4-space is assumed to see a warp factor which ends up at the tiny value $e^{2 A}(r=0) = e^{2 A_0}$. The warped throat is glued to some conformal Calabi-Yau orientifold representing the bulk at $\smash{r^2 = r_{\dot{m}} r^{\dot{m}} = r_{\mathrm{UV}}^2}$, for some $\smash{r_{\mathrm{UV}}}$. See Fig. \ref{brane setup - localised 4-cycle}.

\vspace{-36pt}

To make calculations explicit, it will be assumed that the metric at the tip of the throat corresponds to the geometry $(\mathrm{T}^4/\mathbb{Z}_2) \times \mathrm{T}^2$. Moreover, in analogy with the KS-metric at the throat tip in eqn. (\ref{tip metric}), an overall scaling with the constant $e^{2 A_0}$ is assumed, giving
\begin{equation} \label{tip metric torus}
    g_{m'n'} (\psi, r < r_{\mathrm{UV}}) \overset{r \sim 0}{\sim} g^{(\mathrm{T}^4 / \mathbb{Z}_2)}_{m'n'} \, e^{2 A_0}, \qquad \qquad \qquad g_{3 \bar{3}} (r < r_{\mathrm{UV}}) \overset{r \sim 0}{\sim} g^{(\mathrm{T}^2)}_{3 \bar{3}} \, e^{2 A_0}.
\end{equation}

\paragraph{D7-brane Chiral Superfield}
If the D7-brane wraps a 4-cycle which is entirely localised at the tip of the warped throat, then the metric of the 4-cycle needs to be evaluated at that point in the transverse space. Observing the strong warping condition $e^{-4 A_0} \gg c$, the kinetic term for the D7-brane scalar field takes the form
\begin{equation*}
   \begin{split}
       S_\textrm{kin}^{\textrm{D7-scalar}} = - \dfrac{1}{2 \pi g_s l_s^4} \int\displaylimits_{X_{1,3}} \! \de^4 x \, \sqrt{- g_{4}} \int\displaylimits_{\Sigma_4} \! \de^4 y \, \sqrt{ g^{0}_{\Sigma_4}} \; e^{2\Omega - 4A_0 + \sdil} \, g^{0}_{3 \bar{3}} \, g^{\mu \nu} \, \nabla_\mu \sigma^{3} \nabla_\nu \bar{\sigma}^{3}.
   \end{split} 
\end{equation*}
Because in the current setup neither the warp factor nor the internal metric depend on the 4-cycle coordinates, one can easily observe that such an action reads
\begin{equation*}
    S_\textrm{kin}^{\textrm{D7-scalar}} = - \dfrac{\mathcal{V}_4^0}{2 \pi g_s} \int_{X_{1,3}} \! \de^4 x \, \sqrt{- g_4} \; e^{2\Omega - 4 A_0 + \sdil} \, g^{0}_{3 \bar{3}} \, g^{\mu \nu} \, \nabla_\mu \sigma^{3} \nabla_\nu \bar{\sigma}^{3},
\end{equation*}
where the 4-cycle dimensionless unwarped volume at the tip of the throat is defined as
\begin{equation*}
    \mathcal{V}_4^0 = \dfrac{1}{l_s^4} \int_{\Sigma_4} \de^4 y \, \sqrt{ g^{0}_{\Sigma_4}} \sim e^{4 A_0}.
\end{equation*}
In the end, the K\"{a}hler matter metric has to be
\begin{equation*}
Z_{\sigma^{3} \bar{\sigma}^{3}} = \dfrac{1}{\pi g_s} \, \dfrac{e^{2 \Omega-4A_0}}{[-i (\tau - \bar{\tau})]} \, \mathcal{V}_4^0 \, g^{0}_{3 \bar{3}}.
\end{equation*}
Interestingly, the D7-brane scalar K\"{a}hler matter metric shows two distinct features now that the D7-brane lies at the strongly warped throat-tip rather than extending along the throat:
\begin{itemize}
    \item[-] a dependence on the warp factor, which is reasonable because the whole D7-brane is localised at strong warping;
    \item[-] a dependence on the K\"{a}hler modulus, which means the D7-brane fields are sequestered and effectively very similar to a D3-brane localised at the tip of the throat.
\end{itemize}
Also notice that the matter metric has the effective volume and warp factor scaling $Z_{\sigma^3 \bar{\sigma}^3} \sim e^{2 \Omega + 2 A_0}$, in accord with the result of Ref. \cite{Burgess:2006mn}, following the scaling of the metric $g^{0}_{3 \bar{3}}$ and, correspondingly, of the volume of the 4-cycle at the tip of the throat $\mathcal{V}_4^0$.

Again, the total mass term emerges from the interference of the DBI- and CS-actions, but for the purposes of determining the suppression factors one can simply focus on e.g. the DBI-action, which, in real notation, is of the form
\begin{equation*}
    S_{\textrm{mass}}^{\textrm{D7-scalar}} = - \dfrac{1}{2 \pi g_s} \! \int\displaylimits_{X_{1,3}} \de^4 x \sqrt{- g_{4}} \int\displaylimits_{\Sigma_4} \de^4 y \sqrt{g^{0}_{\Sigma_4}} \; \dfrac{g_s^2 \, e^{4\Omega + 4 A_0 + 2 \sdil}}{8 \pi \Vzero \kappa_4^2 l_s^2} \, G^{0}_{m'n'\dot{r}} \, \bar{G}^{0}_{p'q' \dot{s}} \, g^{m'p'}_{0} \! g^{n'q'}_{0} \sigma^{\dot{r}} \sigma^{\dot{s}}.
\end{equation*}
As the theory at the tip of the throat sees a constant warp factor, one can expand the harmonic mass-sourcing $\smash{(2,1)}$-flux easily.  The supersymmetric mass-sourcing $(2,1)$-flux is still proportional to the harmonic form $\chi_\vartheta = \eta \wedge \de \bar{w}^3$, with $\eta$ the holomorphic $(2,0)$-form of the space $\mathrm{T}^4 / \mathbb{Z}_2$, and can be written as \cite{Camara:2004jj}
\begin{equation*}
    \hat{G}^{0}_3 = f(r=0) \chi_\vartheta,
\end{equation*}
where $f=f(r)$ is a function of the transverse direction (again, the hat denotes the component of the flux that sources a mass term). In terms of the 2-form component, which can be identified as $g_2 = f \, \eta$, the expansion thus reads
\begin{equation*}
    g_2^{0} = \dfrac{1}{\omega_{(0)}^{\Sigma_4} } \, \eta \int_{\Sigma_4} g_2^{0} \wedge \bar{\eta},
\end{equation*}
where $\smash{\omega_{(0)}^{\Sigma_4} = \int_{\Sigma_4} \eta \wedge \bar{\eta}}$. The mass term can thus be expressed as
\begin{equation*}
    S_{\textrm{mass}}^{\textrm{D7-scalar}} = - \dfrac{1}{2 \pi g_s l_s^4} \! \int\displaylimits_{X_{1,3}} \!\! \de^4 x \sqrt{- g_{4}} \int\displaylimits_{\Sigma_4} \!\! \de^4 y \sqrt{g^{0}_{\Sigma_4}} \; \dfrac{1}{8 \pi \Vzero} \, \dfrac{g_s^2}{\kappa_4^2} \, e^{4\Omega + 4 A_0 + 2 \sdil} \, g_2^{0} \, \cdot \, \bar{g}_2^{0} \, \sigma^{3} \bar{\sigma}^{3}.
\end{equation*}
It turns out that the 4-cycle integral can be performed straightforwardly and reads
\begin{equation*}
    \int_{\Sigma_4} \de^4 y \sqrt{\detr{g^{0}_{\Sigma_4}}} \; g_2^{0} \, \cdot \bar{g}_2^{0} = \int_{\Sigma_4} g_2 \wedge \bar{g}_2 = \dfrac{1}{\omega_{(0)}^{\Sigma_4} } \int_{\Sigma_4} g_2^{0} \wedge \bar{\eta} \int_{\Sigma_4} \bar{g}_2^{0} \wedge \eta,
\end{equation*}
so the scalar mass term is simply
\begin{equation*}
        S_{\textrm{mass}}^{\textrm{D7-scalar}} = - \dfrac{1}{2 \pi g_s} \! \int_{X_{1,3}} \!\! \de^4 x \sqrt{- g_{4}} \; \dfrac{1}{8 \pi \Vzero} \, \dfrac{g_s^2 \, e^{4\Omega + 4 A_0 + 2 \sdil}}{\omega_{(0)}^{\Sigma_4}  \kappa_4^2 l_s^4} \int_{\Sigma_4} g_2^{0} \wedge \bar{\eta} \int_{\Sigma_4} \bar{g}_2^{0} \wedge \eta \; \sigma^{3} \bar{\sigma}^{3}.
\end{equation*}
With a pure $(2,1)$-flux background at the tip, such a mass can be generated by means of the superpotential bilinear coupling
\begin{equation}
    \begin{split}
        \bigl[\tilde{\mu}_{\sigma^{3} \sigma^{3}}\bigr]_{(2,1)} & = - \dfrac{e^{-A_0}\Vzero}{\pi \kappa_4 l_s^2} \, \biggl[ g_{3 \bar{3}}^{0} \, \dfrac{ \mathcal{V}_4^0}{\Vzerofour} \biggr]^{\!1/2} \, \biggl[ \der_\tau \der_{u^{\vartheta}} \int_{Y_6} \bigl[ G_3 \wedge \Omega \bigr] \, \delta^{(2)} (r) \biggr]_{(2,1)} \\
        & = \dfrac{e^{-A_0} \Vzero}{\pi \, [-i(\tau - \bar{\tau})] \kappa_4 l_s^2} \, \biggl[ g_{3 \bar{3}}^{0} \, \dfrac{\mathcal{V}_4^0}{\Vzerofour} \biggr]^{\! 1/2} \int_{\Sigma_4} \bar{g}_2^{0} \wedge \eta.
    \end{split} \label{mu-term-(2,1) flux}
\end{equation}
Similarly to the case of eqn. (\ref{bulk D7-brane - mu}), the identification is made assuming the validity of the relationship $\smash{e^{\kappa_4^2 \hat{K}_{\mathrm{cs}}} \Vw = \Vzerofour / \omega_{(0)}^{\Sigma_4}}$. This is not necessarily true in every compactification, in which case an additional factor $\smash{[(\Vzerofour \omega_{(0)})/(\omega_{(0)}^{\Sigma_4} \Vzero))]^{1/2}}$ can be inserted in $\tilde{\mu}_{\sigma^3 \sigma^3}$.

\paragraph*{Comment on Generic Flux Backgrounds}
For generic flux-backgrounds, similar challenges arise as in paragraph \ref{extended 4-cycle: case i}.  However, for ISD-fluxes, if the K\"{a}hler modulus is stabilised by non-perturbative effects, the $(0,3)$-flux is localised away from the tip \cite{Dymarsky:2010mf, Baumann:2010sx}. Therefore, the $(0,3)$-flux does not actually contribute to the integral in $\tilde{\mu}_{\sigma_3\sigma_3}$, and the K\"{a}hler potential coupling $H_{\sigma^3 \sigma^3}$ can also be set to zero.

Notice that, following eqns. (\ref{soft masses}, \ref{B-term}), even if $(0,3)$-flux is present in the bulk, and therefore there is a non-zero F-term for the volume modulus, cancellations hold such that if $\smash{H_{\sigma^3 \sigma^3} = 0}$ then it follows that $\smash{B_{\sigma^3 \sigma^3} = 0}$ and $\smash{m^2_{\sigma^3 \sigma^3, \, \mathrm{soft}} = 0}$, consistently with the fact that the tip of the throat only experiences $(2,1)$-fluxes \cite{Camara:2004jj, Lust:2005bd}.\footnote{Notice, however, the discrepancy between eqns. (3.25, \cite{Camara:2004jj}) and (6.24, \cite{Lust:2005bd}).}

\paragraph*{Warp Factors and Field Redefinitions}
The superpotential bilinear coupling $\tilde{\mu}_{\sigma^{3} \sigma^{3}}$ in eqn. (\ref{mu-term-(2,1) flux}) depends on the warp factor through $e^{-A_0}$, $g^{0}_{3 \bar{3}}$ and $\mathcal{V}_4^0$. It is convenient to make the warp factor dependences explicit. Two possible approaches are now discussed.

One can focus on a highly warped compactification described by means of a K\"{a}hler potential of the form in eqn. (\ref{warped Kaehler potential}). In order to match the D7-brane chiral multiplet kinetic and mass terms with such a structure, first of all one has to redefine the D7-brane scalar field (and consequently its superpartner too) as
\begin{equation} \label{localised D7-brane field redefinition}
    \check{\sigma}^{3} \check{\bar{\sigma}}^{3} = e^{-4A_0} \, \mathcal{V}_4^0 \, g^{0}_{3 \bar{3}} \, \sigma^{3} \bar{\sigma}^{3}.
\end{equation}
In this way, the kinetic and mass terms read
\begin{equation*}
\begin{split}
    S^{\textrm{D7-scalar}} = & - \dfrac{1}{2 \pi g_s} \int_{X_{1,3}} \! \de^4 x \, \sqrt{- g_{4}} \; e^{2\Omega + \sdil} \, g^{\mu \nu} \, \nabla_\mu \check{\sigma}^{\dot{a}} \nabla_\nu \check{\bar{\sigma}}^{\dot{b}} \\
    & - \dfrac{1}{2 \pi g_s} \int_{X_{1,3}} \! \de^4 x \sqrt{- g_{4}} \; \dfrac{g_s^2}{8 \pi \Vzero} \, \dfrac{e^{4\Omega + 2 A_0 + 2 \sdil}}{\omega_{(0)}^{\Sigma_4} \kappa_4^2 l_s^4} \, \dfrac{e^{6 A_0}}{\mathcal{V}_4^0 \, g^{0}_{3 \bar{3}}} \int_{\Sigma_4} g_2^{0} \wedge \bar{\eta} \int_{\Sigma_4} \bar{g}_2^{0} \wedge \eta \; \check{\sigma}^{3} \check{\bar{\sigma}}^3.
\end{split}
\end{equation*}
By relabelling the fields as $\check{\sigma}^{3} \to \sigma^{3}$ for simplicity, one obtains the final action via the K\"{a}hler matter metric
\begin{equation} \label{throat D7-brane - Z}
    Z_{\sigma^3 \bar{\sigma}^3} = \dfrac{1}{\pi g_s} \, \dfrac{e^{2 \Omega}}{[-i (\tau - \bar{\tau})]}
\end{equation}
and the superpotential bilinear coupling
\begin{equation} \label{throat D7-brane - mu}
    \bigl[ \tilde{\mu}_{\sigma^3 \sigma^3} \bigr]_{(2,1)} = \biggl[ \dfrac{e^{6 A_0}}{g^{0}_{3 \bar{3}}} \dfrac{ \Vzerofour}{\mathcal{V}_4^0} \biggr]^{\! 1/2} \, \biggl[ \dfrac{\Vzerotwo}{\pi [-i(\tau - \bar{\tau})] \kappa_4 l_s^2} \int_{\Sigma_4} \bar{g}_2^{0} \wedge \eta \biggr].
\end{equation}
Thanks to the field redefinition and the K\"{a}hler potential shift, the bilinear potential is actually independent of the warp factor.

A second possibility is to replace the original $e^{A_0}$-dependence in the bilinear coupling $\tilde{\mu}_{\sigma^3 \sigma^3}$ with a trilinear term coupling $z^{1/3}$ to the product $\sigma^3 \sigma^3$ \cite{Bena:2018fqc, Dudas:2019pls}, where $z$ is the complex structure modulus fixing the warp factor at the tip as $\langle z \rangle^{1/3} \sim e^{A_0}$, assuming for concreteness a Klebanov-Strassler throat. This will be discussed further below.

\paragraph{D7-brane Gauge Sector}
From the D7-brane DBI-action one can observe the kinetic action for the 4-dimensional gauge field to be
\begin{equation*}
    S_{\textrm{kin}}^{\textrm{D7-vector}} = - \dfrac{1}{8 \pi g_s} \int_{X_{1,3}} \! \de^4 x \, \sqrt{- g_{4}} \; [e^{-4A_0} + c] \, \mathcal{V}_4^0 \, g^{\mu \rho} g^{\nu \sigma} \, F_{\mu \nu} F_{\rho \sigma}
\end{equation*}
and therefore the Yang-Mills coupling is
\begin{equation*}
\dfrac{4 \pi}{g^2_{\mathrm{YM}}} = \mathrm{Im} \, \tau_{\mathrm{YM}} = [e^{-4A_0} + c] \, \dfrac{\mathcal{V}_4^0}{g_s} = \biggl[ e^{-4A_0} - \dfrac{i}{2} \, (\rho - \bar{\rho}) \biggr] \, \dfrac{\mathcal{V}_4^0}{g_s} \sim e^{-4A_0} \, \dfrac{\mathcal{V}_4^0}{g_s},
\end{equation*}
following the condition $\smash{e^{-4 A_0} \gg c}$. One can thus conclude that the gauge kinetic function has to be
\begin{equation} \label{throat D7-brane - f}
f_{\mathrm{D7}} \sim \dfrac{\mathcal{V}_4^0}{2 \pi g_s} \, e^{-4A_0}.
\end{equation}
Notice that, as the volume of the wrapped 4-cycle depends on the warp factor due to the behaviour of the metric eqn. (\ref{tip metric torus}), the term $\mathcal{V}_4^0 \, e^{-4A_0}$ is actually independent of the warp factor.  The subleading term in $f_{\mathrm{D7}}$ instead depends on the warp factor, and, as for the $\tilde{\mu}$-term above, it can be written as a holomorphic contribution in the complex structure modulus $z^{4/3}$ \cite{Bena:2018fqc, Dudas:2019pls}. Also, although the subleading term in $f_{\mathrm{D7}}$ contributes a soft gaugino mass, due to the $e^{4A_0}$ redshift factor it is always suppressed with respect to the anomaly-mediated mass contributions discussed below.

\subsection{D3-/D7-brane Intersecting States} \label{single D3/D7 SUGRA}
Interactions in the low-energy effective action involving D3-/D7-brane intersecting states will now be worked out. Tools other than dimensional reduction need to be used since a higher dimensional effective theory for such states is unknown.

\subsubsection{D3-brane and D7-brane extending from the Throat Tip into the Bulk}
For intersecting D3-/D7-branes, where the D3-brane is at the tip of a warped throat and the D7-brane wraps a 4-cycle extending from the tip into the bulk with the configuration described in paragraph \ref{extended 4-cycle: case iii}, the couplings for the intersecting states in the K\"{a}hler and superpotential of eqns. (\ref{Kaehler potential - D3/D7}, \ref{superpotential - D3/D7}) are as follows.
\begin{itemize}
    \item Following the studies of scattering amplitudes in Refs. \cite{Ibanez:1998rf, Lust:2004cx, Lust:2004fi}, suggests one to define the K\"{a}hler matter metrics for the intersecting D3-/D7-brane states as
    \begin{equation} \label{bulk D3/D7 - Z}
        Z_{\varphi \bar{\varphi}} = Z_{\tilde{\varphi} \tilde{\bar{\varphi}}} = \dfrac{1}{2 \pi g_s} \, e^{2 \Omega}.
    \end{equation}
    The references \cite{Ibanez:1998rf, Lust:2004cx, Lust:2004fi} find the structure $Z_{\varphi\bar{\varphi}} = 1/[-i (\rho - \bar{\rho})]$ in an unwarped compactification, and eqn. (\ref{bulk D3/D7 - Z}) is its natural generalisation.  Further, symmetry arguments reveal that the fields $\varphi$ and $\tilde{\varphi}$ do not have flux-induced masses \cite{Camara:2004jj}.  The resulting no-scale structure implies they can be included within the logarithmic K\"{a}hler potential (together with the other chiral superfields) by defining the $\rho$-term as
    \begin{equation*}
        \kappa_4^2 K = - 3 \, \mathrm{ln} \, \biggl[ 2 \, e^{-2 \Omega} - \dfrac{\kappa_4^2}{3 \pi g_s} \, \varphi \bar{\varphi} \biggr].
    \end{equation*}
    \item As they need to be massless, the intersecting states do not have any bilinear $H$- or $\tilde{\mu}$-coupling. However, one needs to account for a would-be mass term in the case in which the D3- and D7-brane are separated, as explained by Ref. \cite{Camara:2004jj}. As will also be argued in subsubsection \ref{dimensional reduction of intersecting states}, the superpotential term which accounts for this is generated by the Yukawa couplings
    \begin{equation} \label{bulk D3-D7 - Yukawa}
        \tilde{Y}_{\Ds \varphi \tilde{\varphi}} = - \tilde{Y}_{\Dt \varphi \tilde{\varphi}} = \dfrac{1}{g_s} \, \biggl[ \dfrac{2}{\pi} [\Vzero]^3 \biggr]^{1/2} = \tilde{y}.
    \end{equation}
    It will be shown below that such terms are fundamental in order to generate the leading order flux-mediated couplings between the D7-brane and the intersecting states. Notice that the canonically normalised physical Yukawa couplings involving $\sigma^3$ are suppressed by the warp factor, while those involving $\varphi^3$ are not, consistently with their different localisations with respect to $\varphi$ and $\tilde{\varphi}$.
\end{itemize}
The corresponding action has the D-term potential, the F-term potential, and some soft supersymmetry-breaking couplings.
\begin{itemize}
    \item The D-term potential emerges because the intersecting states are charged under the D3- and the D7-brane gauge fields, with couplings
    \begin{equation*}
    g^{-2}_{\mathrm{D3}} = - \dfrac{i}{4 \pi g_s} \, (\tau - \bar{\tau}), \qquad \qquad g^{-2}_{\mathrm{D7}} = - \dfrac{i \, \Vzerofour}{4 \pi g_s} \, (\rho - \bar{\rho} + 2 i c_0) = \dfrac{\Vzerofour}{2 \pi g_s} \, e^{-2 \Omega}.
    \end{equation*}
    It is now easy to infer that the D-term potential for the field $\varphi$ is
    \begin{equation}
    \begin{split}
        V_D^{(\mathrm{susy})} & = \dfrac{1}{2} \, g^2_{\mathrm{D}3} \, (Z_{\varphi \bar{\varphi}} \varphi \bar{\varphi})^2 + \dfrac{1}{2} \, g^2_{\mathrm{D}7} \, (Z_{\varphi \bar{\varphi}} \varphi \bar{\varphi})^2 \\
        & = \dfrac{e^{4 \Omega}}{2 \pi g_s \, [-i (\tau - \bar{\tau})]} \, ( \varphi \bar{\varphi} )^2 + \dfrac{e^{6 \Omega}}{4 \pi g_s \, \Vzerofour} \, ( \varphi \bar{\varphi} )^2,
    \end{split}
    \end{equation}
    and similarly for the field $\tilde{\varphi}$. It is interesting to observe that the specific value of the redshift factor at the tip of the throat does not appear.
    \item On the other hand, in an ISD-background the F-term potential comes from the effective superpotential
    \begin{equation*}
    W_{\mathrm{susy}} = \dfrac{1}{2} \mu_{\Ds \Ds} \Ds \Ds + y \, (\Ds - \Dt) \, \varphi \tilde{\varphi},
    \end{equation*}
    where for the sake of simplicity the trilinear term
    \begin{equation*}
        y = e^{\kappa_4^2 \mathcal{K}/2} \tilde{y}
    \end{equation*}
    has been defined, and reads $V_F^{\mathrm{susy}} = Z^{i \bar{j}} \bigl[\der_i W_{\mathrm{susy}}\bigr] \bigl[\der_{\bar{j}} \bar{W}_{\mathrm{susy}}\bigr]$. This potential gives the redshifted D7-brane supersymmetric mass, but also the couplings between the pure and the intersecting brane states. First of all, one has the cubic interaction
    \begin{equation} \label{bulk D3-D7 - trilinear flux coupling}
    \begin{split}
        V_{\textrm{cubic}}^{(\Ds \varphi \tilde{\varphi})} & = \Bigl[ Z^{\sigma^3 \bar{\sigma}^3} \mu_{\sigma^3 \sigma^3} \bar{y} \sigma^3 \bar{\varphi} \tilde{\bar{\varphi}} + \mathrm{c.c.} \Bigr] \\
        & = - \dfrac{1}{4 \pi \, \kappa_4} \, \dfrac{e^{6 \Omega + 2 A_0}}{[-i (\tau - \bar{\tau})] \, \omega_w^{\Sigma_4}} \biggl[\! \dfrac{2}{\pi \Vzero} \! \biggr]^{\! 1/2} \! \dfrac{\Vwfour}{\Vzerofour} \Biggl[ \! \biggl[ \dfrac{1}{l_s^2} \int_{\Sigma_4} \!\! \bar{g}_2 \wedge \eta \biggr] \Ds \bar{\varphi} \tilde{\bar{\varphi}} + \mathrm{c.c.} \Biggr].
    \end{split}
    \end{equation}
    Additionally, one can observe two distinct quartic interactions which involve only the intersecting states. First of all, there is the standard quartic potential
    \begin{equation} \label{bulk D3-D7 - quartic coupling}
    \begin{split}
        V_{\textrm{quartic}}^{(\varphi \bar{\varphi})} & = Z^{\Ds \bDs} y \bar{y} \, \varphi \tilde{\varphi} \bar{\varphi} \tilde{\bar{\varphi}} + Z^{\Dt \bDt} y \bar{y} \, \varphi \tilde{\varphi} \bar{\varphi} \tilde{\bar{\varphi}} \\
        & = \dfrac{1}{2 \pi g_s} \, \dfrac{e^{6\Omega + 2 A_0}}{\omega_w^{\Sigma_4}} \, \dfrac{\Vwfour}{\Vzerofour} \, \varphi \tilde{\varphi} \bar{\varphi} \tilde{\bar{\varphi}} + \dfrac{\Vzerofour}{\pi g_s} \, \dfrac{e^{4\Omega + 2 A_0}}{[-i (\tau - \bar{\tau})] \, \omega_w^{\Sigma_4}} \, \dfrac{\Vwfour}{\Vzerofour} \, g^{3 \bar{3}}_{0} \, \varphi \tilde{\varphi} \bar{\varphi} \tilde{\bar{\varphi}},
    \end{split}
    \end{equation}
    in which the warp factor redshifts the D7-brane term, but not the D3-brane one due to the cancellation induced by the inverse metric $g^{3 \bar{3}}_0 \sim e^{-2 A_0}$. This does not happen for the D7-brane because its matter metric is determined by the bulk metric $g_{3 \bar{3}}$. Second, there are the quartic interactions that represent the would-be mass terms, namely
    \begin{equation}
    \begin{split}
        V_{\textrm{quartic}}^{(\Ds \varphi \bar{\varphi})} & = y \bar{y} \, Z^{\tilde{\varphi} \tilde{\bar{\varphi}}} \, (\Ds - \Dt ) ( \bDs - \bDt ) \, \varphi \bar{\varphi} \\
        & = \dfrac{1}{\pi g_s} \, \dfrac{e^{4\Omega + 2 A_0} \, \Vwfour}{[-i (\tau - \bar{\tau})] \, \omega_w^{\Sigma_4}} \, (\Ds - \Dt) (\bDs - \bDt) \, \varphi \bar{\varphi}
    \end{split}
    \end{equation}
    and the equivalent term for the field $\tilde{\varphi}$, which are redshifted by the warp factor as must be due to the location of the intersection at the tip of the throat.
    \item In order to determine the supersymmetry-breaking terms for the states $\varphi$ and $\tilde{\varphi}$, instead, it is necessary to determine the Riemann tensor associated to the K\"{a}hler matter metrics. In order to show the general structure of the couplings, in this discussion the possibility of having both $(2,1)$- and $(0,3)$-fluxes is considered.\footnote{Notice that a $(0,3)$-flux does not necessarily affect the supersymmetric couplings: the D3-brane does not have supersymmetric couplings depending on ISD-fluxes, while the D7-brane effective $\mu$-term is correct so long as the conditions around eqn. (\ref{bulk D7-brane - H}) are fulfilled.} One finds the Levi-Civita connection $\Gamma_{\rho \varphi}^{\varphi} = i \, e^{2 \Omega} / 2$, which implies that the only non-vanishing component of the Riemann tensor is
    \begin{equation*}
        R_{\rho \bar{\rho} \varphi \bar{\varphi}} = \dfrac{1}{2 \pi g_s} \dfrac{1}{4} \, e^{6 \Omega}.
    \end{equation*}
    So, as a manifestation of sequestering, in an ISD-background the identity still holds
    \begin{equation*}
    m^2_{\varphi \bar{\varphi}, \, \mathrm{soft}} = \hat{m}^{w}_{3/2} \hat{\bar{m}}^{w}_{3/2} \, Z_{\varphi \bar{\varphi}} - \hat{\mathcal{F}}^\rho \hat{\bar{\mathcal{F}}}^{\bar{\rho}} \, R_{\rho \bar{\rho} \varphi \bar{\varphi}} = 0,
    \end{equation*}
    and the fields $\varphi$ and $\tilde{\varphi}$ stay massless even when supersymmetry is broken by $\rho$. Due to the lack of an $H$- or a $\tilde{\mu}$-term for these fields there is no $B$-coupling either.
    
    Finally, one has to consider the supersymmetry-breaking scalar trilinear couplings, which must be studied with some care. For the couplings to the D7-brane scalar $\Ds$, one finds
    \begin{equation*}
        \nabla_\rho Y_{\Ds \varphi \tilde{\varphi}} = \der_\rho Y_{\Ds \varphi \tilde{\varphi}} + \dfrac{1}{2} \, \kappa_4^2 \hat{K}_\rho Y_{\Ds \varphi \tilde{\varphi}} - 3 \, \Gamma^{l}_{\rho \Ds}  Y_{l \varphi \tilde{\varphi}} = \dfrac{3i}{2} \, e^{2 \Omega} Y_{\Ds \varphi \tilde{\varphi}}
    \end{equation*}
    as a consequence of the fact that, because of the special form of the D7-brane matter metric, its associated Levi-Civita connection vanishes, i.e. $\Gamma^{\Ds}_{\rho \Ds} = 0$. One also finds
    \begin{equation*}
    \begin{split}
        \nabla_\rho Y_{\varphi \tilde{\varphi} \Ds} & = \der_\rho Y_{\varphi \tilde{\varphi} \Ds} + \dfrac{1}{2} \, \kappa_4^2 \hat{K}_\rho \, Y_{\varphi \tilde{\varphi} \Ds} - 3 \, \Gamma^{l}_{\rho \varphi} \, Y_{l \tilde{\varphi} \Ds} = 0, \\
        \nabla_\rho Y_{\tilde{\varphi} \Ds \varphi} & = \der_\rho Y_{\tilde{\varphi} \Ds \varphi} + \dfrac{1}{2} \, \kappa_4^2 \hat{K}_\rho \, Y_{\tilde{\varphi} \Ds \varphi} - 3 \, \Gamma^{l}_{\rho \tilde{\varphi}} \, Y_{l \Ds \varphi} = 0,
    \end{split}
    \end{equation*}
    because in this case the connection is exactly such as to cancel the first two terms. For the couplings with the D3-brane scalar $\Dt$, one finds that all the covariant derivatives vanish too as a consequence of the form of the K\"{a}hler matter metric. Therefore, the only supersymmetry-breaking trilinear coupling is $A_{\Ds \varphi \tilde{\varphi}}$, see eqn. (\ref{A-term}). If one writes the $(0,3)$-flux as $G'_3 = g'_2 (w^3, \bar{w}^3) \wedge \de \bar{w}^3$, with a suitable $(0,2)$-form $g'_2 = g'_2 (w^3, \bar{w}^3)$ on the 4-cycle, then this becomes\footnote{In this calculation the coupling involving the intersecting states is present only if there is a $(0,3)$-flux at the tip of the throat.  This is not necessarily what happens in a fully stabilised model, where the non-perturbative corrections that stabilise the volume modulus localise the $(0,3)$-flux in the bulk. Consistently with this, the F-term of the field $\rho$ is small, so effectively one finds a small $A$-term.}
    \begin{equation} \label{A-term - 37-/73-sector}
    A_{\Ds \varphi \tilde{\varphi}} = \dfrac{3}{4 \pi} \, \dfrac{e^{6 \Omega + 2 A_0}}{[-i (\tau - \bar{\tau})] \, \omega_w^{\Sigma_4} \, \kappa_4} \, \dfrac{\Vwfour}{\Vzerofour} \, \biggl[ \dfrac{2}{\pi \Vzero} \biggr]^{\! 1/2} \dfrac{1}{l_s^2} \int_{\Sigma_4} \bar{g}'_2 \wedge \bar{\eta}.
    \end{equation}
    
    Evidently, in the presence of supersymmetry-breaking imaginary anti-self-dual fluxes, one would obtain mass corrections for the scalars $\varphi$ and $\tilde{\varphi}$ sourced by both the axio-dilaton and the complex structure modulus. Also, one would obtain new trilinear terms coupling these fields to the D3-brane scalar $\Dt$ too.
\end{itemize}

Notice that in an ISD-background the intersecting D3-/D7-brane states couple to the background fluxes only via the mediation of the D7-brane fields as the interactions with the D3-brane fields are protected by the no-scale structure of the latter.

\subsubsection{D3-brane and D7-brane at the Tip of the Throat}
For a system of intersecting D3-/D7-branes where the D7-brane wraps a 4-cycle that is localised at the tip of a warped throat, as in subsubsection \ref{single throat D7-brane SUGRA}, or where the D7-brane wraps a 4-cycle extending through the throat with fields localised at the tip, as in paragraph \ref{extended 4-cycle: case ii}, the intersecting state parameters of the K\"{a}hler and superpotentials of eqns. (\ref{Kaehler potential - D3/D7}, \ref{superpotential - D3/D7}) are as follows:
\begin{itemize}
    \item the K\"{a}hler matter metric is
    \begin{equation} \label{throat D3/D7 - Z}
        Z_{\varphi \bar{\varphi}} = Z_{\tilde{\varphi} \tilde{\bar{\varphi}}} = \dfrac{1}{2 \pi g_s} \, e^{2 \Omega};
    \end{equation}
    \item setting $\beta = e^{-A_0}$, as discussed in subsubsection \ref{dimensional reduction of intersecting states}, the Yukawa couplings are
    \begin{subequations} \label{throat D3-D7 - Yukawa}
        \begin{align}
            \tilde{Y}_{\Ds \varphi \tilde{\varphi}} & = \dfrac{\Vzero}{g_s} \, \biggl[ \dfrac{2}{\pi} \, \Vzerotwo \biggr]^{1/2 \!} \beta = \tilde{y} \beta, \\
            \tilde{Y}_{\Dt \varphi \tilde{\varphi}} & = - \tilde{y}.
        \end{align}
    \end{subequations}
    In this case the canonically normalised physical Yukawa couplings are not redshifted.
\end{itemize}
These account for the sequestered nature of the fields as well as for the presence of the would-be mass term due to any brane separation.

For the intersecting state contributions to the D-term potential, F-term potential and soft supersymmetry-breaking terms, the fact that the D7-brane is localised and therefore has a no-scale-like matter metric (cfr. eqns. (\ref{bulk D7-brane - Z}, \ref{throat D7-brane - Z})) gives rise to particular features.
\begin{itemize}
    \item The D3- and the D7-brane gauge couplings are (neglecting the $\rho$-dependent term for the D7-brane)
    \begin{equation*}
    g^{-2}_{\mathrm{D3}} = - \dfrac{i}{4 \pi g_s} \, (\tau - \bar{\tau}), \qquad \qquad \qquad \qquad g^{-2}_{\mathrm{D7}} = \dfrac{\mathcal{V}_4^0}{2 \pi g_s} \, e^{-4 A_0},
    \end{equation*}
    so the D-term potential for the field $\varphi$ reads
    \begin{equation}
        V_D^{(\mathrm{susy})} = \dfrac{1}{2 \pi g_s \, [-i (\tau - \bar{\tau})]} \, e^{4\Omega} \, (\varphi \bar{\varphi})^2 + \dfrac{1}{4 \pi g_s \, \mathcal{V}_4^0} \, e^{4 \Omega + 4 A_0} \, (\varphi \bar{\varphi} )^2.
    \end{equation}
    The volume dependence is now different for the D7-brane-induced potential. However, the warp factor at the tip of the throat is still effectively missing.
    \item As usual, the F-term potential comes from the effective superpotential and, in addition to the D7-brane supersymmetric mass, there are couplings between the pure and the intersecting brane states. One finds the cubic interaction
    \begin{equation} \label{throat D3-D7 - trilinear flux coupling}
        V_{\textrm{cubic}}^{(\Ds \varphi \tilde{\varphi})} = \dfrac{1}{4 \pi \kappa_4} \dfrac{e^{4 \Omega + A_0}}{[-i (\tau - \bar{\tau})] \, \omega_{(0)}^{\Sigma_4}} \biggl[ \dfrac{2}{\pi \Vzero} \, \dfrac{e^{6 A_0} }{g^{0}_{3 \bar{3}} \mathcal{V}_4^0} \biggr]^{\! 1/2} \Biggl[ \! \biggl[ \dfrac{1}{l_s^2} \int_{\Sigma_4} \bar{g}_2^{0} \wedge \eta \biggr] \, \Ds \bar{\varphi} \tilde{\bar{\varphi}} + \mathrm{c.c.} \Biggr].
    \end{equation}
    Compared to the potential of eqn. (\ref{bulk D3-D7 - trilinear flux coupling}), this potential is less warped down due to the term $\beta = e^{-A_0}$. The pure intersecting states' quartic interactions are
    \begin{equation} \label{throat D3-D7 - quartic coupling}
        V_{\textrm{quartic}}^{(\varphi \bar{\varphi})} = \dfrac{1}{2 \pi g_s \omega_{(0)}^{\Sigma_4}} \, e^{4\Omega} \, \varphi \tilde{\varphi} \bar{\varphi} \tilde{\bar{\varphi}} + \dfrac{1}{\pi g_s} \dfrac{e^{4\Omega + 2 A_0}}{[-i (\tau - \bar{\tau})] \, \omega_{(0)}^{\Sigma_4}} \, g_{0}^{3 \bar{3}} \, \varphi \tilde{\varphi} \bar{\varphi} \tilde{\bar{\varphi}},
    \end{equation}
    where for the D3-brane induced term, the redshift effect is again cancelled by the metric, while for the D7-brane the cancellation arises due to the specific setup with the wrapped 4-cycle at the tip of the throat and the field redefinition of eqn. (\ref{localised D7-brane field redefinition}) (see subsubsection \ref{dimensional reduction of intersecting states}).  There is also the quartic would-be separation mass interaction
    \begin{equation}
        V_{\textrm{quartic}}^{(\Ds \varphi \bar{\varphi})} = \dfrac{1}{\pi g_s} \dfrac{e^{4\Omega + 2 A_0}}{[-i (\tau - \bar{\tau})] \, \omega_{(0)}^{\Sigma_4}} \, (\Ds e^{-A_0} - \Dt) (\bDs e^{-A_0} - \bDt) \, \varphi \bar{\varphi}.
    \end{equation}
    \item For the supersymmetry-breaking terms, it is obvious that in a pure $(2,1)$-flux there cannot be any. In particular, one finds no flux-dependent $A$-coupling for the intersecting D3-/D7-brane states. In fact, even if there were a $(0,3)$-flux, the trilinear scalar coupling $A_{\sigma^3 \varphi \tilde{\varphi}}$ would vanish due to the no-scale structure of the modulus $\smash{\rho}$.
\end{itemize}

\subsubsection{A 6-dimensional Description of the Intersecting States} \label{dimensional reduction of intersecting states}
One can further motivate the form of the K\"{a}hler and superpotential for the D3-/D7-brane intersecting states by a qualitative analysis of their would-be effective field theory.

One can consider the setup in which the branes are separated due to a non-zero difference $\delta Z^3 = \langle \pi^3 \rangle - \langle \phi^3 \rangle$, where $\pi^3$ and $\phi^3$ are the string frame D7- and D3-brane positions in the D7-brane transverse direction, respectively (recall $\sigma^3=\vevc^{3/4} \pi^3$, $\varphi^3 = \vevc^{3/4} \phi^3$, as in appendix \ref{D-brane actions}). A displacement of the D3-brane in the D7-brane longitudinal directions does not induce mass terms, so the intersecting states can be assumed to be 6-dimensional fields living in the non-compact 4-dimensional spacetime as well as in the 2-dimensional compact space which separates the D3- and D7-branes along the transverse complex direction of the latter.  In the string frame, the supersymmetric mass term for the 6-dimensional intersecting states $\theta$ and $\tilde{\theta}$ is
\begin{equation*}
    M^2_{\theta \bar{\theta}} = M^2_{\tilde{\theta} \bar{\tilde{\theta}}} = G_{3 \bar{3}} \delta Z^3 \delta \bar{Z}^3.
\end{equation*}
with $G_{MN}$ the string frame metric, and $\theta, \tilde{\theta}$ will soon be related to the 4-dimensional fields $\varphi, \tilde{\varphi}$.  The kinetic action must be of the form
\begin{equation*}
S_{\textrm{D3/D7}} = - \dfrac{1}{2 \pi l_s^2} \int_{X_{1,3} \times \mathrm{T}^2} \de^6 x \, \sqrt{- G_{6}} \; e^{-n \Phi} \, \biggl[ G^{\mu \nu} \der_\mu \theta \der_\nu \bar{\theta} + \bigl( G_{3 \bar{3}} \delta Z^3 \delta \bar{Z}^3 \bigr) \, \theta \bar{\theta} \biggr],
\end{equation*}
with $n$ a constant representing the fact that usually actions in the string frame are normalised with overall dilaton factors. Then, in the 4-dimensional Einstein frame one obtains
\begin{equation} \label{D3/D7 action}
S_{\textrm{D3/D7}} = - \dfrac{\Vzerotwo}{2 \pi g_s^n} \int_{X_{1,3}} \! \de^4 x \, \sqrt{- g_{4}} \; e^{2 \Omega + (1-n) \sdil} \, \biggl[ g^{\mu \nu} \der_\mu \varphi \der_\nu \bar{\varphi} + e^{2 \Omega + \sdil} \, \bigl( g_{3 \bar{3}} \delta \zeta^{3} \delta \bar{\zeta}^{3} \bigr) \, \varphi \bar{\varphi} \biggr],
\end{equation}
where the brane position moduli have been rescaled as explained in appendix \ref{D-brane actions}, leading to $\delta \zeta^{3} = \vevc^{3/4} \delta Z^3 = \beta \langle \sigma^3 \rangle - \langle \varphi^3 \rangle$, and the same scaling has been performed on the intersecting states, i.e. $\varphi = \vevc^{3/4} \theta$. The factor $\beta$ is
\begin{equation*}
    \beta = \left\lbrace\! \begin{array}{lcl}
        1, & \qquad & \textrm{D7-brane extended from tip to bulk}, \\[0.75ex]
        e^{-A_0}, & & \textrm{D7-brane localised at tip},
    \end{array} \right.
\end{equation*}
where the warp factor emerges only for the case in which the D7-brane multiplet is localised at the tip, following the extra field redefinition (\ref{localised D7-brane field redefinition}). Noticeably, such a construction is compatible with a supersymmetric description, i.e. by means of a $\tilde{\mu}$-tilde coupling, only if the dilaton power takes the value $n=1$ as a different choice cannot reproduce in supergravity the action of eqn. (\ref{D3/D7 action}).

So far, this action applies to any intersecting D3-/D7-brane setup, but it is convenient to specialise to the case in which the D3-brane is located at the tip of a warped throat. As the intersection takes place at the tip of the throat, the action has the form
    \begin{equation*}
    S_{\textrm{D3/D7}} = - \dfrac{1}{2 \pi g_s} \int_{X_{1,3}} \! \de^4 x \, \sqrt{- \detr{g_{4}}} \; \biggl[ e^{2\Omega} \, g^{\mu \nu} \der_\mu \varphi \der_\nu \bar{\varphi} + e^{4 \Omega + 2 A_0 + \sdil} \, \bigl(\delta \zeta^{3} \delta \bar{\zeta}^{3} \bigr) \, \varphi \bar{\varphi} \biggr],
    \end{equation*}
where advantage has been taken of the fact that the internal metric scales as $g^0_{3 \bar{3}} \sim e^{2 A_0}$ and the 2-torus volume factor has been absorbed by the fields. Below, the two distinct scenarios which are of interest are discussed separately, assuming the formulation with the K\"{a}hler potential $2 A_0$-shift.
\begin{itemize}
    \item If the warp factor is a function of only the longitudinal coordinate along the 4-cycle which is wrapped by the D7-brane, the action above can be reproduced in a supersymmetric way by means of the K\"{a}hler and superpotential terms
    \begin{subequations}
    \begin{align}
        Z_{\varphi \bar{\varphi}} & = \dfrac{1}{2 \pi g_s} \, e^{2 \Omega}, \\
        \tilde{\mu}_{\varphi \varphi} & = \dfrac{1}{g_s} \, \biggl[ \dfrac{2}{\pi} \bigl[ \Vzero \bigr]^3 \biggr]^{1/2 \!} \delta \zeta^{3}. \label{D3/D7 mu-term (bulk)}
    \end{align}
    \end{subequations}
    \item If the warp factor is only a function of the transverse direction to the 4-cycle, the K\"{a}hler potential (\ref{closed-strings - Kaehler potential}) is cancelled and it implies that the K\"{a}hler and superpotential couplings are
    \begin{subequations}
    \begin{align}
        Z_{\varphi \bar{\varphi}} & = \dfrac{1}{2 \pi g_s} \, e^{2 \Omega}, \\
        \tilde{\mu}_{\varphi \varphi} & = \dfrac{\Vzero}{g_s} \, \biggl[ \dfrac{2}{\pi} \, \Vzerotwo \biggr]^{1/2 \!} \delta \zeta^{3}, \label{D3/D7 mu-term (throat)}
    \end{align}
    \end{subequations}
\end{itemize}
For definiteness, the calculations have been referred to the case where the whole internal space is factorised as $\Sigma_4 \times \mathrm{T}^2$. The complex structure moduli dependence in the complete mass terms $m^2_{\varphi \bar{\varphi}} = e^{\kappa_4^2 \hat{K}} Z^{\varphi \bar{\varphi}} \tilde{\mu}_{\varphi \varphi} \tilde{\bar{\mu}}_{\bar{\varphi} \bar{\varphi}}$ has not been captured by the dimensional reduction above.

In a theory in which the D3- and D7-brane scalars are dynamical, the terms $\tilde{\mu}_{\varphi \varphi}$ can be used to fix the trilinear couplings as $\tilde{Y}_{\sigma^3 \varphi \varphi} = \tilde{\mu}_{\varphi \varphi}\ab_{\langle \varphi^3 \rangle=0} / \langle \sigma \rangle^3$, $\tilde{Y}_{\varphi^3 \varphi \varphi} = \tilde{\mu}_{\varphi \varphi}\ab_{\langle \sigma \rangle^3=0} / \langle \varphi \rangle^3$. This is a simplified example since it contains only one intersecting field, while in reality there are both the 37- and the 73-states. However, provided a diagonalisation of the states, the structure of the Yukawa couplings is correct. In this way, from the bilinear couplings in eqns. (\ref{D3/D7 mu-term (bulk)}, \ref{D3/D7 mu-term (throat)}) one obtains the trilinear couplings in eqns. (\ref{bulk D3-D7 - Yukawa}, \ref{throat D3-D7 - Yukawa}), respectively.

As commented on in subsubsection \ref{single throat D7-brane SUGRA}, if a complex structure modulus $z$ associated to the tip of the throat controls the warp factor, then one might choose to not use the redefinition eqn. (\ref{localised D7-brane field redefinition}) of the D7-brane scalars at the tip of the warped throat, instead obtaining couplings to $z^p$, with $p \geq 0$.

\section{Warped Anti-D3- and D7-branes} \label{warped anti-D3- and D7-branes}
First, this section overviews the supergravity description of anti-D3-branes in terms of constrained superfields, following the results that have recently been derived in Ref. \cite{Cribiori:2019hod}, which are for a different metric Ansatz to eqn. (\ref{metric}) and outside the regimes of field localisation at the tip, eqns. (\ref{localisation condition}, \ref{supergravity condition}) (see also Ref. \cite{GarciadelMoral:2017vnz}). Second, this section shows how to extend these results to anti-D3-/D7-brane constructions, including in particular the intersecting states, building on results of the previous section for D3-/D7-brane constructions. Finally, considering how these local models may eventually be embedded in global compactifications, the effects of moduli stabilisation and anomaly mediation on the open-string degrees of freedom will be worked out referring to the KKLT scenario for definiteness.  Along with the dimensional reductions in appendix \ref{D-brane actions}, use is made of appendix \ref{LEEFT SUGRA in IIB compactifications with non-linearly realised SUSY}, which derives the supergravity expansions that are suitable in the presence of non-linear supersymmetry.

\subsection{Pure Anti-D3-brane} \label{single anti-D3-brane}
The particle content of D3- and anti-D3-branes is the same, but the couplings with the bulk and other sources are different due to the opposite RR-charges, with implications on the supersymmetry transformations.  This subsection begins with a brief general discussion on anti-D3-brane supersymmetry breaking and their low-energy effective field theory descriptions, then the field content and action are described in detail.

\subsubsection{Anti-D3-brane Supersymmetry Breaking}
In type IIB  Calabi-Yau orientifolds, anti-D3-branes do not preserve the same supersymmetry as the closed-string sector since the orientifold-invariant supersymmetry charge realises supersymmetry only non-linearly on their worldvolume, whereas the supersymmetry charge that would be linearly realised on the brane is projected out. In particular, the gaugino transformation under the surviving supersymmetry takes the form $\sqrt{2} \delta_\epsilon \lambda \sim \epsilon / l^2$, where the factor $\smash{l \sim g_s^{-1/4} \tilde{\tau}_{\mathrm{D3}}^{-1/4}}$ is set by the effective anti-D3-brane tension $\tilde{\tau}_{\mathrm{D3}}$ (i.e. after taking into account any localisation effects).

Because the scale $1/l$ never vanishes, there is no scale at which supersymmetry becomes linearly realised, and there is not the usual F- or D-term whose vacuum expectation value may become zero to restore linear supersymmetry. Nevertheless, because the worldvolume action remains supersymmetric, whilst there is no vacuum in which the anti-D3-brane goldstino has a non-zero supersymmetry transformation, it is effectively a spontaneously broken symmetry. As a consequence of non-linearity, the anti-D3-brane degrees of freedom cannot be encoded in standard $\mathcal{N}_4=1$ multiplets; instead, all the massless degrees of freedom of the anti-D3-brane must be packaged into constrained superfields. Once the tool of constrained supermultiplets is introduced, there is no technical difference with respect to the low energy effective theory describing ``standard'' F-term spontaneous supersymmetry breaking below the supersymmetry breaking scale.

Constrained superfields in global supersymmetry are thoroughly discussed in Ref. \cite{Komargodski:2009rz} as a tool to describe effective theories with broken supersymmetry when the superpartners that become heavy due to the mass-splitting are integrated out. The simplest example is the nilpotent chiral superfield, whose only physical degree of freedom is its fermion playing the role of Volkov-Akulov goldstino for broken supersymmetry \cite{Volkov:1973ix}. A generic treatment of constrained superfields in both global and local supersymmetry can be found in Ref. \cite{DallAgata:2016syy}. As recently discussed in Ref. \cite{Cribiori:2020bgt}, it should be noted that, although the massless degrees of freedom realise non-linear supersymmetry as if their superpartners had been integrated out, above the supersymmetry-breaking scale the full infinite tower of string states is necessary for a consistent supersymmetric theory, and there is no energy scale above which supersymmetry in the usual sense is restored (see Ref. \cite{Cribiori:2020sct} for related discussions).

\subsubsection{Anti-D3-brane Constrained Multiplets}
To place the anti-D3-brane fields in constrained supermultiplets, one matches the non-linear supersymmetry transformations for the brane fields with those of a specific constrained superfield, as done in great detail by Refs. \cite{Bergshoeff:2013pia, Vercnocke:2016fbt, Kallosh:2016aep}.
\begin{itemize}
    \item The gaugino $\lambda$, which plays the role of the goldstino, is described via the fermion component $\psi^X$ of a chiral superfield $X$ that satisfies the nilpotency condition \cite{Rocek:1978nb, Ivanov:1978mx, Lindstrom:1979kq, Casalbuoni:1988xh}
    \begin{equation} \label{X^2 = 0}
        X^2 = 0.
    \end{equation}
    This effectively removes its scalar $\varphi^X$ in favour of the spinor $\psi^X$, indeed implying the identification $\smash{\varphi^X = \psi^X \psi^X / F^X}$, with the auxiliary field $F^X$ being non-vanishing by assumption. At leading order in $l$, i.e. the scale at which the tower of string states enters into play, the gaugino $\lambda$ and the goldstino $\psi^X$ are then related as
    \begin{equation*}
        \lambda \sim \dfrac{1}{2 l^2} \dfrac{\psi^X}{F^X},
    \end{equation*}
    with the non-linear supersymmetry variation $\sqrt{2} \delta_\epsilon \lambda \sim \epsilon / l^2$. If the anti-D3-brane sits at the tip of a warped throat, then this supersymmetry-breaking scale is the warped string scale $\smash{l \sim 1/m_s^w}$. In fact, comparing with the anti-D3-brane energy density uplift below, one can see that the effective anti-D3-brane tension at the tip of the throat scales as $\smash{\tilde{\tau}_{\mathrm{D3}}^{1/4} \sim g_s^{-1/4} m_s^w}$.
    
    As the goldstino is contained in a chiral multiplet, the would-be gaugino D-term breaking is actually described as an F-term breaking. Eventually the gaugino is fixed as $\lambda = 0$ in the unitary gauge. Refs. \cite{Farakos:2013ih, Dudas:2015eha, Bergshoeff:2015tra, Hasegawa:2015bza, Ferrara:2015gta} discuss the supergravity generalisation of this construction.
    \item The Abelian gauge field $A_\mu$ is contained in the vector degrees of freedom of a field-strength chiral multiplet $W_\alpha$ satisfying the constraint \cite{Komargodski:2009rz, Klein:2002vu}
    \begin{equation} \label{X W_alpha = 0}
        X W_\alpha = 0,
    \end{equation}
    which removes the gaugino $\zeta^W$ by making it proportional to the goldstino $\psi^X$.
    \item The so-called modulini $\psi^a$ are described by the fermionic degrees of freedom of three chiral superfields $Y^a$ satisfying the constraints \cite{Brignole:1997pe, DallAgata:2015pdd}
    \begin{equation} \label{X Y^a = 0}
        X Y^a = 0,
    \end{equation}
    which remove the scalars $\varphi^{Y^a}$ by making them proportional to the goldstino $\psi^X$.
    \item The scalars $\varphi^a$ describing position fluctuations are encoded in the scalar degrees of freedom of three chiral superfields $H^a$ satisfying the constraints \cite{Komargodski:2009rz, DallAgata:2016syy}
    \begin{equation} \label{bar X D H^a = 0}
        \bar{X} \mathcal{D}_{\alpha} H^a = 0,
    \end{equation}
    with $\mathcal{D}_\alpha$ the supersymmetry-covariant derivative, which makes both the spinors $\psi^{H^a}$ and the auxiliary fields $F^{H^a}$ proportional to the goldstino $\psi^X$. As it is constrained, the solution to the F-term field equation is not the usual $F^{H^a} = e^{\kappa_4^2 K / 2} K^{H^a \bar{I}} \nabla_{\bar{I}} \bar{W}$, but rather a goldstino-dependent expression which vanishes in the unitary gauge.
\end{itemize}

\subsubsection{Anti-D3-brane Supergravity} \label{single anti-D3-brane SUGRA}
The supergravity formulation of a single anti-D3-brane at the tip of a warped throat in an orientifold compactification with Hodge number $h_+^{1,1} = 1$ is reported below. One can follow the dimensional reductions of Refs. \cite{Camara:2003ku, Grana:2003ek, McGuirk:2012sb, Bergshoeff:2015jxa, Cribiori:2019hod} and adapt them to the metric of eqn. (\ref{metric}).

\paragraph{Anti-D3-brane Uplift Energy}
Anti-D3-branes provide a positive energy uplift to the vacuum energy. Given the warp factor $A_0$ at the anti-D3-brane location, in the 4-dimensional Einstein frame it reads
\begin{equation*}
S_\Lambda^{\overline{\mathrm{D3}}} = - \dfrac{1}{\kappa_4^4} \, \int \de^4 x \, \sqrt{- \detr{g_{4}}} \; \dfrac{g_s^3}{4 \pi [ \Vzero ]^2} \, \dfrac{e^{4 \Omega}}{e^{-4A_0} + c}.
\end{equation*}
In the setup with the anti-D3-brane at the tip of the throat, the warp factor dominates over the volume modulus, so that the effective form of the term above is
\begin{equation*}
S_\Lambda^{\overline{\mathrm{D3}}} = - \dfrac{1}{\kappa_4^4} \, \int \de^4 x \, \sqrt{- \detr{g_{4}}} \; \dfrac{g_s^3}{4 \pi [\Vzero]^2} \, e^{4\Omega + 4A_0}.
\end{equation*}
Such a vacuum energy can be reproduced in supergravity in a very easy way as the F-term potential contribution of the goldstino $X$ by defining the K\"{a}hler and superpotential
\begin{subequations}
    \begin{align}
        \kappa_4^2 \hat{K} & = {\begin{aligned}[t] & - \mathrm{ln} [- i (\tau - \bar{\tau})] - \mathrm{ln} [-i \omega_w] + \mathrm{ln} \biggl[ \dfrac{2}{\pi} \dfrac{\Vw}{[\Vzero]^{3}} \! \biggr] \\
        & - 3 \, \mathrm{ln} \biggl[ 2 \, e^{-2 \Omega} - \dfrac{4 \kappa_4^2}{3 g_s} \dfrac{\Vw}{\Vzero} \dfrac{e^{-2A_0} \, X \bar{X}}{[-i(\tau - \bar{\tau})] [-i \omega_w]} \biggr], \end{aligned}}  \label{closed-strings and goldstino - Kaehler potential} \\
        \kappa_4^3 \hat{W} & = \dfrac{g_s}{l_s^2} \int_{Y_6} G_3 \wedge \Omega + \sqrt{2} \, g_s \kappa_4 X, \label{closed-strings and goldstino - superpotential}
    \end{align}
\end{subequations}
with the actual total K\"{a}hler potential being $\kappa_4^2 \mathcal{K} = 2 A_0 + \kappa_4^2 \hat{K}$. In the unitary gauge, the only change to the closed-string sector effective theory induced by the nilpotent superfield is the anti-D3-brane uplift contribution to the F-term potential (as long as the goldstino is aligned completely with the spinor in $X$ \cite{Ferrara:2014kva}).\footnote{Explicitly, the correction to the F-term potential is $\smash{\Delta \mathcal{V}_F = e^{2 A_0 + \kappa_4^2 \hat{K}} \, \hat{K}^{X \bar{X}} \nabla_X \hat{W} \nabla_{\bar{X}} \hat{\bar{W}}}$, with the terms
\begin{equation*}
\begin{array}{lcl}
    \hat{K}_{X \bar{X}} = \dfrac{2}{g_s} \, \dfrac{e^{2\Omega - 2A_0}}{[-i(\tau - \bar{\tau})] [-i \omega_w]} \, \dfrac{\Vw}{\Vzero}, & \qquad \qquad \qquad \qquad & \displaystyle \kappa_4^3 \nabla_X \hat{W} = \sqrt{2} g_s \kappa_4.
\end{array}
\end{equation*}}

Notice that Ref. \cite{Cribiori:2019hod} does not work with the $2A_0$-shift in the K\"{a}hler potential, as is appropriate in regimes not fulfilling eqn. (\ref{localisation condition}).\footnote{Also, in Ref. \cite{Cribiori:2019hod} the warp factor depends on the brane scalars, i.e. $A_0 = A_0 (H^a, \bar{H}^a)$, which would imply a kinetic term correction for the scalars whenever there is the $2A_0$-shift in the K\"{a}hler potential. In the formulation presented here, the term $A_0$ is independent of the brane scalars.}

\paragraph*{Complex Structure Moduli in Warped Throats}
In type IIB $\smash{\mathcal{N}_4=1}$ compactifications the axio-dilaton and the complex structure moduli are typically stabilised at high energy scales; however, in a KS-throat, the complex structure modulus $z$, which controls the size of 3-sphere at the throat tip, stays in the low-energy effective theory \cite{Bena:2018fqc}. For a dimensionless field $z$, its vacuum expectation value fixes the warp factor at the tip of the throat as \cite{Giddings:2001yu}
\begin{equation} \label{complex structure modulus / warp factor}
    \langle z \bar{z} \rangle^{1/3} = e^{2 A_0} = e^{-4 \pi K / 3 g_s M},
\end{equation}
where $M$ and $K$ are the quantised $F_3$- and $H_3$-fluxes through the conifold 3-sphere and its dual 3-cycle, respectively.

Ref. \cite{Douglas:2007tu} computes the K\"{a}hler metric for the complex structure modulus $z$. Moreover, Ref. \cite{Dudas:2019pls} shows the way to include such a field within the supergravity formulation together with an uplifting anti-D3-brane. Together with the K\"{a}hler modulus shift used here, one can postulate the K\"{a}hler and superpotential
\begin{equation*}
    \begin{split}
        \kappa_4^2 \hat{K} & = - 3 \, \mathrm{ln} \biggl[ 2 \, e^{-2 \Omega} - \dfrac{4 \kappa_4^2}{3 g_s} \dfrac{\Vw}{\Vzero} \dfrac{X \bar{X}}{[-i(\tau - \bar{\tau})] [-i \omega_w]} \biggr] + Z_{z \bar{z}}(z \bar{z}) z \bar{z}, \\
        \kappa_4^3 \hat{W} & = \dfrac{g_s}{l_s^2} \int_{Y_6} G_3 \wedge \Omega + W(z) + \sqrt{2} \, g_s \kappa_4 z^{1/3} X, 
    \end{split}
\end{equation*}
where the K\"{a}hler metric $Z_{z \bar{z}}$ and the superpotential $W(z)$ determine the vacuum expectation value of the field $z$ to be that in equation (\ref{complex structure modulus / warp factor}); for brevity, the constant term and the axio-dilaton and other complex structure moduli have been dropped. Also, one may include the K\"{a}hler potential shift as the extra K\"{a}hler potential coupling
\begin{equation*}
    \kappa_4^2 \delta \hat{K} = \dfrac{1}{3} \, \mathrm{ln} \, z \bar{z} \sim 2 A_0.
\end{equation*}
Such a term does not participate in the K\"{a}hler metric but only in the overall scaling of the energy scales, as it needs to do, and to some scalar and fermionic couplings.

In the KS-throat, the unwarped metric at the tip of the throat scales as $g^{0}_{mn} \sim e^{2 A_0}$, which is crucial as it sets the K\"{a}hler matter metric of the open-string degrees of freedom sitting at the tip of the throat. Therefore, writing the warp factor at the tip in terms of the complex structure modulus leads, for example, to a coupling from the would-be kinetic term of the form
\begin{equation*}
    \delta \mathcal{K} = \dfrac{1}{2 \pi g_s} \, e^{2\Omega} g^0_{a \bar{b}} H^a \bar{H}^b \sim \dfrac{1}{2 \pi g_s} \, (z \bar{z})^{1/3} e^{2\Omega} \delta_{a \bar{b}} H^a \bar{H}^b.
\end{equation*}
It would be interesting to incorporate all such interactions between $z$ and the open-string fields in a complete supergravity description.

Obviously, if the throat is not of the Klebanov-Strassler type, the details of the potentials are different, but by analogy one should expect qualitatively similar results.

\paragraph{Anti-D3-brane Modulini}
For the modulini of an anti-D3-brane, the pure kinetic term reads
\begin{equation*}
S_{\textrm{kin}}^{\overline{\textrm{D3}}\textrm{-modulini}} = - \dfrac{i}{2 \pi g_s} \int_{X_{1,3}} \! \de^4 x \, \sqrt{- \detr{g_4}} \; e^{2\Omega} \, g^{0}_{a \bar{b}} \, \bar{\psi}^b \bar{\sigma}^\mu \nabla_\mu \psi^a.
\end{equation*}
This can be matched with a supergravity formulation by encoding the spinors $\psi^a$ in the constrained multiplets $Y^a$, with $X Y^a = 0$, and using the K\"{a}hler potential
\begin{equation*}
\kappa_4^2 \hat{K} = - 3 \, \mathrm{ln} \, \biggl[ 2 \, e^{-2 \Omega} - \dfrac{\kappa_4^2}{3 \pi g_s} \, g^{0}_{a \bar{b}} \, Y^a \bar{Y}^b - \dfrac{4 \kappa_4^2}{3 g_s} \, \dfrac{e^{-2A_0}}{[-i(\tau - \bar{\tau})] [-i \omega_w]} \, \dfrac{\Vw}{\Vzero} \, X \bar{X} \biggr],
\end{equation*}
or alternatively, after an easy logarithmic expansion, with the K\"{a}hler matter metric
\begin{equation} \label{anti-D3 modulini - Z}
    Z_{Y^a \bar{Y}^b} = \dfrac{1}{2 \pi g_s} \, e^{2 \Omega} \, g^{0}_{a \bar{b}}.
\end{equation}
For the mass term, from the dimensional reduction one finds
\begin{equation*}
S_{\textrm{mass}}^{\overline{\textrm{D3}}\textrm{-modulini}} = - \dfrac{i}{2 \pi g_s} \int_{X_{1,3}} \! \de^4 x \, \sqrt{- \detr{g_4}} \; \bigl[ m_{\psi^a \psi^b} \psi^a \psi^b + \mathrm{c.c.} \bigr],
\end{equation*}
with the mass\footnote{In Ref. \cite{McGuirk:2012sb} the holomorphic 3-form is defined in terms of the gamma-matrices that are suitable for the geometry at the tip of the throat. Given the internal Dirac matrices $\gamma_m$ and the internal spinor $\eta_+$ of positive chirality and norm $\eta_+^\dagger \eta_+ = 1$ which defines the $\mathrm{SU}(3)$-structure of the space, with $\eta_-$ its conjugate, the holomorphic 3-form and the K\"{a}hler form are defined as
\begin{equation*}
    l_s^3 \Omega_{mnp} = \eta_-^\dagger \gamma_{mnp} \eta_+, \qquad \qquad \qquad \qquad \tilde{\omega}_{mn} = i \, \eta_+^\dagger \gamma_{mn} \eta_+.
\end{equation*}
To make estimates in terms of the warp factor scaling, then one needs to consider the qualitative behaviour $l_s^3 \Omega^{0} \sim e^{3 A_0} \, (n^0_f)^{3/2}$, consistently with the metric behaviour. This observation is important for section \ref{analysis of the mass hierarchies}.}
\begin{equation*}
m_{\psi^a \psi^b} = - \dfrac{1}{\bigl[4 \pi \Vzero \bigr]^{1/2}} \, \dfrac{g_s}{4 \kappa_4} \, e^{3\Omega+4A_0+ \sdil/2} \, l_s^4 g^{0}_{\overline{c}(a} \, \Omega^{0}_{b)de} \, (\bar{G}_3^-)_{0}^{\bar{c} d e}.
\end{equation*}
Following the method of Ref. \cite{Cribiori:2019hod}, this mass term can be generated via a K\"{a}hler potential bilinear coupling
\begin{equation} \label{anti-D3 modulini - H}
    H_{Y^a Y^b} = \dfrac{i}{4 \pi g_s^2} \, \dfrac{\Vw^{1/2}}{[-i(\tau - \bar{\tau})][-i \omega_w]^{1/2}} \, e^{2\Omega + A_0} \, l_s^4 \, g^{0}_{\overline{c}(a} \, \Omega^{0}_{b)de} \, (\bar{G}_3^-)_{0}^{\bar{c} d e} \, \kappa_4 \bar{X}.
\end{equation}
Indeed, as required, in an imaginary self-dual background one obtains the effective $\mu$-term
\begin{equation*}
    \mu_{Y^a Y^b} = - \bar{\mathcal{F}}^X \, \der_{\bar{X}} H_{Y^a Y^b} = \dfrac{i}{2 \pi g_s} \, m_{\psi^a \psi^b}.
\end{equation*}
The scale of the canonically normalised mass is \cite{Burgess:2006mn}
\begin{equation*}
    (m_{\overline{\mathrm{D3}}}^w)^2 = \dfrac{g_s^2}{\mathcal{V}^{2/3}} \, \dfrac{1}{\kappa_4^2} \, e^{2 A_0}.
\end{equation*}

\paragraph{Anti-D3-brane Scalars}
The pure kinetic action for the anti-D3-brane scalars takes the form
\begin{equation*}
S_\textrm{kin}^{\overline{\textrm{D3}}\textrm{-scalars}} = - \dfrac{1}{2 \pi g_s} \int_{X_{1,3}} \! \de^4 x \, \sqrt{- \detr{g_4}} \; e^{2 \Omega} \, g^{0}_{a \bar{b}} \, g^{\mu \nu} \, \nabla_\mu \varphi^a \nabla_\nu \bar{\varphi}^b.
\end{equation*}
In order to correctly account for the expected no-scale structure (see paragraph \ref{D3-brane chiral superfields}), one needs to generalise the full K\"{a}hler potential for the K\"{a}hler modulus as
\begin{equation*}
\begin{split}
    \kappa_4^2 \hat{K} = - 3 \, \mathrm{ln} \biggl[ 2 \, e^{-2 \Omega} & - \dfrac{4 \kappa_4^2}{3 g_s} \, \dfrac{e^{-2A_0}}{[-i(\tau - \bar{\tau})] [-i \omega_w]} \, \dfrac{\Vw}{\Vzero} \, X \bar{X} - \dfrac{\kappa_4^2}{3 \pi g_s} \, g^{0}_{a \bar{b}} \, Y^a \bar{Y}^b - \dfrac{\kappa_4^2}{3 \pi g_s} \, g^{0}_{a \bar{b}} \, H^a \bar{H}^b \biggr],
\end{split}
\end{equation*}
where $H^a$ are the constrained chiral multiplets containing the scalars $\varphi^a$. Indeed, in this way the K\"{a}hler matter metric is
\begin{equation} \label{anti-D3 scalars - Z}
    Z_{H^a \bar{H}^b} = \dfrac{1}{2 \pi g_s} \, e^{2 \Omega} \, g^{0}_{a \bar{b}} + \dfrac{\kappa_4^2}{3 \pi g_s^2} \, \dfrac{e^{4 \Omega - 2 A_0}}{[-i (\tau - \bar{\tau})] [-i \omega_w]} \, \dfrac{\Vw}{\Vzero} \, X \bar{X} \, g^{0}_{a \bar{b}}.
\end{equation}
For the scalar masses, from the combination of the relevant parts of the DBI- and CS-actions one finds the term
\begin{equation*}
\begin{split}
    S_{\textrm{mass}}^{\overline{\textrm{D3}}\textrm{-scalars}} = - \dfrac{1}{2 \pi g_s} \int_{X_{1,3}} \! \de^4 x \, \sqrt{- \detr{g_4}} \; \dfrac{e^{4 \Omega}}{4 \pi \Vzero} \, \dfrac{g_s^2}{\kappa_4^2} \, [l_s^2 \nabla_a \nabla_{\bar{b}} (e^{4A} + \alpha)]_{0} \, \varphi^a \bar{\varphi}^b.
\end{split}
\end{equation*}
If only $(2,1)$-flux is present at the tip of the throat, the anti-D3-brane scalar mass-squared trace can be evaluated at leading order thanks to the GKP-equations, which, at a position in the internal space with pure $(2,1)$-flux background, imply the relation \cite{Giddings:2001yu, Camara:2003ku, Bena:2015qfa}\footnote{In the GKP-setup \cite{Giddings:2001yu}, by rearranging the 4-dimensional components of the Einstein equations and the field equation of the 4-form potential, one can show the condition (for the conventions, see appendix \ref{appendix: warped dimensional reduction})
\begin{equation*}
\begin{split}
    \nabla^m \nabla_m \bigl( e^{4 A} + \alpha \bigr) = \dfrac{e^{2 A}}{24 \, \mathrm{Im} \, \tau} \, \bigl[ i G_{mnp} + (*_6 G)_{mnp} \bigr] \, \bigr[ -i \overline{G}^{\hat{m} \hat{n} \hat{p}} + (*_6 \overline{G})^{\hat{m} \hat{n} \hat{p}} \bigr] & \\
    + e^{-6 A} \, \bigl[ \nabla_m \bigl( e^{4 A} + \alpha \bigr) \bigr] \, \bigl[ \hat{\nabla}^{\hat{m}} \bigl( e^{4 A} + \alpha \bigr) \bigr] & \\
    - 2 \hat{\kappa}_{10}^2 \, e^{2 A} \, \Bigr[ \dfrac{1}{4} \, \bigl( \hat{g}^{\mu \nu} T_{\mu \nu} - \hat{g}^{mn} T_{m n} \bigr)_{\textrm{(source)}} - T_3 \, \rho_3^{\textrm{(source)}} \Bigr] &.
\end{split}
\end{equation*}
In a background with ISD-fluxes $G_3 = -i *_6 G_3$ and the condition $e^{4A} = \alpha$, one can observe that:
\begin{itemize}
    \item the source term vanishes for an anti-D3-brane and is subleading in the string length for a D7-brane;
    \item all the flux contributions are expected to have the same functional dependence as the 3-form term.
\end{itemize}
Therefore, in a pure $(2,1)$-flux, one finds the equation in the main text. Obviously a similar result holds for a generic imaginary self-dual $(2,1)$- and $(0,3)$-flux background.}
\begin{equation*}
g^{a \bar{b}} \, \nabla_{a} \nabla_{\bar{b}} e^{4A} = \dfrac{1}{12} \, e^{8A + \sdil} \, G_{2,1}^- \, \cdot \, \bar{G}_{2,1}^-.
\end{equation*}
In accord with Ref. \cite{Camara:2003ku}, in a pure $(2,1)$-flux background the anti-D3-brane supertrace vanishes, and the scalar masses are provided by a $\mu$-term equivalent to that of the modulini. It is then natural to try to generate the $\mu$-term by using an equivalent $H$-coupling to the modulini, that is $H_{H^a H^b} = H_{Y^a Y^b}$ given in eqn. (\ref{anti-D3 modulini - H}).  Some care is needed, as the constrained superfield $H^a$ does not have an independent F-term, and so its couplings in the supergravity expansions are different to the standard case, as shown in appendix \ref{LEEFT SUGRA in IIB compactifications with non-linearly realised SUSY}.  It turns out that the coupling $H_{H^a H^b} = H_{Y^a Y^b}$, is still able to generate a mass
\begin{equation*}
    m^2_{\varphi^a \bar{\varphi}^b} = 2 \, Z^{H^c \bar{H}^d} \hat{\mathcal{F}}^M \hat{\bar{\mathcal{F}}}^N H_{H^a H^c, \bar{N}} \bar{H}_{\bar{H}^b \bar{H}^d, M}, 
\end{equation*}
but it will be seen around the derivaton of eqn. (\ref{anti-D3 scalars - extra H}) that this choice also originates unwanted bilinear couplings. An alternative way to describe the mass term is to use a coupling $H_{Y^a H^b}$, with $H_{Y^a H^b} = H_{Y^a Y^b}$.\footnote{\label{footnote} For a coupling $H_{Y^a H^b}$, expanding the F-term scalar potential, one finds that the scalar mass term for the multiplet $H^a$ (without independent F-term) is given by eqn. (\ref{susy mass}) (it is generated by the F-term of $Y^a$), while the scalar mass term for the multiplet $Y^a$ is given by eqn. (\ref{no F-term mass}).} From the F-term of the multiplet $Y^a$ one now obtains the scalar mass
\begin{equation*}
    m^2_{\varphi^a \bar{\varphi}^b} = Z^{Y^c \bar{Y}^d} \hat{\mathcal{F}}^M \hat{\bar{\mathcal{F}}}^N H_{H^a Y^c, \bar{N}} \bar{H}_{\bar{H}^b \bar{Y}^d, M}
\end{equation*}
and it also turns out that the unwanted bilinear interactions are avoided. Such an $H$-term also contributes a coupling $\smash{m^2_{Y^a \bar{Y}^b} Y^a \bar{Y}^b}$, but this is actually a fermionic term that vanishes in the unitary gauge. In conclusion, one reproduces the scalar mass by means of the K\"{a}hler potential bilinear coupling
\begin{equation} \label{anti-D3 scalars - H}
    H_{H^a Y^b} = \dfrac{i}{4 \pi g_s^2} \, \dfrac{\Vw^{1/2}}{[-i(\tau - \bar{\tau})][-i \omega_w]^{1/2}} \, e^{2\Omega + A_0} \, l_s^4 \, g^{0}_{\overline{c}(a} \, \Omega^{0}_{b)de} \, (\bar{G}_3^-)_{0}^{\bar{c} d e} \, \kappa_4 \bar{X},
\end{equation}
with the supersymmetric scalar mass being $\smash{m^2_{\varphi^a \bar{\varphi}^b} = Z^{Y^c \bar{Y}^d} \, \mu_{H^a Y^c} \bar{\mu}_{\bar{H}^b \bar{Y}^d}}$.

The analysis of subsection \ref{LEEFT SUGRA in IIB compactifications with non-linearly realised SUSY} shows that in general there is also a would-be soft supersymmetry-breaking coupling mass
\begin{equation*}
    \begin{split}
        m^2_{\varphi^a \bar{\varphi}^b, \, \mathrm{soft}} = \kappa_4^2 \hat{\mathcal{V}}_F Z_{H^a \bar{H}^b} - \hat{\mathcal{F}}^M \hat{\bar{\mathcal{F}}}^N \bigl[ Z_{H^a \bar{H}^b, M \bar{N}} - 2 \, \Gamma^{H^c}_{M H^a} \, Z_{H^c \bar{H}^d} \, \bar{\Gamma}^{\bar{H}^d}_{\bar{N} \bar{H}^b} \bigr] \\
        + \bigl[ \hat{m}^{w}_{3/2} \hat{\mathcal{F}}^M Z_{H^a \smash{\bar{H}}^b, M} + \smash{\hat{\bar{m}}}^{w}_{3/2} \smash{\hat{\bar{\mathcal{F}}}}^N Z_{H^a \smash{\bar{H}}^b, N} \bigr] &.
    \end{split}
\end{equation*}
In a $(2,1)$-flux background, the only contribution is from the $X$-field F-term, which gives
\begin{equation}
    \begin{split}
        m^2_{\varphi^a \bar{\varphi}^b, \, \mathrm{soft}} = \dfrac{2}{3} \, \kappa_4^2 V_{\overline{\mathrm{D3}}} \, Z_{H^a \bar{H}^b}.
    \end{split}
\end{equation}
This term can be seen to emerge in the dimensional reduction as follows. In the presence of the anti-D3-brane scalars, the volume is shifted and the total Weyl-factor should be such that \cite{Giddings:2005ff, Baumann:2007ah}
\begin{equation} \label{Weyl factor shift}
    e^{- 2 \Omega'} = e^{- 2 \Omega} - \dfrac{\kappa_4^2}{6 \pi g_s} \,  g_{a \bar{b}}^0 H^a \bar{H}^b,
\end{equation}
with the actual uplift energy $V_{\overline{\mathrm{D3}}}' = 4 \pi \vevc^3 \, e^{4\Omega'+4A_0}/g_s l_s^4$. If one expands this energy in $H^a$, then what is obtained is exactly the sought-after factor, being
\begin{equation*}
    V_{\overline{\mathrm{D3}}}'(e^{2\Omega'}) =  V_{\overline{\mathrm{D3}}} (e^{2\Omega}) \, \biggl[ 1 + \dfrac{2}{3} \, \kappa_4^2 Z_{H^a \bar{H}^b} H^a \bar{H}^b \biggr].
\end{equation*}

If a non-zero $(0,3)$-flux were present at the tip of the throat too, the scalar masses would receive extra contributions in the dimensional reduction. This cannot be added as a would-be supersymmetric $\smash{\mu}$-term, since an $F^X$-induced extra contribution gives cross-terms between $(2,1)$- and $(0,3)$-fluxes in the scalar mass-squared trace, which are not seen from the dimensional reduction \cite{Camara:2003ku}, and an $F^\rho$-induced would-be $\mu$-coupling cannot work either because it is impossible to find a scaling $H_{ab} \propto e^{n\Omega}$ giving a mass $m^2_{a \bar{b}} \propto e^{4 \Omega}$. Instead, the matching can be achieved via a would-be soft-breaking term, by adding an extra $X \bar{X}$-term in the K\"{a}hler metric in eqn. (\ref{anti-D3 scalars - Z}). Notice that, even in the presence of a non-vanishing $F^\rho$-term, the scalar masses are still partially protected by a no-scale cancellation
\begin{equation*}
    Z_{H^a \bar{H}^b, \, \rho \bar{\rho}} - 2 \, \Gamma^{H^c}_{\rho H^a} \, Z_{H^c \bar{H}^d} \, \bar{\Gamma}^{\bar{H}^d}_{\bar{\rho} \bar{H}^b} = 0.
\end{equation*}
This is a specific feature of the constrained-superfield would-be supersymmetry-breaking mass expression, since the usual soft supersymmetry-breaking mass vanishes in the presence of a logarithmic structure but due to a different cancellation involving the gravitino mass. However, there is an extra $F^\rho$-induced term giving an unwanted mass $m^2_{a \overline{b}} \propto e^{6 \Omega}$: this is a common issue for highly-warped setups if working with just one K\"{a}hler modulus, but, if needed, it can be cancelled by an extra $X \overline{X}$-term. In the main scenario considered, only a $(2,1)$-flux is present at the tip of the throat, so the $(0,3)$-flux induced must vanish. 

From the dimensional reduction one also obtains bilinear and trilinear couplings. For an Abelian anti-D3-brane, the bilinear coupling is
\begin{equation*}
\begin{split}
    S_{\textrm{bilinear}}^{\overline{\mathrm{D3}}\textrm{-scalars}} = - \dfrac{1}{2 \pi g_s} \int_{X_{1,3}} \! \de^4 x \, \sqrt{- \detr{g_4}} \; \dfrac{e^{4 \Omega}}{8 \pi \Vzero} \, \dfrac{g_s^2}{\kappa_4^2} \, \biggl( [l_s^2 \nabla_a \nabla_{b} (e^{4A} + \alpha)]_{0} \, \varphi^a \varphi^b + \mathrm{c.c.} \biggr),
\end{split}
\end{equation*}
whilst there are no trilinear couplings. The description within supergravity follows from the discussions in subsections \ref{LEEFT SUGRA in IIB compactifications} and \ref{LEEFT SUGRA in IIB compactifications with non-linearly realised SUSY}. As there are no bilinear $\tilde{\mu}$-couplings, for a term $H_{H^a H^b}$ the generic $B$-coupling would be
\begin{equation*}
    \begin{split}
        B_{\varphi^a \varphi^b} = \kappa_4^2 \hat{\mathcal{V}}_F H_{H^a H^b} + \hat{\bar{m}}^{w}_{3/2} \hat{\bar{\mathcal{F}}}^M \der_{\bar{M}} H_{H^a H^b} + \hat{m}^{w}_{3/2} \hat{\mathcal{F}}^M \hat{\nabla}_M H_{H^a H^b} \\
        - \hat{\mathcal{F}}^M \hat{\bar{\mathcal{F}}}^{\bar{N}} \, \bigl( H_{H^a H^b, M \bar{N}} - 4 \, \Gamma^{l}_{Mi} H_{H^a H^b, \bar{N}} \bigr) & .
    \end{split}
\end{equation*}
One can now observe that if a term $\smash{H_{H^a H^b} \propto \bar{X} \, e^{2 \Omega}}$ were used to generate the mass term $m_{\varphi^a\bar{\varphi}^b}$, it would also give a $B$-term scaling as $\smash{B_{\varphi^a \varphi^b} \propto e^{6\Omega + 4 A_0}}$, which is not present in the dimensional reduction.  Although this could be cancelled by a suitable counter-term $\smash{H'_{H^a H^b} \propto X \bar{X} \, e^{4 \Omega + A_0}}$, it is simpler to instead obtain the mass term via the coupling $\smash{H_{H^a Y^b}}$, as chosen in eqn. (\ref{anti-D3 scalars - H}); this only generates a bilinear term $\smash{B_{\varphi^a Y^b}}$, which is not a scalar coupling and vanishes in the unitary gauge. Finally, the required coupling $\smash{B_{\varphi^a \varphi^b}}$ above can be obtained by defining an extra $H$-term
\begin{equation} \label{anti-D3 scalars - extra H}
    H'_{H^a H^b} = \dfrac{1}{2 \pi g_s} \, \dfrac{e^{4 \Omega}}{8 \pi \Vzero} \, \dfrac{g_s^2}{\kappa_4^2} \, [l_s^2 \nabla_a \nabla_{b} (e^{4A} + \alpha)]_{0} \, \dfrac{X \bar{X}}{\hat{\mathcal{F}}^X \hat{\bar{\mathcal{F}}}^X},
\end{equation}
which only affects the $B$-term because this is the only term scaling as a second $X$-derivative of the $H$-term.

\paragraph{Anti-D3-brane Gauge Field}
Compared to the D3-brane gauge field, the anti-D3-brane gauge field is described by the same DBI-action but by an opposite CS-action, which results in the 4-dimensional action
\begin{equation*}
    S_{\textrm{kin}}^{\overline{\textrm{D3}}\textrm{-vector}} = - \dfrac{1}{4 \pi g_s} \int_{X_{1,3}} e^{-\sdil} \, F_2 \wedge * F_2 - \dfrac{1}{4 \pi g_s} \int_{X_{1,3}} C_0 \, F_2 \wedge F_2.
\end{equation*}
Of course, the gauge kinetic function cannot be $\smash{f_{\overline{\mathrm{D3}}} = i \bar{\tau} / 2 \pi g_s}$ as it is not holomorphic in the axio-dilaton. A solution to this issue is given in Ref. \cite{Cribiori:2019hod}, which finds
\begin{equation*}
    f_{\overline{\mathrm{D3}}} = \bigl(\bar{\mathcal{D}}^2 - 8 \mathcal{R} \bigr) \, \biggl( \dfrac{\bar{X} \bar{f}_{\mathrm{D3}}(\bar{\tau})}{\bar{\mathcal{D}}^2 \bar{X}} \biggr),
\end{equation*}
with $\mathcal{D}_\alpha$ the supergravity fermionic derivative and $\mathcal{R}$ the gravity multiplet. This function is holomorphic thanks to the projectors but at the same time has a superspace expansion
\begin{equation} \label{anti-D3-brane - f}
    f_{\overline{\mathrm{D3}}} = \dfrac{i \bar{\tau}}{2 \pi g_s} + \mathcal{O}(X).
\end{equation}
Because $X$ is the nilpotent superfield, all the extra terms are proportional to the goldstino and therefore vanish in the unitary gauge.

\subsection{Anti-D3-/D7-brane Intersecting States} \label{single anti-D3/D7 SUGRA}
For intersecting anti-D3-/D7-branes systems, the pure anti-D3- and pure D7-states have been described in the previous subsections.  It is also possible to provide a supergravity formulation of anti-D3-/D7-brane intersecting states:
\begin{itemize}
    \item on the one hand, one can infer the scaling factors for the kinetic and interaction terms of anti-D3-/D7-brane intersecting states using the D3-/D7-brane system discussed in subsection \ref{single D3/D7 SUGRA};
    \item on the other hand, the tools of constrained superfields allow one to formulate the low-energy theory in the language of supergravity.
\end{itemize}

\subsubsection{Anti-D3-/D7-brane Constrained Superfields and Couplings}
The strings stretching between the anti-D3- and the D7-brane give two scalar fields $\varphi$ and $\tilde{\varphi}$ as well as two Weyl spinors $\psi$ and $\tilde{\psi}$; in particular, the fields $(\varphi, \psi)$ and $(\tilde{\varphi}, \tilde{\psi})$ are in conjugate representations of the gauge groups.

Similarly to the pure anti-D3-brane states, as the anti-D3-/D7-brane intersecting states do not respect the supersymmetry of Calabi-Yau orientifold compactification, the natural tool to describe them is constrained superfields. It is impossible to identify the constrained superfields for the intersecting states by comparing supersymmetry variations because the latter are unknown as they cannot be inferred from a dimensional reduction.  However, one can postulate the following ones:
\begin{enumerate}[(i)]
    \item the scalar fields $\varphi$ and $\tilde{\varphi}$ belong to the chiral superfields $H$ and $\tilde{H}$ satisfying the spinor-removing constraints
    \begin{equation} \label{X barX D_alpha H = 0}
        X \bar{X} \, \mathcal{D}_\alpha H = 0, \qquad \qquad \qquad X \bar{X} \, \mathcal{D}_\alpha \tilde{H} = 0;
    \end{equation}
    \item the Weyl spinors $\psi$ and $\tilde{\psi}$ belong to the chiral superfields $Y$ and $\tilde{Y}$ satisfying the scalar-removing constraints
    \begin{equation} \label{X Y = 0}
        X Y = 0, \qquad \qquad \qquad \qquad X \tilde{Y} = 0.
    \end{equation}
\end{enumerate}
These constraints have been chosen because they are the easiest way \cite{DallAgata:2016syy} to remove the undesired degrees of freedom from the effective theory below the anti-D3-brane supersymmetry-breaking scale. In particular, notice that the constraint for the scalar fields is such as to leave an independent F-term \cite{DallAgata:2015zxp}.

In the strongly-warped regimes set by eqns. (\ref{localisation condition}, \ref{supergravity condition}), the K\"{a}hler potential contains the $2A_0$-shift as in eqn. (\ref{warped Kaehler potential}). Given the closed-string and anti-D3-brane goldstino potentials $\hat{K}$ and $\hat{W}$ of eqns. (\ref{closed-strings and goldstino - Kaehler potential}, \ref{closed-strings and goldstino - superpotential}), one can argue that the total K\"{a}hler potential and superpotential are
\begin{subequations}
\begin{gather}
    \begin{split}
        K = \hat{K} & + Z_{Y^a \bar{Y}^b} Y^a \bar{Y}^b + \dfrac{1}{2} \bigl[ H_{H^a H^b} Y^a Y^b + \mathrm{c.c.} \bigr] \\
        & + Z_{H^a \bar{H}^b} H^a \bar{H}^b + \dfrac{1}{2} \bigl[ H_{Y^a H^b} Y^a H^b + \mathrm{c.c.} \bigr] \\
        & + Z_{\Ds \bDs} \Ds \bDs + \dfrac{1}{2} \bigl[ H_{\Ds \Ds} \Ds \Ds + \mathrm{c.c.} \bigr] \\
        & + Z_{H \bar{H}} H \bar{H} + Z_{Y \bar{Y}} Y \bar{Y} + Z_{\tilde{H} \tilde{\bar{H}}} \tilde{H} \tilde{\bar{H}} + Z_{\tilde{Y} \tilde{\bar{Y}}} \tilde{Y} \tilde{\bar{Y}},
    \end{split} \label{Kaehler potential - anti-D3/D7} \\[1.5ex]
    \begin{split}
        W = \hat{W} & + \dfrac{1}{2} \, \tilde{\mu}_{\Ds \Ds} \Ds \Ds + \tilde{y} (\beta \Ds - Y^3 - H^3) Y \tilde{Y} \\
        & + \tilde{y} (\beta \Ds - Y^3 - H^3) H \tilde{Y} + \tilde{y} (\beta \Ds - Y^3 - H^3) Y \tilde{H}, 
    \end{split} \label{superpotential - anti-D3/D7}
\end{gather}
\end{subequations}
The pure anti-D3- and D7-brane terms follow from those discussed in subsubsections \ref{single bulk D7-brane SUGRA}, \ref{single throat D7-brane SUGRA}, \ref{single anti-D3-brane SUGRA}, and their theory is the same except for the anti-D3-brane uplift effect on the D7-brane to be discussed. The other terms represent the intersecting states and will be discussed below.

The field $H, Y$, and $\tilde{H}, \tilde{Y}$ have charges $q_{\overline{\mathrm{D3}}}=1,-1$ and $q_{\mathrm{D7}}=-1,1$, respectively, under the anti-D3- and D7 gauge groups.

\subsubsection{Anti-D3-brane with D7-brane from the Throat Tip into the Bulk} \label{single anti-D3/D7-bulk SUGRA}
In the setup in which the anti-D3-brane sits at the tip of the warped throat and the D7-brane wraps a 4-cycle extending from the throat tip into the bulk, the couplings for the intersecting states in eqns. (\ref{Kaehler potential - anti-D3/D7}, \ref{superpotential - anti-D3/D7}) are as follows.
\begin{itemize}
    \item Because the kinetic terms are not affected by the flux-induced supersymmetry breaking, for anti-D3-/D7-brane intersecting states one can make use of the same K\"{a}hler matter metric terms as for the D3-/D7-brane case. The logarithmic structure that is equivalent to eqn. (\ref{bulk D3/D7 - Z}) for D3-/D7-branes is generaralised to
    \begin{equation*}
        \kappa_4^2 K = - 3 \, \mathrm{ln} \, \biggl[ 2 \, e^{-2 \Omega} - \dfrac{4 \kappa_4^2}{3 g_s} \dfrac{\Vw}{\Vzero} \dfrac{e^{-2A_0} \, X \bar{X}}{[-i(\tau - \bar{\tau})] [-i \omega_w]} - \dfrac{\kappa_4^2}{3 \pi g_s} \, \varphi \bar{\varphi} \biggr].
    \end{equation*}
    so the matter metrics for anti-D3-/D7-branes are defined to be
    \begin{subequations} \label{bulk anti-D3/D7 - Z}
        \begin{align}
            Z_{H \bar{H}} & = \dfrac{1}{2 \pi g_s} \, e^{2 \Omega} + \dfrac{\kappa_4^2}{3 \pi g_s^2} \, \dfrac{e^{4 \Omega - 2 A_0}}{[-i (\tau - \bar{\tau})] [-i \omega_w]} \, \dfrac{\Vw}{\Vzero} \, X \bar{X}, \\
            Z_{Y \bar{Y}} & = \dfrac{1}{2 \pi g_s} \, e^{2 \Omega}.
        \end{align}
    \end{subequations}
    This is consistent with the intersecting states not acquiring flux-induced masses \cite{Camara:2004jj} due to similar cancellations to those discussed for the anti-D3-brane scalars.
    \item For the trilinear couplings in the superpotential, further explanations are required, as two related but distinct features from the higher dimensional setup need considering.
    \begin{enumerate}[(i)]
        \item Using the internal space symmetries of the flux-dependent couplings, Ref. \cite{Camara:2004jj} shows that the anti-D3-/D7-brane intersecting states couple only to the pure anti-D3-brane states and not to the pure D7-brane states. The coupling 3-form flux can be written as $G''_3 = g''_2 \wedge \de w^3$, where $g''_2 = g''_2 (w^3, \bar{w}^3)$ is a combination of $(1,1)$-forms on the 4-cycle, and the scalar trilinear couplings are of the kind
        \begin{equation*}
            t_{\alpha \beta \gamma} = \dfrac{1}{\kappa_4} \, u (e^{2\Omega}, e^{2A_0}) \, c_{\alpha \beta \gamma},
        \end{equation*}
        where (see appendix \ref{appendix: throat geometry} for the explicit expressions of the $(1,1)$-forms, $\zeta_i$)
        \begin{equation} \label{bulk anti-D3-/D7-brane CIU flux-dependent scalar couplings}
        \begin{split}
            \displaystyle c_{H^3 \bar{H} \tilde{\bar{H}}} & = \dfrac{1}{l_s^2} \int_{\Sigma_4} g''_2 \wedge \zeta_1, \\
            c_{H^3 H \tilde{H}} & = \dfrac{1}{l_s^2} \int_{\Sigma_4} g''_2 \wedge \zeta_2, \\
            c_{H^3 \tilde{H} \tilde{\bar{H}}} & = \dfrac{1}{l_s^2} \int_{\Sigma_4} g''_2 \wedge (\zeta_3 + \zeta_4) = c_{H^3 H \bar{H}},
        \end{split}
        \end{equation}
        with the overall factor
        \begin{equation*}
            u (e^{2\Omega}, e^{2A_0}) = \dfrac{1}{4 \pi} \, \dfrac{e^{7 \Omega + 3 A_0}}{[-i (\tau - \bar{\tau})]^{1/2} \, [-i \omega_w]} \, \biggl[ \dfrac{1}{\pi \Vzero} \biggr]^{1/2} \dfrac{\Vwfour}{\Vzerofour}.
        \end{equation*}
        A $(2,1)$-flux sources the coupling, but it is not the same flux that sources the D7-brane mass. Ref. \cite{Camara:2004jj} identifies the flux components that the couplings depend on, while the overall scaling $u$ has been inferred from the D3-/D7-brane case (see eqns. (\ref{bulk D3-D7 - trilinear flux coupling}, \ref{A-term - 37-/73-sector}), and note that the matching is done in terms of canonically normalised fields).
        \item Also, one needs to account for the mass due to the brane separation in a supersymmetric way since both the scalars and the spinors acquire the same separation mass. A way to do that is via a trilinear coupling in the superpotential.
    \end{enumerate}
    A natural guess to implement both these facts in the 4-dimensional effective theory is a generalisation of the trilinear coupling in eqn. (\ref{bulk D3-D7 - Yukawa}), with all the permutations accounting for the fact that now scalars and spinors are in different multiplets. Because for ISD-fluxes both the anti-D3- and the D7-brane have an effective superpotential bilinear coupling, though, such a term would again generate a coupling of the D7-brane state $\sigma^3$ with the intersecting states. A way to avoid it is to exclude the coupling\footnote{The removal of the term $\Delta_1 W = \tilde{y} \Ds H \tilde{H}$ prevents the couplings with the D7-brane, the absence of the term $\Delta_2 W = - \tilde{y} H^3 H \tilde{H}$ prevents the repetition of quartic couplings of the anti-D3-brane with the intersecting states already generated by the other terms -- which however also generate the would-be separation mass terms in an elegant way including the D7-brane scalar too --  and the absence of the term $\smash{\Delta_3 W = - \tilde{y} Y^3 H \tilde{H}}$ prevents the coupling $\smash{\bar{y} \mu_{33} H^3 \bar{H} \tilde{\bar{H}}}$, which is also forbidden by the symmetry arguments of Ref. \cite{Camara:2004jj}.}
    \begin{equation*}
        \Delta W = \tilde{y} ( \Ds - H^3 - Y^3 ) H \tilde{H}.
    \end{equation*}
    As a matter of fact the trilinear couplings of the proposed superpotential in eqn. (\ref{superpotential - anti-D3/D7}), namely
    \begin{subequations}
        \begin{align}
            & \tilde{Y}_{\Ds Y \tilde{Y}} = \tilde{Y}_{\Ds H \tilde{Y}} = \tilde{Y}_{\Ds Y \tilde{H}} = \tilde{y}, \\
            & \tilde{Y}_{Y^3 Y \tilde{Y}} = \tilde{Y}_{Y^3 H \tilde{Y}} = \tilde{Y}_{Y^3 Y \tilde{H}} = - \tilde{y}, \\
            & \tilde{Y}_{H^3 Y \tilde{Y}} = \tilde{Y}_{H^3 H \tilde{Y}} = \tilde{Y}_{H^3 Y \tilde{H}} = - \tilde{y},
        \end{align}
    \end{subequations}
    are enough to generate the desired couplings apart from a couple, which however will be dealt with in paragraph \ref{cftvp couplings}.
\end{itemize}

\paragraph{Standard Supergravity Terms}
One now needs to determine the effective D- and F-term potentials as well as the soft would-be supersymmetry-breaking couplings. Most of the terms have already been worked out in the earlier discussions on anti-D3- and D7-brane states, so one can focus on the interplay between the branes and on the new terms from intersecting states.
\begin{itemize}
    \item For the D7-brane, most of the calculations hold as in the analysis of the pure D7-brane in subsubsection \ref{single bulk D7-brane SUGRA}, as now summarised.
    
    For the supersymmetric terms, the effective $\mu$-coupling and the corresponding supersymmetric mass is exactly the same as discussed in subsubsection \ref{single bulk D7-brane SUGRA}. On the other hand, the effective superpotential couplings follow straightforwardly from the superpotential and are
    \begin{equation} \label{bulk anti-D3/D7 - effective Yukawas - D7-brane}
        Y_{\Ds Y \tilde{Y}} = Y_{\Ds H \tilde{Y}} = Y_{\Ds Y \tilde{H}} = y.
    \end{equation}
    Notice that the superpotential gives exactly the same (and no extra) Yukawa couplings as the D3-/D7-brane construction, since only the terms with one scalar and two spinors generate proper Yukawa terms.
    
    For the supersymmetry-breaking terms, assuming that the K\"{a}hler metric and the $H$-term do not depend on $X$ since they come from a deformation of the axio-dilaton K\"{a}hler potential, from the general expression one can observe the soft-breaking mass (where $m^{\mathrm{flux}}_{\Ds \bDs, \, \mathrm{soft}}$ represents the flux-induced soft-breaking terms)
    \begin{equation*}
        \begin{split}
            m^2_{\Ds \bDs, \, \mathrm{soft}} & = \bigl( \hat{m}^{w}_{3/2} \hat{\bar{m}}^{w}_{3/2} + \kappa_4^2 \mathcal{V}_F \bigr) \, Z_{\Ds \bDs} - \mathcal{F}^M \bar{\mathcal{F}}^{\bar{M}} \, R_{M \bar{N} \Ds \bDs} \\
            & = ( m^{\mathrm{flux}}_{\Ds \bDs, \, \mathrm{soft}} )^2 + \delta m^2_{\Ds \bDs, \, \mathrm{soft}},
        \end{split}
    \end{equation*}
    which clearly has an uplifting contribution due to the supersymmetry breaking by the anti-D3-brane, with
    \begin{equation}
        \delta m^2_{\Ds \bDs, \, \mathrm{soft}} = \kappa_4^2 V_{\overline{\mathrm{D3}}} \, Z_{\Ds \bDs} = \biggl[ \dfrac{g_s}{2 \pi \Vzero}\biggr]^2 \, \dfrac{e^{4\Omega + 4 A_0} \, \Vzerofour}{\kappa_4^2 [-i (\tau - \bar{\tau})]}.
    \end{equation}
     
    The effective $B$-term follows a similar destiny since it can be seen to read
    \begin{equation}
            B_{\Ds \Ds} = B^{\mathrm{flux}}_{\Ds \Ds} + \kappa_4^2 V_{\overline{\mathrm{D3}}} H_{\Ds \Ds}.
    \end{equation}
    Finally, the trilinear $A$-terms do not really generate any scalar trilinear coupling as the trilinear terms of eqn. (\ref{bulk anti-D3/D7 - effective Yukawas - D7-brane}) never involve three scalars due to the constraints, which means that the would-be scalar trilinear coupling is actually a fermionic interaction.
    \item For the anti-D3-brane, there is no substantial difference with respect to the analysis of subsubsection \ref{single anti-D3-brane SUGRA} since there are no new bilinear couplings in the K\"{a}hler potential or in the superpotential. One also has the superpotential trilinear couplings
    \begin{subequations} \label{bulk anti-D3/D7 - effective Yukawas - anti-D3-brane}
    \begin{align}
        & Y_{Y^3 Y \tilde{Y}} = Y_{Y^3 H \tilde{Y}} = Y_{Y^3 Y \tilde{H}} = -y, \\
        & Y_{H^3 Y \tilde{Y}} = Y_{H^3 H \tilde{Y}} = Y_{H^3 Y \tilde{H}} = -y.
    \end{align}
    \end{subequations}
    Evidently, these terms just add couplings between the anti-D3-brane and the intersecting states, but do not cause any particular modification to the pure anti-D3-brane action. Again, the superpotential also gives exactly the same Yukawa couplings as in the D3-/D7-brane construction.
    \item For the anti-D3-/D7-brane intersecting states, because their K\"{a}hler potential and superpotential expansion terms do not involve bilinear terms apart from the K\"{a}hler matter metric, one simply has the trilinear superpotential couplings discussed above and the soft-breaking masses
    \begin{equation*}
        m^2_{\varphi \bar{\varphi}, \, \mathrm{soft}} = \bigl( \hat{m}^{w}_{3/2} \hat{\bar{m}}^{w}_{3/2} \, Z_{H \bar{H}} - \mathcal{F}^\rho \bar{\mathcal{F}}^{\rho} \, R_{\rho \bar{\rho} H \bar{H}} \bigr) + \bigl( \kappa_4^2 V_{\overline{\mathrm{D3}}} \, Z_{H \bar{H}} - \mathcal{F}^X \bar{\mathcal{F}}^{X} \, R_{X \bar{X} H \bar{H}} \bigr),
    \end{equation*}
    and similarly for the counterpart $\tilde{\varphi}$. The first contribution vanishes in an ISD-background before non-perturbative corrections kick in, but the second one does not and reads
    \begin{equation}
        \delta m^2_{\varphi \bar{\varphi}, \, \mathrm{soft}} = \dfrac{2}{3} \, \kappa_4^2 V_{\overline{\mathrm{D3}}} \, Z_{H \bar{H}} = \biggl[ \dfrac{g_s}{2 \pi \Vzero}\biggr]^2 \dfrac{e^{6 \Omega + 4 A_0}}{3 \kappa_4^2}.
    \end{equation}
    Because these fields have no pure bilinear and trilinear couplings in the K\"{a}hler and superpotential, they do not have soft-breaking bilinear and trilinear terms either.
    \item To conclude, one must consider the complete effective D- and F-term potentials.
    
    First of all, for the D-term potential, one has again the positive semi-definite quartic self-interaction terms (and similarly for the corresponding field $\tilde{\varphi}$)
    \begin{equation}
    \begin{split}
        V_D^{(\mathrm{susy})} & = \dfrac{1}{2} \, g^2_{\mathrm{D}3} \, (Z_{H \bar{H}} \varphi \bar{\varphi})^2 + \dfrac{1}{2} \, g^2_{\mathrm{D}7} \, (Z_{H \bar{H}} \varphi \bar{\varphi})^2 \\
        & = \dfrac{1}{2 \pi g_s \, [-i (\tau - \bar{\tau})]} \, e^{4\Omega} \, (\varphi \bar{\varphi})^2 + \dfrac{1}{4 \pi g_s \, \Vzerofour} \, e^{6 \Omega} \, (\varphi \bar{\varphi})^2.
    \end{split}
    \end{equation}

    Second, for the F-term potential, most of the terms that are generated are actually fermionic interactions and not scalar couplings. Taking into account the effective bilinear terms from the pure D7- and anti-D3-branes as well as the Yukawa couplings in eqns. (\ref{bulk anti-D3/D7 - effective Yukawas - D7-brane}, \ref{bulk anti-D3/D7 - effective Yukawas - anti-D3-brane}), one obtains the effective superpotential
    \begin{equation*}
    \begin{split}
        W_{\mathrm{susy}} = \dfrac{1}{2} \, \mu_{\Ds \Ds} \Ds \Ds + \dfrac{1}{2} \, \mu_{Y^a Y^b} Y^a Y^b + \mu_{Y^a H^b} Y^a H^b + y \, (\Ds - Y^3 - H^3) Y \tilde{Y} \\
        + y \, (\Ds - Y^3 - H^3) H \tilde{Y} + y \, (\Ds - Y^3 - H^3) Y \tilde{H} &.
    \end{split}
    \end{equation*}
    Therefore, the effective F-term potential takes the form\footnote{For ease of notation, only the non-fermionic terms have been reported. Denoting the fermionic terms that one would have as $\mathcal{O}(X)$, the actual expression one finds is
    \begin{equation*}
    \begin{split}
        V_F^{(\mathrm{susy})} = Z^{\Ds \bDs} \bigl( \mu_{\Ds \Ds} \Ds + \mathcal{O}(X) \bigr) \bigl( \bar{\mu}_{\bDs \bDs} \bDs + \mathcal{O}(\bar{X}) \bigr) + Z^{Y^a \bar{Y}^b} \bigl( \mu_{Y^a H^c} \varphi^c + \mathcal{O}(X) \bigr) \bigl( \bar{\mu}_{\bar{Y}^b \bar{H}^d} \bar{\varphi}^d + \mathcal{O}(\bar{X}) \bigr) \\
        + Z^{Y \bar{Y}} \bigl[ y \bigl(\Ds - \varphi^3 \bigr) \tilde{\varphi} + \mathcal{O}(X) \bigr] \bigl[ \bar{y} \bigl(\bDs - \bar{\varphi}^3 \bigr) \tilde{\bar{\varphi}} + \mathcal{O}(\bar{X}) \bigr] \\[0.25ex]
        + Z^{\tilde{Y} \tilde{\bar{Y}}} \bigl[ y \bigl(\Ds - \varphi^3 \bigr) \varphi + \mathcal{O}(X) \bigr] \bigl[ \bar{y} \bigl(\bDs - \bar{\varphi}^3 \bigr) \bar{\varphi} + \mathcal{O}(\bar{X}) \bigr] & .
    \end{split}
    \end{equation*}}
    \begin{equation}
    \begin{split}
        V_F^{(\mathrm{susy})} = Z^{\Ds \bDs} \, \mu_{\Ds \Ds} \bar{\mu}_{\bDs \bDs} \Ds \bDs + Z^{Y^a \bar{Y}^b} \, \mu_{Y^a H^c} \bar{\mu}_{\bar{Y}^b \bar{H}^d} \varphi^c \bar{\varphi}^d \\
        + Z^{Y \bar{Y}} \bigl[ y \bigl(\Ds - \varphi^3 \bigr) \tilde{\varphi} \bigr] \bigl[ \bar{y} \bigl(\bDs - \bar{\varphi}^3 \bigr) \tilde{\bar{\varphi}} \bigr] \\
        + Z^{\tilde{Y} \tilde{\bar{Y}}} \bigl[ y \bigl(\Ds - \varphi^3 \bigr) \varphi \bigr] \bigl[ \bar{y} \bigl(\bDs - \bar{\varphi}^3 \bigr) \bar{\varphi} \bigr] & .
    \end{split}
    \end{equation}
    One immediately recognises the D7-brane supersymmetric mass, the anti-D3-brane scalar mass and the would-be separation mass for the anti-D3-/D7-brane intersecting states, with the same volume scaling as for the D3-/D7-brane case.
    
    The constrained multiplets $H^a$ have constrained F-terms, but they always appear in mixed $H^a Y^b$-, $H^a H$- and $H^a \tilde{H}$-couplings.  Therefore they both contribute the non-standard couplings discussed in appendix \ref{LEEFT SUGRA in IIB compactifications with non-linearly realised SUSY}, which turn out be fermionic and vanishing in the unitary gauge, and standard couplings via the effect of $Y^b$, $H$, $\tilde{H}$, which have unconstrained F-terms and end up providing bosonic terms in the action (see footnote \ref{footnote}).
\end{itemize}

\paragraph{$X\bar{X}$-dependent Interaction Terms} \label{cftvp couplings}
The supergravity formulation described so far incorporates all expected couplings, except the trilinear flux couplings in eqn. (\ref{bulk anti-D3-/D7-brane CIU flux-dependent scalar couplings}) and an anti-D3-/D7-brane version of the D3-/D7-brane quartic potential (\ref{bulk D3-D7 - quartic coupling}).

These couplings can be obtained by considering a specific class of supersymmetric terms, introduced in Refs. \cite{Cribiori:2017laj, Cribiori:2019hod}. This involves the nilpotent goldstino field in such a way as to only contribute bosonic terms to the component action, with the fermionic terms vanishing in the unitary gauge. Indeed, the coupling in eqn. (\ref{bulk anti-D3-/D7-brane CIU flux-dependent scalar couplings}) can be described by adding to the K\"{a}hler potential in eqn. (\ref{Kaehler potential - anti-D3/D7}) the deformation
\begin{equation} \label{bulk anti-D3-/D7-brane - Kaehler potential trilinear}
    \delta \mathcal{K} = \dfrac{2 [\Vwfour]^2}{g_s^4 \Vzerofour} \, \biggl[ \dfrac{1}{\pi} \, \Vzero \biggr]^{1/2} \, \dfrac{\kappa_4^2 X \bar{X} \, e^{5 \Omega - 3 A_0}}{[-i(\tau - \bar{\tau})]^{3/2} \, [-i \omega_w]^2} \, \Bigl[ \kappa_4 \, c_{\alpha \beta \gamma} H^\alpha H^\beta H^\gamma + \mathrm{c.c.} \Bigr].
\end{equation}
The only modification that this induces in the bosonic action comes from the second derivative with respect to $X$ and $\bar{X}$, namely $\delta V_F = \delta \mathcal{K}_{X \bar{X}} \mathcal{F}^X \bar{\mathcal{F}}^X$ as all the other terms contain the scalar component of $X$, which is proportional to the goldstino. One can similarly include the coupling of eqn. (\ref{bulk D3-D7 - quartic coupling}).

\subsubsection{Anti-D3-brane and D7-brane at the Tip of the Throat} \label{single anti-D3/D7-throat SUGRA}
If the anti-D3-brane and the D7-brane are localised at the tip of the warped throat, the supergravity couplings for the intersecting states in eqns. (\ref{Kaehler potential - anti-D3/D7}, \ref{superpotential - anti-D3/D7}) are given explicitly as follows.
\begin{itemize}
    \item As in subsubsection \ref{single anti-D3/D7-bulk SUGRA}, the matter metric terms for the anti-D3-/D7-brane case read  
    \begin{subequations} \label{throat anti-D3/D7 - Z}
        \begin{align}
            Z_{H \bar{H}} & = \dfrac{1}{2 \pi g_s} \, e^{2 \Omega} + \dfrac{\kappa_4^2}{3 \pi g_s^2} \, \dfrac{e^{4 \Omega - 2 A_0}}{[-i (\tau - \bar{\tau})] [-i \omega_w]} \, \dfrac{\Vw}{\Vzero} \, X \bar{X}, \\
            Z_{Y \bar{Y}} & = \dfrac{1}{2 \pi g_s} \, e^{2 \Omega}.
        \end{align}
    \end{subequations}
    \item For the cubic superpotential term, one can again follow subsubsection \ref{single anti-D3/D7-bulk SUGRA}. For a localised D7-brane there is no $(0,3)$-flux mediated coupling for the intersecting D3-/D7-brane states, so, following the tangent space symmetry arguments \cite{Camara:2004jj} and the scaling factors determined therein, the trilinear scalar couplings are still of the form
    \begin{equation*}
        t_{\alpha \beta \gamma} = \dfrac{1}{\kappa_4} \, u (e^{2\Omega}, e^{2A_0}) \, c_{\alpha \beta \gamma},
    \end{equation*}
    where the flux and index structure is
    \begin{equation} \label{throat anti-D3-/D7-brane CIU flux-dependent scalar couplings}
    \begin{split}
        \displaystyle c_{H^3 \bar{H} \tilde{\bar{H}}} & = \dfrac{1}{l_s^2} \int_{\Sigma_4} g''_2 \wedge \zeta_1, \\
        c_{H^3 \tilde{H} \tilde{\bar{H}}} & = \dfrac{1}{l_s^2} \int_{\Sigma_4} g''_2 \wedge (\zeta_2 + \zeta_3) = c_{H^3 H \bar{H}},
    \end{split}
    \end{equation}
    but with an overall factor
    \begin{equation*}
        u (e^{2\Omega}, e^{2A_0}) = \dfrac{1}{4 \pi} \, \dfrac{e^{4 \Omega + 2 A_0}}{[-i (\tau - \bar{\tau})]^{1/2} \, [-i \omega_{(0)}^{\Sigma_4}]} \, \biggl[ \dfrac{1}{\pi \Vzero} \, \dfrac{e^{6 A_0}}{g^{0}_{3 \bar{3}} \mathcal{V}_4^0} \biggr]^{\! 1/2}.
    \end{equation*}
    The matching with the scaling for the D3-/D7-brane coupling in  eqn. (\ref{throat D3-D7 - trilinear flux coupling}) is done in terms of the canonically normalised fields. Anyway, as in subsubsection \ref{single anti-D3/D7-bulk SUGRA}, the Yukawa couplings are still simply
    \begin{subequations}
        \begin{align}
            & \tilde{Y}_{\Ds Y \tilde{Y}} = \tilde{Y}_{\Ds H \tilde{Y}} = \tilde{Y}_{\Ds Y \tilde{H}} = \tilde{y} \beta, \\
            & \tilde{Y}_{Y^3 Y \tilde{Y}} = \tilde{Y}_{Y^3 H \tilde{Y}} = \tilde{Y}_{Y^3 Y \tilde{H}} = - \tilde{y}, \\
            & \tilde{Y}_{H^3 Y \tilde{Y}} = \tilde{Y}_{H^3 H \tilde{Y}} = \tilde{Y}_{H^3 Y \tilde{H}} = - \tilde{y},
        \end{align}
    \end{subequations}
    with $\beta = e^{-A_0}$, from the discussion of subsubsection \ref{dimensional reduction of intersecting states}. 
\end{itemize}

\paragraph{Standard Supergravity Terms}
Again, one can study the interactions term by term.
\begin{itemize}
    \item For the D7-brane, the results of subsubsection \ref{single throat D7-brane SUGRA} still hold with the further anti-D3-brane contribution to the soft-breaking mass\footnote{Since the K\"{a}hler metric now contains a factor $\smash{e^{2\Omega} / \mathrm{Im} \, \tau}$, it is now ambiguous whether it comes from a shift in the axio-dilaton K\"{a}hler potential or the K\"{a}hler modulus one. In the latter case, the D7-brane K\"{a}hler metric acquires an $X\bar{X}$-dependence, and there is an additional contribution to $\smash{ \delta m^2_{\Ds \bDs, \, \mathrm{soft}}}$, which results in an overall factor $f=2/3$ in the total expression.}
    \begin{equation*}
        \delta m^2_{\Ds \bDs, \, \mathrm{soft}} = \kappa_4^2 V_{\overline{\mathrm{D3}}} \, Z_{\Ds \bDs} = \biggl[ \dfrac{g_s}{2 \pi \Vzero} \biggr]^2 \, \dfrac{e^{6 \Omega + 4 A_0}}{\kappa_4^2 [-i (\tau - \bar{\tau})]}
    \end{equation*}
    and the $B$-term
    \begin{equation}
        B_{\Ds \Ds} = \kappa_4^2 V_{\overline{\mathrm{D3}}} H_{\Ds \Ds}.
    \end{equation}
    Further, now one has the effective superpotential couplings
    \begin{equation} \label{throat anti-D3/D7 - effective Yukawas - D7-brane}
        Y_{\Ds Y \tilde{Y}} = Y_{\Ds H \tilde{Y}} = Y_{\Ds Y \tilde{H}} = y \, e^{-A_0}.
    \end{equation}
    Finally, the trilinear $A$-terms do not generate any scalar trilinear coupling since in fact they correspond to fermionic interactions.
    \item For the anti-D3-brane, the same results as in subsubsection \ref{single anti-D3-brane SUGRA} hold identically. Further, there are the superpotential trilinear couplings
    \begin{subequations} \label{trhoat anti-D3/D7 - effective Yukawas - anti-D3-brane}
    \begin{align}
        & Y_{Y^3 Y \tilde{Y}} = Y_{Y^3 H \tilde{Y}} = Y_{Y^3 Y \tilde{H}} = -y, \\
        & Y_{H^3 Y \tilde{Y}} = Y_{H^3 H \tilde{Y}} = Y_{H^3 Y \tilde{H}} = -y.
    \end{align}
    \end{subequations}
    \item For the anti-D3-/D7-brane intersecting states, once again the only thing to add is the soft-breaking mass
    \begin{equation}
        \delta m^2_{\varphi \bar{\varphi}, \, \mathrm{soft}} = \dfrac{2}{3} \kappa_4^2 V_{\overline{\mathrm{D3}}} \, Z_{H \bar{H}} = \biggl[ \dfrac{g_s}{ 2 \pi \Vzero} \biggr]^2 \, \dfrac{e^{6 \Omega + 4 A_0}}{3 \kappa_4^2}.
    \end{equation}
    \item To conclude, one must discuss the effective D- and F-term potentials. For the D-term potential, one has again
    \begin{equation}
        V_D^{(\mathrm{susy})} = \dfrac{e^{4\Omega}}{2 \pi g_s \, [-i (\tau - \bar{\tau})]} \, (\varphi \bar{\varphi})^2 + \dfrac{e^{4 \Omega + 4 A_0}}{4 \pi g_s \, \mathcal{V}_4^0} \, (\varphi \bar{\varphi})^2.
    \end{equation}
    For the F-term potential, from the effective superpotential
    \begin{equation*}
    \begin{split}
        W_{\mathrm{susy}} = \dfrac{1}{2} \, \mu_{\Ds \Ds} \Ds \Ds + \dfrac{1}{2} \, \mu_{Y^a Y^b} Y^a Y^b + \mu_{Y^a H^b} Y^a H^b + y \, (\Ds e^{-A_0} - Y^3 - H^3) Y \tilde{Y} \\
        + y \, (\Ds e^{-A_0} - Y^3 - H^3) H \tilde{Y} + y \, (\Ds e^{-A_0} - Y^3 - H^3) Y \tilde{H} & ,
    \end{split}
    \end{equation*}
    so that the effective F-term potential reads as usual
    \begin{equation*}
    \begin{split}
        V_F^{(\mathrm{susy})} = Z^{\Ds \bDs} \, \mu_{\Ds \Ds} \bar{\mu}_{\bDs \bDs} \Ds \bDs + Z^{Y^a \bar{Y}^b} \, \mu_{Y^a H^c} \bar{\mu}_{\bar{Y}^b \bar{H}^d} \varphi^c \bar{\varphi}^d \\
        + Z^{Y \bar{Y}} \bigl[ y (\Ds e^{-A_0} - \varphi^3) \tilde{\varphi} \bigr] \bigl[ \bar{y} (\bDs e^{-A_0} - \bar{\varphi}^3) \tilde{\bar{\varphi}} \bigr] \\
        + Z^{\tilde{Y} \tilde{\bar{Y}}} \bigl[ y (\Ds e^{-A_0} - \varphi^3) \varphi \bigr] \bigl[ \bar{y} (\bDs e^{-A_0} - \bar{\varphi}^3) \bar{\varphi} \bigr] & .
    \end{split}
    \end{equation*}
\end{itemize}

\paragraph{$X\bar{X}$-dependent Interaction Terms}
For completeness, one has to include in the theory the flux-dependent trilinear couplings between the anti-D3-brane and the intersecting states in eqn. (\ref{bulk anti-D3-/D7-brane CIU flux-dependent scalar couplings}). Again, one can do so by means of the K\"{a}hler potential
\begin{equation} \label{throat anti-D3-/D7-brane - Kaehler potential trilinear}
        \delta \mathcal{K} = \dfrac{2 \Vzerofour}{g_s^4} \, \dfrac{\kappa_4^2 X \bar{X} \, e^{2 \Omega - 4 A_0}}{[-i(\tau - \bar{\tau})]^{3/2} \, [-i \omega_{(0)}^{\Sigma_4}]^2} \, \biggl[ \dfrac{\Vzero}{\pi} \dfrac{e^{6 A_0}}{g^{0}_{3 \bar{3}} \mathcal{V}_4^0} \biggr]^{\! 1/2} \, \Bigl[ \kappa_4 \, c_{\alpha \beta \gamma} H^\alpha H^\beta H^\gamma + \mathrm{c.c.} \Bigr].
\end{equation}
One can do the same for the quartic coupling in eqn. (\ref{throat D3-D7 - quartic coupling}).

\subsection{Moduli Stabilisation and Anomaly Mediation} \label{moduli stabilisation + anomaly mediation}
The scenario presented so far provides a toy model towards quasi-realistic constructions with non-linear supersymmetry in which most scalars are massive. However, the volume modulus is a runaway direction due to the anti-D3-brane uplift and its stabilisation affects the other fields of the theory. Moreover, as will also be discussed, some fields receive non-negligible mass contributions from anomaly mediation effects.

\subsubsection{Moduli Stabilisation via Perturbative and Non-Perturbative Corrections}
Due to the no-scale structure of the theory, tree-level type IIB flux compactifications lack the stabilisation of the K\"{a}hler modulus; nevertheless, this can be fixed once $\alpha'$- and non-perturbative corrections are included.

For concreteness, the KKLT scenario \cite{Kachru:2003aw} for the K\"{a}hler modulus stabilisation will be considered here, but analogous computations could be performed for the Large Volume Scenario \cite{Conlon:2005ki}. Two important modifications to the closed string $\hat{K}$ and $\hat{W}$ for the present analysis are the following.
\begin{enumerate}[(i)]
    \item In the KKLT approach, the K\"{a}hler modulus potential receives non-perturbative corrections from effects such as D7-brane gaugino condensation\footnote{This mechanism and its stability after the anti-D3-brane uplift have been scrutinised carefully in the literature and, despite the criticisms, there is no clear proof for it to be inconsistent. For the most recent discussions, see for instance Refs. \cite{Hamada:2019ack, Carta:2019rhx, Gautason:2019jwq, Bena:2019mte, Kachru:2019dvo}.} or Euclidean D3-brane instantons. Both effects can be described in the low-energy supergravity theory by means of a superpotential of the form
    \begin{equation*}
        \delta \hat{W}_{\mathrm{np}} = \dfrac{1}{\kappa_4^3} \, A \, e^{a i \rho},
    \end{equation*}
    where $A$ and $a$ are parameters whose details depend on the origin of the non-perturbative effects. This correction against a non-vanishing flux superpotential stabilises the volume modulus and, together with the anti-D3-brane uplift, it may give a 4-dimensional non-supersymmetric de Sitter vacuum.
    \item The $\alpha'$-corrections modify the K\"{a}hler potential for the volume modulus as \cite{Becker:2002nn}
    \begin{equation*}
        \kappa_4^2 \hat{K} = - 2 \, \mathrm{ln} \, \biggl[ \Bigl(2 \, e^{-2 \Omega} \Bigr)^{3/2} + \dfrac{1}{2} \, \hat{\xi} \biggr],
    \end{equation*}
    where, given the parameter $\xi = - \zeta(3) \chi / 16 \pi^3$, with $\smash{\zeta = \zeta(s)}$ the Riemann $\zeta$-function and the Euler number $\chi = 2 \, (h^{1,1} - h^{2,1})$ taken to be positive, the deformation is
    \begin{equation*}
        \hat{\xi} = \hat{\xi} (\tau, \bar{\tau}) = [-i ( \tau - \bar{\tau})]^{3/2} \xi.
    \end{equation*}
    Although $\alpha'$-corrections to a KKLT-setup with anti-D3-brane uplift do not qualitatively modify the stabilisation of $\rho$, as they are subleading in the volume suppression, they turn out to provide leading order contributions to some open-string masses, specifically the intersecting scalars.
\end{enumerate}
Note that, as discussed in subsubsection \ref{generalities on warped 4-dimensional SUGRAs}, in strongly warped scenarios, the effects of supergravity corrections are warped down in the scalar potential due to localisation effects, leading to a modification of the usual scales. This stems from the $2A_0$-shift in the K\"{a}hler potential (see eqn. (\ref{warped Kaehler potential})).

Typically the axio-dilaton and complex structure moduli are fixed at higher energy scales than the K\"{a}hler modulus and the open-string degrees of freedom, determining the flux background to be imaginary self-dual. This happens also in highly warped compactifications, as discussed in subsection \ref{field localisation}, so in the low-energy effective field theory they can be regarded as constant terms. An exception may be the complex structure moduli associated to the throat base at the strongly warped end \cite{Bena:2018fqc, Dudas:2019pls}. 

For the open-string sector, the anti-D3-brane scalars receive leading-order flux-induced mass contributions, so non-perturbative and $\alpha'$-corrections would give at most subleading corrections. A similar reasoning applies to the D7-brane scalars. Spinors are less affected than the scalars since they do not get soft-breaking contributions. On the other hand, the intersecting states do not have flux-induced masses, so such corrections play a relevant role.

Including the perturbative and non-perturbative corrections, the relevant terms in the supergravity theory for the volume modulus $\rho$ and the anti-D3-/D7-brane intersecting state scalars $\varphi$ are (the additional constant terms, including the $2A_0$-shift in the K\"{a}hler potential, as in eqn. (\ref{warped Kaehler potential}), will be included later)
\begin{subequations}
\begin{align}
    \kappa_4^2 K & = - 2 \, \mathrm{ln} \biggl[ \biggl( [-i (\rho - \bar{\rho}) + 2 c_0 ] - \dfrac{\kappa_4^2}{3} \, \g X \bar{X} - \dfrac{\kappa_4^2}{3} \, g_{\varphi \bar{\varphi}} \varphi \bar{\varphi} \biggr)^{3/2} + \dfrac{\hat{\xi}}{2} \biggr], \label{Kaehler potential + string-loop and non-perturbative corrections} \\
    \kappa_4^3 W & = W_0 + A \, e^{a i \rho} + \kappa_4 \s X, \label{superpotential + string-loop and non-perturbative corrections}
\end{align}
\end{subequations}
where, recall,  $e^{-2\Omega} = \mathrm{Im} \rho + c_0$ with $c_0 = {\mathcal V}_w/\mathcal{V}_{(0)}$.  The function $g_{\varphi \bar{\varphi}}$ can be read off from eqns. (\ref{bulk anti-D3/D7 - Z}, \ref{throat anti-D3/D7 - Z}), while the definitions of the constant GVW-term and of the anti-D3-brane parameters $W_0$, $\g$ and $\s$, respectively, can be extracted from eqns. (\ref{closed-strings and goldstino - Kaehler potential}, \ref{closed-strings and goldstino - superpotential}) and read
\begin{equation*}
    W_0 = \dfrac{g_s}{l_s^2} \biggl\langle \int_{Y_6} G_3 \wedge \Omega \biggr\rangle, \qquad \qquad \g = \dfrac{4}{g_s} \dfrac{\Vw}{\Vzero} \dfrac{e^{-2A_0}}{\langle -i(\tau - \bar{\tau}) \rangle \langle -i \omega_w \rangle}, \qquad \qquad \s = \sqrt{2} g_s.
\end{equation*}
The contributions from the vacuum expectation values of the axio-dilaton and the complex structure moduli as well as the constant terms have not been reported in the K\"{a}hler potential for brevity, but they will be reinserted when discussing physical scales.

Although the underlying string construction is different, as far as the scalar fields are concerned the supergravity theory of eqns. (\ref{Kaehler potential + string-loop and non-perturbative corrections}, \ref{superpotential + string-loop and non-perturbative corrections}) is formally equivalent to the one studied in detail in Ref. \cite{Aparicio:2015psl},\footnote{In Ref. \cite{Aparicio:2015psl}, the matter sector is modelled using a D3-brane in the bulk, with supersymmetry broken by a distant anti-D3-brane. For the scalar fields, this turns out to have an analogous supergravity formulation, the only differences being the $2 A_0$-shift to the K\"{a}hler potential and the $c_0$-shift to the K\"{a}hler modulus.} so this subsubsection mostly summarises the main results. After a standard calculation, the F-term potential of this model can be written as
\begin{equation*}
    V_F = \hat{V}_F (\rho, \bar{\rho}) + \Delta V_F (\varphi, \bar{\varphi}),
\end{equation*}
where $\hat{V}_F$ is the K\"{a}hler modulus potential, as a consequence of the breaking of the no-scale structure by the corrections and uplift term, while $\Delta V_F$ is the scalar potential for the scalar field $\varphi$, generating a mass term among other interactions.

\paragraph{K\"{a}hler Modulus Stabilisation and Minkowski Vacuum}
On the one hand, one can show that the leading order hidden-sector supersymmetry-breaking F-term potential reads
\begin{equation}
    \hat{V}_F = V_F^{\mathrm{KKLT}+\alpha'} + V_F^{\overline{\mathrm{D3}}+\alpha'},
\end{equation}
where the $\alpha'$-corrected KKLT-potential and uplift energy respectively read
\begin{equation*}
    \begin{split}
        V_F^{\mathrm{KKLT}+\alpha'} & = \dfrac{1}{\kappa_4^4} \biggl[ \dfrac{a^2 A \bar{A} \, e^{i a (\rho - \bar{\rho})}}{3 \, [ - i (\rho - \bar{\rho}) + 2 c_0]} + \dfrac{a ( W_0 \bar{A} \, e^{ -i a \bar{\rho}} + \bar{W}_0 A \, e^{i a \rho})}{[ - i (\rho - \bar{\rho}) + 2 c_0]^2} \biggr] + \delta_{\alpha'} V_F^{\mathrm{KKLT}}, \\
        V_F^{\overline{\mathrm{D3}}+\alpha'} & = \dfrac{\s^2}{\g [-i (\rho - \bar{\rho}) + 2 c_0]^2 \kappa_4^4} + \delta_{\alpha'} V_F^{\overline{\mathrm{D3}}},
    \end{split}
\end{equation*}
with the $\alpha'$-corrections being
\begin{equation*}
    \begin{split}
        \delta_{\alpha'} V_F^{\textrm{KKLT}} & = \dfrac{\hat{\xi}}{2 \kappa_4^4} \biggl[ \dfrac{1}{6} \dfrac{a^2 A \bar{A} \, e^{i a (\rho - \bar{\rho})}}{[-i (\rho - \bar{\rho}) \!+\! 2 c_0]^{5/2}} \!-\! \dfrac{a [W_0 \bar{A} \, e^{ -i a \bar{\rho}} + \bar{W}_0 A \, e^{i a \rho}]}{2 [ - i (\rho - \bar{\rho}) + 2 c_0]^{7/2}} \!+\! \dfrac{3 \, W_0 \bar{W}_0}{2 [ - i (\rho - \bar{\rho}) + 2 c_0]^{9/2}} \biggr], \\
        \delta_{\alpha'} V_F^{\overline{\mathrm{D3}}} & = - \dfrac{\hat{\xi}}{2 \kappa_4^4} \, \dfrac{\s^2}{\g [-i (\rho - \bar{\rho}) \!+\! 2 c_0]^{7/2}}.
    \end{split}
\end{equation*}
By parametrising the superpotential constants as $W_0 = \ab W_0 \ab \, e^{i \theta}$ and $A = \ab A \ab \, e^{i \alpha}$, given the definition of the K\"{a}hler modulus
\begin{equation*}
    \rho = \chi + i \, c,
\end{equation*}
one finds that at leading order the axion $\chi$ is minimised as $a \langle \chi \rangle = \theta - \alpha + n \pi$. Then, the leading order $c$-dependent scalar potential is
\begin{equation} \label{c-field scalar potential}
    V (c) = \dfrac{1}{\kappa_4^4} \, \dfrac{a \ab A \ab}{2} \, \biggl[ \dfrac{1}{3} \, \dfrac{a \ab A \ab \, e^{-2 a c}}{[c + c_0]} - \dfrac{\ab W_0 \ab \, e^{-a c}}{[c + c_0]^2} \biggr] + \dfrac{1}{\kappa_4^4} \, \dfrac{\s^2}{4 \, \g [c + c_0]^2}.
\end{equation}
Defining the shifted variables $c' = c + c_0$ and $\ab B \ab = \ab A \ab \, e^{a c_0}$ \cite{Frey:2008xw}, one obtains results which are formally equivalent to those of Ref. \cite{Aparicio:2015psl}. In the large volume regime, in which $c \gg 1$, the stationary condition $\der V / \der c = 0$ gives the solution
\begin{equation} \label{KKLT+anti-D3 vev}
    \ab W_0 \ab = \dfrac{2}{3} \, \langle a [c + c_0] \rangle \ab A \ab e^{- \langle a c \rangle} + \dfrac{1}{a \, \g} \, \dfrac{\s^2}{\langle a  [c + c_0] \rangle \ab A \ab} e^{\langle a c \rangle}.
\end{equation}
Further, a Minkowski vacuum $\langle \hat{V}_F \rangle = 0$ can be obtained if the parameter $s$ fulfils the leading order equality
\begin{equation} \label{KKLT+anti-D3 Minkowksi vacuum condition}
\s^2 = \dfrac{2}{3} \, a \, \g \, \langle a  [c + c_0] \rangle \ab A \ab^2 e^{- 2 \langle a c \rangle}.
\end{equation}
Of course one might want to impose a de Sitter vacuum, but anyway the vacuum energy has to be small. The $\alpha'$-corrections would modify the vacuum conditions only at subleading order in the volume.

One can write the vacuum expectation value (\ref{KKLT+anti-D3 vev}) in view of the Minkowksi vacuum condition (\ref{KKLT+anti-D3 Minkowksi vacuum condition}) as $\ab W_0 \ab = (2/3) \, (\langle a [c + c_0] \rangle + 1) \, \ab A \ab \, e^{- \langle a c \rangle}$, or, more conveniently and at leading order in the volume, as
\begin{equation*}
    \ab W_0 \ab^2 = \dfrac{2 \s^2}{3 \, \g} \, \langle [c + c_0] \rangle.
\end{equation*}
By taking this into account, the gravitino mass, namely $\smash{\kappa_4^2 \hat{m}^2_{3/2} = \langle e^{\kappa_4^2 \hat{K}} \, \hat{W} \hat{\bar{W}} \rangle}$, at leading order in the volume is
\begin{equation*}
    \hat{m}^2_{3/2} = \dfrac{1}{\kappa_4^2} \, \dfrac{\s^2}{12 \, \g \langle [c + c_0] \rangle^2}.
\end{equation*}
Similarly, one can see that the not-yet canonically normalised K\"{a}hler modulus mass is
\begin{equation*}
    \hat{m}^2_{cc} = \dfrac{1}{2} \dfrac{\der^2 V}{\der c^2} \biggr\ab_{c = \langle c \rangle} = \dfrac{1}{\kappa_4^4} \, \dfrac{a^2 \s^2}{4 \, \g \langle [c + c_0] \rangle^2}.
\end{equation*}
    
Finally, the combination of fluxes, non-perturbative corrections and anti-D3-brane uplift induces a non-zero F-term for the field $\rho$, along with the one for $X$. In the Minkowski vacuum of eqns. (\ref{KKLT+anti-D3 vev}, \ref{KKLT+anti-D3 Minkowksi vacuum condition}), at leading order in the volume one finds\footnote{In the presence of perturbative and non-perturbative corrections (and an anti-D3-brane), the axio-dilaton F-term becomes non-zero too. However, it is small compared to the F-terms for $X$ and $\rho$ \cite{Aparicio:2015psl}.}
\begin{equation*}
    \hat{F}^X = \biggl[\dfrac{6}{\g} \, (\langle c \rangle + c_0) \biggr]^{1/2} \dfrac{\hat{\bar{m}}_{3/2}}{\kappa_4} , \qquad \qquad \qquad \qquad \hat{F}^\rho = \dfrac{i}{a} \, \hat{\bar{m}}_{3/2}.
\end{equation*}
This means that the goldstino $\psi_{\textrm{g}}$ is now a linear combination of the anti-D3-brane gaugino and of the K\"{a}hler modulino (see e.g. Refs. \cite{Antoniadis:2014oya, Hasegawa:2015bza} for progress in the couplings between the gravitino and $\psi^X$). The unitary gauge does not exactly set to zero the spinor component $\psi^X$ of the nilpotent superfield, but rather the goldstino. This means that the anti-D3-brane models in this section have a plethora of interactions between the fields coupled to $X$ and/or $\rho$ and the linear combination of $\psi^X$ and $\psi^\rho$ that is orthogonal to the goldstino. This spinor $\psi_{\textrm{g}}'$ is massive, with a mass of at least the same order as the K\"{a}hler modulus mass. However, from the scalar potential, one can see that the scales at which each F-term comes into play have a different volume suppression, being \cite{Dudas:2019pls}
\begin{equation} \label{fx and frho}
    f_X = \bigl[ K_{X \bar{X}} F^X \bar{F}^X \bigr]^{1/2\!} \sim \dfrac{\hat{m}_{3/2}}{\kappa_4}, \qquad \qquad \qquad f_\rho = \bigl[ K_{\rho \bar{\rho}} F^\rho \bar{F}^\rho \bigr]^{1/2\!} \sim \dfrac{1}{\mathcal{V}^{2/3}} \, \dfrac{\hat{m}_{3/2}}{\kappa_4}.
\end{equation}
This suggests that, due to the hierarchically smaller volume suppression, the anti-D3-brane still provides the dominant contribution to the goldstino $\psi_{\textrm{g}}$, thus not changing drastically the scenario compared to the case where the goldstino is provided by the anti-D3-brane alone.

\paragraph{Open-string Mass Terms}
In order to write the open-string scalar potential in a convenient way it is helpful to consider the complete canonical normalisation of the scalar field, including the $\alpha'$-corrections. At the end of the day, one finds the $\varphi$-field scalar potential
\begin{equation}
    \Delta V_F = \biggl[ \dfrac{2 \kappa_4^2}{3} \, \bigl( V_F^{\mathrm{KKLT}+\alpha'} + V_F^{\overline{\mathrm{D3}}+\alpha'} \bigr) + \Theta_F \biggr] \dfrac{(1 + \delta_{Z_{\varphi \bar{\varphi}}}) \, g_{\varphi \bar{\varphi}} \varphi \bar{\varphi}}{[-i (\rho - \bar{\rho}) + 2 c_0]}.
\end{equation}
where the correction to the field normalisation is $\smash{\delta_{Z_{\varphi \bar{\varphi}}} = - \hat{\xi}/2 [-i (\rho - \bar{\rho}) + 2 c_0]^{3/2}}$. In this form, it is easy to impose the vacuum solutions. The $\Theta_F$-term reads
\begin{equation*}
    \Theta_F = \Theta_F^{\mathrm{KKLT}+\alpha'} + \Theta_F^{\overline{\mathrm{D3}}+\alpha'},
\end{equation*}
with the KKLT- and uplift-like terms
\begin{equation*}
\begin{split}
    \Theta_F^{\mathrm{KKLT}+\alpha'} & = \dfrac{5 \hat{\xi}}{72 \kappa_4^2} \biggl[ \dfrac{a^2 A \bar{A} \, e^{a i (\rho - \bar{\rho})}}{[-i (\rho - \bar{\rho}) + 2 c_0]^{5/2}} \!+\! \dfrac{3 a (\bar{A} W_0 e^{-a i \bar{\rho}} + A \bar{W}_0 e^{ai \rho})}{[-i (\rho - \bar{\rho}) + 2 c_0]^{7/2}} \!+\! \dfrac{9 W_0 \bar{W}_0}{[-i (\rho - \bar{\rho}) + 2 c_0]^{9/2}} \biggr], \\
    \Theta_F^{\overline{\mathrm{D3}}+\alpha'} & = \dfrac{\hat{\xi}}{12 \kappa_4^2} \, \dfrac{\s^2}{\g [-i (\rho - \bar{\rho}) + 2 c_0]^{7/2}}.
\end{split}
\end{equation*}
In the Minkowski vacuum of eqns. (\ref{KKLT+anti-D3 vev}, \ref{KKLT+anti-D3 Minkowksi vacuum condition}), only $\Theta_F$ contributes to the scalar masses. At leading order, its KKLT-like term happens to vanish, so the potential is fixed by its uplift-like term and it is positive definite. In particular, one finds the mass term
\begin{equation}\label{uplift-mass-term}
    \Delta V_F\ab_{\rho = \langle \rho \rangle} = \dfrac{\s^2 g_{\varphi \bar{\varphi}}}{12 \g [-i \langle \rho - \bar{\rho} \rangle + 2 c_0]^{9/2}} \, \dfrac{\hat{\xi}}{\kappa_4^2} \, \varphi \bar{\varphi}.
\end{equation}

\paragraph{Complete Scalar Potential and Mass Terms}
For a fully-fledged calculation, one must insert the axio-dilaton and complex structure modulus K\"{a}hler potentials and the constant term, as in eqns. (\ref{closed-strings and goldstino - Kaehler potential}, \ref{closed-strings and goldstino - superpotential}). Further, the $2 A_0$-shift in $K$ also needs to be included, as in eqn. (\ref{warped Kaehler potential}), and the consequent redshift will be indicated by the superscript `$w$', in line with the notation in the rest of the article. Finally, recall that hatted quantities mean they are purely determined by the supersymmetry-breaking hidden-sector potentials. 

Developing the observations made at the end of subsubsection \ref{generalities on warped 4-dimensional SUGRAs} on the redshifting of non-perturbative contributions to the scalar potential in strongly warped scenarios, notice that the $2A_0$-shift in the K\"{a}hler potential does not change qualitatively the shape of the scalar potential in the presence of KKLT-like non-perturbative corrections and anti-D3-brane uplift, but it affects it quantitatively. Indeed, the uplift term from the anti-D3-brane is scaled by the usual factor $e^{4A_0}$, but the pure closed-string sector term, which is usually unwarped, is now also scaled down by a factor $e^{2A_0}$. The moduli stabilisation is thus somewhat more delicate, as the uplift from the anti-D3-brane should not be too large with respect to the close string stabilisation so as to cause a runaway. Also, all the masses are now redshifted by an extra factor $e^{2A_0}$.

In detail, in the closed-string sector, the gravitino mass and the non-canonically normalised K\"{a}hler modulus mass read, respectively,
\begin{align}
    (\hat{m}^{w}_{3/2})^2 & = \dfrac{1}{\kappa_4^2} \dfrac{g_s^3}{12 \pi [\Vzero]^2} \, e^{4 \Omega + 4 A_0} \sim \dfrac{g_s^3}{\mathcal{V}^{4/3}} \, \dfrac{1}{\kappa_4^2} e^{4A_0}, \label{gravitino mass} \\[1.0ex]
    (\hat{m}^{w}_{cc})^2 & = \dfrac{3 a^2}{\kappa_4^2} (\hat{m}^{w}_{3/2})^2 \sim \dfrac{g_s^3}{\mathcal{V}^{4/3}} \, \dfrac{1}{\kappa_4^4} e^{4A_0}. \label{volume modulus mass}
\end{align}
Notice that two factors contribute to make the gravitino mass highly suppressed, i.e. the $e^{2A_0}$-redshift and the small bulk $(0,3)$-flux, which in the tuning towards a de Sitter/Minkowski vacuum ends up providing a lower volume- but enhanced warp factor-suppression. 

Moreover, if the open-string scalars are the intersecting state fields $\varphi$ and $\tilde{\varphi}$, then in terms of the gravitino mass their non-canonically normalised mass is
\begin{equation} \label{intersecting state scalar mass}
    m^2_{\varphi \bar{\varphi}} = \dfrac{\xi}{8 \pi g_s} \, e^{5 \Omega - 3 \sdil / 2} \, (\hat{m}^{w}_{3/2})^2 \sim \dfrac{\xi g_s^2}{\mathcal{V}^{3}} \, \dfrac{1}{\kappa_4^2} e^{4A_0},
\end{equation}
and similarly for the field $\tilde{\varphi}$. Such a mass is quite small due to a large volume suppression and the effect of warping, but it is necessarily positive definite. Notice that it vanishes in the absence of the $\alpha'$-corrections, namely if one sets $\xi = 0$.

Further, for D7-branes extending from the bulk to the throat the gauge kinetic function is determined by the volume modulus (see eqn. (\ref{bulk D7-brane - f})) and one finds the $F^\rho$-induced gaugino mass
\begin{equation} \label{F-term induced D7-brane gaugino mass}
    m_{1/2}^{\mathrm{D7}} = \dfrac{e^{2 \Omega}}{2 a} \, \hat{m}^w_{3/2} \sim \dfrac{g_s^3}{\mathcal{V}^{2}} \, \dfrac{1}{\kappa_4^2} e^{4A_0}.
\end{equation}
For D7-branes at the tip of the throat, there is a dependence on the volume modulus but it is highly redshifted (see eqn. (\ref{throat D7-brane - f})).

\paragraph{Corrections to Pure Anti-D3- and D7-brane Couplings}
The effect of the K\"{a}hler modulus stabilisation on the masses and the couplings of the pure anti-D3- and D7-brane states can also be worked out using supergravity, as will now be summarised. It is useful to note that the F-term for the volume modulus $\rho$ has an extra volume-suppression in the presence of non-perturbative corrections, while the F-term for the goldstino $X$ is unchanged. A key observation will be that the non-perturbative corrections induce scales that are never bigger than the flux-induced ones discussed before, so in the end the orders of magnitude for masses and couplings are unchanged.

For the pure D7- and anti-D3-brane chiral multiplets, the canonically normalised would-be supersymmetric masses are $e^{A_0}m_{\mathrm{D7}}$ or $m_{\mathrm{D7}}^w$ for the D7-brane fields localised in the bulk or at the tip, respectively (see ss. \ref{single bulk D7-brane SUGRA}, \ref{single throat D7-brane SUGRA}), and $\smash{m_{\overline{\mathrm{D3}}}^w}$ for the anti-D3-brane (see ss. \ref{single anti-D3-brane SUGRA}). For such fields, the $\rho$-field F-term does not participate in the effective $\mu$-terms, leaving these would-be supersymmetric masses unchanged. The would-be soft-breaking masses can be seen to be never bigger than these flux-induced terms, being at most of the order of the gravitino mass. Indeed, after canonically normalising, the would-be soft-breaking terms are at most of order
\begin{equation}
    m^i_{\mathrm{soft}} \sim \hat{m}_{3/2}^w.
\end{equation}
For both bulk- and throat-localised D7-branes, one finds canonically normalised soft-break\-ing masses $\smash{m_{\mathrm{soft}}^{77} \sim \hat{m}_{3/2}^w}$, with $\smash{e^{A_0} m_{\mathrm{D7}} \sim \hat{m}_{3/2}^w}$ (assuming $\theta \sim \theta'$) and $\smash{m_{\mathrm{D7}}^w \gg \hat{m}_{3/2}^w}$. For anti-D3-branes, one similarly finds $\smash{(m_{\mathrm{soft}}^{\overline{3} \overline{3}})^2 \sim - (\hat{m}_{3/2}^w)^2}$, where, at leading order in the volume, the key role is played by the F-term of the goldstino $X$. The $B$-terms are unaffected for the anti-D3-branes, coming from an $X \bar{X}$-term, while they receive normalised contributions for the D7-branes of order $\smash{B_i \sim (e^{A_0} m_{\mathrm{D7}} + \hat{m}_{3/2}^w) \hat{m}_{3/2}^w}$ or $\smash{B_i \sim m_{\mathrm{D7}}^w \hat{m}_{3/2}^w}$, for bulk or tip localisation, respectively. In the former case, the soft-breaking corrections compete with the flux-induced ones, but do not dominate, while in the latter the corrections are irrelevant for the mass eigenvalues. Finally, notice that the trilinear soft-breaking couplings with the intersecting states are inserted via the $X \bar{X}$-coupling and are thus unaffected.

As has been mentioned, the non-perturbative effects do not directly affect the open-string sector $X \bar{X}$-couplings. However, one may expect corrections for all the couplings, with a scale set by $\smash{\hat{m}_{3/2}^w}$. For the pure anti-D3-brane, such corrections would be irrelevant, as $\smash{m_{\overline{\mathrm{D3}}}^w \gg \hat{m}_{3/2}^w}$. On the other hand, considering the counter-part D3-/D7-branes, the soft-breaking trilinear coupling depends on the $\rho$-field F-term and is thus suppressed in the presence of non-perturbative corrections. All these changes must be implemented by hand, modifying the scalings in the $X \bar{X}$-terms.

In all these couplings, the $\alpha'$-corrections may only contribute at most with volume-suppressed terms and are thus irrelevant for fixing the orders of magnitude. An intuitive explanation for this can be seen in the fact that they do not participate in the stabilisation of the K\"{a}hler modulus and they are subleading in the F-terms.

\subsubsection{Anomaly Mediation}
In supersymmetric theories with a hidden sector, anomaly mediation provides a one-loop contribution to gaugino masses and trilinear scalar couplings, and a two-loop contribution to charged scalar masses \cite{Randall:1998uk, Giudice:1998xp}. Again, this is discussed in a setup similar to the current one in Ref. \cite{Aparicio:2015psl}, so only an essential review is reported below.

In the case of a diagonalisable K\"{a}hler matter metric, one can show that the anomaly-mediated gaugino masses, the scalar masses and the trilinear couplings read \cite{Randall:1998uk, Giudice:1998xp, Bagger:1999rd, Binetruy:2000md, Everett:2008ey, DEramo:2013dzi}
\begin{subequations}
\begin{align}
    m^a_{1/2}\Bigr|_{\mathrm{anom}} & = \dfrac{\beta_{g_a}}{g_a} \, \biggl[ \hat{m}^{w}_{3/2} - \dfrac{\kappa_4^2}{3} \, \hat{\mathcal{F}}^M \hat{K}_M \biggr], \label{anomaly-mediated gaugino mass} \\
    m^2_i\Bigr|_{\mathrm{anom}} & = \dfrac{1}{2} \, \beta_{h} \dfrac{\der \gamma^i}{\der h} \, \biggl[ \hat{m}^{w}_{3/2} - \dfrac{\kappa_4^2}{3} \, \hat{\mathcal{F}^M} \hat{K}_M \biggr] \biggl[ \hat{\bar{m}}^{w}_{3/2} - \dfrac{\kappa_4^2}{3} \, \hat{\bar{\mathcal{F}}^M} \hat{K}_{\bar{M}} \biggr], \label{anomaly-mediated scalar mass} \\
    A_{ijk}\Bigr|_{\mathrm{anom}} & = \dfrac{1}{2} \, y_{ijk} \, (\gamma^i + \gamma^j + \gamma^k) \biggl[ \hat{m}^{w}_{3/2} - \dfrac{\kappa_4^2}{3} \, \hat{\mathcal{F}^M} \hat{K}_M \biggr], \label{anomaly-mediated trilinear coupling}
\end{align}
\end{subequations}
where $y_{ijk}$ are the canonically normalised Yukawa couplings, $h$ represents any running coupling and $\beta_h$ is the corresponding beta-function, with $\gamma^i$ the $i$-field anomalous dimension. These expressions refer to the canonically normalised fields, with indices lowered and raised by Kronecker deltas $\smash{\delta_{i \bar{j}}}$ and $\smash{\delta^{i \bar{j}}}$.
\begin{itemize}
    \item Given the quadratic Casimir invariant in the adjoint representation $C_2 (G)$ and the generator normalisation $C (r_G)$ for the representation $r_G$, respectively, the beta-functions for the gauge couplings $g$ read
    \begin{equation*}
        \beta_g = - \dfrac{g^3}{16 \pi^2} \, b,
    \end{equation*}
    where $b$ is the coefficient
    \begin{equation*}
        b = \dfrac{11}{3} \, C_2 (G) - \dfrac{2}{3} \, n_f \, C(r^f_G) - \dfrac{1}{3} \, n_s \, C(r^s_G),
    \end{equation*}
    with $n_f$ and $n_s$ being the spinors and scalars in the representations $r^f_G$ and $r^s_G$ of the gauge group $G$, respectively. For the special unitary group $\mathrm{SU}(n)$, with $n > 1$, one has the set of values
    \begin{table}[H] \[
        \begin{array}{ccccc}
        \toprule
        \textrm{particle representation} & \quad \quad & C & \qquad & C_2 \\
        \midrule
        \boldsymbol{n} & & \dfrac{1}{2} & & \dfrac{n^2 - 1}{2n} \\[2.0ex]
        (\boldsymbol{n}, \boldsymbol{\bar{n}}) & & n & & n \\
        \midrule
    \end{array}
    \] \end{table} \vspace{-10pt}
    and for an Abelian group $\mathrm{U}(1)$ one finds $C(y) = y^2$ and $C_2(y) = 0$, where $y$ is the particle charge.
    \item One can write schematically the beta-functions for the Yukawa couplings $y_{ijk}$ as
    \begin{equation*}
        \beta_{y_{ijk}} = \tensor{f}{^l_{i}} (g,y) \, \tensor{y}{_{ljk}} + \tensor{f}{^l_{j}} (g,y) \, \tensor{y}{_{ilk}} + \tensor{f}{^l_{k}} (g,y) \, \tensor{y}{_{ijl}},
    \end{equation*}
    where $\tensor{f}{^i_j} (g,y)$ are functions generally scaling as $f (g,y) \sim b' g^2 + b'' y \overline{y}$, for some model-dependent coefficients $b'$ and $b''$, where the details of the index structure can be ignored for simplicity for the present purposes of determining just parametric dependences \cite{Randall:1998uk}.
    \item Finally, the anomalous dimension $\gamma^i$ can be written as
    \begin{equation*}
        \gamma^i = \dfrac{1}{16 \pi^2} \, \biggl( \dfrac{1}{2} \, \sum_{j,k} \tensor{Y}{_{ijk}} \tensor{\bar{Y}}{_{\bar{i}\bar{j}\bar{k}}} - 2 \sum_{a} g^2_a C(r^i_{G_a}) \biggr).
    \end{equation*}
\end{itemize}
The relevant mass scales are worked out below for intersecting anti-D3-/D7-branes. For single branes, the only non-neutral fields of the model are in the intersecting sector, which is thus the only one receiving corrections. More realistic non-Abelian models with multiple branes have a larger non-neutral spectrum, but the mass scales, being fixed by the gauge couplings, are analogous. In particular, the $b$-coefficients are typically negative due to the large number of degrees of freedom.
\begin{itemize}
    \item For a D7-brane wrapping a 4-cycle extending along the throat, the anomaly-mediated gaugino mass is slightly more suppressed than the volume modulus F-term contribution, being
    \begin{equation} \label{anomaly-mediated bulk D7-gaugino mass}
        m_{1/2}^{\mathrm{D7}}\Bigr|_{\mathrm{anom}} \simeq - \dfrac{g^2_{\mathrm{D7}}}{16 \pi^2} \, b_{\mathrm{D7}} \, \hat{m}^{w}_{3/2} = - \dfrac{g_s b_{\mathrm{D7}}}{8 \pi \Vzerofour} \, e^{2 \Omega} \, \hat{m}^{w}_{3/2},
    \end{equation}
    Instead, if the D7-brane wraps a 4-cycle that is localised at the infrared end of the throat, the anomaly-mediated mass is
    \begin{equation} \label{anomaly-mediated throat D7-gaugino mass}
        m_{1/2}^{\mathrm{D7}}\Bigr|_{\mathrm{anom}} \simeq - \dfrac{g^2_{\mathrm{D7}}}{16 \pi^2} \, b_{\mathrm{D7}} \, \hat{m}^{w}_{3/2} = - \dfrac{g_s b_{\mathrm{D7}}}{8 \pi} \, \hat{m}^{w}_{3/2}.
    \end{equation}
    In the presence of non-Abelian anti-D3-branes, there are extra would-be gaugini apart from the goldstino and their anomaly-mediated mass is\footnote{If one considers the effects of a non-zero axio-dilaton F-term, the gaugino mass contribution is at most of order $\smash{m_{1/2}^{\overline{\mathrm{D3}}} \sim (3 \hat{m}^{w}_{3/2} / 2 a \mathcal{V}^{2/3})}$ \cite{Aparicio:2015psl}, so it is usually subleading with respect to the anomaly-mediated one.}
    \begin{equation} \label{anomaly-mediated anti-D3-gaugino mass}
        m_{1/2}^{\overline{\mathrm{D3}}}\Bigr|_{\mathrm{anom}} \simeq - \dfrac{g^2_{\overline{\mathrm{D3}}}}{16 \pi^2} \, b_{\overline{\mathrm{D3}}} \, \hat{m}^{w}_{3/2} = - \dfrac{g_s b_{\overline{\mathrm{D3}}}}{4 \pi  [-i (\tau - \bar{\tau})]} \, \hat{m}^{w}_{3/2}.
    \end{equation}
    \item For the intersecting state scalars, which classically are vanishing, in principle the full anomaly-mediated mass term is
    \begin{equation*}
            m^2_{\varphi}\Bigr|_{\mathrm{anom}} \simeq \dfrac{1}{2} \Bigl[g_{\mathrm{D7}}^4 b_{\mathrm{D7}} C(r^\varphi_{\mathrm{D7}}) + g_{\overline{\mathrm{D3}}}^4 b_{\overline{\mathrm{D3}}} C(r^\varphi_{\overline{\mathrm{D3}}}) \Bigr] \biggl(\dfrac{\hat{m}^{w}_{3/2}}{8 \pi^2}\biggr)^2 + \delta_y \dot{m}^2_{\varphi},
    \end{equation*}
    where $\smash{\delta_y \dot{m}^2_{\varphi}}$ represents the Yukawa coupling-dependent contribution. This scales as $\smash{\delta_y \dot{m}^2_{\varphi} \sim y \overline{y} (b' g^2 + b'' y \overline{y})}$. For anti-D3-branes and localised D7-branes, one finds the scalings $g^2 \sim g_s$ and $y \sim g_s^{1/2}$, while for extended D7-branes the anti-D3-brane terms, unchanged, are the dominating ones. So the leading gauge coupling- and Yukawa coupling-dependent corrections have the same parametric dependence, and in the following the focus is going to be on the former for simplicity. In particular, notice that these tend to be negative-definite in quasi-realistic constructions with $b<0$, and they therefore compete with the positive-definite $\alpha'$-induced correction. So, for a D7-brane wrapping a 4-cycle extending along the throat, the leading order anomaly-mediated scalar mass is dominated by the anti-D3-brane contribution and reads
    \begin{equation} \label{anomaly-mediated intersecting state scalar mass - bulk D7-brane}
        m^2_{\varphi}\Bigr|_{\mathrm{anom}} \simeq \dfrac{g_s^2 b_{\overline{\mathrm{D3}}} C(r^\varphi_{\overline{\mathrm{D3}}})}{8 \pi^2 [-i (\tau - \bar{\tau})]^2} (\hat{m}^{w}_{3/2})^2 + \delta_y \dot{m}^2_{\varphi}.
    \end{equation}
    On the other hand, for a D7-brane wrapping a 4-cycle localised at the tip of the throat, the leading order term is
    \begin{equation} \label{anomaly-mediated intersecting state scalar mass - throat D7-brane}
        m^2_{\varphi}\Bigr|_{\mathrm{anom}} \simeq \biggl[ \dfrac{g_s^2 b_{\overline{\mathrm{D3}}} C(r^\varphi_{\overline{\mathrm{D3}}})}{8 \pi^2 [-i (\tau - \bar{\tau})]^2} + \dfrac{g_s^2 b_{\mathrm{D7}} C(r^\varphi_{\mathrm{D7}})}{32 \pi^2} \biggl( \dfrac{e^{4A_0}}{\mathcal{V}_4^0} \biggr)^2 \biggr] (\hat{m}^{w}_{3/2})^2 + \delta_y \dot{m}^2_{\varphi}.
    \end{equation}
    Such masses are negative definite as long as the $b$-coefficients are negative. These contributions are in close competition with the $\alpha'$-induced terms and the tachyonic terms might dominate, leading to an instability. 
    \item The contributions to the trilinear couplings are again determined in view of the gauge coupling terms and read
    \begin{equation*}
        A_{ijk}\Bigr|_{\mathrm{anom}} \simeq y_{ijk} \sum_a \dfrac{1}{b_a} \sum_{l=i,j,k} C_2(r^l_{G_a}) \, m^a_{1/2}\Bigr|_{\mathrm{anom}} + \delta_y \dot{A}_{ijk},
    \end{equation*}
    where the Yukawa-coupling contribution is of order $\smash{\delta_y \dot{A}_{ijk} \sim y_{ijk} \, y \overline{y} \, \hat{m}^{w}_{3/2}}$. This means that such trilinear terms are of up to order $\smash{A\bigr|_{\mathrm{anom}} \sim g_s \, y \, \hat{m}_{3/2}^w}$. Compared to the pure flux-induced terms, one can see that these tend to be leading for D7-branes extending along the throat and subleading for D7-branes wrapping 4-cycles at the tip of the throat.
\end{itemize}

\section{Overview on the Extension to Non-Abelian Theories} \label{overview on the extension to non-Abelian theories}
So far, the focus has been only on single anti-D3- and D7-branes. This section outlines a way to extend the previous results to multiple coincident branes at orbifold singularities, which provide quasi-realistic models with non-Abelian gauge groups and matter fields in bifundamental representations. The identification of the non-Abelian sectors with appropriate constrained superfields is worked out, and the new supergravity interactions are found, first for anti-D3-brane stacks, then for anti-D3-/D7-brane systems. Finally, the low-energy effective field theory corresponding to anomaly-free combinations of anti-D3-/D7-branes on orbifold singularities within flux compactifications is spelled out in some detail. 

Although an explicit realisation of a Calabi-Yau orientifold with orbifold-like singularities is beyond the scope of this paper, the results in sections \ref{warped D3- and D7-branes} and \ref{warped anti-D3- and D7-branes} hold in any such construction. In particular, the consequences of the orbifolding are in the richer array of gauge group representations particles may fall into, as reviewed below, but the gauge couplings and masses computed in earlier sections continue to hold in general. At the same time, there is a very interesting interplay between the orbifolding and supersymmetry breaking by anti-D3-branes, whereby, after the orbifolding, the bifundamental matter stretching between anti-D3-branes and D7-branes will have scalars and fermions in different gauge representations. Other minor differences, due for instance to orbifold symmetries projecting out certain background fluxes, are commented on explicitly. In a complete construction, local and global RR-tadpole cancellation would restrict the combinations of fluxes, anti-D3-branes and wrapped D7-branes appearing at each fixed point of the geometry.

\subsection{Non-Abelian Anti-D3-branes} \label{non-Abelian anti-D3-branes - constrained superfields}
First of all it is necessary to describe a stack of coincident anti-D3-branes in the language of $\mathcal{N}_4=1$ supergravity by extending its constrained superfields to the non-Abelian framework and adding a few new couplings which are non-zero only in the non-Abelian case.

\subsubsection{Particle Content}
The gauge group of a stack of $n$ coincident anti-D3-branes at a smooth point in the internal space is the non-Abelian group $\mathrm{U}(n)$. The group $\mathrm{U}(n)$ fulfils the isomorphism
\begin{equation*}
    \mathrm{U}(n) \simeq \mathrm{SU}(n) \times \mathrm{U}(1) / \mathbb{Z}_n,
\end{equation*}
so its generators $t_{I}$, with $I = 0, i$, consist of the $n$-dimensional identity  $t_0 = 1_n$ and of the $n$-dimensional Hermitean generators $t_i$ of the group $\mathrm{SU}(n)$, with $i=1,\dots,n^2-1$.

The particle content contains the following degrees of freedom:
\begin{itemize}
    \item a non-Abelian gauge vector, i.e.
    \begin{equation*}
        \hat{A}_\mu = \hat{A}^I_{\mu} t_I = A_{\mu} 1_n + A^i_{\mu} t_i;
    \end{equation*}
    \item a gaugino in the adjoint representation, i.e.
    \begin{equation*}
        \hat{\lambda} = \hat{\lambda}^I t_I = \lambda 1_n + \lambda^i t_i;
    \end{equation*}
    \item three complex scalars in the adjoint representation, i.e.
    \begin{equation*}
        \hat{\varphi}^a = \hat{\varphi}^{a I} t_I = \varphi^a 1_n + \varphi^{a i} t_i;
    \end{equation*}
    \item three modulini in the adjoint representation, i.e.
    \begin{equation*}
        \hat{\psi}^a = \hat{\psi}^{a I} t_I = \psi^a 1_n + \psi^{a i} t_i.
    \end{equation*}
\end{itemize}
The field $A_\mu$ gauges the $\mathrm{U}(1)$-component and the fields $A^i_\mu$ gauge the non-Abelian $\mathrm{SU}(n)$-component. Also, the fields $\lambda$, $\varphi^a$ and $\psi^a$ are netural under the Abelian group and singlets of the $\mathrm{SU}(n)$-component, whereas the fields $\lambda^i$, $\varphi^{a i}$ and $\psi^{a i}$ are neutral under the Abelian group and in the adjoint representation of the $\mathrm{SU}(n)$-component.

As it is a singlet under all the gauge groups, the spinor $\lambda$ is the goldstino of the theory. Therefore, it can be placed in a nilpotent chiral superfield $X$ just as in eqn. (\ref{X^2 = 0}), with
\begin{equation} \label{non-Abelian anti-D3-brane - nilpotent superfield}
    X^2 = 0.
\end{equation}
Being a singlet, the nilpotent superfield is sufficient to define the other constraints in a similar fashion as for a single anti-D3-brane, thanks to the linearity of their solutions \cite{Komargodski:2009rz}.
\begin{itemize}
    \item The non-Abelian gaugini $\lambda^i$ can be packaged in the chiral superfield
    \begin{equation*}
        \tilde{X} = X^i t_i,
    \end{equation*}
    which is neutral under the $\mathrm{U}(1)$- and in the adjoint of the $\mathrm{SU}(n)$-component of the gauge group, with the scalars removed by a constraint like the one in eqn. (\ref{X Y^a = 0}), i.e.
    \begin{equation}
        X \tilde{X} = 0.
    \end{equation}
    \item Similarly, the full gauge vector can be described by the field-strength chiral superfield
    \begin{equation*}
        \hat{W}_{\alpha} = W_\alpha + \tilde{W}_\alpha,
    \end{equation*}
    with $W_\alpha = W_\alpha 1_n$ and $\tilde{W}_\alpha = W^i_{\alpha} \, t_i$, where the spinor components are removed by the constraints\footnote{In addition to the constraint, there may be a modified Wess-Zumino gauge condition, as discussed in the Abelian case by Ref. \cite{Komargodski:2009rz}, which easily extends to the non-Abelian case.} (generalising that of eqn. (\ref{X W_alpha = 0}))
    \begin{subequations}
    \begin{align}
        X W_\alpha & = 0, \\
        X \tilde{W}_\alpha & = 0.
    \end{align}
    \end{subequations}
    As the nilpotent superfield $X$ is a singlet, these constraints are gauge invariant.\footnote{Notice that, if the constraint reads $X \tilde{W} = 0$, then, given the gauge transformation induced by the chiral superfield $\Lambda$, the constraint $X [e^{i\Lambda} \tilde{W} e^{-i\Lambda}]  = e^{i\Lambda} X \tilde{W} e^{-i\Lambda} = 0$ holds too.} Also notice that the condition $X \hat{W}_\alpha = 0$ is equivalent to the two constraints written above.
    \item For the modulini, one can define the chiral superfields
    \begin{equation*}
        \hat{Y}^a = Y^a + \tilde{Y}^a,
    \end{equation*}
    with $Y^a = Y^a 1_n$ and $\tilde{Y}^a = Y^{a i} t_i$, and remove the scalar components by means of the constraints (generalising the ones in eqn. (\ref{X Y^a = 0}))
    \begin{subequations}
    \begin{align}
        X Y^a & = 0, \\
        X \tilde{Y}^a & = 0.
    \end{align}
    \end{subequations}
    Again, gauge invariance is preserved and an equivalent condition is $X \hat{Y}^a = 0$.
    \item Finally, the scalars can again be encoded in the chiral superfields
    \begin{equation*}
        \hat{H}^a = H^a + \tilde{H}^a,
    \end{equation*}
    with $H^a = H^a 1_n$ and $\tilde{H}^a = H^{a i} t_i$, with the spinor and auxilary field components removed by constraints (generalising those of eqn. (\ref{bar X D H^a = 0}))
    \begin{subequations}
    \begin{align}
        \bar{X} \mathcal{D}_\alpha H^a & = 0; \\
        \bar{X} \mathcal{D}_\alpha \tilde{H}^a & = 0.
    \end{align}
    \end{subequations}
    These are gauge-invariant for gauge transformations with a chiral superfield $\Omega$ such that $\bar{X} (\Omega - \bar{\Omega}) = 0$, which implies the constraint $\bar{X} \mathrm{D}_\alpha \Omega = 0$ and is consistent with the gauge-fixing choice $X V = 0$ (see ref. \cite{Komargodski:2009rz} for more details). Again, one can simply write the condition $X \mathcal{D}_\alpha \hat{H}^a = 0$.
\end{itemize}

\subsubsection{Supergravity Formulation} \label{non-Abelian anti-D3-brane supergravity}
Given the superfield spectrum above, one needs to extend the $\mathcal{N}_4=1$ description of subsubsection \ref{single anti-D3-brane SUGRA} to a non-Abelian theory. Adapting the existing Abelian couplings to their non-Abelian version is straightforward. Moreover, to match the dimensionally-reduced effective action of Refs. \cite{Grana:2003ek, McGuirk:2012sb} one needs to generate a further cubic and quartic scalar interaction as well as some Yukawa couplings.

Quite remarkably, one can verify that the only extra terms which need to be included in the supergravity theory are those in the trilinear superpotential
\begin{equation} \label{non-Abelian anti-D3-brane superpotential}
    \delta \hat{W} = \dfrac{\upsilon}{4 \pi g_s} \, l_s^3 \Omega^{0}_{abc} \, \mathrm{tr} \, \hat{Y}^a \hat{Y}^b \hat{H}^c + \dfrac{\upsilon}{4 \pi g_s} \, l_s^3 \Omega^{0}_{abc} \, \mathrm{tr} \, \hat{Y}^a \hat{H}^b \hat{H}^c,
\end{equation}
where the normalisation constant is $\upsilon^2 = 4 \pi \, e^{-2 A_0} \, [\Vzero]^{3}$. One could account for the warp factor by considering the throat complex structure modulus \cite{Bena:2018fqc, Dudas:2019pls}.
\begin{itemize}
    \item Since it contains two spinors and one scalar, the first term in the superpotential only represents a Yukawa coupling between the modulini $\hat{\psi}^a$ and the scalars $\hat{\varphi}^a$ of the form
    \begin{equation*}
        y_{\hat{\psi}^a \hat{\psi}^b \hat{\varphi}^c} = y_{\hat{Y}^a \hat{Y}^b \hat{H}^c} = \dfrac{e^{3 \Omega}}{2 \pi g_s [-i (\tau - \bar{\tau})]^{1/2}} \, \dfrac{\Vw^{1/2}}{[-i \omega_w]^{1/2}} \, l_s^3 \Omega^{0}_{abc},
    \end{equation*}
    which corresponds to the couplings in Refs. \cite{Grana:2003ek, McGuirk:2012sb}, provided the insertion of the complex structure moduli in $\omega_w$ (not captured explicitly in the dimensional reduction).
    \item In a similar way as for D3-branes, the Yukawa terms also generate the quartic scalar potential and part of the cubic potential \cite{Grana:2003ek}. Indeed, now one has the effective anti-D3-brane superpotential
    \begin{equation*}
        \begin{split}
            W^{\overline{\mathrm{D3}}}_{\mathrm{susy}} = \, & \dfrac{1}{2} \, \mu_{\hat{Y}^a \hat{Y}^b} \mathrm{tr} \, \hat{Y}^a \hat{Y}^b + \dfrac{1}{2} \, \mu_{\hat{Y}^a \hat{H}^b} \mathrm{tr} \, \hat{Y}^a \hat{H}^b \\
            + & \dfrac{1}{2} \, y_{\hat{Y}^a \hat{Y}^b \hat{Y}^c} \mathrm{tr} \, \hat{Y}^a \hat{Y}^b \hat{H}^c + \dfrac{1}{2} \,  y_{\hat{Y}^a \hat{H}^b \hat{H}^c} \mathrm{tr} \, \hat{Y}^a \hat{H}^b \hat{H}^c,
        \end{split}
    \end{equation*}
    which in the unitary gauge generates the F-term scalar potential
    \begin{equation*}
        V_F^{(\mathrm{susy})} = Z^{\hat{Y}^a \hat{\bar{Y}}^b} \, \mathrm{tr} \, \bigl( \mu_{\hat{Y}^a \hat{H}^c} \hat{\varphi}^c + y_{\hat{Y}^a \hat{H}^c \hat{H}^d} \hat{\varphi}^c \hat{\varphi}^d \bigr) \bigl( \bar{\mu}_{\hat{\bar{Y}}^b \hat{\bar{H}}^e} \hat{\bar{\varphi}}^e + \bar{y}_{\hat{\bar{Y}}^b \hat{\bar{H}}^e \hat{\bar{H}}^f} \hat{\bar{\varphi}}^e \hat{\bar{\varphi}}^f \bigr).
    \end{equation*}
    Further, the D-term potential now reads
    \begin{equation*}
        V_D = \dfrac{1}{2} \, g_{\overline{\mathrm{D3}}}^2 \, \mathrm{tr} \, ( Z_{\hat{H}^a \hat{\bar{H}}^b} \hat{\varphi}^a \hat{\bar{\varphi}}^b) (Z_{\hat{H}^c \hat{\bar{H}}^d} \hat{\varphi}^c \hat{\bar{\varphi}}^d).
    \end{equation*}
    Obviously, the quadratic term in the F-term potential is the usual anti-D3-brane mass term. Then, consistently with the results of Refs. \cite{Grana:2003ek, McGuirk:2012sb}, the cubic term reads\footnote{In the presence of $(0,3)$-flux at the tip of the throat, there would be a further soft-breaking contribution to the trilinear scalar potential.}
    \begin{equation*}
        V^{\overline{\mathrm{D3}}}_{\textrm{cubic}} = - \biggl[ \dfrac{e^{4 \Omega + 4 A_0}}{16 \pi [-i (\tau -\bar{\tau})] \kappa_4} \, \biggl[ \dfrac{1}{2 \pi \omega} \dfrac{\Vw}{\Vzero} \biggr]^{1/2 \!} l_s (\bar{G}_3^{+ 0})_{a \bar{b} \bar{c}} \, \hat{\varphi}^a \hat{\bar{\varphi}}^b \hat{\bar{\varphi}}^c + \mathrm{c.c.} \biggr],
    \end{equation*}
    while the D-term potential and the quartic term of the F-term potential combine to give the usual would-be $\mathcal{N}_4 = 4$ scalar potential
    \begin{equation*}
        V^{\overline{\mathrm{D3}}}_{\textrm{quartic}} = \dfrac{e^{4 \Omega}}{8 \pi g_s [-i (\tau - \bar{\tau})]} \, \dfrac{\Vw}{\omega} \, g^{0}_{a \bar{b}} g^{0}_{c \bar{d}} \, \mathrm{tr} \, \biggl[ [\hat{\varphi}^a, \hat{\varphi}^c] [\hat{\bar{\varphi}}^b, \hat{\bar{\varphi}}^d] + [\hat{\varphi}^a, \hat{\bar{\varphi}}^d] [\hat{\bar{\varphi}}^b, \hat{\varphi}^c] \biggr],
    \end{equation*}
    which concludes the discussion of the consistency with the dimensional reductions in Refs. \cite{Grana:2003ek, McGuirk:2012sb}.
\end{itemize}

\subsection{Non-Abelian Anti-D3-/D7-brane Systems} \label{non-Abelian anti-D3-/D7-branes - constrained superfields}
As a further step toward quasi-realistic constructions, one can add a stack of $w$ intersecting D7-branes to the system with $n$ anti-D3-branes. The new states are as follows.
\begin{itemize}
    \item The D7-brane worldvolume is enhanced to a non-Abelian $\mathrm{U}(w)$-theory, where the gauge group is factorisable as $\mathrm{U}(w) \simeq \mathrm{SU}(w) \times \mathrm{U}(1) / \mathbb{Z}_w$, with $K=0,k$, for $k=1,\dots,w^2-1$. The degrees of freedom are then:
    \begin{itemize}
        \item a non-Abelian gauge vector and a spinor in the adjoint representation, i.e.
        \begin{equation*}
            \hat{B}_{\mu} = B_{\mu} 1_w + B^k_{\mu} \, t_k, \qquad \qquad \qquad \hat{\zeta} = \zeta 1_w + \zeta^k \, t_k;
        \end{equation*}
        \item a scalar and another spinor in the adjoint representation, i.e.
        \begin{equation*}
            \hat{\sigma}^3 = \sigma^3 1_w + \sigma^{3 \, k} \, t_k, \qquad \qquad \qquad \hat{k}^3 = \eta^3 1_w + \eta^{3 \, k} \, t_k.
        \end{equation*}
    \end{itemize}
    As D7-branes do not break supersymmetry, these fields make up standard multiplets. In particular, there are an Abelian vector superfield $Y_\alpha$, containing $B_\mu$ and $\zeta$, a non-Abelian $\mathrm{SU}(w)$ vector superfield $\tilde{Y}_\alpha$, containing $B^k_\mu$ and $\zeta^k$, a neutral chiral multiplet $\sigma^3$, containing $\sigma^3$ and $\eta^3$, and a chiral multiplet $\tilde{\sigma}^3$, containing $\sigma^{3 \, k}$ and $\eta^{3 \, k}$.
    \item For the anti-D3-/D7-brane intersecting states, the situation does not differ too much from the setup with single branes. The degrees of freedom are:
    \begin{itemize}
        \item two scalar fields $\varphi$ and $\tilde{\varphi}$ from the $\bar{3} 7$- and $7 \bar{3}$-sectors, respectively, with the former in the fundamental representation of the group $\mathrm{U}(n)$ and in the antifundamental of $\mathrm{U}(w)$, and the latter in the conjugate representation;
        \item two spinor fields $\psi$ and $\tilde{\psi}$ from the $\bar{3} 7$- and $7 \bar{3}$-sectors, respectively, with the former in the fundamental representation of the group $\mathrm{U}(n)$ and in the antifundamental of $\mathrm{U}(w)$, and the latter in the conjugate representation.
    \end{itemize}
    As usual, these fields cannot be packaged in standard supermultiplets with respect to the closed-string sector supersymmetry, but rather in constrained superfields.
    \begin{itemize}
        \item The scalars can be encoded in the chiral superfields $H$ and $\tilde{H}$ such as to remove their spinor components, generalising eqns. (\ref{X barX D_alpha H = 0}), i.e.
        \begin{subequations}
        \begin{align}
            X \bar{X} \, \mathcal{D}_\alpha H & = 0, \\
            X \bar{X} \, \mathcal{D}_\alpha \tilde{H} & = 0.
        \end{align}
        \end{subequations}
        \item The spinors can be encoded in the chiral superfields $Y$ and $\tilde{Y}$ such as to remove their scalar components, generalising eqns. (\ref{X Y = 0}), i.e.
        \begin{subequations}
        \begin{align}
            X Y & = 0, \\
            X \tilde{Y} & = 0.
        \end{align}
        \end{subequations}
    \end{itemize}
    Again, thanks to the linearity of the constraints, their solutions are simple generalisations of the Abelian ones. Notice that a superfield in the fundamental representation of a group $\mathrm{U}(p)$ has a charge $q = +1$ under the corresponding Abelian subgroup and is in the fundamental representation of the $\mathrm{SU}(p)$-subgroup, and correspondingly for the antifundamental representation.
\end{itemize}

\subsection{Anti-D3-/D7-branes at Orbifold Singularities}
An interesting class of model-building setups is the one with anti-D3-branes and D7-branes at orbifold singularities, as introduced by Ref. \cite{Aldazabal:2000sa} and implemented by Ref. \cite{Cascales:2003wn} in a more complete quasi-realistic flux setup (see also Ref. \cite{Marchesano:2004yn}).

The fact that the branes sit at an orbifold singularity breaks each gauge group $\mathrm{U}(m)$ into several subgroups $\mathrm{U}(m_i)$.  Interestingly, the anti-D3-/D7-intersecting scalars and spinors now transform in different representations of the unbroken gauge groups, and so have no semblance to being superpartners.

\subsubsection{Outline of the Gauge Group Breaking and Massless Spectrum}
One considers a 10-dimensional spacetime of the kind $X_{1,9} = \mathbb{R}^{1,3} \times Y_6$, where $Y_6$ is the 6-dimensional orbifold $O^6 = T^6 / \mathbb{Z}_N$. The action $\theta$ of the $\mathbb{Z}_N$-twist on the complex internal coordinates is
\begin{equation*}
    z^a \; \overset{\mathbb{Z}_N}{\longrightarrow} \; \alpha^{l_a} z^a,
\end{equation*}
with the definition $\alpha = e^{2 \pi i / N}$ and the bulk supersymmetry  condition $\sum_{a=1}^3 l_a = 0 \; \mathrm{mod} \, N$; for simplicity, only the case where $l_3$ is even is discussed. The action of the $\mathbb{Z}_N$-twist on the massless degrees of freedom of a stack of $n$ anti-D3-branes is then as follows.
\begin{itemize}
    \item Because they are orthogonal to the orbifolded directions, the action of the $\mathbb{Z}_N$-twist on the anti-D3-brane gauge vector fields is simply
    \begin{equation*}
        \hat{A}_\mu \; \overset{\mathbb{Z}_N}{\longrightarrow} \; \Gamma_{\theta, \bar{3}} \, \hat{A}_\mu \, \Gamma_{\theta, \bar{3}}^{-1},
    \end{equation*}
    where, given $N$ arbitrary integers $n_i$, with $i=0,1,\dots,N-1$, such that $\sum_{i=0}^{N-1} n_i = n$, the representation of the orbifold matrix is chosen to be
    \begin{equation*}
    \Gamma_{\theta, \bar{3}} = \mathrm{diag} \, \bigl( 1_{n_0}, \alpha 1_{n_1}, \dots, \alpha^{N-1} 1_{n_{N-1}} \bigr).
    \end{equation*}
    Therefore, it is not difficult to infer that the invariant generators generate the subgroup
    \begin{equation*}
        G_{\mathbb{Z}_N} =\bigotimes_{i=0}^{N-1} \mathrm{U}(n_i).
    \end{equation*}
    \item The three complex scalars $\hat{\varphi}^a$ transform under the orbifold twist $\theta$ as
    \begin{equation*}
        \hat{\varphi}^a \; \overset{\mathbb{Z}_N}{\longrightarrow} \; \alpha^{l_a} \, \Gamma_{\theta, \bar{3}} \, \hat{\varphi}^a \, \Gamma_{\theta, \bar{3}}^{-1},
    \end{equation*}
    which implies that the orbifold-invariant scalar fields fall into the representations
    \begin{equation*}
        \sum_{a=1}^3 \sum_{i=0}^{N-1} (\boldsymbol{n}_i,\overline{\boldsymbol{n}}_{i+l_a}).
    \end{equation*}
    \item The four Weyl spinors are associated to the states $\ket{\lbrace s_m \rbrace}_{m=1}^4$, where the half-integers $s_m=\pm 1/2$ define their chirality \cite{BLT}, and compatibly with the GSO-projection can be labelled as $\hat{\psi}^{r}$, with $r=0$ corresponding to the would-be gaugino $\hat{\lambda}$ and $r=a=1,2,3$ corresponding to the would-be modulini $\hat{\psi}^a$. The orbifold twist takes the form
    \begin{equation*}
        \hat{\psi}^{r} \; \overset{\mathbb{Z}_N}{\longrightarrow} \; \alpha^{s_m k_m} \, \Gamma_{\theta, \bar{3}} \, \hat{\psi}^{r} \, \Gamma_{\theta, \bar{3}}^{-1},
    \end{equation*}
    where $k_m$ are integers defining the orbifold action on the fermions, with $\sum_{m=1}^4 k_m = 0 \; \mathrm{mod} \, N$ and $l_1 = k_3+k_4$, $l_2 = k_2+k_4$ and $l_3 = k_2+k_3$, and the calculations show that the orbifold-invariant subset of the spinor $\hat{\lambda}$ transform in the representation
    \begin{equation*}
        \sum_{i=0}^{N-1} (\boldsymbol{n}_i,\overline{\boldsymbol{n}}_{i}),
    \end{equation*}
    while from the would-be modulini $\hat{\psi}^a$ one obtains the representations
    \begin{equation*}
    \sum_{a=1}^3 \sum_{i=0}^{N-1} (\boldsymbol{n}_i,\overline{\boldsymbol{n}}_{i+l_a}).
    \end{equation*}
\end{itemize}
In the presence of D7-branes, the reasoning is analogous. Just as for the action of the orbifold twist on the anti-D3-brane degrees of freedom, one defines the matrix
\begin{equation*}
    \Gamma_{\theta,7} = \mathrm{diag} \, \bigl( 1_{w_0} , \alpha 1_{w_1}, \dots, \alpha^{N-1} 1_{w_{N-1}} \bigr),
\end{equation*}
and essentially follows the same reasoning as above. The description of the orbifold action on the anti-D3-/D7-brane intersecting states can also be worked out in a similar way. \\

The full spectrum is summarised below.
\begin{itemize}
    \item The $\bar{3} \bar{3}$-sector provides a simple would-be supersymmetric massless spectrum.
    \begin{enumerate}[(i)]
        \item The vector fields and adjoint Weyl spinors transform in identical representations of the group $\bigotimes_{i=0}^{N-1} \mathrm{U}(n_i)$, i.e. in particular:
        \begin{equation} \label{bar3bar3-vector}
        \begin{array}{lcl}
            \textrm{$\bar{3} \bar{3}$-sector vectors:} & \qquad &\displaystyle \boldsymbol{r^{(\bar{3}\bar{3})}_v} = \sum_{i=0}^{N-1} (\boldsymbol{n}_i,\overline{\boldsymbol{n}}_{i}); \\[3.5ex]
            \textrm{$\bar{3} \bar{3}$-sector Weyl spinors:} & &\displaystyle \boldsymbol{r^{(\bar{3}\bar{3})}_{W_0}} = \sum_{i=0}^{N-1} (\boldsymbol{n}_i,\overline{\boldsymbol{n}}_{i}).
        \end{array}
        \end{equation}
        \item The $3 N$ complex scalar fields and the remaining $3 N$ Weyl spinors transform in identical bi-fundamental representations of the group $\bigotimes_{i=0}^{N-1} \mathrm{U}(n_i)$, namely:
        \begin{equation} \label{bar3bar3-chiral}
        \begin{array}{lcl}
            \textrm{$\bar{3} \bar{3}$-sector scalars:} & \quad &\displaystyle \boldsymbol{r^{(\bar{3}\bar{3})}_s} = \sum_{a=1}^3 \sum_{i=0}^{N-1} (\boldsymbol{n}_i,\overline{\boldsymbol{n}}_{i+l_a}); \\[3.5ex]
            \textrm{$\bar{3} \bar{3}$-sector Weyl spinors:} & &\displaystyle \boldsymbol{r^{(\bar{3}\bar{3})}_{W}} = \sum_{a=1}^3 \sum_{i=0}^{N-1} (\boldsymbol{n}_i,\overline{\boldsymbol{n}}_{i+l_a}).
        \end{array}
        \end{equation}
    \end{enumerate}
    \item The $7 \bar{3}$- and $\bar{3} 7$-sectors provide the following non-supersymmetric massless spectrum, transforming in distinct bifundamental representations:
    \begin{enumerate}[(i)]
        \item two sets of $N$ scalar fields:
        \begin{subequations}
        \begin{align}
            \textrm{$7 \bar{3}$-sector scalars:} \quad \quad \quad \quad \boldsymbol{r^{(7\bar{3})}_s} & = \sum_{i=0}^{N-1} (\overline{\boldsymbol{n}}_i,\boldsymbol{w}_i), \label{7bar3-scalar} \\
            \textrm{$\bar{3} 7$-sector scalars:} \quad \quad \quad \quad \boldsymbol{r^{(\bar{3}7)}_s} & = \sum_{i=0}^{N-1} (\boldsymbol{n}_i,\overline{\boldsymbol{w}}_i); \label{bar37-scalar}
        \end{align}
        \end{subequations}
        \item two sets of $N$ Weyl spinors: 
        \begin{subequations}
        \begin{align}
            \textrm{$7 \bar{3}$-sector Weyl spinors:} \quad \quad \quad \quad \boldsymbol{r^{(7\bar{3})}_W} & = \sum_{i=0}^{N-1} (\overline{\boldsymbol{n}}_i,\boldsymbol{w}_{i-l_3/2}), \label{7bar3-Weyl} \\
            \textrm{$\bar{3} 7$-sector Weyl spinors:} \quad \quad \quad \quad \boldsymbol{r^{(\bar{3}7)}_W} & = \sum_{i=0}^{N-1} (\boldsymbol{n}_{i-l_3/2},\overline{\boldsymbol{w}}_i). \label{bar37-Weyl}
        \end{align}
        \end{subequations}
    \end{enumerate}
    \item Finally, in the $77$-sector, one has a supersymmetric spectrum, as follows:
    \begin{enumerate}[(i)]
        \item the vector fields and a class of Weyl spinors form a number $N$ of $\mathcal{N}_4=1$ vector multiplets:
        \begin{equation}
            \textrm{$77$-sector vector multiplets:} \quad \quad \quad \quad \boldsymbol{r^{(77)}_V} = \sum_{i=0}^{N-1} (\boldsymbol{w}_i,\overline{\boldsymbol{w}}_i);
        \end{equation}
        \item the scalars fields and the Weyl spinors form a number $N$ of $\mathcal{N}_4=1$ chiral multiplets:
        \begin{equation} \label{77-chiral}
        \textrm{$77$-sector chiral multiplets:} \quad \quad \quad \quad \boldsymbol{r^{(77)}_C} = \sum_{i=0}^{N-1} (\boldsymbol{w}_i,\overline{\boldsymbol{w}}_{i+l_3}).
        \end{equation}
    \end{enumerate}
\end{itemize}
Such representations factorise according to the factorisation of the groups $\mathrm{U}(p_i)$, for instance if a field is in the representation $\boldsymbol{p}_i$ with respect to the group $\mathrm{U}(p_i)$, it has charge $q=1$ under its $\mathrm{U}(1)$-component and is in the representation $\boldsymbol{p}_i$ of the $\mathrm{SU}(p_i)$-component.

As models with anti-D3- and D7-branes at orbifold singularities contain chiral fermions in fundamental representations of the gauge groups, the theory is anomalous unless special cancellations occur, which is usually guaranteed by RR-tadpole cancellation \cite{Angulo:2002wf}. The specific configurations which make the theory anomaly-free are spelled out below and amount to the combinations of the sets of integers $\lbrace n_i \rbrace_{i=0}^{N-1}$ and $\lbrace w_i \rbrace_{i=0}^{N-1}$ that happen to give a theory in which all the anomalous Feynman diagrams add up to zero.
\begin{enumerate}[(i)]
    \item The condition that cancels out all the non-Abelian anomalies that arise from the $\mathrm{SU}(n_i)$- and $\mathrm{SU}(w_i)$-subgroups is \cite{Aldazabal:2000sa, Cascales:2003wn, Angulo:2002wf}
    \begin{equation} \label{anomaly-free_condition}
        4 \biggl[ \prod_{a=1}^{3} \mathrm{sin} \, \Bigl( \dfrac{\pi k l_a}{N} \Bigr) \biggr] \, \mathrm{tr} \, \Gamma_{\theta^k, \bar{3}} - \mathrm{sin} \Bigl( \dfrac{\pi k l_3}{N} \Bigr) \, \mathrm{tr} \, \Gamma_{\theta^k,7} = 0.
    \end{equation}
    \item Under the condition above, the mixed Abelian/non-Abelian diagrams are pseudo-anomalous, which implies that the Abelian factors actually acquire a mass via the Green-Schwarz mechanism, apart from the linear combination\footnote{Actually, this combination exists as long as all the integers $n_i$ are non-zero. Moreover, some $\mathbb{Z}_N$-orbifolds might have further anomaly-free linear combinations. An explanation to this is in Ref. \cite{Aldazabal:2000sa}, ss. 2.3.} \cite{Aldazabal:2000sa, Ibanez:2001nd, Cascales:2003wn}
    \begin{equation} \label{non-anomalous Abelian group}
        Q = \sum_{i=0}^{N-1} \dfrac{Q_{n_i}}{n_i}.
    \end{equation}
    Depending on the model, there may be additional non-anomalous combinations.
\end{enumerate}
In principle, the gauge fields in the multiplets from the spectrum reported in eqn. (\ref{bar3bar3-vector}) are the vectors\footnote{The notation should not be misleading: for instance, $\smash{\tilde{A}^{(i)}_\mu}$ denotes the gauge field for the $\smash{\mathrm{SU}(n_i)}$-component, and it can be expanded as $\smash{\tilde{A}^{(i)}_\mu = \tilde{A}^{(i) \, k}_\mu t^{(i)}_k}$, with $\smash{t^{(i)}_k}$ the Hermitean generators of $\smash{\mathrm{SU}(n_i)}$.} $\smash{\hat{A}^{(i)}_\mu = A^{(i)}_\mu + \tilde{A}^{(i)}_\mu}$, one for each different $\smash{\mathrm{U}(n_i)}$-subgroup, and similarly for the $\mathrm{U}(w_i)$-subgroups. However:
\begin{enumerate}[(i)]
    \item the non-Abelian gauge fields $\tilde{A}^{(i)}_\mu$ of the $\mathrm{SU}(n_i)$-components are non-anomalous if the condition in eqn. (\ref{anomaly-free_condition}) is satisfied, and similarly for the $\mathrm{SU}(w_i)$-components;
    \item all the Abelian gauge vectors $A_\mu^{(i)}$ are anomalous and hence disappear from the low-energy effective theory, apart from the linear combination given in eqn. (\ref{non-anomalous Abelian group}), i.e.
    \begin{equation*}
        V_\mu = \sum_{i=0}^{N-1} \dfrac{A_\mu^{(i)}}{n_i}.
    \end{equation*}
\end{enumerate}
Additional anti-D3-branes at other fixed points are also included in order to cancel the D7-brane anomaly induced there. Even though the corresponding new $\mathrm{U}(1)$-factors are anomaly-free, they still acquire a mass via the St\"{u}ckelberg coupling \cite{Ibanez:2001nd, Cascales:2003wn}.

\subsubsection{Outlook on a Supergravity Formulation} \label{outlook of anti-D3-/D7-brane at orbifold singularities SUGRA}
Given the massless spectrum of anti-D3-/D7-branes at orbifold singularities, one can now describe the effective theory in the language of $\mathcal{N}_4=1$ supergravity. In particular, one needs to identify the goldstino and understand how to encode the remaining degrees of freedom in supermultiplets.

If the anti-D3-brane sits at an orbifold singularity, the goldstino survives (see eqn. (\ref{bar3bar3-vector})) and the same supersymmetry breaking takes place as if it is at a smooth point (a similar breaking also happens for anti-D3-branes sitting at an orientifold singularity, as in Ref. \cite{Kallosh:2014wsa}). With multiple anti-D3-branes, the following reasoning holds.
\begin{enumerate}[(i)]
    \item At a smooth point, the anti-D3-brane goldstino would be the neutral singlet contained in the $\mathrm{U}(n)$-gaugino $\hat{\lambda}$. At an orbifold singularity, the original $\mathrm{U}(n)$-gaugino $\hat{\lambda}$ suffers the orbifold projection
    \begin{equation*}
        \hat{\lambda} \; \overset{\mathbb{Z}_N}{\longrightarrow} \; \Gamma_{\theta, \bar{3}} \, \hat{\lambda} \, \Gamma_{\theta, \bar{3}}^{-1},
    \end{equation*}
    which singles out several diagonal components as several gaugini $\hat{\lambda}^{(i)}$ for each of the subgroups $\mathrm{U}(n_i)$. For each of these, one extracts a neutral singlet $\lambda^{(i)}$ under the $\mathrm{U}(1)_i$ and $\mathrm{SU}(n_i)$ subgroups.
    \item Only one linear combination of the gaugini and their would-be vector superpartners is actually massless, with orthogonal combinations acquiring a mass via the Green-Schwarz mechanism \cite{Cascales:2003wn}. In accordance with eqn. (\ref{non-anomalous Abelian group}), the goldstino of the theory is thus the linear combination
    \begin{equation*}
        \psi_{\textrm{g}} = \sum_{i=0}^{N-1} \dfrac{\lambda^{(i)}}{n_i}
    \end{equation*}
    since it is the only massless gauge-neutral spinor on the anti-D3-brane worldvolume.
\end{enumerate}
The goldstino can be encoded as usual in a nilpotent superfield $X$. After the identification of the goldstino, one can easily infer the main characteristics of the supergravity effective field theory of the remaining fields in the massless spectrum. Details are below.
\begin{itemize}
    \item In the $\bar{3} \bar{3}$-sector, the situation is as follows.
    \begin{enumerate}[(i)]
        \item The vectors and the neutral Weyl spinors that transform in the adjoint representations $\smash{\boldsymbol{r^{(\bar{3}\bar{3})}_v} = \boldsymbol{r^{(\bar{3}\bar{3})}_{W_0}}}$ are the orbifold-invariant blocks of the fields $\tilde{A}_\mu$ and $\tilde{\lambda}$, plus the non-anomalous Abelian component $V_\mu$ and the goldstino $\psi_{\textrm{g}}$. Therefore, they belong to the orbifold-invariant blocks from the constrained superfields $\tilde{W}_\alpha$, $\tilde{X}$, $W_\alpha$ and $X$, respectively.
        
        The vectors are massless and provide the standard-like model gauge fields, with the goldstino being set to zero in the unitary gauge. On the other hand, the would-be non-Abelian gaugini are extra massless degrees of freedom which can be made massive via non-trivial effects such as anomaly mediation.
        \item The complex scalars and Weyl spinors transforming in the bifundamental representations $\smash{\boldsymbol{r^{(\bar{3}\bar{3})}_s} = \boldsymbol{r^{(\bar{3}\bar{3})}_W}}$ are the orbifold-invariant blocks of the fields $\varphi^a$, $\tilde{\varphi}^a$, $\psi^a$ and $\tilde{\psi}^a$, and therefore belong to the orbifold-invariant blocks from the constrained superfields $H^a$, $\tilde{H}^a$, $Y^a$ and $\tilde{Y}^a$, respectively.
        
        All these fields are massive in the presence of $(2,1)$-flux at the anti-D3-brane location. Scalars receive further subleading contributions originating from perturbative and non-perturbative corrections to the theory. Notice that not all the orbifold singularities allow for $(2,1)$-fluxes, in which case the corrections become leading.\footnote{For a supersymmetric $\mathbb{Z}_N$-twist, a necessary condition for the $(2,1)$-flux to survive the orbifold projection is that at least one of the $l_a$-coefficients be $l_a = N/2$, which is not satisfied e.g. by a $\mathbb{C}^3/\mathbb{Z}_3$-singularity, but it is for instance by $\mathbb{C}^3/\mathbb{Z}_4$; the flux can also be preserved for singularities of the form $(\mathbb{C}^2/\mathbb{Z}_N) \times \mathbb{C}$, $\mathbb{C}^3/[\mathbb{Z}_M \!\times \mathbb{Z}_N]$ and $\mathbb{C}^3/[\mathbb{Z}_M \!\times \mathbb{Z}_N \!\times \mathbb{Z}_K]$ \cite{Aldazabal:2000sa, Camara:2003ku, Camara:2004jj}.
        
        Moreover, depending on the orbifold action, the specific flux components that render the modulini massive \cite{Camara:2003ku} might be projected out. The trace condition allows this situation while keeping the scalars massive.}
    \end{enumerate}
    \item In the $7 \bar{3}$- and $\bar{3}7$-sectors, the situation is as follows.
    \begin{enumerate}[(i)]
        \item The scalars transforming in the bifundamental representations $\smash{\boldsymbol{r^{(\bar{3}7)}_s}}$ and $\smash{\boldsymbol{r^{(7\bar{3})}_s}}$ are the orbifold-invariant blocks of the fields $\varphi$ and $\tilde{\varphi}$, and therefore belong to the corresponding blocks from the constrained superfields $H$ and $\tilde{H}$, respectively.
        
        Such fields are massive after supersymmetry breaking and receive contributions from anomaly mediation.  Anomaly-mediated mass contributions can be negative and lead to tachyonic instabilities, but they may be balanced by other effects such as the $\alpha'$-corrected uplift contribution.
        \item The spinors belonging to the bifundamental representations $\smash{\boldsymbol{r^{(\bar{3}7)}_W}}$ and $\smash{\boldsymbol{r^{(7\bar{3})}_W}}$ are the orbifold-invariant blocks of the fields $\psi$ and $\tilde{\psi}$, and therefore belong to the corresponding blocks from the constrained superfields $Y$ and $\tilde{Y}$, respectively.
        
        Such fields are always massless and represent the matter content of the standard-like model extension built at the orbifold singularity.
    \end{enumerate}
    \item In the $77$-sector, the situation is the following.
    \begin{enumerate}[(i)]
        \item The fields in the vector multiplets in the adjoint representations $\boldsymbol{r^{(77)}_V}$ are the invariant blocks from the fields $B_\mu$ (if anomaly-free), $\tilde{B}_\mu$, $\zeta$ and $\tilde{\zeta}$, and therefore belong to the corresponding blocks of the vector multiplets $Y_\alpha$ and $\tilde{Y}_\alpha$.
        
        Such gauge fields are chosen to correspond to interactions in a hidden sector. In a pure-flux background, the gaugini are massive only in the presence of $(0,3)$-flux, which is not present at the tip of the throat. However, for bulk-extended D7-branes they acquire masses from a non-zero volume modulus F-term and even for throat-localised D7-branes they acquire a mass from the anomaly mediation mechanism.
        \item The fields in the chiral multiplets in the bifundamental representations $\boldsymbol{r^{(77)}_C}$ are the invariant blocks of the fields $\sigma^3$, $\tilde{\sigma}^3$, $\eta^3$ and $\tilde{\eta}^3$, and therefore belong to the corresponding blocks of the chiral superfields $\sigma^3$ and $\tilde{\sigma}^3$.
        
        All these fields are massive in the presence of $(2,1)$-flux, with further contributions from perturbative and non-perturbative corrections to the theory.
    \end{enumerate}
\end{itemize}
Now that the supermultiplets have been identified, given the $\mathcal{N}_4 = 1$ supergravity formulation of a system with intersecting anti-D3- and D7-branes at a smooth point in the internal space, in order to describe the theory of intersecting anti-D3- and D7-branes at an orbifold singularity one can simply reduce the original superfields to the subset that is invariant under the orbifold twist.

Notice that there exist singularities with further massless would-be vector superfields. In particular, this is a feature of orbifolds which leave invariant at least one of the complex directions \cite{Aldazabal:2000sa}. In this case this gives rise to extra massless Abelian gauge fields and neutral spinor fields.

\section{Analysis of the Mass Hierarchies} \label{analysis of the mass hierarchies}
Together, sections \ref{warped D3- and D7-branes}, \ref{warped anti-D3- and D7-branes} and \ref{overview on the extension to non-Abelian theories} provide the tools to formulate the supergravity theory for chiral gauge theories from intersecting anti-D3-/D7-branes on warped orbifold singularities in type IIB Calabi-Yau orientifold flux compactifications. The physical mass scales that emerge in such constructions are now discussed, with a view towards quasi-realistic standard-like models. In the scenario considered:
\begin{itemize}
    \item the localisation condition of eqn. (\ref{localisation condition}) is assumed, implying that closed-string sector fields, apart from the gravitino, tend to localise near the redshifted end of the throat;
    \item the hierarchy of eqn. (\ref{supergravity condition}) between the gravitino mass-sourcing fluxes is assumed, implying that the gravitino is localised in the bulk and a low-energy supergravity description is consistent.
\end{itemize}
It is also assumed that only $(2,1)$-fluxes are present at the tip of the throat. For ease of notation, the normalisation $\smash{\Vzero = 1}$ is considered in the rest of this section.

\subsection{Pure D7- and Anti-D3-brane States}
Pure D7- and anti-D3-brane states are discussed first, as their masses are essentially determined by the dimensional reduction of the worldvolume actions. In particular, except for some of the gaugini, the $77$- and $\bar{3} \bar{3}$-states are not critically dependent on the interplay between each other, neither on the way in which the K\"{a}hler modulus is stabilised nor on anomaly mediation effects.
\begin{itemize}
    \item For D7-branes that wrap 4-cycles extending from the tip of a warped throat into the bulk, the fate of the hidden matter chiral multiplets can be one of two possibilities, in accord with subsubsection \ref{single bulk D7-brane SUGRA}.
    \begin{itemize}
        \item If the mass-sourcing fluxes do not have specific hierarchies, then the D7-brane chiral superfield is localised near the tip of the throat with a mass of the order of the flux-induced axio-dilaton one, that is, from the normalisation induced by the matter metric and the $\mu$-coupling in eqns. (\ref{localised bulk D7-brane - Z}, \ref{localised bulk D7-brane - mu}), a canonical supersymmetric mass
        \begin{equation} \label{localised bulk D7-brane canonical mass}
            m^2_{77} \sim (m_{\mathrm{D7}}^w)^2 \sim \dfrac{g_s^2}{\mathcal{V}^{2/3}} \, \dfrac{1}{\kappa_4^2} \, e^{2 A_0},
        \end{equation}
        which is of the same order of magnitude as the warped Kaluza-Klein scale $m^{w}_{\mathrm{KK}}$ of eqn. (\ref{throat KK-mass}), above the cutoff scale of the theory.
        \item If the fluxes are such that the D7-brane chiral multiplet does not localise near the tip, then, from the matter metric and the $\mu$-coupling in eqns. (\ref{bulk D7-brane - Z}, \ref{bulk D7-brane - mu}), the canonically normalised supersymmetric mass is
        \begin{equation} \label{bulk D7-brane canonical mass}
            m^2_{77} \sim e^{2A_0} m_{\mathrm{D7}}^2 \sim \dfrac{{\theta'}^2 g_s^2}{\mathcal{V}^2} \, \dfrac{1}{\kappa_4^2} \, e^{2 A_0},
        \end{equation}
        where $\theta'$ is a small number representing the small bulk flux. In this case, the chiral multiplet survives the warped Kaluza-Klein cutoff.
    \end{itemize}
    Again following subsubsection \ref{single bulk D7-brane SUGRA}, given the gauge kinetic function of eqn. (\ref{bulk D7-brane - f}), the hidden-sector gauge couplings are of order
    \begin{equation} \label{bulk D7-brane gauge coupling}
        g^2_{\mathrm{D7}} \sim \dfrac{g_s}{\mathcal{V}^{2/3}}.
    \end{equation}
    In the absence of $(0,3)$-flux, if there are no supersymmetry-breaking or anomaly-mediation effects, the D7-brane gaugino is massless.
    \item For D7-branes that wrap 4-cycles localised at the tip of a warped throat, from the discussion in subsubsection \ref{single throat D7-brane SUGRA} with the matter metric and the $\mu$-coupling of eqns. (\ref{throat D7-brane - Z}, \ref{throat D7-brane - mu}), the hidden chiral matter multiplets acquire the canonical mass \cite{Burgess:2006mn}
    \begin{equation} \label{throat D7-brane canonical mass}
        m^2_{77} \sim (m_{\mathrm{D7}}^w)^2 \sim \dfrac{g_s^2}{\mathcal{V}^{2/3}} \, \dfrac{1}{\kappa_4^2} \, e^{2 A_0}.
    \end{equation}
    This means that the fields do not survive the cutoff unless the mass-sourcing $(2,1)$-flux is parametrically smaller than other fluxes in the throat that generate the warped Kaluza-Klein scale. Also, subsubsection \ref{single throat D7-brane SUGRA}, thanks to the gauge kinetic function of eqn. (\ref{throat D7-brane - f}), indicates that the hidden gauge couplings scale as
    \begin{equation} \label{throat D7-brane gauge coupling}
        g^2_{\mathrm{D7}} \sim g_s.
    \end{equation}
    Again, the gaugino is massless in the absence of supersymmetry-breaking or anomaly-mediation effects.
    \item For anti-D3-branes, the modulini and scalar exotics have masses of the same order of magnitude, as discussed in subsubsection \ref{single anti-D3-brane SUGRA}. From the matter metrics of eqns. (\ref{anti-D3 modulini - Z}, \ref{anti-D3 scalars - Z}) and the $H$-couplings of eqns. (\ref{anti-D3 modulini - H}, \ref{anti-D3 scalars - H}), one finds once again that a $(2,1)$-flux sources a canonical mass \cite{Burgess:2006mn}
    \begin{equation} \label{anti-D3-brane canonical mass}
        m^2_{\bar{3} \bar{3}} \sim (m_{\overline{\mathrm{D3}}}^w)^2 \sim \dfrac{g_s^2}{\mathcal{V}^{2/3}} \, \dfrac{1}{\kappa_4^2} \, e^{2 A_0}.
    \end{equation}
    Further, given the gauge kinetic function in eqn. (\ref{anti-D3-brane - f}), the gauge coupling scales as
    \begin{equation} \label{anti-D3-brane gauge coupling}
        g^2_{\overline{\mathrm{D3}}} \sim g_s.
    \end{equation}
    As the anti-D3-brane gaugino is the goldstino of the theory, it is always massless. For non-Abelian branes, there can be anomaly-mediation effects, otherwise the gaugini are always massless in the models under consideration.
\end{itemize}

\subsection{Anti-D3-/D7-brane intersecting states and Stable K\"{a}hler Modulus}
In the presence of intersecting anti-D3- and D7-branes, following subsection \ref{single anti-D3/D7 SUGRA}, both the scalars and the spinors from the $\bar{3}7$- and $7\bar{3}$-sectors are massless if one ignores perturbative and non-perturbative corrections. However, the string perturbative and non-perturbative effects are crucial for both stabilising the K\"{a}hler modulus and for making the intersecting scalars massive, as discussed in subsection \ref{moduli stabilisation + anomaly mediation}. In contrast, for the pure anti-D3- and D7-brane states, these only induce suppressed extra contributions which are only significant for some of the gaugini. A relevant role can also be played by anomaly mediation.

As discussed in subsection \ref{moduli stabilisation + anomaly mediation}, following eqn. (\ref{gravitino mass}), the interplay between perturbative and non-perturbative corrections imply that the gravitino acquires a mass of order
\begin{equation} \label{gravitino canonical mass}
    (\hat{m}^{w}_{3/2})^2 \sim \dfrac{g_s^{3}}{\mathcal{V}^{4/3}} \dfrac{1}{\kappa_4^2} \, e^{4 A_0}.
\end{equation}
Roughly, this can be written in terms of the warped Kaluza-Klein scale and the condition of eqn. (\ref{localisation condition}) shows that this mass is bounded above as
\begin{equation*}
    (\hat{m}^{w}_{3/2})^2 \lesssim \dfrac{g_s}{\mathcal{V}^2} \, (m^{w}_{\mathrm{KK}})^2.
\end{equation*}
This means that the gravitino is well within the cutoff of the theory. Also, the K\"{a}hler modulus is stabilised and, from eqn. (\ref{volume modulus mass}), its canonically normalised\footnote{For ease of notation, although this is the mass of the canonically normalised modulus (which requires taking into account both the K\"{a}hler metric and inserting the appropriate dimension for a 4-dimensional field), the symbol $\mathcal{V}$ is maintained from now on since the volume is what is controlled by the canonically normalised field $c$.} mass is of order
\begin{equation} \label{Kaehler modulus canonical mass}
    (\hat{m}^{w}_{\mathcal{V}})^2 \sim \mathcal{V}^{4/3} \, (\hat{m}^{w}_{3/2})^2,
\end{equation}
with the upper bound
\begin{equation*}
    (\hat{m}^{w}_{\mathcal{V}})^2 \lesssim \dfrac{g_s}{\mathcal{V}^{2/3}} \, (m^{w}_{\mathrm{KK}})^2,
\end{equation*}
leaving it well within the warped Kaluza-Klein cutoff too. Finally, from the matter metrics in eqns. (\ref{bulk anti-D3/D7 - Z})/(\ref{throat anti-D3/D7 - Z}) and the mass of eqn. (\ref{intersecting state scalar mass}), the canonical masses for the $\bar{3}7$-/$7 \bar{3}$-states visible scalar are of order
\begin{equation} \label{intersecting scalar canonical mass}
    m^2_{\bar{3}7} \sim m^2_{7\bar{3}} \sim \dfrac{\xi}{\mathcal{V}} \, (\hat{m}^{w}_{3/2})^2.
\end{equation}
Again, one can easily verify that these fields survive the 4-dimensional cutoff, being
\begin{equation*}
    m^2_{\bar{3}7} \sim m^2_{7\bar{3}} \lesssim \dfrac{g_s \xi}{\mathcal{V}^{3}} \, (m^{w}_{\mathrm{KK}})^2.
\end{equation*}

As discussed in subsection \ref{moduli stabilisation + anomaly mediation}, moduli stabilisation has effects on the gaugini, and anomaly mediation affects both the gaugini and the intersecting states.
\begin{itemize}
    \item For D7-branes wrapping a 4-cycle extended from the throat tip into the bulk, from eqn. (\ref{F-term induced D7-brane gaugino mass}), the non-zero volume F-term induces D7-brane hidden gaugini masses of order
    \begin{equation} \label{bulk D7-brane gaugino canonical mass}
        m_{1/2}^{\mathrm{D7}} \sim \dfrac{1}{\mathcal{V}^{2/3}} \, \hat{m}^{w}_{3/2},
    \end{equation}
    An anomaly-mediated contribution is also there but it has an extra string coupling suppression, as can be seen in eqn. (\ref{anomaly-mediated bulk D7-gaugino mass}). Further, from eqn. (\ref{anomaly-mediated intersecting state scalar mass - bulk D7-brane}), the anti-D3-/D7-brane visible sector intersecting scalars anomaly-mediated mass contribution is
    \begin{equation} \label{anomaly-mediated scalar mass - bulk D7-brane}
        \delta m^2_{\bar{3}7} \sim \delta m^2_{7\bar{3}} \sim - g_s^2 (\hat{m}^{w}_{3/2})^2,
    \end{equation}
    which competes with the $\alpha'$-induced contribution. One generally also has a Yukawa coupling-induced term of the same order of magnitude.
    \item If the D7-brane wraps a 4-cycle which is localised at the tip of the warped throat, then eqn. (\ref{anomaly-mediated throat D7-gaugino mass}) indicates that the D7-brane hidden gaugino acquires an anomaly mediated mass of order 
    \begin{equation} \label{throat D7-brane gaugino canonical mass}
        m_{1/2}^{\mathrm{D7}} \sim g_s \, \hat{m}^{w}_{3/2}.
    \end{equation}
    Further, from eqn. (\ref{anomaly-mediated intersecting state scalar mass - throat D7-brane}), the anti-D3-/D7-brane visible sector intersecting scalars' anomaly-mediated contribution is
    \begin{equation} \label{anomaly-mediated scalar mass - throat D7-brane}
        \delta m^2_{\bar{3}7} \sim \delta m^2_{7\bar{3}} \sim - g_s^2 \, (\hat{m}^{w}_{3/2})^2.
    \end{equation}
    This dominates the term induced by the $\alpha'$-induced contribution, generating an instability, unless the volume and the string coupling are properly tuned. A Yukawa coupling-induced term of the same order of magnitude is also generally there.
    \item Anomaly mediation also generates masses for the anti-D3-brane visible sector would-be gaugini apart from the goldstino, which are present for non-Abelian anti-D3-branes. In this case, thanks to eqn. (\ref{anomaly-mediated anti-D3-gaugino mass}), the order of magnitude is
    \begin{equation} \label{anti-D3-brane gaugino canonical mass}
        m_{1/2}^{\overline{\mathrm{D3}}} \sim g_s \, \hat{m}^{w}_{3/2}.
    \end{equation}
\end{itemize}

An interesting scenario is the one in which the mass-sourcing $(2,1)$-flux is such that the pure anti-D3- and D7-brane chiral multiplets are heavier than the cutoff scale. Since their positions are stabilised at the expectation values $\langle \varphi^a \rangle = 0$ and $\langle \sigma^3 \rangle = 0$, the trilinear couplings disappear. One is left with an effective theory in which the 4-dimensional degrees of freedom are:
\begin{itemize}
    \item the non-anomalous visible and hidden gauge vectors, which are massless, and the gaugini, which are massless if Abelian and with masses of the order of magnitude in eqns. (\ref{bulk D7-brane gaugino canonical mass})/(\ref{throat D7-brane gaugino canonical mass}) and (\ref{anti-D3-brane gaugino canonical mass}) otherwise;
    \item the intersecting states, namely some standard-like model spinors and exotic scalars in fundamental representations of the gauge groups, where the spinors are massless and the scalars have masses of the order of magnitude in eqn. (\ref{intersecting scalar canonical mass});
    \item the graviton, which is obviously massless, and a gravitino with a mass of the order of magnitude in eqn. (\ref{gravitino canonical mass});
    \item the K\"{a}hler modulus and its superpartner, with masses of the order in eqn. (\ref{Kaehler modulus canonical mass}), which constitute the lightest closed-string hidden-sector particles after the gravitino.
\end{itemize}
In models at orbifold singularities, the intersecting states are generally such that the scalars and the spinors are in different representations of the gauge groups, meaning that they do not even have would-be superpartners, but rather represent just a bunch of different charged spin-0 and spin-1/2 particles.

\subsection{Sample Mass Scales} \label{sample mass scales}
A qualitative spectrum that summarises the typical mass scales in models with  intersecting anti-D3- and D7-branes for strongly warped compactifications, i.e. satisfying the condition in eqn. (\ref{localisation condition}), and in the limit where the bulk $(0,3)$-flux is sufficiently small that a 4-dimensional supergravity formulation is allowed, i.e. satisfying eqn. (\ref{supergravity condition}), is reported below.

In detail, Fig. \ref{localised open-string sector - energy scales} reports a qualitative sample spectrum, in units of the reduced Planck mass $m_P$, in the case where the anti-D3-brane sits at the tip of the warped throat and the D7-brane wraps a 4-cycle extending from the throat tip into the bulk, with its chiral multiplet localised at the tip (see ss. \ref{extended 4-cycle: case ii}, \ref{single anti-D3-brane SUGRA}, \ref{single anti-D3/D7-throat SUGRA}). A similar spectrum emerges if the D7-brane wraps a 4-cycle localised at the throat tip (see ss. \ref{single throat D7-brane SUGRA}, \ref{single anti-D3-brane SUGRA}, \ref{single anti-D3/D7-throat SUGRA}), with only minor changes in the gauge sector. Instead, if the D7-brane wraps a 4-cycle extending from the bulk into the tip, with the chiral multiplet localised in the bulk, the only difference is in the smaller mass of the latter (see ss. \ref{extended 4-cycle: case iii}, \ref{single anti-D3-brane SUGRA}, \ref{single anti-D3/D7-bulk SUGRA}).

The volume modulus is stabilised by KKLT-like non-perturbative corrections and $\alpha'$-corrections are inserted too (see ss. \ref{moduli stabilisation + anomaly mediation}). The sample values are $g_s = 5 \cdot 10^{-2}$ and $e^{2 A_0} = 10^{-8}$  as well as $a = 0.1$, $\ab A \ab=1$ and $\ab W_0 \ab = 10^{-5}$, with $\langle \mathrm{Im} \, \tau \rangle = 1$, $\langle - i \omega_w \rangle = 1$, $\Vw/\Vzero = 1$ and $c_0=1$, which, for the scalar potential in eqn. (\ref{c-field scalar potential}), give a volume vacuum expectation value $\langle \mathcal{V} \rangle = 1.6 \cdot 10^3$ and a minimum energy $\Lambda \sim 10^{-26} m_P^4$ (which can as usual be adjusted with further fine-tuning). As usual, these parameters have been tuned to ensure the volume modulus stabilisation (for recent progress towards a top-down understanding of the KKLT parameter space see e.g. Refs. \cite{Blumenhagen:2019qcg, Demirtas:2019sip, Bena:2018fqc, Bena:2019mte}). In particular, the values chosen here are close to the original Ref. \cite{Kachru:2003aw} and satisfy the assumptions of the current setup, but are only one example in a vast parameter space. Along with the Minkowski vacuum condition of eqns. (\ref{KKLT+anti-D3 vev}, \ref{KKLT+anti-D3 Minkowksi vacuum condition}), the most stringent bounds are:
\begin{itemize}
    \item the localisation condition in eqn. (\ref{localisation condition}), which requires a small enough volume, compared to the warp factor, such that $\smash{\langle \mathcal{V} \rangle^{2/3} \lesssim e^{-A_0}}$;
    \item a small GVW-superpotential $\ab W_0 \ab$, which is necessary for the KKLT-vacuum but also to accomplish the supergravity condition in eqn. (\ref{supergravity condition});
    \item a string coupling that is large enough to be a reasonable gauge coupling in the visible sector, being $g_{\mathrm{vis}}^2 \sim 2 \pi g_s$, but also sufficiently small, compared to the volume $\langle \mathcal{V} \rangle$, as to satisfy the inequality $\xi/(g_s^2 \langle \mathcal{V}\rangle) \gtrsim 1$, which prevents tachyons in the intersecting sector.
\end{itemize}
Roughly, in order to have reasonable gauge couplings and to avoid open-string tachyons, the string coupling has to be of order $g_s \sim 10^{-2}$ and the volume is thus forced to be roughly at most of the order of magnitude $\langle \mathcal{V} \rangle \sim 10^3$. Therefore, the gravitino mass in eqn. (\ref{gravitino canonical mass}) - to which all the other 4-dimensional effective masses are proportional - indicates that what really suppresses the masses is the redshift factor $e^{A_0}$. In particular, the parameters chosen here place the gravitino mass and scalar exotics just above the current observational bounds. However, by stretching the parameters of the non-perturbative superpotential correction, one may achieve scenarios where the redshift $e^{A_0}$ is small enough to make the gravitino - and consequently all the other low-energy fields - arbitrarily light. On the other hand, bigger values of the redshift $e^{A_0}$ are also possible and provide masses that can be a few orders of magnitude larger.

\begin{figure}[H]
\centering
\begin{tikzpicture}[scale=0.60]

\node at (0,-1){};

\def \w {20}
\def \h {5.2}
\def \v {14.8}
\def \hb {4.5}
\def \vg {4.25}
\def \vo {3.75}
\def \d {1.5}
\def \G {20}
\def \legend {19.5}

\def \mP {18.4}
\def \ms {15.5}
\def \mKK {14.8}
\def \mus {12.0}
\def \muKK {11.0}
\def \mEW {2.1}
\def \cutoff {10.3}

\def \mflux {13.9}
\def \muflux {11.5}
\def \gravitino {6.3}
\def \Kaehler {8.4}
\def \brane {11.5}
\def \intersecting {4.5}
\def \hgaugino {4.1}
\def \gaugino {5.2}

\coordinate (O) at (0,0);
\coordinate (mP) at (0,\mP);
\coordinate (ms) at (0,\ms);
\coordinate (mKK) at (0,\mKK);
\coordinate (mus) at (0,\mus);
\coordinate (muKK) at (0,\muKK);
\coordinate (cutoff) at (0,\cutoff);
\coordinate (gravitino1) at (0,\gravitino);
\coordinate (mEW) at (0,\mEW);

\coordinate (mflux) at (\h,\mflux);
\coordinate (muflux) at (\h,\muflux);
\coordinate (gravitino) at (\h,\gravitino);
\coordinate (Kaehler) at (\h,\Kaehler);
\coordinate (brane) at (\v,\brane);
\coordinate (intersecting) at (\v,\intersecting);
\coordinate (hgaugino) at (\h,\hgaugino);
\coordinate (gaugino) at (\v,\gaugino);
\coordinate (hgauge) at (\h,0);

\coordinate (G) at (0,\G);

\coordinate (O') at (\w,0);
\coordinate (ms') at (\w,\ms);
\coordinate (mP') at (\w,\mP);
\coordinate (mKK') at (\w,\mKK);
\coordinate (mus') at (\w,\mus);
\coordinate (muKK') at (\w,\muKK);
\coordinate (cutoff') at (\w,\mP);
\coordinate (gravitino2) at (\w,\gravitino);
\coordinate (mEW') at (\w,\mEW);

\coordinate (hidden) at (\h,\legend);
\coordinate (visible) at (\v,\legend);

\draw[->] (O) -- (G) node [left] {$E$};
\fill[fill=red, opacity=0.1] (cutoff) rectangle (cutoff');

\node [left] at (O) {$0$};
\node [left] at (mP) {$m_P$};
\node [left] at (ms) {$m_s$};
\node [left] at (mKK) {$m_{KK}$};
\node [left] at (mus) {$m^{w}_{s}$};
\node [left] at (muKK) {$m^{w}_{\mathrm{KK}}$};
\node [left] at (mEW) {$m_{EW}$};

\draw [loosely dotted] (mP) -- (mP');
\draw [loosely dotted] (ms) -- (ms');
\draw [loosely dotted] (mKK) -- (mKK');
\draw [loosely dotted] (mus) -- (mus');
\draw [dotted] (muKK) -- (muKK') node {};
\draw [loosely dotted] (gravitino1) -- (gravitino2);
\draw [loosely dotted] (mEW) -- (mEW');
\draw [loosely dotted] (O) -- (O');

\draw[->, gray] (\w,0) -- (\w,\G) node [right] {$m / m_P$};
\node[gray] at (\w+1.25,18.4) {$1$};
\draw[fill] (\w,18.4) circle (0.025);
\node[gray] at (\w+1.25,15.4) {$10^{-3}$};
\draw[fill] (\w,15.4) circle (0.025);
\node[gray] at (\w+1.25,14.4) {$10^{-4}$};
\draw[fill] (\w,14.4) circle (0.025);
\node[gray] at (\w+1.25,12.4) {$10^{-6}$};
\draw[fill] (\w,12.4) circle (0.025);
\node[gray] at (\w+1.25,11.4) {$10^{-7}$};
\draw[fill] (\w,11.4) circle (0.025);
\node[gray] at (\w+1.25,8.4) {$10^{-10}$};
\draw[fill] (\w,8.4) circle (0.025);
\node[gray] at (\w+1.25,6.4) {$10^{-12}$};
\draw[fill] (\w,6.4) circle (0.025);
\node[gray] at (\w+1.25,4.4) {$10^{-14}$};
\draw[fill] (\w,4.4) circle (0.025);
\node[gray] at (\w+1.25,2.4) {$10^{-16}$};
\draw[fill] (\w,2.4) circle (0.025);

\node at (hidden) {\emph{hidden sectors}};
\node at (visible) {\emph{visible sector}};

\node at (mflux) {};
\node at (muflux) {$\! m^{w}_{\tau}, \, m^{w}_{u^\alpha}, \, m_{77}$};
\node[fill=white] at (gravitino) {$\hat{m}^{w}_{3/2}$};
\node at (Kaehler) {$\hat{m}^{w}_{\mathcal{V}}$};
\node at (brane) {$m_{\bar{3} \bar{3}}$};
\node at (intersecting) {$m_{\bar{3}7}^{\mathrm{scalar}}, \, m_{7\bar{3}}^{\mathrm{scalar}}$};
\node at (gaugino) {$m^{\overline{\mathrm{D3}}}_{1/2}$};
\node at (hgaugino) {$m^{\mathrm{D7}}_{1/2}$};

\draw[fill=white] (\v-\vg,0) rectangle (\v+\vg,2.25);
\fill[fill=green, opacity=0.15] (\v-\vg,0) rectangle (\v+\vg, 2.25);
\node[align=center] at (\v, 1.1) {observed standard model};

\draw[thick, draw=orange, fill=white!70!orange] (\v-\vo,-0.55) rectangle (\v+\vo,0.55);
\node [align=center] at (\v,0) {$\bar{3} \bar{3}$-vectors, $\bar{3}7$-/$7 \bar{3}$-spinors};

\draw[thick, draw=cyan, fill=white!85!cyan] (\h-\hb,-0.55) rectangle (\h+\hb,0.55);
\node [align=center] at (\h,0) {$77$-vectors, Abelian $77$-gaugini};

\end{tikzpicture}
\caption{A qualitative sample of the mass scales in models with intersecting anti-D3- and D7-branes in highly warped compactifications, i.e. such that $\smash{\langle \mathcal{V} \rangle^{2/3} \leq e^{-A_0}}$, with KKLT-like non-perturbative corrections and $\alpha'$-corrections, and a small bulk $(0,3)$-flux such that the gravitino localises in the bulk. Where the spin is not indicated, the masses refer to the supermultiplet as the soft-breaking corrections do not dominate. The observed standard model energy range and the relevant scales above the cutoff are shown explicitly. The graph refers to an anti-D3-brane sitting at the throat tip and a D7-brane wrapping a 4-cycle extending from the tip into the bulk, with the D7-brane chiral multiplet localised at the tip, where the gauge couplings are $g^2_{\overline{\mathrm{D3}}} \sim 0.3$ and $g^2_{\mathrm{D7}} \sim 2 \cdot 10^{-3}$. A similar spectrum emerges if the D7-brane wraps a 4-cycle localised at the tip, with then the D7-brane scales similar to the anti-D3-brane scales, so $g^2_{\mathrm{D7}} \sim g^2_{\overline{\mathrm{D3}}}$ and $\smash{m^{\mathrm{D7}}_{1/2} \sim m^{\overline{\mathrm{D3}}}_{1/2}}$. If the D7-brane wraps a 4-cycle extending into the bulk and the mass-sourcing $(2,1)$-fluxes are such that the D7-brane chiral multiplet localises in the bulk, then the latter approaches the gravitino mass scale, $m_{77} \sim \hat{m}^{w}_{3/2}$.}
\label{localised open-string sector - energy scales}
\end{figure}

Although a detailed exploration of the phenomenological implications of such scenarios is not the main aim of this article, a few comments are due. Notice that in the mass scales all the numerical factors have been dropped and only the parametric dependences on $e^{A_0}$, $g_s$ and $\langle \mathcal{V} \rangle$ have been taken into account.
\begin{itemize}
    \item From the cosmological perspective, the models do not present the cosmological moduli problem \cite{Banks:1993en, deCarlos:1993wie, Banks:1995dt, Nakamura:2006uc} since all the hidden moduli are heavier than the visible scalars. Whether or not there is a gravitino problem depends on the decay channels and abundances, but, in any case, the gravitino, with mass of order $\smash{\hat{m}_{3/2}^w \sim 8 \cdot 10^{-13} m_P}$, is sufficiently heavy to decay soon enough as not to spoil the BBN-physics, with a lower bound at roughly $\smash{m_{3/2}^{\mathrm{min}} \sim 10^{-13} m_P}$ \cite{Kawasaki:2004qu, Endo:2006zj, Nakamura:2006uc}. The models also contain some massless hidden $\mathrm{U}(1)$-gaugini and some heavy non-Abelian gaugini from the 77-sector, with masses $\smash{m_{1/2}^{\mathrm{D7}} \sim 6 \cdot 10^{-15} m_P}$ for a wrapped 4-cycle extending into the bulk, with a very small gauge coupling of order $g_{\mathrm{hid}}^2 \sim 2 \cdot 10^{-3}$, or $\smash{m_{1/2}^{\mathrm{D7}} \sim 4 \cdot 10^{-14} m_P}$ for a wrapped 4-cycle at the throat tip, with  coupling $g_{\mathrm{hid}}^2 \sim 0.3$. If the D7-brane chiral multiplet localises near the tip, its mass scale is above the cutoff, while if its mass-sourcing bulk flux is small enough and it stays in the bulk, then its mass is comparable to the gravitino one, i.e. $m_{77} \sim \hat{m}_{3/2}^w$.
    \item From the particle physics point of view, the visible sector consists of one Abelian and a few non-Abelian gauge groups plus some charged massless spinors in bifundamental representations as well as some heavy charged bifundamental scalars and a few slightly heavier non-Abelian gaugini. All the gauge couplings are of order $g_{\mathrm{vis}}^2 \sim 0.3$. For a gravitino with a mass of order $\smash{\hat{m}^w_{3/2} \sim 8 \cdot 10^{-13} m_P}$, these scalar masses are of order $\smash{m^{\mathrm{scalar}}_{\bar{3} 7} \sim m^{\mathrm{scalar}}_{7 \bar{3}} \sim 2 \cdot 10^{-14} m_P}$, while for the gaugini they are $\smash{m_{1/2}^{\overline{\mathrm{D3}}} \sim 4 \cdot 10^{-14} m_P}$. These values are consistent with the observational bounds \cite{Tanabashi:2018oca}.
\end{itemize}

It is important to discuss the scale at which the supersymmetry-breaking mass splittings come into play. Indeed, whilst there is no scale at which superpartners emerge for the $\bar{3}\bar{3}$- and $\bar{3}7$-/$7\bar{3}$-states, closed-string and $77$-multiplets do have supersymmetry-breaking mass splittings, and $\bar{3}\bar{3}$-states and $\bar{3}7$-/$7\bar{3}$-scalars also acquire soft mass contributions from supersymmetry breaking effects. The breaking of supersymmetry by the anti-D3-branes takes place at the warped-string scale $\smash{m_s^w}$, where  the full tower of string states comes into play \cite{Cribiori:2020bgt}. However, the relevant mass scale for supersymmetry breaking in the low-energy theory is instead controlled by the gravitino mass scale $\smash{\hat{m}^w_{3/2}}$, as will now be explained. In a near-Minkowski vacuum, the orders of magnitude of the contributions to the F-term scalar potential are fixed by the scales \cite{Aparicio:2015psl, Dudas:2019pls}
\begin{equation*}
    f_X = \bigl[ \mathcal{K}_{X \bar{X}} \mathcal{F}^X \bar{\mathcal{F}}^X \bigr]^{1/2\!} \sim \hat{m}_{3/2}^w m_P, \qquad \qquad \qquad f_\rho = \bigl[ \mathcal{K}_{\rho \bar{\rho}} \mathcal{F}^\rho \bar{\mathcal{F}}^\rho \bigr]^{1/2\!} \sim \dfrac{1}{\mathcal{V}^{2/3}} \, \hat{m}_{3/2}^w m_P,
\end{equation*}
although the anti-D3-brane uplift energy and the KKLT-like K\"{a}hler modulus potentials combine non-trivially with the gravitino mass-dependent contribution to give a near-zero cosmological constant. One may then define a supersymmetry-breaking scale in the low-energy theory as $\smash{m_{\mathrm{SUSY}} \sim f_X^{1/2}}$. Nevertheless, for both the K\"{a}hler modulus and the open-string sector, the orders of magnitude of the mass splittings read
\begin{equation*}
    \hat{m}^{w}_{\mathcal{V}} \sim \mathcal{V}^{2/3} \, \hat{m}^{w}_{3/2}, \qquad \qquad \qquad m^{\mathrm{open}}_{\mathrm{soft}} \sim \hat{m}_{3/2}^w.
\end{equation*}
So, even though there is no order parameter able to restore supersymmetry for the anti-D3-brane, the mass-splittings are not at the scale $\smash{m_s^w}$ or $m_{\mathrm{SUSY}}$, but rather they are fixed by the gravitino mass $\smash{\hat{m}_{3/2}^w}$ in the stabilised model: as usual, the canonical normalisation in physical units sets the volume modulus mass at a slightly volume-enhanced gravitino scale, whereas for the open-string contributions the scale is immediately set at the scale $\smash{m_{\mathrm{soft}} \sim m_{\mathrm{SUSY}}^2 / m_P \sim \hat{m}_{3/2}^w}$ by the mediation of gravity. Moreover, for the low-energy bifundamental scalars, this scale is further reduced by cancellations at leading order and they are the lightest (exotic) visible particles.

To end, it is worthwhile to stress that the particle spectra discussed here represent the generic low-energy effective theory corresponding to intersecting anti-D3-/D7-branes at an orbifold-like singularity, located at the tip of a strongly warped throat in a Calabi-Yau orientifold flux compactification, with the K\"{a}hler modulus stabilised in a KKLT-like framework. An explicit and globally consistent realisation of such constructions is left for future work.

\section{Conclusions} \label{conclusions}

This article has developed the supergravity description for the low-energy effective field theory of intersecting anti-D3-/D7-brane systems on orbifold singularities at the tip of warped throats, in stabilised type IIB Calabi-Yau orientifold flux compactifications.  Such string configurations could plausibly provide a realisation of the gauge and matter sectors of the Standard Model of Particle Physics, along with a rich hidden sector, with a geometric origin for large hierarchies of scales and a non-standard realisation of supersymmetry breaking. The anti-D3-brane degrees of freedom realise the bulk $\mathcal{N}_4=1$ supersymmetry only non-linearly, and thus break supersymmetry spontaneously, with the goldstino corresponding to the neutral massless gaugino that is always present.  When the branes are placed on orbifold singularities, moreover, the anti-D3-/D7-brane intersecting fermions and bosons transform in different bifundamental representations of the gauge groups; thus they in no way resemble superpartners.  New descriptions are therefore necessary, namely non-linear supergravity using constrained superfields. The focus of the article has been on the main distinctive features of these novel non-supersymmetric scenarios and their low energy descriptions, while the realisation of globally consistent concrete models is left for future studies.

The paper began by reviewing the properties of warped flux compactifications in section \ref{warped IIB closed-string sector}.  In particular, for strongly warped throats and bulk volumes that are not too large, i.e. satisfying eqn. (\ref{localisation condition}), bulk fields tend to dynamically localise near the tip of the throat, where energy scales are suppressed due to a gravitational redshift.  In order to have a 4-dimensional gravitino localised in the bulk, with Planck-suppressed couplings to match those of the graviton, as expected in supergravity, special fluxes satisfying condition (\ref{supergravity condition}) have also been assumed. The strong warping can eventually be captured in the low-energy supergravity theory describing degrees of freedom at the bottom of the throat via a constant shift the K\"{a}hler potential by the redshift logarithm $\mathrm{ln} \, e^{2 A_0} = 2 A_0$ \cite{Burgess:2006mn}. 

Taking this highly warped flux background, the low-energy effective theory for a supersymmetric D3-/D7-brane system was reviewed in section \ref{warped D3- and D7-branes}. Two qualitatively different scenarios were considered: first with the D7-brane wrapping a 4-cycle extending from the tip along the throat into the bulk, second with the wrapped 4-cycle localised at the tip.  Moreover, in the first case, the D7-brane chiral supermultiplet may be localised in the bulk or at the tip, depending on its mass-sourcing fluxes.  The possibility of six-dimensional integrals being dominated by the warped throat or the bulk were also both considered. For the $33$- and $77$-states, the effective field theory for the light degrees of freedom can be found by simply matching the 4-dimensional interactions found via dimensional reduction with those obtained in linear supergravity (including soft-breaking terms in the presence of supersymmetry-breaking fluxes).  For the $37$- and $73$-states, further tools are necessary, in particular the allowed interactions can be inferred using the internal space symmetries \cite{Camara:2004jj}. The power of linear supergravity is that, having identified the K\"{a}hler potential, superpotential, gauge kinetic functions and Fayet-Iliopoulos terms by matching with a few dimensionally reduced interactions, the complete action necessary for supersymmetry can be inferred, including couplings to bulk moduli.

With these preparations, the low-energy description of anti-D3-/D7-branes at the bottom of warped throats in supersymmetric warped flux compactifications was worked out, first for Abelian setups in section \ref{warped anti-D3- and D7-branes} and then for non-Abelian stacks of branes on orbifold singularities in section \ref{overview on the extension to non-Abelian theories}. Despite supersymmetry breaking, the non-linear supergravity construction provides a useful framework for the low-energy theory, including the couplings with bulk fields. After identifying the appropriate constrained superfields to encapsulate the low-energy fields, their interactions were worked out, building on both the single anti-D3-brane case \cite{Cribiori:2019hod} and the supersymmetric D3-/D7-brane cases above.  Most of the interactions can be described within standard supergravity expansions with hidden-sector supersymmetry breaking and soft-breaking terms.  However, in the presence of constrained superfields where the constraint also fixes the auxiliary field in the multiplet in terms of the goldstino, the supergravity expansions are non-standard, and are computed in appendix \ref{LEEFT SUGRA in IIB compactifications with non-linearly realised SUSY}. Another consequence of the anti-D3-brane supersymmetry breaking is a few couplings involving intersecting states, which would follow from analogy with the supersymmetric D3-/D7-brane case, but do not appear to fit in to the non-linear supergravity expansions.  These can instead be realised via a new interaction proportional to the nilpotent goldstino superfield, i.e. an $X\bar{X}$-term \cite{Cribiori:2017laj}, which provides each coupling term by term, plus further interactions proportional to the goldstino and vanishing in the unitary gauge.  Although this somewhat weakens the power of the supergravity formulation, at least in the current understanding, the latter allows an embedding of bottom-up open string scenarios with brane supersymmetry breaking into fully stabilised compactifications, including perturbative and non-perturbative effects.  This is essential to understand their phenomenology and cosmology. 

To this end, the D-brane setups were embedded in the KKLT-scenario, with the anti-D3-branes providing both gauge and matter sectors as well as the anti-D3-brane uplift to Minkowski/de Sitter vacuum energy.  Attractively, the small bulk $(0,3)$-flux backgrounds, necessary to balance against non-perturbative effects and stabilise the K\"{a}hler modulus, also help satisfy condition (\ref{supergravity condition}) allowing a supergravity description \cite{Burgess:2006mn}. The technology developed can easily be applied to other moduli stabilisation scenarios, and less warped scenarios, outside the validity of eqns. (\ref{localisation condition}, \ref{supergravity condition}).

The low-energy effective actions thus found have several interesting features.  The complex structure, axio-dilaton, $77$-, and $\bar{3}\bar{3}$-sector chiral multiplets acquire would-be supersymmetric mass terms from $(2,1)$-fluxes, at a scale above the cut-off (as well as subleading soft-breaking masses from the anti-D3-brane supersymmetry breaking). Physically, this means that the open-string moduli corresponding to brane positions are stabilised at the tip of the throat.  Instead, fermionic $\bar{3}7$- and $7\bar{3}$-states remain massless and could provide the standard-like model visible sector, whilst scalar visible sector exotic $\bar{3}7$- and $7\bar{3}$-states -- in distinct bifundamental representations – always have (would-be) soft-breaking masses, due to the anti-D3-brane and volume modulus supersymmetry breaking.  Because the latter is suppressed by no-scale-like cancellations, $\alpha'$-corrections (positive-definite masses) and anomaly mediation (tachyonic masses) can set the scale of the exotic scalar masses \cite{Aparicio:2015psl}, and which contribution wins depends on the parameter choices.  Moduli stabilisation and anomaly mediation also provide mass terms for the $77$- and $\bar{3}\bar{3}$-sector gaugini. As well as the mass scales, the leading supersymmetric and soft-breaking bilinear and trilinear couplings have all been computed. The visible $\bar{3}\bar{3}$- and hidden $77$-sector gauge couplings are fixed by the string coupling. All this is spelled out in section \ref{analysis of the mass hierarchies}.

As well as the light visible sector (standard-like model gauge fields and fermions, and scalar exotics), and a light hidden gauge sector plus matter, when embedding in KKLT-like scenarios for K\"{a}hler modulus stabilisation, the volume modulus and gravitino remain in the effective field theory, whereby cosmological bounds on the gravitino constrain the parameter space. Notice that the KKLT small parameter $\ab W_0 \ab$ implies a small gravitino mass, which is then further reduced by warping. Although the precise mass scales are model-dependent, the pattern of masses and their parametric dependence on the warp factor, volume and string-coupling are fairly universal within the KKLT scenario. Whilst a thorough phenomenological study, including renormalisation-group flows of the scales, is beyond the scope of this paper, if the warping is too strong, the gravitino mass $\smash{\hat{m}_{3/2}^w \sim (g_s^{3/2} e^{A_0} / \mathcal{V}^{2/3}) \, m_P \, e^{A_0}}$ may be so suppressed as to be ruled out by the observational bounds that confirm the BBN-physics, while the exotic scalar masses $\smash{m_{\bar{3}7} \sim \hat{m}_{3/2}^w / \mathcal{V}^{1/2}}$ may be ruled out by observation in accelerators.  Conversely, weaker warping allows scales to be pushed far beyond current experimental bounds. \\

This work leads to several interesting and important open questions. First and foremost is a rigorous understanding of the extent to which non-linearly realised supersymmetry and strong warping can help resolve hierarchy problems like the gauge hierarchy. The presence of spontaneous supersymmetry breaking, and yet no scale at which the usual superpartners appear, is an intriguing feature of these scenarios. Recently there has been a great deal of interest towards non-supersymmetric constructions in string theory (see e.g. Refs. \cite{Blaszczyk:2014qoa, *Abel:2015oxa, *Ashfaque:2015vta, *Blaszczyk:2015zta, *Nibbelink:2015vha, *Nibbelink:2016lzi, *Florakis:2016ani, *Abel:2017rch, *GrootNibbelink:2017luf, *Abel:2017vos, *McGuigan:2019gdb, *Faraggi:2020wej}) and it is very compelling to understand the relation between the D-brane supersymmetry breaking considered here and other approaches in the literature.  See Ref. \cite{Cribiori:2020sct} for an upcoming work in this direction. 

From a model building point of view, it would be essential to build warped throats that allow viable singularities at their tip, and the presence of simple 4-cycles (like for instance the $\mathrm{K}3$-surface or the 4-torus $\mathrm{T}^4$) at their tip or along their length would then allow easy explicit dimensional reductions. Geometric constructions with warped throats hosting a 4-torus $\mathrm{T}^4$ at the tip and $\mathbb{Z}_3$-singularities are built in Ref. \cite{Cascales:2003wn}. It would be fruitful to extend the present work to anti-D3/D7-brane systems on more general toric singularities, such as in Refs. \cite{Verlinde:2005jr, Conlon:2008wa, Krippendorf:2010hj, Dolan:2011qu, Cicoli:2012vw}, at the tip of warped throats. Related work on the construction of throats with branes at singularities can already be found e.g. in Refs. \cite{Franco:2005fd, Krippendorf:2010hj, Dolan:2011qu, Cicoli:2012vw, Franco:2014hsa, Kallosh:2015nia, Garcia-Etxebarria:2015lif, Retolaza:2016alb} and on throats with wrapped D7-branes in Refs. \cite{Ouyang:2003df, Kuperstein:2004hy, Chen:2008jj}. Ultimately, globally consistent compactifications, with appropriate singularities, cycles and sources that fulfil RR-tadpole cancellation, should be constructed, to which the results presented here would apply.

Various possible instabilities arising from anti-D3-branes in flux backgrounds should also be explored, since this work has completely neglected the brane backreaction and the details of the complex structure modulus that governs the warp factor at the throat tip. In particular, as shown by Ref. \cite{Kachru:2002gs}, $p$ anti-D3-branes in the flux background of the KS-throat with $M$ units of RR-flux are metastable and long-lived for sufficiently small ratio $p/M$, with brane-flux decay occurring non-perturbatively via brane polarisation \`{a} la Myers \cite{Myers:1999ps} (for an overview of past debates on this picture, see Ref. \cite{Blaback:2019ucp}). Also, Ref. \cite{Bena:2018fqc} has shown that for a KS-throat the anti-D3-branes may induce a complex structure instability, depending on the amount of flux relative to the branes. It would be interesting to investigate these dynamics in other relevant throats and in the presence of orbifold singularities. Additionally, so far, world-volume fluxes on the D7-branes have been neglected for simplicity, though they can contribute interesting D-terms and F-terms. 

Once globally consistent, realistic constructions, approaching the standard model of particle physics have been identified, detailed phenomenological and cosmological studies would be possible.

\section*{Acknowledgments}
It is a pleasure to thank Niccolò Cribiori, Fernando Quevedo, Radu Tatar, Timm Wrase and Ivonne Zavala for helpful comments and discussions at several stages of this project.

\appendix

\section{Dimensional Reduction in Warped Compactifications} \label{appendix: warped dimensional reduction}
This section reviews the dimensional reduction of closed- and open-string sectors in warped compactifications.\footnote{See also Ref. \cite{Burgess:2020qsc} for a recent discussion of the scaling properties of the closed- and open-string effective theories in string compactifications.} It is meant to set the notation for the main text and to provide a review of how the scaling factors are obtained in warped dimensional reductions.

\subsection{Warped Closed-String Sector in Type IIB String Theory}
In type IIB compactifications, in principle the theory is formulated in terms of the string-frame metric $ds_{10}^2 = G_{MN} \, \de x^M \de x^N$. Given the gravitational coupling $2 \check{\kappa}_{10}^2 = l_s^8 / 2 \pi$, where the string length is $l_s = 2 \pi \sqrt{\alpha'}$, the 10-dimensional massless bosonic action reads \cite{Polchinski}
\begin{equation*}
\begin{split}
    S^{\mathrm{boson}}_{\mathrm{IIB}} = \dfrac{1}{2 \check{\kappa}_{10}^2} \int_{X_{1,9}} \biggl[ e^{- 2 \Phi} \biggl( R_{10} \star 1 + 4 \, \de \Phi \wedge \star \, \de \Phi - \dfrac{1}{2} \, H_3 \wedge \star \, H_3 \biggr) & \biggr] \\
    + \dfrac{1}{2 \check{\kappa}_{10}^2} \int_{X_{1,9}} \biggl[ - \dfrac{1}{2} \, F_1^{s} \wedge \star \, F_1^{s} - \dfrac{1}{2} \, \tilde{F}_3^{s} \wedge \star \, \tilde{F}_3^{s} - \dfrac{1}{4} \, \tilde{F}_5^{s} \wedge \star \, \tilde{F}_5^{s} & \biggr] \\
    - \dfrac{1}{4 \check{\kappa}_{10}^2} \int_{X_{1,9}} C_4^{s} \wedge H_3 \wedge \, \tilde{F}_3^{s} &,
\end{split}
\end{equation*}
where the NSNS- and RR-sector field-strength tensors are respectively defined as $H_3 = \de B_2$ and $\tilde{F}^{s} = \de C^{s} - H_3 \wedge C^{s}$. Also, the string coupling is $g_s = e^{\langle \Phi \rangle}$, where $\Phi$ is the dilaton. Then, the 10-dimensional Einstein frame metric is defined as
\begin{equation*} \label{string frame}
\hat{g}_{MN} = e^{- (\Phi - \langle \Phi \rangle)/2} G_{MN},
\end{equation*}
which can be expressed more easily in terms of the shifted dilaton $\sdil = \Phi - \langle \Phi \rangle$. In this way, in a more compact notation, the low-energy effective action can eventually be written as \cite{Polchinski:2000uf}
\begin{equation}
    \begin{split}
        S^{\mathrm{boson}}_{\mathrm{IIB}} = \dfrac{1}{2 \hat{\kappa}_{10}^2} \int_{X_{1,9}} \biggl[ \hat{R}_{10} \hat{*} 1 - \dfrac{\de \tau \wedge \hat{*} \, \de \bar{\tau}}{2 \, (\mathrm{Im} \, \tau)^2} - \dfrac{G_3 \wedge \hat{*} \, \bar{G}_3}{2 \, \mathrm{Im} \, \tau} - \dfrac{1}{4} \, \tilde{F}_5 \wedge \hat{*} \, \tilde{F}_5 & \biggr] \\
        - \dfrac{1}{4 \hat{\kappa}_{10}^2} \int_{X_{1,9}} C_4 \wedge H_3 \wedge \tilde{F}_3 & ,
    \end{split}
\end{equation}
where the physical $10$-dimensional gravitational coupling is $2 \hat{\kappa}_{10}^2 = g_s^2 l_s^8 / 2 \pi$ and the RR-fields have been rescaled as $C = g_s C^{s}$. Further, the axio-dilaton and the complexified 3-form flux have been defined as $\tau = C_0 + i \, e^{-\sdil}$ and $G_3 = \tilde{F}_3 - i \, e^{-\sdil} H_3$, respectively.

In a Calabi-Yau orientifold compactification with non-zero background fluxes, the field equations imply a non-trivial warp factor \cite{Giddings:2001yu, DeWolfe:2002nn}. Following Refs. \cite{Giddings:2005ff, Frey:2008xw}, the volume-controlling real K\"{a}hler modulus $c = c (x)$ appears as a shift in the warp factor $e^A = e^{A(y)}$, leading to the definition of the generalised warp factor
\begin{equation*}
    e^{-4 A[c(x),y]} = e^{-4A(y)} + c(x),
\end{equation*}
with the 10-dimensional Einstein-frame metric taking the form
\begin{equation*}
    ds_{10}^2 = \dfrac{1}{[e^{-4A} + c]^{1/2}} \, \breve{g}_{\mu \nu} \, \de x^\mu \de x^\nu + [e^{-4A} + c]^{1/2} \, \breve{g}_{m n} \, \de y^m \de y^n.
\end{equation*}
As discussed by Ref. \cite{Frey:2008xw}, one can Weyl rescale this to the 4-dimensional Einstein frame, while also introducing a compensator field $b=b(y)$ that is necessary to solve the Einstein equations, with the full metric reading
\begin{equation} \label{4-dimensional Einstein frame metric}
    ds_{10}^2 = \dfrac{e^{2 \Omega} \, \vevc^{3/2}}{[e^{-4A} + c]^{1/2}} \, (g_{\mu \nu} \, \de x^\mu \de x^\nu + 2 \der_\mu c \, \der_m b \, \de x^\mu \de y^m ) + [e^{-4A} + c]^{1/2} \, g_{mn} \, \de y^m \de y^n.
\end{equation}
In particular, in the Weyl rescaling one has the K\"{a}hler modulus-dependent factor
\begin{equation*}
    e^{2 \Omega} = \dfrac{\displaystyle \int_{Y_6} \de^6 y \, \sqrt{g_{6}}}{\displaystyle \int_{Y_6} \de^6 y \, \sqrt{g_{6}} \; [e^{-4 A} + c]}
\end{equation*}
and for generality also an arbitrary constant $\vevc^{3/2}$ has been introduced, which in this case will be chosen as $\vevc \sim \langle c \rangle$ \cite{Conlon:2005ki}.\footnote{\label{constant Weyl factor} Notice that the canonically normalised masses in Planck units are independent of constant Weyl rescalings and most references work with $\vevc=1$.} The warp factor has the following behaviours:
\begin{itemize}
    \item in the infrared region of the throat $\tau_6$, the background warp factor is much larger than the volume modulus, that is $e^{-4A}(y \in \tau_6) \gg \langle c \rangle \gg 1$ so that
    \begin{equation*}
        e^{-\langle 4A[c] \rangle} \sim e^{-4A}, \qquad \textrm{for $y \in \tau_6$};
    \end{equation*}
    \item in the bulk region of the compact space, the background warp factor is negligible, that is  $e^{-4A}(y \in Y_6 \backslash \tau_6) \ll c$, so
    \begin{equation*}
        e^{-\langle 4A[c] \rangle} \sim \langle c \rangle, \qquad \textrm{for $y \in Y_6 \backslash \tau_6$}.
    \end{equation*}
\end{itemize}

The dimensional reduction of the closed-string sector action, to find the 4-dimensional low-energy effective theory corresponding to the flux compactification, is now reviewed for the most relevant degrees of freedom. Following the very definition of the 4-dimensional Einstein frame, the type IIB Einstein-Hilbert action becomes
\begin{equation*}
\begin{split}
    S^{\mathrm{IIB}}_{\mathrm{EH}} = \dfrac{1}{2 \hat{\kappa}_{10}^2} \int_{X_{1,9}} \! \de^{10} x \, \sqrt{-\detr{\hat{g}_{10}}} \; \hat{R}_{10} = \dfrac{1}{2 \kappa_{4}^2} \int_{X_{1,3}} \! \de^4 x \, \sqrt{- \detr{g_{4}}} \; R_{4} + \delta S^{\mathrm{IIB}}_{\mathrm{EH}},
\end{split}
\end{equation*}
with the 4-dimensional gravitational coupling defined as
\begin{equation}
    2 \kappa_4^2 = \dfrac{2 \hat{\kappa}_{10}^2}{\vevc^{3/2} l_s^6 \Vzero} =  \dfrac{g_s^2 l_s^2}{2 \pi \vevc^{3/2} \Vzero}
\end{equation}
and the term $\delta S^{\mathrm{IIB}}_{\mathrm{EH}}$ standing for the internal curvature and other derivative terms, emerging from the remainder of the Ricci scalar, which provide contributions to the kinetic terms and the scalar potential for the geometric moduli. In particular, the K\"{a}hler modulus kinetic term is reproduced by means of the K\"{a}hler potential \cite{Frey:2008xw}
\begin{equation*}
\kappa_{4}^2 \hat{K}(\rho, \bar{\rho}) = - 3 \, \mathrm{ln} \, \bigl[ - i (\rho - \bar{\rho}) + 2 c_0 \bigr],
\end{equation*}
with $c_0 = \Vw/ \Vzero$, where the complexified K\"{a}hler modulus $\rho$ is defined as
\begin{equation*}
    \rho(x) = \chi(x) + i \, c (x),
\end{equation*}
with $\chi$ being the 4-form axion. The description of the other closed-string sector fields follows with specific features determined by warping effects \cite{DeWolfe:2002nn, Douglas:2007tu}.
\begin{itemize}
    \item For the axio-dilaton $\tau$, it is immediate to check that the kinetic term is
    \begin{equation*}
        \begin{split}
            S_{\textrm{axio-dilaton}} & = \dfrac{1}{2 \hat{\kappa}_{10}^2} \int_{X_{1,9}} \! \de^{10} x \, \sqrt{-\detr{\hat{g}_{10}}} \; \biggl[ - \dfrac{1}{2 \, (\mathrm{Im} \, \tau)^2} \, \hat{g}^{MN} \der_M \tau \der_N \bar{\tau} \biggr] \\
            & = \dfrac{1}{2 \kappa_{4}^2} \int_{X_{1,3}} \! \de^4 x \, \sqrt{- \detr{g_{4}}} \biggl[ - \dfrac{1}{2 \, (\mathrm{Im} \, \tau)^2} \, g^{\mu \nu} \, \der_\mu \tau \der_\nu \bar{\tau} \biggr],
        \end{split}
    \end{equation*}
    which is reproduced by the usual K\"{a}hler potential $\kappa_4^2 \hat{K} (\tau, \bar{\tau}) = - \mathrm{ln} \, [-i (\tau - \bar{\tau})]$.
    \item For the complex structure moduli $u^\alpha$, with $\alpha=1,\dots,h_-^{2,1}$, the dimensional reduction is more involved. In particular, one needs the quantities
    \begin{equation*}
    \omega_w = \int_{Y_6} e^{-4A} \Omega \wedge \bar{\Omega}, \qquad \qquad \qquad \qquad \hat{K}_{\alpha \bar{\beta}} = -\dfrac{1}{\omega_w} \int_{Y_6} e^{-4A} \, \chi_\alpha \wedge \bar{\chi}_\beta,
    \end{equation*}
    which provide the warped version of the complex structure moduli K\"{a}hler potential, $\kappa_4^2 \hat{K} (u , \bar{u}) = - \mathrm{ln} \, [-i \omega_w]$, and the explicit K\"{a}hler metric \cite{Candelas:1990pi, DeWolfe:2002nn}, where $\Omega$ and $\chi_\alpha$ are the unwarped harmonic 3-form and $(2,1)$-form basis, respectively.
\end{itemize}

To have a complete supergravity formulation, one must also match the scalar potential that arises from the dimensional reduction. The following calculation only captures the axio-dilaton and complex structure moduli potential as it neglects the details of the coupling with the warp factor, the volume modulus and the compensator field. It is just meant to argue the emergence of the GVW-superpotential \cite{Gukov:1999ya} and to fix the overall constants. The functional dependence of the scalar potential is set by the 3-form term as the remaining terms from the Einstein-Hilbert and 5-form actions can be combined with the 3-form action, cancelling the contribution from imaginary self-dual fluxes $G_3^-$ and leaving pure imaginary anti-self-dual fluxes $G_3^+$ \cite{Giddings:2001yu, DeWolfe:2002nn}, with
\begin{equation*}
    G_3^\pm = \dfrac{1}{2} \, (1 \pm i *_6) G_3.
\end{equation*}
Refs. \cite{Giddings:2005ff, Frey:2008xw} show that if the warp factor $e^{-4A}$ solves the field equations, so does the shifted warp factor $e^{-4A}+c$. Assuming then for simplicity the background value for the volume $\langle c \rangle$, one can express this 10-dimensional potential in terms of the 4-dimensional Einstein-frame metric, i.e.
\begin{equation*}
\begin{split}
    S_{\textrm{3-form}} & = \dfrac{1}{2 \hat{\kappa}_{10}^2} \int_{X_{1,9}} \! \de^{10} x \, \sqrt{-\detr{\hat{g}_{10}}} \; \biggl[ - \dfrac{1}{12 \, \mathrm{Im} \, \tau} \, G_3^+ \, \hat{\cdot} \, \bar{G}_3^+ \biggr] \\
    & = \dfrac{\vevc^3}{2 \hat{\kappa}_{10}^2} \int_{X_{1,3}} \! \de^4 x \, \sqrt{- \detr{g_{4}}} \int_{Y_6} \de^6 y \, \sqrt{\detr{g_{6}}} \; \biggl[ - \dfrac{e^{\langle 4\Omega \rangle + \langle 4 A[c] \rangle}}{12 \, \mathrm{Im} \, \tau} \, G_3^+ \cdot \bar{G}_3^+ \biggr].
\end{split}
\end{equation*}
The most interesting case to consider is the one where integrations are dominated by the throat region $\tau_6$, in which $e^{-\langle 4A[c] \rangle} \sim e^{-4A}$. Because the GKP field equations require the imaginary anti-self-dual 3-forms $e^{4A} G_3^+$ to be harmonic \cite{Giddings:2001yu, DeWolfe:2002nn}, without loss of generality one can focus on the $(3,0)$-component and expand it as
\begin{equation*}
e^{4A} \, G_{(3,0)} = \dfrac{1}{\omega_w} \, \Omega \int_{\tau_6} G_3 \wedge \bar{\Omega},
\end{equation*}
so that the action can be written as
\begin{equation*}
    S_{\textrm{3-form}} = \dfrac{\vevc^3}{2 \hat{\kappa}_{10}^2} \int_{X_{1,3}} \!\! \de^4 x \sqrt{- \detr{g_{4}}} \int_{\tau_6} \Biggl[ - \dfrac{i}{2} \, \dfrac{e^{\langle 4 \Omega \rangle - 4 A}}{\mathrm{Im} \, \tau \, \omega_w^2} \, \Omega \wedge \bar{\Omega} \, \biggl[ \int_{\tau_6} G_3 \wedge \bar{\Omega} \biggr] \! \biggl[ \int_{\tau_6} \bar{G}_3 \wedge \Omega \biggr] \Biggr].
\end{equation*}
The integral over the internal space is now easily seen to be
\begin{equation} \label{lambda}
    \lambda \sim \int_{Y_6} e^{-4A} \, \Omega \wedge \bar{\Omega} = \omega_w \sim \omega_w \, \dfrac{\Vw}{\Vzero} \, e^{\langle 2\Omega \rangle},
\end{equation}
where an approximate unit factor has been introduced in the final relation, for convenience in the comparison with the supergravity action below.  At the end of the day, the 3-form action is (the numerical factor can be determined by properly taking into account the axio-dilaton and 5-form contributions to the scalar potential \cite{DeWolfe:2002nn})
\begin{equation*}
\begin{split}
    S_{\textrm{3-form}} & = \dfrac{\vevc^3}{2 \hat{\kappa}_{10}^2} \int_{X_{1,3}} \! \de^4 x \, \sqrt{- \detr{g_{4}}} \; \Biggl[ - \dfrac{i \, e^{\langle 6 \Omega \rangle}}{\mathrm{Im} \, \tau \, \omega_w} \, \dfrac{\Vw}{\Vzero} \, \biggl[ \int_{Y_6} G_3 \wedge \bar{\Omega} \biggr] \biggl[\int_{Y_6} \bar{G}_3 \wedge \Omega \biggr] \Biggr] \\
    & = \dfrac{1}{2 \kappa_{4}^4} \dfrac{g_s^2}{4 \pi} \int_{X_{1,3}} \! \de^4 x \, \sqrt{- \detr{g_{4}}} \; \Biggl[ - \dfrac{i \, e^{\langle 6 \Omega \rangle}}{\mathrm{Im} \, \tau \, \omega_w} \, \dfrac{\Vw}{[\Vzero]^3} \, \dfrac{1}{l_s^4} \, \biggl[ \int_{Y_6} G_3 \wedge \bar{\Omega} \biggr] \biggl[ \int_{Y_6} \bar{G}_3 \wedge \Omega \biggr] \Biggr]
\end{split}
\end{equation*}
The last step takes into account the definition of the 4-dimensional Planck units while keeping the bulk integrals scaled with the appropriate string length factors (recalling the scalings $G_3 \sim l_s^2$ and $\Omega \sim 1$).  This result gives a way to understand how to insert the volume and warped-volume factors in the effective supergravity formulation whereby the K\"{a}hler and superpotential of eqns. (\ref{closed-strings - Kaehler potential}, \ref{closed-strings - superpotential}) reproduce it exactly.

A similar analysis can be done with the opposite approximation that bulk integrals dominate over throat integrals, which leads to the unwarped limit. The calculation follows analogously but it is easier since the warping in the integrations is irrelevant, i.e. $\Vw \sim \Vzero$ and $\omega_w \sim \omega_{(0)} = \int_{Y_6} \Omega \wedge \bar{\Omega}$. In more detail, one may start from the 10-dimensional potential written above noticing the identities $e^{4A[c]}=1/c=e^{-4u}$ and $e^{2\Omega} = 1/c = e^{-4u}$, and reduce it along the same lines, with the 3-form flux $G_3^+$ being harmonic. Alternatively, formally this limit can be found by setting $e^{4A} = 1$ in all the final integrated expressions, so that $\Vw = \Vzero$ and $\omega_w = \omega_{(0)}$. One obtains the famous results of Refs. \cite{Giddings:2001yu, Jockers:2004yj}. The warped expressions are always kept in the main text for the sake of generality.

\subsection{General D-brane Action} \label{D-brane actions}
As is well-known, the uncompactified $(p+1)$-dimensional worldvolume theory of a stack of $n$ coincident D$p$- or anti-D$p$-branes consists of the following massless degrees of freedom:
\begin{itemize}
    \item from the NS-sector, a vector $A_{\alpha}$ which gauges the non-Abelian gauge group $\mathrm{U}(n)$ and $9-p$ scalars $\phi^{\dot{m}}$ in the adjoint representation of the group $\mathrm{U}(n)$, with the indices $\alpha$ and $\dot{m}$ respectively running over the worldvolume longitudinal and transverse directions, meaning $\alpha = 0,\dots,p$ and $\dot{m}=p+1,\dots,9$;
    \item from the R-sector, some spinors $\psi^A$ in the adjoint representation of the group $\mathrm{U}(n)$, where the family index $A$ counts the number of $(p+1)$-dimensional spinors descending from a single 10-dimensional Majorana-Weyl spinor.
\end{itemize}
The difference between branes and anti-branes is their charge under the RR-fields, which is $q=1,-1$, respectively.

The effective action describing the massless degrees of freedom of coincident D-branes is a non-Abelian generalisation of the effective action describing a single D-brane \cite{Myers:1999ps}. In detail, it is the summation of a Dirac-Born-Infeld and a Chern-Simons action, i.e.
\begin{equation}
    S^{\mathrm{D}p} = S^{\mathrm{D}p}_{\mathrm{DBI}} + S^{\mathrm{D}p}_{\mathrm{CS}}.
\end{equation}
The string frame description will be reviewed first as this is what is usually suitable for deriving effective actions, and then everything will be re-expressed in the Einstein frame.  The embedding function of the D$p$-brane worldvolume $\mathcal{W}_{1,p}$ into the $10$-dimensional spacetime $X_{1,9}$ will be represented by
\begin{equation*}
\varphi: \;\; \mathcal{W}_{1,p} \, \hookrightarrow \, X_{1,9},
\end{equation*}
where the pull-back of a $10$-dimensional vector $v = v_M \, \de x^M$ is defined as $v_{\alpha} = (\varphi_* v)_\alpha = v_M \, \der_\alpha x^M$, and similarly for tensors of arbitrary rank.

For brevity, only the bosonic action is discussed below. An analysis of the general D$p$-brane fermionic action can be found in Refs. \cite{Marolf:2003ye, Marolf:2003vf, Martucci:2005rb} (see also Ref. \cite{Grana:2002tu}).

\subsubsection{Dirac-Born-Infeld Action}
In the string frame, the Dirac-Born-Infeld term for a stack of D$p$-branes at a generic smooth point in the internal manifold takes the form\footnote{Notice that one can define the brane tension as $\tau_{\mathrm{D}p} = T_p/g_s$ as a direct consequence of writing the dilaton factor in the action in terms of the shifted dilaton field $\sdil = \Phi - \langle \Phi \rangle$.}
\begin{equation*}
S_{\mathrm{DBI}}^{\mathrm{D}p} = - T_p \int_{\mathcal{W}_{1,p}} \!\! \de^{p+1} \xi \; \mathrm{str} \biggl[ \; e^{-\Phi} \, \sqrt{- \mathrm{det} \, \bigl[ \Gamma_{\alpha\beta} \bigr] \cdot \mathrm{det} \, \bigl[ \tensor{Q}{^{\dot{n}}_{\dot{m}}} \bigr]} \biggr],
\end{equation*}
where $T_p = 2 \pi / l_s^{p+1}$ is the D$p$-brane tension. Also, one has the rank-2 tensor
\begin{equation*}
\Gamma_{\alpha\beta} \equiv E_{\alpha \beta} + E_{\alpha {\dot{m}}} \, \tensor{\bigl( Q^{-1} - 1 \bigr)}{^{\dot{m}}_{\dot{l}}} \, E^{{\dot{l}}{\dot{n}}} \, E_{{\dot{n}} \beta} + 2 \pi \alpha' F_{\alpha \beta},
\end{equation*}
along with the combination of the string frame metric tensor and the 2-form NSNS-field, $E_{MN} = G_{MN} + B_{MN}$, with $E_{\alpha\beta}$ being its pull-back on the worldvolume, as well as the purely non-Abelian rank-$(1,1)$ tensor
\begin{equation*}
\tensor{Q}{^{\dot{n}}_{\dot{m}}} = \delta^{\dot{n}}_{\dot{m}} + 2 \pi i \alpha' \, \bigl[ \phi^{\dot{n}}, \phi^{\dot{k}} \bigr] \, E_{{\dot{k}}{\dot{m}}}.
\end{equation*}
The determinant `$\mathrm{det}$' is with respect to spacetime indices, while the trace `$\mathrm{str}$' is the symmetrised trace over the gauge group indices such that the Lie matrix-valued terms $F_{\alpha \beta}$, $D_\alpha \phi^{\dot{m}}$ and $[\phi^{\dot{m}}, \phi^{\dot{n}}]$ are treated as commuting (no other terms are treated as commuting).

One can write the action in the Einstein frame by redefining the metric and NSNS-field combination as $\hat{e}_{MN} = e^{-\sdil/2} \, E_{MN} = \hat{g}_{MN} + e^{-\sdil/2} \, B_{MN}$. Elementary operations then reveal the action to take the form
\begin{equation} \label{DBI_action}
S_{\mathrm{DBI}}^{\mathrm{D}p} = - \tau_{\mathrm{D}p} \int_{\mathcal{W}_{1,p}} \!\! \de^{p+1} \xi \; \mathrm{str} \, \biggl[ e^{\, (p-3) \, \sdil/ 4} \, \sqrt{- \mathrm{det} \, \bigl[ \hat{\gamma}_{\alpha\beta} \bigr] \cdot \mathrm{det} \, \bigl[ \tensor{Q}{^{\dot{n}}_{\dot{m}}} \bigr]} \biggr],
\end{equation}
where the physical D$p$-brane tension turns out to be $\tau_{\mathrm{D}p} = 2 \pi / g_s l_s^{p+1}$. Also, one redefines the rank-2 tensor as
\begin{equation*}
\hat{\gamma}_{\alpha\beta} \equiv \hat{e}_{\alpha \beta} + \hat{e}_{\alpha {\dot{m}}} \, \tensor{\bigl( Q^{-1} - 1 \bigr)}{^{\dot{m}}_{\dot{l}}} \, \hat{e}^{{\dot{l}{\dot{n}}}} \, \hat{e}_{{\dot{n}} \beta} + 2 \pi \alpha' \, e^{-\sdil/2} \, F_{\alpha \beta},
\end{equation*}
whilst the rank-$(1,1)$ tensor is still
\begin{equation*}
\tensor{Q}{^{\dot{n}}_{\dot{m}}} = \delta^{\dot{n}}_{\dot{m}} + 2 \pi i \alpha' \, e^{\sdil/2} \, \bigl[ \phi^{\dot{n}}, \phi^{\dot{k}} \bigr] \, \hat{e}_{{\dot{k}}{\dot{m}}}.
\end{equation*}

\subsubsection{Chern-Simons Action}
The Chern-Simons action is the same both in the string and the 10-dimensional Einstein frame up to the rescaling of the RR-fields and it takes the form
\begin{equation} \label{CS_action}
S_{\mathrm{CS}}^{\mathrm{D}p} = q \, \tau_{\mathrm{D}p} \int\displaylimits_{\mathcal{W}_{1,p}} \mathrm{str} \, \biggl\lbrace \biggl[\varphi_* \biggl( e^{2 \pi i \alpha' \mathrm{i}_{\dot{\phi}} \mathrm{i}_{\dot{\phi}}} \biggl( \sum_{l=0}^{4} C_{2l} \wedge e^{B_2} \biggr) \biggr) \biggr] \wedge e^{2 \pi \alpha' F_2} \biggr\rbrace,
\end{equation}
where $\mathrm{i}_{\dot{\phi}}$ denotes the interior product with the vector field $\phi^{\dot{m}}$, i.e. for a general $n$-form
\begin{equation*}
    \mathrm{i}_{\dot{\phi}} A_n = \dfrac{1}{(p-1)!} \phi^{\dot{m}} \, A_{{\dot{m}} \, M_1 \dots M_{n-1}} \, \de x^{M_1} \wedge \dots \de x^{M_{n-1}}.
\end{equation*}

\subsubsection{Further Remarks}
One typically chooses to work in the so-called static gauge, in which, given the expansion parameter $\al = l_s^2/2\pi$ for ease of notation, the brane position is parametrised as
\begin{equation*}
    X^{\dot{\mu}} (\xi) = \delta^{\dot{\mu}}_\alpha \, \xi^\alpha, \quad \quad \quad \quad \quad \quad Y^{\dot{m}} (\xi) = y_0^{\dot{m}} + \al \phi^{\dot{m}} (\xi).
\end{equation*}
In detail, $y_0^{\dot{m}}$ are the background brane positions in the Dirichlet directions while the terms $\delta Y^{\dot{m}} = \al \phi^{\dot{m}}$ represent fluctuations thereof. Moreover, the notation is such that:
\begin{itemize}
    \item indices $\dot{\mu}$ span both the 4-dimensional spacetime and the $p-3$ internal directions wrapped by the D$p$-brane, i.e. $\dot{\mu}=\mu, m'$, with $m'=4,\dots,p$;
    \item indices $\dot{m}$ span the internal directions which are not wrapped, i.e. $\dot{m}=p+1,\dots,9$.
\end{itemize}
Under these premises, one has to take care of the following facts.
\begin{itemize}
    \item The DBI- and CS-actions involve pull-backs of 10-dimensional fields onto the brane worldvolume: these are a generalised version of the standard pull-back as they involve non-Abelian fields. For instance the non-Abelian pull-back on the worldvolume of a 1-form $v = v_M \, \de x^M$ is
    \begin{equation*}
    \varphi_* v = v_{\dot{\mu}} \, \delta^{\dot{\mu}}_\alpha \, \de \xi^\alpha + \al \, \nabla_\alpha \phi^{\dot{m}} \, v_{\dot{m}} \, \de \xi^\alpha,
    \end{equation*}
    where $\nabla_\alpha$ is the standard gauge covariant derivative, as a generalisation of the standard pull-back expression involving $\der_\alpha y^m$. Generalisations to $n$-forms are immediate.
    \item Fields on the brane worldvolume must be expressed as functions of the coordinates $\xi^\alpha$, of course. A generic 10-dimensional function $f = f(x^M)$ can be written as a non-Abelian Taylor expansion on the worldvolume, i.e.
    \begin{equation*}
    f (x^{\dot{\mu}},y^{\dot{m}}) = \sum_{k=0}^{+\infty} \dfrac{\al^k}{k!} \, \phi^{{\dot{m}}_1} \phi^{{\dot{m}}_2} \dots \phi^{{\dot{m}}_k} \, \der_{{\dot{m}}_1} \der_{{\dot{m}}_2} \dots \der_{{\dot{m}}_k} \, f (x^{\dot{\mu}}, y^{\dot{m}}_0),
    \end{equation*}
    which accounts for the fluctuations of the D$p$-brane in terms of the non-Abelian displacements from the original position $y_0^{\dot{m}}$.
\end{itemize}

\subsubsection{D3, anti-D3- and D7-brane Kinetic and Mass Terms}
In the probe approximation, an explicit dimensional reduction of the D3- and anti-D3-brane action has been performed in Refs. \cite{Camara:2003ku, Grana:2003ek, McGuirk:2012sb, Bergshoeff:2015jxa, Cribiori:2019hod}, while the study of the D7-brane action can be found in Refs. \cite{Camara:2004jj, Jockers:2004yj, Marchesano:2008rg}. Most references work with the metric form
\begin{equation*}
    ds_{10}^2 = e^{2 A[c]} \, \breve{g}_{\mu \nu} \, \de x^\mu \de x^\nu + e^{-2 A[c]} \, \breve{g}_{m n} \, \de y^m \de y^n.
\end{equation*}
In this subsubsection the results are taken directly from such references. For a 4-dimensional theory, the worldvolume degrees of freedom must be reduced, and they are sensitive to the details of the wrapped $(p-3)$-cycle. It is also convenient to combine pairs of real scalars into single complex scalars as $\smash{\phi^{a} = \phi^{\dot{m}=2a+2} + i \, \phi^{\dot{m}=2a+3}}$, and similarly for the modulini.
\begin{itemize}
    \item For D3- and anti-D3-branes, the dimensional reduction proceeds in the same way except for the different interference between the DBI- and CS-actions due to the different RR-charge. All the terms evaluated at the brane location carry a symbol '$0$'.
    
    First of all one finds the cosmological constant contribution
    \begin{equation*}
        S_\Lambda^{\mathrm{D3}_q} = - (1-q) \tau_{\mathrm{D3}} \int \de^4 x \, \sqrt{- \breve{g}_{4}} \; e^{4A_0[c]},
    \end{equation*}
    which explains the anti-D3-brane uplift energy.
    
    Further, the pure scalar kinetic and mass terms turn out to be (there are also bilinear $\phi^a \phi^b$-couplings with the same scaling as the mass terms)
    \begin{equation*}
    S^{\mathrm{D3}_q}_{\textrm{scalars}} = - \tau_{\mathrm{D3}} \al^2 \int_{X_{1,3}} \! \de^4 x \, \sqrt{- \breve{g}_{4}} \; \biggl[ \breve{g}_{a \bar{b}}^{0} \, \breve{g}^{\mu \nu} \, \breve{\nabla}_\mu \phi^a \breve{\nabla}_\nu \bar{\phi}^b + [\nabla_a \nabla_{\bar{b}} (e^{4A[c]} - q \alpha)]_{0} \, \phi^a \bar{\phi}^b \biggr].
    \end{equation*}
    Following the GKP-equations \cite{Giddings:2001yu, DeWolfe:2002nn,Giddings:2005ff}, the anti-D3-brane scalars are massive for imaginary self-dual $(2,1)$- and $(0,3)$-fluxes, whereas for D3-branes they are massless.
    
    For the modulini, one finds the kinetic and mass action\footnote{The dimensional reduction of the 10-dimensional Majorana-Weyl spinor to the 4-dimensional Weyl spinors is the same as in Ref. \cite{McGuirk:2012sb} since $\smash{e^{-4 A_0[c]} \sim e^{- 4 A_0}}$ for branes at the tip of the throat.}
    \begin{equation*}
        S^{\mathrm{D3}_q}_{\textrm{modulini}} = - i \tau_{\mathrm{D3}} \al^2 \int_{X_{1,3}} \! \de^4 x \, \sqrt{- \breve{g}_{4}} \; \biggl[ \breve{g}^{0}_{a \bar{b}} \, \breve{\bar{\psi}}^b \breve{\bar{\sigma}}^\mu \breve{\nabla}_\mu \breve{\psi}^a + \bigl( m^{(q)}_{\breve{\psi}^a \breve{\psi}^b} \breve{\psi}^a \breve{\psi}^b + \mathrm{c.c.} \bigr) \biggr].
    \end{equation*}
    For anti-D3-branes, the modulini masses are purely sourced by $(2,1)$-fluxes and read
    \begin{equation*}
        m^{(q=-1)}_{\breve{\psi}^a \breve{\psi}^b} = - \dfrac{1}{4} \, e^{4A_0[c] + \sdil/2} \, \breve{g}^{0}_{\overline{c}(a} \, l_s^3 \breve{\Omega}^{0}_{b)de} \, (\bar{G}_3^-)_{0}^{\breve{\bar{c}}\breve{d}\breve{e}},
    \end{equation*}
    while for D3-branes they are sourced by imaginary anti-self-dual $(1,2)$-fluxes. 
    
    One also finds the gauge vector action
    \begin{equation*}
        S^{\mathrm{D3}_q}_{\textrm{gauge}} = - \dfrac{\tau_{\mathrm{D3}} \al^2}{2} \int_{X_{1,3}} e^{-\sdil} \, F_2 \wedge \breve{*} F_2 + \dfrac{q \tau_{\mathrm{D3}} \al^2}{2} \int_{X_{1,3}} C_0 \, F_2 \wedge F_2.
    \end{equation*}
    The gaugino mass is sourced by $(0,3)$- and $(3,0)$-fluxes for anti-D3- and D3-branes, respectively. 
    \item For D7-branes, the reduction to a 4-dimensional action depends on the wrapped internal 4-cycle, so only the general features of bosons will be discussed. Let the 4-cycle be spanned by the coordinates $(z^1,z^2)$ and let $z^3$ be transverse direction.
    
    For the transverse scalar $\pi^3=\phi^3$, the pure kinetic action is
    \begin{equation*}
        S_\textrm{kin}^{\textrm{D7-scalar}} = - \tau_{\mathrm{D7}} \al^2 \int_{X_{1,3}} \! \de^4 x \, \sqrt{- \breve{g}_{4}} \int_{\Sigma_4} \! \de^4 y \, \sqrt{ \breve{g}_{\Sigma_4}} \; [e^{-4A} + c] \, e^{\sdil} \, \breve{g}_{3 \bar{3}} \, \breve{g}^{\mu \nu} \, \breve{\nabla}_\mu \pi^{3} \breve{\nabla}_\nu \bar{\pi}^{3}.
    \end{equation*}
    The total mass term emerges from the interference of the DBI- and CS-actions, with the terms adding up or cancelling out. The full expression is complicated, but the scalings can be read from the DBI-term and the mass action has the form
    \begin{equation*}
        S_{\textrm{mass}}^{\textrm{D7-scalar}} = - \dfrac{\tau_{\mathrm{D7}} \al^2}{2} \int_{X_{1,3}} \! \de^4 x \sqrt{- \breve{g}_{4}} \int_{\Sigma_4} \! \de^4 y \sqrt{ \breve{g}_{\Sigma_4}} \; \dfrac{e^{2 \sdil}}{e^{-4A} + c} \, G_{m'n'\dot{r}} \, \tensor{\bar{G}}{^{\breve{m}'} ^{\breve{n}'} _{\!\dot{s}}} \, \pi^{\dot{r}} \pi^{\dot{s}},
    \end{equation*}
    in real notation. As D7-branes preserve the same supersymmetry as the orientifold, the supersymmetric mass is sourced by a $(2,1)$-flux (but IASD-fluxes source supersymmetry-breaking masses as well). For the theory to have no Freed-Witten anomalies \cite{Freed:1999vc}, the 2-form $B_2$ must be constant over the 4-cycle and in this case the supersymmetric mass is sourced specifially by the flux $G_{12 \bar{3}}$. 
    
    One also finds the gauge vector kinetic action
    \begin{equation*}
        S_{\textrm{kin}}^{\textrm{D7-vector}} = - \dfrac{\tau_{\mathrm{D7}} \al^2}{4} \int_{X_{1,3}} \! \de^4 x \, \sqrt{- \detr{\breve{g}}_{4}} \int_{\Sigma_4} \de^4 y \, \sqrt{ \detr{\breve{g}_{\Sigma_4}}} \; [e^{-4A} + c] \, \breve{g}^{\mu \rho} \breve{g}^{\nu \sigma} \, F_{\mu \nu} F_{\rho \sigma},
    \end{equation*}
    with gaugino masses sourced by $(0,3)$-fluxes.
\end{itemize}

In order to switch to the 4-dimensional Einstein frame defined in eqn. (\ref{metric}), which is necessary to single out the leading order K\"{a}hler modulus couplings, one can make the identifications
\begin{equation*}
    \breve{g}_{\mu \nu} = e^{2\Omega} \, \vevc^{3/2} \, g_{\mu \nu}, \qquad \qquad \qquad \qquad \qquad \breve{g}_{mn} = g_{mn}.
\end{equation*}
Notice that one also needs to transform the Pauli matrix 4-vector as $\breve{\sigma}^\mu = e^{-\Omega} \, \vevc^{-3/4} \, \sigma^\mu$ and to rescale the spinors as $\breve{\psi} = e^{-\Omega/2} \vevc^{-3/8} \tilde{\psi}$ (for similar calculations, see e.g. Refs. \cite{Cremmer:1978hn, Grana:2003ek}).

It is also convenient to renormalise the fields in such a way as to remove the $\vevc$-factors, which turns out to be very helpful in order to obtain 4-dimensional quantities expressed in the appropriate (string coupling, volume and/or warp factor suppressed) Planck units. So for the D3- and anti-D3-branes one has
\begin{equation*}
    \varphi^a = \vevc^{3/4} \phi^a, \qquad \qquad \qquad \qquad \psi^a = \vevc^{3/4} \tilde{\psi}^a,
\end{equation*}
while for D7-branes one has
\begin{equation*}
    \sigma^3 = \vevc^{3/4} \pi^3.
\end{equation*}
Further couplings that arise from the redefinition of the volume modulus are given in the main text (see eqn. (\ref{Weyl factor shift})). A complete analysis including the compensator field (see eqn. (\ref{4-dimensional Einstein frame metric}) is beyond the scope of this paper but for progresses in that direction see Ref. \cite{Cownden:2016hpf}, where it is shown that cancellations occur such that the D3-brane kinetic term is unaffected. Worldvolume fluxes are also not considered.

\section{Soft Terms for Linear and Non-Linear Supersymmetry} \label{LEEFT SUGRA expansions}
This section overviews the structure of the $\mathcal{N}_4=1$ low-energy effective theories of type IIB compactifications with hidden-sector supersymmetry breaking: first it reviews the well-known results for standard multiplets, then it discusses the modifications that occur in the presence of constrained superfields.

\subsection{Classification of Superfields in IIB Low-Energy Supergravity}
A convenient way to study the low-energy effective $\mathcal{N}_4=1$ theory of type IIB Calabi-Yau orientifold compactifications starts from observing that the degrees of freedom of the model are divided in three groups.
\begin{itemize}
    \item Chiral superfields $\phi^M$ that are gauge-neutral and may acquire a non-zero expectation value and/or a non-zero F-term. These constitute the hidden sector responsible for the breaking of supersymmetry and typically correspond to the closed-string moduli but may also include open-string fields.
    \item Chiral superfields $\varphi^i$ that, in order to preserve the gauge symmetries, necessarily have vanishing vacuum expectation values and F-terms, meaning they do not directly break supersymmetry either. These are typically open-string degrees of freedom and constitute the matter sector.
    \item Vector multiplets $W^A$ which come from both the closed- and the open-string sectors and provide both hidden and observable gauge sectors.
\end{itemize}
In the main text, the breaking of supersymmetry is described as an F-term breaking, so the vector superfields play quite a marginal role. Also, the terms in the action with a number $n$ of $\varphi^i$-fields correspond to order-$n$ couplings as these have zero vacuum expectation values, which motivates the expansion of their theory around the vacuum defined by the fields $\phi^M$.

From the expansion of the F-term potential, one can compute the couplings of the theory for all the chiral multiplets in the theory. To start, it is convenient to express the total K\"{a}hler potential $K$ and the total superpotential $W$ of the theory in the form
\begin{subequations}
\begin{align}
    K & = \hat{K}(\phi,\bar{\phi}) + Z_{i \bar{j}}(\phi,\bar{\phi}) \, \varphi^i \bar{\varphi}^{\bar{j}} + \dfrac{1}{2} \, \bigl( H_{ij}(\phi,\bar{\phi}) \varphi^i \varphi^j + \mathrm{c.c.} \bigr), \label{total K} \\
    W & = \hat{W}(\phi) + \dfrac{1}{2} \, \tilde{\mu}_{ij}(\phi) \, \varphi^i \varphi^j + \dfrac{1}{3} \, \tilde{Y}_{ijk}(\phi) \, \varphi^i \varphi^j \varphi^k, \label{total W}
\end{align}
\end{subequations}
along with the gauge kinetic functions
\begin{equation} \label{total f}
    f_{AB} = f_{AB} (\phi),
\end{equation}
where the K\"{a}hler potential $\hat{K}$ and the superpotential $\hat{W}$ describe the pure supersymmetry-breaking hidden sector, while the gauge kinetic functions $f_{AB}$ and the expansion parameters $Z_{i \bar{j}}$, $H_{ij}$, $\tilde{\mu}_{ij}$ and $\tilde{Y}_{ijk}$ describe their couplings to the fluctuations $\varphi^i$. The gauge kinetic functions are always assumed to be block-diagonal.

Then, from an analysis of the general $\mathcal{N}_4=1$ supergravity action \cite{WB} for the theory (\ref{total K}, \ref{total W}) and (\ref{total f}), one finds the standard low-energy effective component action for the supersymmetry-breaking hidden sector $\phi^M$ and just a few relevant couplings involving the matter sector $\varphi^i$. In detail, denoting all the chiral multiplets of the theory with the indices $I = M, i$, one can simply insert the potentials in eqns. (\ref{total K}, \ref{total W}) into the F-term scalar potential
\begin{equation*}
    V_F = K_{I \bar{J}} F^{I} \bar{F}^{\bar{J}} - 3 \kappa_4^2 \, e^{\kappa_4^2 K} \, W \bar{W},
\end{equation*}
where the F-terms are fixed by their algebraic field equations to be $F^{I} = e^{\kappa_4^2 K/2} \, K^{I \bar{J}} \nabla_{\bar{J}} \bar{W}$, with $\nabla_I W = \der_I W + (\kappa_4^2 \der_{I} K) W$. Fermionic interactions can be discussed in a similar way, and a similar analysis applies for the gauge sectors in eqn. (\ref{total f}). A spontaneous breaking of supersymmetry taking place in the hidden sector is also transmitted to the matter sector with the emergence of mass splittings and certain softly non-supersymmetric couplings.

\subsection{Theories with Linearly Realised Supersymmetry} \label{LEEFT SUGRA in IIB compactifications}
If all the fields realise supersymmetry linearly, then all the degrees of freedom are encoded within standard chiral and vector superfields and the expansions are lengthy but straightforward. This subsection summarises the results of Refs. \cite{Kaplunovsky:1993rd, Brignole:1993dj, Grana:2003ek}.

\begin{itemize}
    \item All the hatted quantities represent the pure $\phi^M$-field terms generated by the K\"{a}hler and superpotential $\hat{K}$ and $\hat{W}$, namely the F-term scalar potential 
    \begin{equation*}
    \hat{V}_F = e^{\kappa_4^2 \hat{K}} \, (\hat{K}^{M \bar{N}} \hat{\nabla}_M \hat{W} \hat{\bar{\nabla}}_{\bar{N}} \hat{\bar{W}} - 3 \kappa_4^2 \, \hat{W} \hat{\bar{W}}),
    \end{equation*}
    with the K\"{a}hler-covariant derivative $\hat{\nabla}_M \hat{W} = \der_M \hat{W} + (\kappa_4^2 \der_M \hat{K}) \hat{W}$, the auxilary fields
    \begin{equation*}
    \hat{F}^M = e^{\kappa_4^2 \hat{K}/2} \, \hat{K}^{M \bar{N}} \, \hat{\bar{\nabla}}_{\bar{N}} \hat{\bar{W}}
    \end{equation*}
    and the gravitino mass
    \begin{equation*}
    \hat{m}_{3/2} = e^{\kappa_4^2 \hat{K}/2} \kappa_4^2 \hat{W}.
    \end{equation*}
    As explained above, the pure supersymmetry-breaking hidden-sector effective theory is the same independently of the matter sector. In particular, in the absence of cancellations, the F-term scalar potential $\hat{V}_F$ sets the supersymmetry-breaking scale at the order of magnitude $\smash{m_{\mathrm{SUSY}} \sim [\hat{K}_{M \bar{N}} \hat{F}^M \hat{\bar{F}}^N]^{1/4} \sim [\hat{m}_{3/2} m_P]^{1/2}}$.
    \item As far as the bosonic interactions are concerned, one can see that the theory generates a low-energy theory described by the Lagrangian
    \begin{equation*}
    \mathcal{L}_{\textrm{$\varphi$-bosons}} = - Z_{i \bar{j}} \, \der_\mu \varphi^i \der^\mu \bar{\varphi}^{\bar{j}} - V_{\mathrm{susy}} - V_{\mathrm{soft}},
    \end{equation*}
    where $V_{\mathrm{susy}}$ and $V_{\mathrm{soft}}$ are the $\varphi^i$-sector supersymmetric and soft supersymmetry-breaking potentials, respectively, given by
    \begin{subequations}
    \begin{align}
        V_{\mathrm{susy}} &= \dfrac{1}{2} \, D^2 + Z^{i \bar{j}} \, \der_i W_{\mathrm{susy}} \der_{\bar{j}} \bar{W}_{\mathrm{susy}}, \label{effective susy scalar potential} \\
        V_{\mathrm{soft}} & = m^2_{i \bar{j}, \, \mathrm{soft}} \, \varphi^i \bar{\varphi}^{\bar{j}} + \biggl( \dfrac{1}{2} \, B_{ij} \, \varphi^i \varphi^j + \dfrac{1}{3} \, A_{ijk} \, \varphi^i \varphi^j \varphi^k + \mathrm{c.c.} \biggr). \label{effective susy-breaking potential}
    \end{align}
    \end{subequations}
    In detail, one can conveniently define the effective superpotential as
    \begin{equation*}
    W_{\mathrm{susy}} = \dfrac{1}{2} \, \mu_{ij} \, \varphi^i \varphi^j + \dfrac{1}{3} \, Y_{ijk} \, \varphi^i \varphi^j \varphi^k,
    \end{equation*}
    where the effective supersymmetric couplings read
    \begin{subequations}
    \begin{align}
        \mu_{ij} & = e^{\kappa_4^2 \hat{K}/2} \, \tilde{\mu}_{ij} + \hat{m}_{3/2} H_{ij} - \hat{\bar{F}}^{\bar{N}} \der_{\bar{N}} H_{ij}, \label{effective mu} \\
        Y_{ijk} & = e^{\hat{K}/2} \, \tilde{Y}_{ijk} \label{effective Y}.
    \end{align}
    \end{subequations}
    In particular this generates the supersymmetric masses
    \begin{equation} \label{susy mass}
        m^2_{i \bar{j}} = Z^{k \bar{l}} \, \mu_{ik} \bar{\mu}_{\bar{j} \bar{l}}
    \end{equation}
    as well as supersymmetric trilinear and supersymmetric quartic couplings. Another supersymmetric term is the D-term potential determined by
    \begin{equation*}
    D = - g \, Z_{i \bar{j}} \, \varphi^i \bar{\varphi}^{\bar{j}},
    \end{equation*}
    with the gauge coupling being
    \begin{equation} \label{gauge coupling}
        g^{-2} = \dfrac{1}{2} \, ( f + \bar{f} ).
    \end{equation}
    Second, one finds the soft supersymmetry-breaking terms
    \begin{subequations}
    \begin{align}
        m^2_{i \bar{j}, \, \mathrm{soft}} & = (\hat{m}_{3/2}\hat{\bar{m}}_{3/2} + \kappa_4^2 \hat{V}_F) \, Z_{i \bar{j}} - \hat{F}^M \hat{\bar{F}}^{\bar{N}} \, R_{M \bar{N} i \bar{j}}, \label{soft masses} \\
        B_{ij} & = {\begin{aligned}[t] (2 \, \hat{m}_{3/2} \hat{\bar{m}}_{3/2} + \kappa_4^2 \hat{V}_F) \, H_{ij} - \hat{\bar{m}}_{3/2} \, \hat{\bar{F}}^{\bar{M}} \, \der_{\bar{M}} H_{ij} + \hat{m}_{3/2} \, \hat{F}^{M} \hat{\nabla}_M H_{ij} \\
        - \hat{F}^M \hat{\bar{F}}^{\bar{N}} \, \hat{\nabla}_M \der_{\bar{N}} H_{ij} - e^{\kappa_4^2 \hat{K}/2} \, \tilde{\mu}_{ij} \, \hat{\bar{m}}_{3/2} + \hat{F}^M \hat{\nabla}_M (e^{\kappa_4^2 \hat{K}/2} \tilde{\mu}_{ij}) &, \end{aligned}} \label{B-term} \\
        A_{ijk} & = \hat{F}^M \hat{\nabla}_M Y_{ijk}, \label{A-term}
    \end{align}
    \end{subequations}
    where, given the Levi-Civita connection of the K\"{a}hler metric $Z_{i \bar{j}}$, i.e. $\Gamma^{j}_{Mi} = Z^{j \bar{k}} \der_M Z_{i \bar{k}}$, the Riemann tensor reads
    \begin{equation*}
    R_{M \bar{N} i \bar{j}} = \der_M \der_{\bar{N}} Z_{i \bar{j}} - \Gamma^{k}_{Mi} \, Z_{k \bar{l}} \, \bar{\Gamma}^{\bar{l}}_{\bar{N} \bar{j}}, \qquad \qquad 
    \end{equation*}
    while the K\"{a}hler-covariant derivatives are
    \begin{equation*}
    \begin{split}
        \hat{\nabla}_M (e^{\kappa_4^2 \hat{K}/2} \tilde{\mu}_{ij}) & = \der_M (e^{\kappa_4^2 \hat{K}/2} \tilde{\mu}_{ij}) + \dfrac{1}{2} \, \kappa_4^2 \hat{K}_M \, e^{\kappa_4^2 \hat{K}/2}\tilde{\mu}_{ij} - 2 \, \Gamma^{k}_{Mi} \, e^{\kappa_4^2 \hat{K}/2}\tilde{\mu}_{kj}, \\
        \hat{\nabla}_M Y_{ijk} & = \der_M Y_{ijk} + \dfrac{1}{2} \, \kappa_4^2 \hat{K}_M \, Y_{ijk} - 3 \, \Gamma^{l}_{Mi} \, Y_{ljk},
    \end{split}
    \end{equation*}
    as well as $\hat{\nabla}_M H_{ij} = \der_M H_{ij} - 2 \, \Gamma^{k}_{Mi} \, H_{kj}$ and $\hat{\nabla}_M H_{ij, \bar{N}} = \der_M H_{ij, \bar{N}} - 2 \, \Gamma^{k}_{Mi} \, H_{kj, \bar{N}}$. Unless there are further suppressions due to a cancellation, the order of magnitude of the canonically normalised matter soft-breaking terms is set by the scale $\smash{m_{\mathrm{soft}} \sim m_{\mathrm{SUSY}}^2 / m_P \sim \hat{m}_{3/2}}$.
    \item As far as fermionic interactions are concerned, the relevant terms are the $\psi^i$-field fermionic masses $m^{\mathrm{f}}_{ij}$ and Yukawa couplings $y_{ijk}$ from the supersymmetric Lagrangian
    \begin{equation*}
    \mathcal{L}_{\textrm{$\psi$-fermions}} = - Z_{i \bar{j}} \, \bar{\psi}^{\bar{j}} \bar{\sigma}^\mu \der_\mu \psi^i - \biggl( \dfrac{1}{2} \, m^{\mathrm{f}}_{ij} \, \psi^i \psi^k + \dfrac{1}{3} \, y_{ijk} \, \varphi^i \psi^j \psi^k + \mathrm{c.c.} \biggr),
    \end{equation*}
    which turn out to be
    \begin{subequations}
    \begin{align}
        m^{\mathrm{f}}_{ij} & = \mu_{ij}, \label{fermionic masses} \\
        y_{ijk} & = Y_{ijk} \label{Yukawa couplings}.
    \end{align}
    \end{subequations}
    Also, the supersymmetry-breaking gaugino masses read
    \begin{equation} \label{gaugino mass}
        m_{1/2} = \hat{F}^M \der_M \, \mathrm{ln} ( f+\bar{f} ).
    \end{equation}
\end{itemize}

\subsection{Theories with Linearly and Non-Linearly Realised Supersymmetry} \label{LEEFT SUGRA in IIB compactifications with non-linearly realised SUSY}
If the theory also contains fields that realise supersymmetry non-linearly, then it is necessary to describe such degrees of freedom using constrained supermultiplets. This is the case for instance of type IIB orientifold models with anti-D3-branes.

Non-linearly realised supersymmetry comes in by means of a nilpotent chiral superfield $X$, whose scalar is constrained to be $\smash{\phi^X = \psi^X \psi^X/2 F^X}$ by the nilpotency condition $X^2=0$: such a multiplet has a non-zero F-term and therefore must be included in the supersymmetry-breaking hidden sector. Other fields may realise supersymmetry non-linearly due to similar constraints with similar solutions, but usually they do not have non-zero F-terms and thus are not in this sector. Anyway, for all such constrained multiplets, there are two distinct scenarios.

\begin{itemize}
    \item If the constraint does not fix the F-term of the multiplet, the usual supergravity expansions of subsection \ref{LEEFT SUGRA in IIB compactifications} still hold and the constraint only fixes either its bosonic or fermionic dynamical degrees of freedom in the final action. In the unitary gauge the fixed components vanish.
    \item If the constraint also fixes the F-term, then the expansions of subsection \ref{LEEFT SUGRA in IIB compactifications} do not hold anymore since they are derived by expanding the F-term too. If the fields $\varphi^i$ correspond to chiral multiplets without independent spinor and auxiliary fields, then the calculation proceeds as follows:
    \begin{itemize}
        \item[-] in principle, the full F-term potential is $V_F = K_{I \bar{J}} F^I \bar{F}^J - 3 \kappa_4^2 \, e^{\kappa_4^2 K} \, W \bar{W}$, with the auxiliary fields given by the well-known solutions to their algebraic field equations, $\smash{\bar{F}^{\bar{J}} = e^{\kappa_4^2 K/2} K^{I\bar{J}} D_I W}$;
        \item[-] however, the constraints on the $\varphi^i$-multiplet auxiliary fields make them purely fermionic terms before algebraically fixing them, so that the actual F-term potential is just $V_F = K_{M \bar{N}} \dot{F}^M \smash{\dot{\bar{F}}}^{\bar{N}} - 3 \kappa_4^2 \, e^{\kappa_4^2 K} \, W \bar{W}$, with $\smash{\dot{\bar{F}}}^{\bar{N}} = e^{\kappa_4^2 K/2} K^{M\bar{N}} \nabla_M W$.
    \end{itemize}
   By performing an expansion as in equations (\ref{total K}, \ref{total W}), one can show that the scalar potential for the fields $\varphi^i$ is of the form
    \begin{equation*}
        V = m^2_{i \bar{j}} \varphi^i \bar{\varphi}^{\bar{j}} + m^2_{i \bar{j}, \, \mathrm{soft}} \, \varphi^i \bar{\varphi}^{\bar{j}} + \Bigl( \dfrac{1}{2} \, B_{ij} \, \varphi^i \varphi^j + \dfrac{1}{3} \, A_{ijk} \, \varphi^i \varphi^j \varphi^k + \mathrm{c.c.} \Bigr).
    \end{equation*}
    Obviously there is no distinction between supersymmetric and supersymmetry-breaking terms, but the notation is meant to emphasise the differences with respect to the standard case. In particular, the two mass contributions read
    \begin{subequations} \label{no F-term mass}
    \begin{align}
        m^2_{i \bar{j}} & = 2 Z^{l \bar{k}} \hat{\bar{F}}^{\bar{N}} \hat{F}^M H_{il, \bar{N}} \bar{H}_{\bar{j} \bar{k}, M}, \\
        m^2_{i \bar{j}, \, \mathrm{soft}} & = \begin{aligned}[t] \kappa_4^2 \hat{V}_F Z_{i \bar{j}} - \hat{F}^M \hat{\bar{F}}^{\bar{N}} \bigl[ Z_{i \bar{j}, M \bar{N}} - 2 \, \Gamma^{k}_{Mi} \, Z_{k \bar{l}} \, \bar{\Gamma}^{\bar{l}}_{\bar{N} \bar{j}} \bigr] \\[-0.5ex]
        + \bigl[ \hat{m}_{3/2} \hat{F}^M Z_{i \bar{j}, M} + \hat{\bar{m}}_{3/2} \smash{\hat{\bar{F}}}^{\bar{N}} Z_{i \bar{j}, N} \bigr] &. \end{aligned}
    \end{align}
    \end{subequations}
    Instead the bilinear $B$-coupling reads
    \begin{equation} \label{no F-term B-term}
        \begin{split}
            B_{ij} = \kappa_4^2 \hat{V}_F H_{ij} + e^{\kappa_4^2 \hat{K}/2} \hat{F}^M \hat{\nabla}_M \tilde{\mu}_{ij} + \hat{\bar{m}}_{3/2} \hat{\bar{F}}^{\bar{N}} H_{ij, \bar{N}} + \hat{m}_{3/2} \hat{F}^M H_{ij, M} \\[-0.5ex]
            - \hat{F}^M \hat{\bar{F}}^{\bar{N}} \, \bigl( H_{ij, M \bar{N}} - 4 \, \Gamma^{l}_{Mi} H_{lj, \bar{N}} \bigr) - 3 \, \hat{\bar{m}}_{3/2} \tilde{\mu}_{ij} &.
        \end{split}
    \end{equation}
    As for the trilinear terms, one only has the would-be supersymmetry-breaking term
    \begin{equation} \label{no F-term A-term}
        A_{ijk} = e^{\kappa_4^2 \hat{K}/2} \, \bigl[\hat{F}^M \hat{\nabla}_{M} \tilde{Y}_{ijk} - 3 \, \hat{\bar{m}}_{3/2} \tilde{Y}_{ijk} \bigr]. 
    \end{equation}
    The K\"{a}hler-covariant derivatives are defined as $\smash{\hat{\nabla}_M \tilde{\mu}_{i j} = \der_M \tilde{\mu}_{i j} + (\kappa_4^2 \hat{K}_M) \tilde{\mu}_{i j}}$ and $\smash{\hat{\nabla}_M \tilde{Y}_{i j k} = \der_M \tilde{Y}_{i j k} + (\kappa_4^2 \hat{K}_M) \tilde{Y}_{i j k}}$. Noticeably, although the structure of all the coupling terms is different, one can observe that the theory is still invariant under the usual K\"{a}hler transformations as all the terms are individually covariant. The case where the scalar and the F-term components of a multiplet are constrained may be discussed in a similar fashion. It is not encountered in the main text and thus left for future study.
\end{itemize}

\section{Geometry of Warped 4-cycles} \label{appendix: throat geometry}
This appendix contains a few observations about the geometry of a 4-cycle wrapped by a D7-brane in the two setups discussed in the main text.

\subsection{Products of 2- and 4-cycles}
In the main text, whenever it is necessary to consider the cycles wrapped by the D7-branes explicitly, as in e.g. subsubsections \ref{single bulk D7-brane SUGRA} and \ref{single throat D7-brane SUGRA}, they are assumed to be (conformally) a 4-dimensional orbifold $O_4 = \mathrm{T}^4 / \mathbb{Z}_2$, and the 6-dimensional space is locally assumed to be (conformally) the product of the orbifold $O_4$ and the 2-torus $\mathrm{T}^2$.

To be concrete, following Refs. \cite{Lust:2004fi, Lust:2004fi, Lust:2005bd}, one considers the 4-dimensional orbifold $O_4$ spanned by the coordinates $(z^1,z^2)$ and the 2-torus $\mathrm{T}^2$ spanned by $z^3$, with $w^a = z^a/l_s$ the dimensionless coordinates. Then: 
\begin{itemize}
    \item on the 4-cycle $O_4 = \mathrm{T}^4 / \mathbb{Z}_2$, the untwisted $(2,0)$- and $(1,1)$-forms are
    \begin{equation*}
        \eta = \de w^1 \wedge \de w^2,
    \end{equation*}
    and
    \begin{equation*}
    \begin{array}{ccc}
        \zeta_1 = \de w^1 \wedge \de \bar{w}^2, & \qquad & \zeta_2 = \de \bar{w}^1 \wedge \de w^2,  \\[0.5ex]
        \zeta_3 = \de w^1 \wedge \de \bar{w}^1, & & \zeta_4 = \de w^2 \wedge \de \bar{w}^2;
    \end{array}
    \end{equation*}
    \item the untwisted harmonic 3-forms on the 6-dimensional space $(\mathrm{T}^4 / \mathbb{Z}_2) \times \mathrm{T}^2$ are then the holomorphic 3-form
    \begin{equation*}
        \Omega = \eta \wedge \de w^3 = \de w^1 \wedge \de w^2 \wedge \de w^3,
    \end{equation*}
    and the $(2,1)$-forms
    \begin{equation*}
            \chi_1 = \dfrac{\de w^1 \wedge \de \bar{w}^2 \wedge \de w^3}{[-i (u^2 - \bar{u}^2)]}, \qquad \chi_2 = \dfrac{\de \bar{w}^1 \wedge \de w^2 \wedge \de w^3}{[-i (u^1 - \bar{u}^1)]}, \qquad \chi_\vartheta = \dfrac{\de w^1 \wedge \de w^2 \wedge \de \bar{w}^3}{[-i (u^3 - \bar{u}^3)]},
    \end{equation*}
    as well as (ignoring the off-diagonal complex structure moduli)
    \begin{equation*}
        \chi_3 = \de w^1 \wedge \de \bar{w}^1 \wedge \de w^3, \qquad \qquad \chi_4 = \de w^2 \wedge \de \bar{w}^2 \wedge \de w^3,
    \end{equation*}
    where the complex structure moduli $u^a$ have been introduced into the relevant elements of the basis, with the definition $\de z^a = \de y^a + u^a \de y^{a +3}$, for $a = 1,2,3$.
\end{itemize}
Also, there are extra moduli corresponding to blown-up singularities which are ignored. Moreover, one can show that the unwarped complex structure K\"{a}hler potential reads
\begin{equation*}
    \hat{K}^{(0)}_{\mathrm{cs}} = - \mathrm{ln} \, \Bigl[ - i \int_{Y_6} \Omega \wedge \bar{\Omega} \Bigr] = - \mathrm{ln} \, \Bigl( [-i (u^1 - \bar{u}^1)] [-i (u^2 - \bar{u}^2)] [-i (u^3 - \bar{u}^3)] \Bigr) - \mathrm{ln} \, \Vzero.
\end{equation*}
In warped scenarios, if the identification of the bulk complex structure moduli still holds, one finds analogus results with the substitution of the unwarped volume with $\Vw$.

\subsection{Complex Structure K\"{a}hler Metrics}
It is convenient to collectively label the basis of the harmonic $(1,1)$-forms on the orbifold $O_4 = \Sigma_4$ as $\zeta_{i}$, with $i=1,\dots,4$, and the basis of harmonic $(2,1)$-forms on the 6-dimensional product $O_4 \times \mathrm{T}^2$ as $\chi_\alpha$, with $\alpha = 1, \dots, 4, \vartheta$. Further there are the harmonic $(2,0)$-form $\eta$ and the harmonic $(3,0)$-form $\Omega$. The explicit complex structure moduli factors $[-i(u^a - \bar{u}^a)]$ will be ignored for brevity. It is then possible to observe the following equivalences.
\begin{itemize}
    \item If the wrapped 4-cycle is extended in the bulk and the warp factor does not vary over the transverse space, then one can observe the identities
    \begin{equation*}
    \omega_w = \int_{Y_6} e^{-4A} \Omega \wedge \bar{\Omega} = \Vzerotwo \int_{\Sigma_4} e^{-4A} \eta \wedge \bar{\eta}
    \end{equation*}
    and
    \begin{equation*}
        \begin{split}
            \int_{Y_6} e^{-4A} \chi_{\alpha} \wedge \bar{\chi}_\beta = \Vzerotwo \, \biggl[ \delta^{i}_{\alpha} \delta^{j}_{\beta} \int_{\Sigma_4} e^{-4A} \zeta_{i} \wedge \bar{\zeta}_{j} - \delta^{\vartheta}_{\alpha} \delta^{\vartheta}_{\beta} \int_{\Sigma_4} e^{-4A} \eta \wedge \bar{\eta} \biggr].
        \end{split}
    \end{equation*}
    This implies that the complex structure moduli metric can be written as
    \begin{equation*}
        \hat{K}_{\alpha \bar{\beta}} = - \dfrac{1}{\omega_w} \int_{Y_6} e^{-4A} \chi_{\alpha} \wedge \bar{\chi}_\beta = \delta^{i}_{\alpha} \delta^{j}_{\beta} \, \hat{K}_{i \bar{j}} + \delta^{\vartheta}_{\alpha} \delta^{\bar{\vartheta}}_{\bar{\beta}},
    \end{equation*}
    with the definitions
    \begin{equation*}
    \hat{K}_{i \bar{j}} = - \dfrac{1}{\omega^{\Sigma_4}_w} \int_{\Sigma_4} e^{-4A} \zeta_{i} \wedge \bar{\zeta}_{j}, \qquad \qquad \qquad \omega^{\Sigma_4}_w = \int_{\Sigma_4} e^{-4A} \eta \wedge \bar{\eta}.
    \end{equation*}
    \item In a setup with the wrapped 4-cycle being localised at the tip of a warped throat, i.e. with the warp factor varying only along the 2-torus, the analysis of the complex structure moduli is also easy. Then, one can observe the identities
    \begin{equation*}
    \omega_w = \int_{Y_6} e^{-4A} \Omega \wedge \bar{\Omega} = \Vwtwo \int_{\Sigma_4} \eta \wedge \bar{\eta}
    \end{equation*}
    and
    \begin{equation*}
        \int_{Y_6} e^{-4A} \chi_{\alpha} \wedge \bar{\chi}_\beta = \Vwtwo \, \biggl[ \delta^{i}_{\alpha} \delta^{j}_{\beta} \int_{\Sigma_4} \zeta_{i} \wedge \bar{\zeta}_{j} - \delta^{\vartheta}_{\alpha} \delta^{\vartheta}_{\beta} \int_{\Sigma_4} \eta \wedge \bar{\eta} \biggr]
    \end{equation*}
    so that the warped version of the complex structure moduli metric is the same as the unwarped one, i.e.
    \begin{equation*}
        \hat{K}_{\alpha \bar{\beta}} = - \dfrac{1}{\omega_w} \int_{Y_6} e^{-4A} \chi_{\alpha} \wedge \bar{\chi}_\beta = \delta^{i}_{\alpha} \delta^{j}_{\beta} \, \hat{K}^{(0)}_{i \bar{j}} + \delta^{\vartheta}_{\alpha} \delta^{\bar{\vartheta}}_{\bar{\beta}},
    \end{equation*}
    with the definitions
    \begin{equation*}
    \hat{K}_{i \bar{j}}^{(0)} = - \dfrac{1}{\omega^{\Sigma_4}_{(0)}} \int_{\Sigma_4} \zeta_i \wedge \bar{\zeta}_j, \qquad \qquad \qquad \omega^{\Sigma_4}_{(0)} = \int_{\Sigma_4} \eta \wedge \bar{\eta}.
    \end{equation*}
\end{itemize}
The explicit complex structure moduli factors may be inserted by following straightforwardly the definitions above. For instance, one can find
\begin{equation*}
\begin{split}
    \omega_w & = \int_{Y_6} e^{-4A} \Omega \wedge \bar{\Omega} = [-i (u^3 - \bar{u}^3)] \Vzerotwo \int_{\Sigma_4} e^{-4A} \eta \wedge \bar{\eta}, \\
    \hat{K}_{\vartheta \bar{\vartheta}} & = - \dfrac{1}{\omega_w} \int_{Y_6} e^{-4A} \chi_{\vartheta} \wedge \bar{\chi}_\vartheta = \dfrac{1}{[-i(u^3 - \bar{u}^3)]^2},
\end{split}
\end{equation*}
as follows directly from the definition $\smash{\chi_{\vartheta} = \eta \wedge \de \bar{w}^3 / [-i(u^3 - \bar{u}^3)]}$, with the identification $\smash{\int_{\mathrm{T}^2} \de w^3 \wedge \de \bar{w}^3 = -i [-i(u^3 - \bar{u})^3] \Vzerotwo}$. As an example, by defining the 2-form $g_2$ via the identification $G_3 = g_2 \wedge \de \bar{w}^3$, given the 3-form expansion $\smash{e^{4A} G_3 = - \hat{K}^{\vartheta \bar{\vartheta}} \chi_{\vartheta} \int_{Y_6} G_3 \wedge \bar{\chi}_{\vartheta} / \omega_w}$, one finds the same expansion that is used in the main text, i.e.
\begin{equation*}
    e^{4A} g_2 = \dfrac{1}{\omega_w^{\Sigma_4}} \, \eta \int_{\Sigma_4} g_2 \wedge \bar{\eta}.
\end{equation*}

\vspace{8pt}

\phantomsection

\addcontentsline{toc}{section}{References}


\bibliographystyle{JHEP}
\bibliography{report}

\end{document}